\documentclass[twocolumn,longtable]{aastex631}
\usepackage[caption=false]{subfig}
\DeclareCaptionLabelFormat{continued}{Continuation of #1~#2}
\begin{document}
\sloppy
\newcommand{\vdag}{(v)^\dagger}
\newcommand\aastex{AAS\TeX}
\newcommand\latex{La\TeX}
\newcommand{\ky}{2002\,KY$_{14}$}
\newcommand{\kylong}{(250112)\,2002\,KY$_{14}$}
\newcommand{\plshort}{2010\,PL$_{66}$}
\newcommand{\pllong}{(499522)\,2010\,PL$_{66}$}
\newcommand{\ph}{2013\,PH$_{44}$}
\newcommand{\phlong}{(471931)\,2013\,PH$_{44}$}
\newcommand{\gx}{2010\,GX$_{34}$}
\newcommand{\gxlong}{2010\,GX$_{34}$}
\newcommand{\yd}{2009\,YD$_7$}
\newcommand{\ydlong}{(353222)\,2009\,YD$_{7}$}
\newcommand{\aeshort}{2016\,AE$_{193}$}
\newcommand{\aelong}{(514312)\,2016\,AE$_{193}$}
\newcommand{\cxshort}{2017\,CX$_{33}$}
\newcommand{\cxlong}{(523798)\,2017\,CX$_{33}$}
\newcommand{\fzshort}{2015\,FZ$_{117}$}
\newcommand{\fz}{2015\,FZ$_{117}$}
\newcommand{\fzlong}{(472760)\,2015\,FZ$_{117}$}
\newcommand{\fcshort}{2008\,FC$_{76}$}
\newcommand{\fclong}{(281371)\,2008\,FC$_{76}$}
\newcommand{\azshort}{2013\,AZ$_{60}$}
\newcommand{\azlong}{2013\,AZ$_{60}$}
\newcommand{\mruncorr}{m$_{11}^\mathrm{R}$}
\newcommand{\mrcorr}{m$_{110}^\mathrm{R}$}
\newcommand{\sn}{1998\,SN$_{165}$}
\newcommand{\de}{1999\,DE$_9$}
\newcommand{\cm}{2000\,CM$_{105}$}
\newcommand{\ok}{2000\,OK$_{67}$}
\newcommand{\hz}{2001\,HZ$_{58}$}
\newcommand{\kn}{2001\,KN$_{76}$}
\newcommand{\qb}{2001\,QB$_{298}$}
\newcommand{\qj}{2001\,QJ$_{298}$}
\newcommand{\qt}{2001\,QT$_{322}$}
\newcommand{\qx}{2001\,QX$_{297}$}
\newcommand{\xd}{2001\,XD$_{255}$}
\newcommand{\xp}{2001\,XP$_{254}$}
\newcommand{\xu}{2001\,XU$_{254}$}
\newcommand{\yh}{2001\,YH$_{140}$}
\newcommand{\cy}{2002\,CY$_{224}$}
\newcommand{\gj}{2002\,GJ$_{32}$}
\newcommand{\gv}{2002\,GV$_{31}$}
\newcommand{\vs}{2002\,VS$_{130}$}
\newcommand{\vt}{2002\,VT$_{130}$}
\newcommand{\vu}{2002\,VU$_{130}$}
\newcommand{\fe}{2003\,FE$_{128}$}
\newcommand{\felong}{(469505)~2003\,FE$_{128}$}
\newcommand{\gh}{2003\,GH$_{55}$}
\newcommand{\ghlong}{(385437)~2003\,GH$_{55}$}
\newcommand{\qa}{2003\,QA$_{92}$}
\newcommand{\ys}{2003\,YS$_{179}$}
\newcommand{\fc}{2003\,FC$_{128}$}
\newcommand{\qw}{2003\,QW$_{90}$}
\newcommand{\qwlong}{(307616)~2003\,QW$_{90}$}
\newcommand{\xr}{2004\,XR$_{190}$}
\newcommand{\tf}{2004\,TF$_{282}$}
\newcommand{\tv}{2004\,TV$_{357}$}
\newcommand{\eo}{2005\,EO$_{304}$}
\newcommand{\ez}{2005\,EZ$_{296}$}
\newcommand{\tb}{2005\,TB$_{190}$}
\newcommand{\rs}{2005\,RS$_{43}$}
\newcommand{\jj}{2007\,JJ$_{43}$}
\newcommand{\orten}{2007\,OR$_{10}$}
\newcommand{\cs}{2008\,CS$_{190}$}
\newcommand{\yg}{2009\,YG$_{19}$}
\newcommand{\ep}{2010\,EP$_{65}$}
\newcommand{\et}{2010\,ET$_{65}$}
\newcommand{\fcten}{2010\,FC$_{49}$}
\newcommand{\jf}{2011\,JF$_{31}$}
\newcommand{\dw}{2012\,DW$_{98}$}
\newcommand{\apthirteen}{2013\,AP$_{183}$}
\newcommand{\at}{2013\,AT$_{183}$}
\newcommand{\fa}{2013\,FA$_{28}$}
\newcommand{\jv}{2013\,JV$_{65}$}
\newcommand{\am}{2014\,AM$_{55}$}
\newcommand{\df}{2014\,DF$_{143}$}
\newcommand{\djfourteen}{2014\,DJ$_{143}$}
\newcommand{\ezfourteen}{2014\,EZ$_{51}$}
\newcommand{\gdfourteen}{2014\,GD$_{54}$}
\newcommand{\gefourteen}{2014\,GE$_{45}$}
\newcommand{\gjfourteen}{2014\,GJ$_{54}$}
\newcommand{\hf}{2014\,HF$_{200}$}
\newcommand{\jq}{2014\,JQ$_{80}$}
\newcommand{\ls}{2014\,LS$_{28}$}
\newcommand{\wanine}{2014\,WA$_{509}$}
\newcommand{\waninelong}{(535018)~2014\,WA$_{509}$}
\newcommand{\waten}{2014\,WA$_{510}$}
\newcommand{\watenlong}{(535028)~2014\,WA$_{510}$)}
\newcommand{\wj}{2014\,WJ$_{510}$}
\newcommand{\wo}{2014\,WO$_{509}$}
\newcommand{\ws}{2014\,WS$_{510}$}
\newcommand{\ye}{2014\,YE$_{50}$}
\newcommand{\yj}{2014\,YJ$_{50}$}
\newcommand{\bbfifteen}{2015\,BB$_{519}$}
\newcommand{\bc}{2015\,BC$_{519}$}
\newcommand{\bz}{2015\,BZ$_{518}$}
\newcommand{\da}{2015\,DA$_{225}$}
\newcommand{\lele}[3]{{#1}\,$\le$\,{#2}\,$\le$\,{#3}}
\newcommand{\tauav}[1]{$\tau_{#1}$/A$_{\rm V}$}
\newcommand{\iav}[1]{I$_{#1}$/A$_{\rm V}$}
\newcommand{\fblue}[1]{F$_{70}$\,}
\newcommand{\fgreen}[1]{F$_{100}$\,}
\newcommand{\fred}[1]{F$_{160}$\,}
\def\absmag{H$_\mathrm{V}$}
\def\tss{$T_\mathrm{ss}$}
\def\rh{$r_\mathrm{h}$}
\def\espec{$\epsilon_\nu$}
\def\ebol{$\epsilon_\mathrm{bol}$}
\def\deq{$D_\mathrm{eq}$}
\def\geomalb{p$_\mathrm{V}$}
\def\qg{2001\,QG$_{298}$}
\def\deff{D$_\mathrm{eff}$}
\def\gcc{g\,cm$^{-3}$}
\def\tiunit{$\mathrm J\,m^{-2}\,s^{-1/2}K^{-1}$}

\received{\today}
\revised{\today}
\accepted{\today}
\submitjournal{ApJS}

\shorttitle{K2 TNOs}
\shortauthors{Kecskem\'ethy et al.}

\graphicspath{{./}{figures/}}

\title{Light curves of transneptunian objects from the K2 mission of the Kepler Space Telescope}

\correspondingauthor{Csaba Kiss}
\email{pkisscs@konkoly.hu}

\author[0000-0002-9511-0901]{Vikt\'oria Kecskem\'ethy}
\affiliation{Konkoly Observatory, Research Centre for Astronomy and Earth Sciences, E\"otv\"os Lor\'and Research Network, \\ Konkoly Thege Mikl\'os \'ut 15-17, H-1121 Budapest, Hungary}
\affiliation{CSFK, MTA Centre of Excellence, Budapest, Konkoly Thege Miklós út 15-17, H-1121, Hungary}
\affiliation{Leiden Observatory, Leiden University, Niels Bohrweg 2, 2333 CA Leiden, The Netherlands}

\author[0000-0002-8722-6875]{Csaba Kiss}
\affiliation{Konkoly Observatory, Research Centre for Astronomy and Earth Sciences, E\"otv\"os Lor\'and Research Network, \\ Konkoly Thege Mikl\'os \'ut 15-17, H-1121 Budapest, Hungary}
\affiliation{CSFK, MTA Centre of Excellence, Budapest, Konkoly Thege Miklós út 15-17, H-1121, Hungary}
\affiliation{E\"otv\"os Lor\'and University, Institute of Physics, P\'azm\'any P\'eter s\'et\'any 1/A, H-1171 Budapest, Hungary}

\author[0000-0002-1698-605X]{R\'obert Szak\'ats}
\affiliation{Konkoly Observatory, Research Centre for Astronomy and Earth Sciences, E\"otv\"os Lor\'and Research Network, \\ Konkoly Thege Mikl\'os \'ut 15-17, H-1121 Budapest, Hungary}
\affiliation{CSFK, MTA Centre of Excellence, Budapest, Konkoly Thege Miklós út 15-17, H-1121, Hungary}

\author[0000-0001-5449-2467]{Andr\'as P\'al}
\affiliation{Konkoly Observatory, Research Centre for Astronomy and Earth Sciences, E\"otv\"os Lor\'and Research Network, \\ Konkoly Thege Mikl\'os \'ut 15-17, H-1121 Budapest, Hungary}
\affiliation{CSFK, MTA Centre of Excellence, Budapest, Konkoly Thege Miklós út 15-17, H-1121, Hungary}
\affiliation{E\"otv\"os Lor\'and University, Faculty of Science, P\'azm\'any P\'eter s\'et\'any 1/A, H-1171 Budapest, Hungary}

\author[0000-0002-0606-7930]{Gyula M. Szab\'o}
\affiliation{ELTE E\"otv\"os Lor\'and University, Gothard Astrophysical Observatory, Szombathely, Hungary}
\affiliation{MTA-ELTE Exoplanet Research Group, 9700 Szombathely, Szent Imre h. u. 112, Hungary}

\author[0000-0002-8159-1599]{L\'aszl\'o Moln\'ar}
\affiliation{Konkoly Observatory, Research Centre for Astronomy and Earth Sciences, E\"otv\"os Lor\'and Research Network, \\ Konkoly Thege Mikl\'os \'ut 15-17, H-1121 Budapest, Hungary}
\affiliation{CSFK, MTA Centre of Excellence, Budapest, Konkoly Thege Miklós út 15-17, H-1121, Hungary}
\affiliation{E\"otv\"os Lor\'and University, Institute of Physics, P\'azm\'any P\'eter s\'et\'any 1/A, H-1171 Budapest, Hungary}
\affiliation{MTA CSFK Lend\"ulet Near-Field Cosmology Research Group}

\author[0000-0003-0926-3950]{Kriszti\'an S\'arneczky}
\affiliation{Konkoly Observatory, Research Centre for Astronomy and Earth Sciences, E\"otv\"os Lor\'and Research Network, \\ Konkoly Thege Mikl\'os \'ut 15-17, H-1121 Budapest, Hungary}
\affiliation{CSFK, MTA Centre of Excellence, Budapest, Konkoly Thege Miklós út 15-17, H-1121, Hungary}

\author[0000-0001-8764-7832]{J\'ozsef Vink\'o}
\affiliation{Konkoly Observatory, Research Centre for Astronomy and Earth Sciences, E\"otv\"os Lor\'and Research Network, \\ Konkoly Thege Mikl\'os \'ut 15-17, H-1121 Budapest, Hungary}
\affiliation{CSFK, MTA Centre of Excellence, Budapest, Konkoly Thege Miklós út 15-17, H-1121, Hungary}
\affiliation{E\"otv\"os Lor\'and University, Institute of Physics, P\'azm\'any P\'eter s\'et\'any 1/A, H-1171 Budapest, Hungary}

\author[0000-0002-3258-1909]{R\'obert Szab\'o}
\affiliation{Konkoly Observatory, Research Centre for Astronomy and Earth Sciences, E\"otv\"os Lor\'and Research Network, \\ Konkoly Thege Mikl\'os \'ut 15-17, H-1121 Budapest, Hungary}
\affiliation{CSFK, MTA Centre of Excellence, Budapest, Konkoly Thege Miklós út 15-17, H-1121, Hungary}
\affiliation{E\"otv\"os Lor\'and University, Institute of Physics, P\'azm\'any P\'eter s\'et\'any 1/A, H-1171 Budapest, Hungary}
\affiliation{MTA CSFK Lend\"ulet Near-Field Cosmology Research Group}


\author[0000-0002-1326-1686]{G\'abor Marton}
\affiliation{Konkoly Observatory, Research Centre for Astronomy and Earth Sciences, E\"otv\"os Lor\'and Research Network, \\ Konkoly Thege Mikl\'os \'ut 15-17, H-1121 Budapest, Hungary}
\affiliation{CSFK, MTA Centre of Excellence, Budapest, Konkoly Thege Miklós út 15-17, H-1121, Hungary}

\author[0000-0001-5531-1381]{Anik\'o Farkas-Tak\'acs}
\affiliation{Konkoly Observatory, Research Centre for Astronomy and Earth Sciences, E\"otv\"os Lor\'and Research Network, \\ Konkoly Thege Mikl\'os \'ut 15-17, H-1121 Budapest, Hungary}
\affiliation{CSFK, MTA Centre of Excellence, Budapest, Konkoly Thege Miklós út 15-17, H-1121, Hungary}
\affiliation{E\"otv\"os Lor\'and University, Faculty of Science, P\'azm\'any P\'eter s\'et\'any 1/A, H-1171 Budapest, Hungary}

\author[0000-0002-1663-0707]{Csilla E. Kalup}
\affiliation{Konkoly Observatory, Research Centre for Astronomy and Earth Sciences, E\"otv\"os Lor\'and Research Network, \\ Konkoly Thege Mikl\'os \'ut 15-17, H-1121 Budapest, Hungary}
\affiliation{CSFK, MTA Centre of Excellence, Budapest, Konkoly Thege Miklós út 15-17, H-1121, Hungary}
\affiliation{E\"otv\"os Lor\'and University, Faculty of Science, P\'azm\'any P\'eter s\'et\'any 1/A, H-1171 Budapest, Hungary}

\author[0000-0002-3234-1374]{L\'aszl\'o L. Kiss}
\affiliation{Konkoly Observatory, Research Centre for Astronomy and Earth Sciences, E\"otv\"os Lor\'and Research Network, \\ Konkoly Thege Mikl\'os \'ut 15-17, H-1121 Budapest, Hungary}
\affiliation{CSFK, MTA Centre of Excellence, Budapest, Konkoly Thege Miklós út 15-17, H-1121, Hungary}
\affiliation{Sydney Institute for Astronomy, School of Physics A29, University of Sydney, NSW 2006, Australia}



\begin{abstract}
The K2 mission of the Kepler Space Telescope allowed the observations of light curves of small solar system bodies throughout the whole Solar system. In this paper we present the results of a collection of K2 transneptunian object observations, between Campaigns C03 (November 2014 -- February 2015) to C19 (August -- September, 2018), which includes 66 targets. Due to the faintness of our targets the detectability rate of a light curve period is $\sim$56\%, notably lower than in the case of other small body populations, like Hildas or Jovian trojans. We managed to obtain light curve periods with an acceptable confidence for 37 targets; the majority of these cases are new identifications. We were able to give light curve amplitude upper limits for the other 29 targets. Several of the newly detected light curve periods are longer than $\sim$24\,h, in many cases close to $\sim$80\,h, i.e., these targets are slow rotators. This relative abundance of slowly rotating objects is similar to that observed among Hildas, Jovian trojans and Centaurs in the K2 mission, and also among main belt asteroids measured with the TESS Space Telescope. Transneptunian objects show notably higher light curve amplitudes at large (D\,$\gtrsim$\,300\,km) sizes than that found among large main belt asteroids, in contrast to the general expectation that due to their lower compressive strength they reach hydrostatic equlibrium at smaller sizes than their inner solar system counterparts.  
\end{abstract}


\section{Introduction} \label{Sect:intro}

Despite almost three decades of observations the number of trans-Nepunian objects (TNOs) with known rotational properties -- at least rotation period and light curve amplitude -- is still rather low. While amplitude estimates exist for a larger number of objects \citep[see e.g.][]{Showalter2021}, the latest version of the Light Curve Database \citep[LCDB,][]{Warner2009} contains only 124 TNOs with known rotation periods. In many cases, even these periods are not very reliable, and different authors derive different solutions based on their own data \citep[see e.g.][]{Sheppard2008,Duffard2009,T14}. 

Light curves hold information on the formation and evolution of individual objects and on the collisional evolution of the asteroid populations. In the main belt large, D\,$\geq$\,40\,km asteroids have likely reached collisional equilibrium and the corresponding Maxwellian spin frequency distribution, while 
the rotation of smaller asteroids are modified by non-collisional and non-gravitational effects as well \citep[see e.g.][and references therein]{Pravec2002}.
In the transneptunian region smaller bodies with diameters D\,$\leq$\,100\,km are likely collisional fragments with their rotation states erased and heavily overwritten, while intermediate sized objects (100\,km\,$\leq$\,D\,$\leq$\,200\,km) have probably been stable to catastrophic breakups, but their spin states are likely strongly influenced by impacts. On the other hand, the largest objects (D\,$\geq$\,200\,km) likely kept their spin states inherited from the formation era in the early Solar system \citep[see][and references therein]{Sheppard2008}. 

Binaries, especially contact and semi-contact systems, however, may alter this picture significantly, as they may end up in fully synchronised spin-orbit states \citep{Noll2020} or become collapsed binaries \citep{Nesvorny2020} at the end of their tidal evolution, potentially increasing the number of slow rotators \citep[][]{Marton2020}. 
While there are a large number of minor planets with known light curves below the collisional decoupling limit in the main belt or in the Hilda and Jovian trojan populations, the vast majority of TNOs with known rotational characteristics are larger than $\sim$100\,km. This means that their rotational properties are expected to be chiefly determined by their formation circumstances and/or by their tidal evolution in multiple systems. 

Earlier reviews of transneptunian light curves reported mean rotation periods of P\,=\,7-8\,h \citep[e.g.][]{Duffard2009}, however, it was also found that the binary transneptunian population rotates slower \citep{T14}, and objects in the cold classical population have larger variability and rotate slower than the non-cold classical TNOs \citep{Benecchi2013,TS19}. In the latter case the mean rotation rates were within the uncertainties, P\,=\,9.47$\pm$1.53\,h, and P\,=\,8.45$\pm$0.58\,h, respectively, for the two samples. In the case of Centaurs -- which are believed to be originated from the transneptunian region -- \citet{Marton2020} identified several targets with long, P\,$\ge$20\,h rotation periods, indicating that the previous ground-based light curve samples might have missed slowly rotating objects.  

While ground-based observations have obvious limitations in detecting long-period light curves \citep[see e.g.][]{Marciniak2015,Marciniak2018}, the K2 mission of the Kepler Space Telescope \citep{K2} allowed long (up to $\sim$80 days), uninterrupted observations of many Solar system objects, including main belt asteroids \citep{Szabo2015,Szabo2016,Molnar2018}, Hildas \citep{Szabo2020}, Jovian trojans \citep{Szabo2017,Ryan2017,Kalup2021}, 
and also the irregular satellites of giant planets \citep{Kiss2016,Farkas2017}. Light curves were also published for a few, selected transneptunian objects based on K2 observations \citep{Pal2015,Pal2016}. 
A common outcome of the studies of larger samples, across all dynamical classes, was the identification of an increased number of targets with long rotation periods compared to previous ground-based studies. A similar trend is observed among the data of nearly 10\,000 main belt asteroids obtained by the TESS Space Telescope \citep{Pal2020}, and asteroids with long rotation periods were identified in other surveys like the Asteroid Terrestrial-impact Last Alert System (ATLAS), the Zwicky Transient Facility 
\citep[ZTF,][]{Erasmus2021}, and the All-Sky Automated Survey for Supernovae \citep[ASAS-SN,][]{Hanus2021}. 

In this paper we present the analysis of light curve data for 66 transneptunian objects obtained by the K2 mission of the Kepler Space Telescope. The observations and the main data reduction steps are summarised in Sect.~\ref{sect:obs}, results and comparison with the light curve characteristics of other populations and datasets are presented in Sect.~\ref{sect:results}, and conclusions are given in Sect.~\ref{sect:conclusions}. 
Large tables of observational data, light curve figures, and the description of electronically available light curve data can be found in the Appendix. 


\section{Observations and data reduction \label{sect:obs}} 

\subsection{Basic data reduction \label{sect:basicdatareduction}}

\begin{figure}
\includegraphics[width=0.5\textwidth]{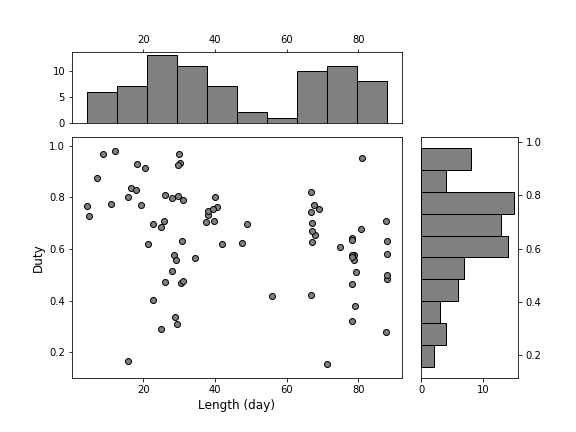}
\caption{Duty cycle versus the full length of the observations. Histograms of duty cycle and observation length are shown on the right and the top of the figure, respectively. \label{fig:duty}}
\end{figure}

The Kepler Space Telescope observed numerous Solar System objects during the K2 mission. The observing strategy and data reduction steps followed in this paper are analogous to other TNO, Centaur and asteroid targets published in previous papers \citep[see][for a summary]{Kiss2020}. Since \textit{Kepler} observed only selected pixels during each 60--80\,d long campaign, pixels over $\sim 30$\,d long arcs of the apparent trajectories were allocated for each target. 
We processed the \textit{Kepler} observations with the \textit{fitsh} software package \citep{Pal2012}. First, we assembled the individual Target Pixel Files of both the track of the target and that of the nearby stars into mosaic images. Astrometric solutions were derived for every mosaic frame in the campaign, using the Full Frame Images (acquired once per campaign) as initial hints, and the frames were registered into the same reference system. We then enlarged the images by roughly 3 times, and rotated them into RA-Dec directions. This enlargement helped to decrease the fringing of the residual images in the next step, where we subtracted a median image from all frames. The median was created from a subset of frames that did not contain the target. We applied simple aperture photometry to the differential images based on the ephemeris (aperture central position in RA and Dec) provided by the {\sl ephemd} tool \citep{Pal2020}. 
We then discarded data points that were contaminated by the residuals of the stellar images, saturation columns and crosstalk patterns from the camera electronics  -- the frequency of incidence of these data points is characterised by the duty cycle, the ratio of the number of frames used for the final light curve derivation and the total number of frames on which the target was theoretically visible. In the case of four targets, \xp, \at, \ws\, and \bc\, the targets were observed in two campaigns, C16 and C18. For these targets we used the merged C16 and C18 data. 
The main characteristics of the observations are listed in Table~\ref{table:orbital} in the Appendix. The distribution of the length of the observations and the respective duty cycle values are shown in Fig.~\ref{fig:duty}. 

\subsection{Period determination \label{sect:perioddetermination}}

The light curves obtained were analysed with a residual minimization algorithm \citep{Pal2016,Molnar2018}. In this method we fit the data with a function $A + B \cos(2\pi f \Delta t) + C \sin(2\pi f \Delta t)$, where $f$ is the trial frequency, $\Delta t = T - t$, where T is the approximate center of the time series, and A, B, and C are parameters to be determined. We search for the minimum in the dispersion curves for each frequency. As demonstrated in \citet{Molnar2018} the best fit frequencies obtained with this method are identical to the results of Lomb–Scargle periodogram or fast Fourier transform analyses, with a notably smaller general uncertainty in the residuals. 

Nevertheless, our period determination was cross-checked with the Fourier spectra obtained with the {\sl Period04} package \citep{P04}, and we also used these spectra to obtain an amplitude upper limit in the case of period non-detections. We typically accepted the period with the strongest minimum in the residual spectrum. It has already been reported in previous K2 data analysis \citep[see e.g.][]{Marton2020} that there is increased noise towards lower frequencies, typically below f\,$\leq$\,1\,{cycle/day} (c/d). We observed the same phenomenon in our TNO data as well. To cope with this increased noise we calculated a local r.m.s.\ noise in the Fourier spectrum for each frequency using a running box with sigma clipping filtering to exclude the spectral peaks from the noise calculation. This r.m.s.\ noise, $\sigma_f$, is presented as a blue curve in the figures in \autoref{sect:nondetect}. Similar noise curves ($\sigma_r$) have been derived for the normalised residual spectra. 

We evaluated the sensitivity of the Fourier method by testing it on two kinds of { synthetic} light curves, typically considered for asteroids and transneptunian objects: a rotating tri-axial ellipsoid with axes $a$\,$\geq$\,$b$\,$\geq$\,$c$, rotating around axis $c$, and a contact binary, with two spherical bodies {of} radius $r_1$\,$\geq$\,$r_2$. As expected, the Fourier method is less sensitive to non-sinusoidal light curves as in these cases the spectral power, or equivalently the light curve amplitude is distributed over several peak frequencies, most prominently the harmonics
of the primary frequency. We found a factor of $\sim$2 decrease of Fourier sensitivity in the case of contact binary light curves, and a decrease of 1.2--1.5 in the case of triaxial ellipsoids compared with sinusoidal light curves with the same amplitude. 

Considering this decreased Fourier sensitivity we accepted the peak in the residual spectrum if at the peak frequency the normalised residual ($A_r$) in the normalised residual spectrum $A_r$\,$\geq$\,5$\sigma_r$ \emph{and} the amplitude in the Fourier spectrum at the same frequency $A_f$\,$\geq$\,3$\sigma_f$. A frequency is considered to be \emph{tentative} if $A_r$\,$\geq$\,5$\sigma_r$ \emph{and} 2.5$\sigma_f$\,$\leq$\,$A_f$\,$\leq$\,3$\sigma_f$. We consider the frequency with the highest $A_r/\sigma_r$ value as the {\it primary} light curve frequency of the target, if the requirements above are met. In some cases there were multiple frequencies with similarly high $A_r/\sigma_r$ values found in the residual spectrum. We also list these {\it secondary} frequencies in Table~\ref{table:freqs}. 

We also investigated our light curves using a second-order Fourier method \citep{Harris1989} to distinguish between single and double peak light curves, comparing the amplitudes of the first and second harmonics, A$_1$ and A$_2$, respectively. A single peak light curve was accepted when A$_1$\,$\geq$\,A$_2$, and a double peak light curve was accepted when A$_1$\,$<$\,A$_2$. In some cases it was difficult to decide between the first and second harmonics as the two amplitudes were close to each other. In these cases we listed both periods in Table~\ref{table:freqs}, with the preferred one listed first and marked as `primary' period. We used these preferred periods later in the subsequent analysis in this paper. 

For targets with a detected light curve period we used the maximum difference between the data points of the folded and binned light curve, and the data in the individual bins to calculate the uncertainty in the amplitude, using standard error propagation. 
In those cases when no clear light curve period could have been identified we used the r.m.s.\ noise curves to characterise the detectable amplitudes -- for each target a single value, $\sigma_f^1$, obtained at f\,=\,1\,cycle\,day$^{-1}${,} is used in general.



\section{Results \label{sect:results}}

\subsection{Absolute magnitudes \label{sect:absmag}}

We determined the absolute magnitudes of the targets, transformed from the K2 observations to the USNO B1.0 R--band photometric system as described above, in the same way as in \citet{Marton2020}. We did not correct for the light curve variations (which are discussed in Sect.~\ref{sect:lc}) but we used simple averaging of the observed brightness values when we calculated the phase angle uncorrected absolute magnitudes (\mruncorr), after a correction for the heliocentric and observer distances. As the phase angle ranges of the observations were rather small, a phase angle correction could not be reliably obtained from these data to calculate the phase angle corrected absolute magnitudes (\mrcorr). Therefore{,} we used a $\beta_R$ linear phase coefficient from \citet{Ayala}, when it was available for a specific target, and in all other cases we used $\beta_R$\,=\,0.176$\pm$0.132\,mag\,deg$^{-1}$, the mean of the $\beta_R$-s of TNOs. These \mruncorr\, and \mrcorr\, values are listed in Table~\ref{table:orbital}. 

\subsection{Light curves and detectability \label{sect:lc}}

Table~\ref{table:freqs} contains the main frequencies identified for our targets. We were able to identify light curve frequencies for 37 targets (see Table~\ref{table:freqs}), of which 10 are considered to be tentative. 
The residual spectrum and the light curve(s) folded with the main characteristic frequency/frequencies are presented in Appendix~\ref{sect:detect}. 
In the case of 29 targets no peaks in the residual and Fourier spectra met the criteria defined above. In these cases we used the $\sigma_f^1$ values to characterise the amplitude upper limits, as listed in Table~\ref{table:freqs}. In Appendix~\ref{sect:nondetect} we present the unfolded light curves and the Fourier spectra of these targets. 


\startlongtable
\begin{deluxetable*}{p{3cm}p{1cm}p{1cm}p{1cm}p{1cm}p{1cm}p{1cm}p{0.7cm}p{5.5cm}}
\tabletypesize{\scriptsize}
\tablecaption{Main frequencies identified in the light curves of our targets. The columns are: 
(1) objects name (provisional designation);
(2-3) frequency and uncertainty (day$^{-1}$);
(4-5) period and uncertainty (h, redundant with frequency);
(6-7) light curve amplitude and uncertainty (mag); upper limits are 1$\sigma$ values based on the mean Fourier amplitude throughout the spectrum. Note that usually there is a notable (up to a factor of two) increase in r.m.s.\ amplitude towards low frequencies, typically below $\sim$1\,c/d;
(8) Flag: P -- primary period; S -- secondary period;  
T -- tentative (2.5\,$<$\,$\sigma_f$\,$<$\,3 in the Fourier spectrum and $\sigma$\,$>$\,5 in the residual spectrum); 
W -- uncertainties in frequency/period include the uncertainty due to nearby minimum in the residual spectrum; 1/2: single peak/double peak period. 
(9) comments and literature data;  
References: BS13 -- \citet{BS13}; K06 -- \citet{Kern2006}; LL06 -- \citet{LL06}; 
P02 -- \citet{P02}; SJ02 -- \citet{SJ02};
T10 -- \citet{T10}; T12 -- \citet{T12}; 
T14 -- \citet{T14}; TS18 -- \citet{TS18}; TS19 -- \citet{TS19}; 
}
\tablehead{\colhead{Name}  &  \colhead{f}  &  \colhead{$\delta$f}  &  \colhead{P}  &  \colhead{$\delta$P}  & \colhead{$\Delta$m}  &  \colhead{$\delta$($\Delta$m)} & 
\colhead{flag} &
\colhead{comments}}
\colnumbers
\startdata
\object{(26375)\,\de} & 1.046 & 0.006 & 22.947 & 0.130 & 0.081 & 0.017 & P & $\Delta$m\,$<$\,0.10, P\,$>$\,12\,h (SJ02) \\ 
    & 1.883 & 0.007 & 12.745 & 0.047 & 0.100 & 0.019 & S &  \\ \hline
\object{(35671)\,\sn} & 4.251 & 0.014 & 5.646 & 0.019 & 0.106 & 0.027 & P/SP & $\Delta$m\,=\,0.15$^{+0.022}_{-0.030}$, P\,=\,10.0$\pm$0.8\,h (P02) \\ 
    & 8.520 & 0.014 & 2.817 & 0.005 & 0.080 & 0.027 & S/DP & $\Delta$m\,=\,0.16$\pm$0.01, P\,=\,8.84\,h or 8.70\,h (LL06) \\ \hline
\object{(66652)\,Borasisi} & 1.208 &  0.002 & 19.868 & 0.032 & 0.216 & 0.057 & P & P=6.4$\pm$0.1 (K06) \\ \hline  
\object{(80806)\,\cm}   & --    & --    & --     & --    & $<$0.045 & --      & & $\Delta$m\,$<$\,0.14 (TS19)    \\ \hline
\object{(119878)\,\cy} &  -- & -- & -- & -- & $<$0.194 & -- & --      \\ \hline
\object{(126154)\,\yh}       & 0.876 & 0.006 & 27.397  & 0.172 & 0.229 & 0.019  & P &  $\Delta$m\,=\,0.13$\pm$0.05, P\,=\,13.2\,h (T10)    \\ 
          & 1.751 & 0.005 & 13.705  & 0.039 & 0.095 & 0.019  & S &    \\ \hline
\object{(127871) \fc} & 0.867 & 0.006 & 27.674 & 0.194 & 0.161 & 0.039 & & -- \\ \hline
(135182) \qt &  -- & -- & -- & -- & $<$0.067 & -- & --  &   \\ \hline
(138537) \ok & 0.710 & 0.038 & 33.803 & 1.809 & 0.155    & 0.042   & T/W & P$>$6h (TS19)     \\ \hline
(145480) \tb & --  & -- & -- & -- & $<$0.025 & -- & & $\Delta$m=\,0.12$\pm$0.01, P\,=\,12.68\,h (T12) \\ \hline
(149348) \vs & 2.778 & 0.023 & 8.639  & 0.072 & 0.217 & 0.060 & T & $\Delta$m\,$\approx$\,0.1 (TS19) \\ \hline
(160147) \kn  & --    & --    & --     & --    & $<$0.043 & --    & --  &   \\ \hline
(182934) \gj &  -- & -- & -- & -- & $<$0.044 & -- & --  &   \\ \hline
(307463) \vu & -- & -- & -- & -- & $<$0.019 & --  & -- &      \\ \hline
(307616)  \qw &  0.278 & 0.011 & 86.331 & 3.515 & 0.145 & 0.025 & P &     \\ 
    &  0.540 & 0.010 & 44.444 & 0.789 & 0.147 & 0.028 & S & \\ \hline  
(308379) \rs &  -- & -- & -- & -- & $<$0.017 & -- & --  &     \\ \hline
(312645) \ep & 1.932 & 0.006 & 12.422 & 0.039 & 0.122 & 0.019  & P & $\Delta$m=\,0.17$\pm$0.03, P$_s$\,=\,7.48\,h (BS13) \\ \hline
(385266) \qb & --    & --    & --     & --    & $<$0.030 & --    & -- &    \\ \hline
(385437) \gh &  0.114 & 0.002 & 210.526 & 3.693 & 0.402 & 0.060 & P &  \\ 
         &  2.562 & 0.003 & 9.3683 & 0.01062 & 0.165 & 0.034 & S/T & \\ \hline 
(408832) \qj &  -- & -- & -- & -- & $<$0.038 & -- & --  &   \\ \hline
(420356) Praamzius & 0.352 & 0.004 & 68.182 & 0.775 & 1.433 & 0.181 & P &  \\ 
          & 9.108 & 0.002 & 2.635 & 0.001 & 0.536 & 0.218 & S/T & \\ \hline
(469420) \xp &  0.514 & 0.005 & 46.719 & 0.422 & 0.282 & 0.056 & P/T &      \\ 
             &  0.205 & 0.004 & 117.157 & 2.328 & 0.275 & 0.056 & S/T &      \\ \hline
(469421) \xd &  -- & -- & -- & -- & $<$0.039 & -- & -- &     \\ \hline
(469505)\,\fe & 3.106 & 0.004 & 7.727  & 0.011 & 0.230 & 0.046 & P & $\Delta$m=\,0.50$\pm$0.14, P\,=\,5.85$\pm$0.15\,h (K06) \\ 
    & 0.582 & 0.005 & 41.203 & 0.360 & 0.292 & 0.047 & S/T & \\ \hline
(469704)\,\ez &  5.074 & 0.005 & 4.730  & 0.005 & 0.431 & 0.093 & P & \\
    &  0.596 & 0.006 & 40.268 & 0.404 & 0.514 & 0.083 & S & \\ \hline 
(470523)\,\cs & --  & -- & -- & -- & $<$0.022 & -- & -- &   \\ \hline
(471137)\,\et & 0.724 & 0.005 & 33.149 & 0.217 & 0.124 & 0.020  & P & $\Delta$m=\,0.13$\pm$0.02, P$_s$\,=\,3.94\,h (BS13) \\ 
    & 1.452 & 0.006 & 16.529 & 0.064 & 0.105 & 0.019  & S & \\ \hline
(471150)  \fcten & -- & -- & -- & -- & $<$0.010 & -- & & \\ \hline
(471318) \jf & 9.030 & 0.006 & 2.658 & 0.002 & 0.487 & 0.045 & P &   \\ 
    & 4.515 & 0.003 & 5.316 & 0.003 & 0.443 & 0.044 & S &  \\ \hline
(472235) \gefourteen & 2.386 & 0.036 & 10.059 & 0.151 & 0.501 & 0.074 & T &     \\ \hline
(508869) \vt & -- & --    & --     & --    & $<$0.034 & --  & -- & $\Delta$m\,$\approx$\,0.21 (T14) \\ \hline
(523658) \dw & 0.790 & 0.005 & 30.380 & 0.184 & 0.241 & 0.051 & T & --   \\ \hline
(523687) \df &  -- & -- & -- & -- & $<$0.015 & -- & --  & -- \\ \hline
(523692) \ezfourteen & 7.500 & 0.005 & 3.200 & 0.002 & 0.145 & 0.026 &  -- &  \\ \hline
(523698) \gdfourteen & 0.542 & 0.007 & 44.280 & 0.563 & 0.262 & 0.050 & T & --  \\ \hline
(523706) \hf & 4.808 & 0.016 & 4.992 & 0.016 & 0.516 & 0.064 &  -- & --   \\ \hline
(523769) \ws & 0.960 & 0.004 & 25.009 & 0.100 & 0.374 & 0.060 & -- & --  \\ \hline
(525462) \eo &  0.658 & 0.006 & 36.474 & 0.306 & 0.515 & 0.094 & -- & --   \\ \hline
(533207) \djfourteen &  -- & -- & -- & -- & $<$0.040 & -- & -- & --    \\ \hline 
(533562) \jq & 1.974 & 0.005 & 12.158 & 0.032 & 0.394 & 0.137 & P & $\Delta$m\,=\,0.76$\pm$0.04, P\,=\,12.16\,h (TS18) \\ 
    & 3.948 & 0.010 & 6.079 & 0.015 & 0.403 & 0.101 & S  \\ \hline
(533676) \ls  &  -- & --    & --     & --    & $<$0.041 & --      & -- & $\Delta$m\,=\,0.35$\pm$0.03, P\,=\,5.52/11.04\,h (TS19) \\ \hline
(535018) \wanine &  -- & -- & -- & -- & $<$0.054 & -- & --  & -- \\ \hline
(535023) \wo &  -- & -- & -- & -- & $<$0.017 & -- & --  & --   \\ \hline
(535028) \waten &  -- & -- & -- & -- & $<$0.021 & -- & -- & --    \\ \hline
(535030) \wj & 0.592 & 0.009 & 40.541 & 0.648 & 0.442 & 0.094 & T/SP & --   \\ \hline
(535228) \ye & 0.421 & 0.009 & 56.970 & 1.236 & 0.414 & 0.083 & T & --     \\ \hline
(535231) \yj &    -- & -- & -- & -- & $<$0.058 & -- & --  & --   \\ \hline
(542258) \apthirteen & 4.792 & 0.010 & 5.008 & 0.010 & 0.440 & 0.078 & T & --      \\ \hline
\hz & 4.671 & 0.005 & 5.138 & 0.005 & 0.457 & 0.100 &  -- & --     \\ \hline
\qx &  -- & -- & -- & -- & $<$0.073 & -- & --   & --   \\ \hline
\xu & 1.360  & 0.005 & 17.654 & 0.064 & 0.523 & 0.102 & -- & --     \\ 
    & 0.680  & 0.003 & 35.294 & 0.128 & 0.681 & 0.099 & -- & --      \\ \hline
\qa &   -- & -- & -- & -- & $<$0.045 & -- & --  & --    \\ \hline
\ys &  1.244 & 0.035 & 19.293 & 0.542 & 1.278 & 0.197 & T & -- \\ \hline
\tf &  -- & -- & -- & -- & $<$0.036 & -- & --  & --     \\ \hline
\tv &  -- & -- & -- & -- & $<$0.041 & -- & --  & --     \\ \hline
\xr &  -- & -- & -- & -- & $<$0.026 & -- & --  & --     \\ \hline
\yg & 6.328 & 0.048 & 3.793 & 0.029 & 0.360 & 0.045 & P & -- \\ 
    & 3.166 & 0.046 & 7.581 & 0.111 & 0.365 & 0.046 & S & -- \\ \hline
\at & 0.305 & 0.002 & 78.788 & 0.346 & 0.402 &  0.051 & P & --  \\ \hline
\fa & -- & -- & -- & -- & $<$0.030 & -- & -- & $\Delta$m\,$\approx$\,0.1 (TS19) \\ \hline
\jv & 3.982 & 0.004 & 6.027 & 0.006 & 0.291 & 0.071 & -- & --    \\ \hline
\am &  -- & -- & -- & -- & $<$0.032 & -- & -- & --    \\ \hline
\gjfourteen &   -- & -- & -- & -- & $<$0.047 & -- & -- & --    \\ \hline
\bbfifteen & 0.328 & 0.008 & 73.202 & 1.801 & 0.774 & 0.113 &      \\ \hline
\bc &  2.697 & 0.005 & 8.899 & 0.006 & 0.330 & 0.061 & P & --     \\  \hline
\bz &  0.760 & 0.004 & 31.579 & 0.083 & 0.392 & 0.079 & P & --      \\  
          &  1.476 & 0.004 & 16.260 & 0.022 & 0.373 & 0.076 & S & --  \\ \hline
\da       &  6.7961 & 0.0034 & 3.531 & 0.002 &  0.511 & 0.128 & P & -- \\ 
          &  0.668 & 0.004 & 35.928 & 0.188 & 0.484 & 0.119 & S/T & -- \\ \hline \hline


\enddata
\label{table:freqs}
\end{deluxetable*}

\begin{figure}[ht!]
\begin{center}
\includegraphics[width=0.47\textwidth]{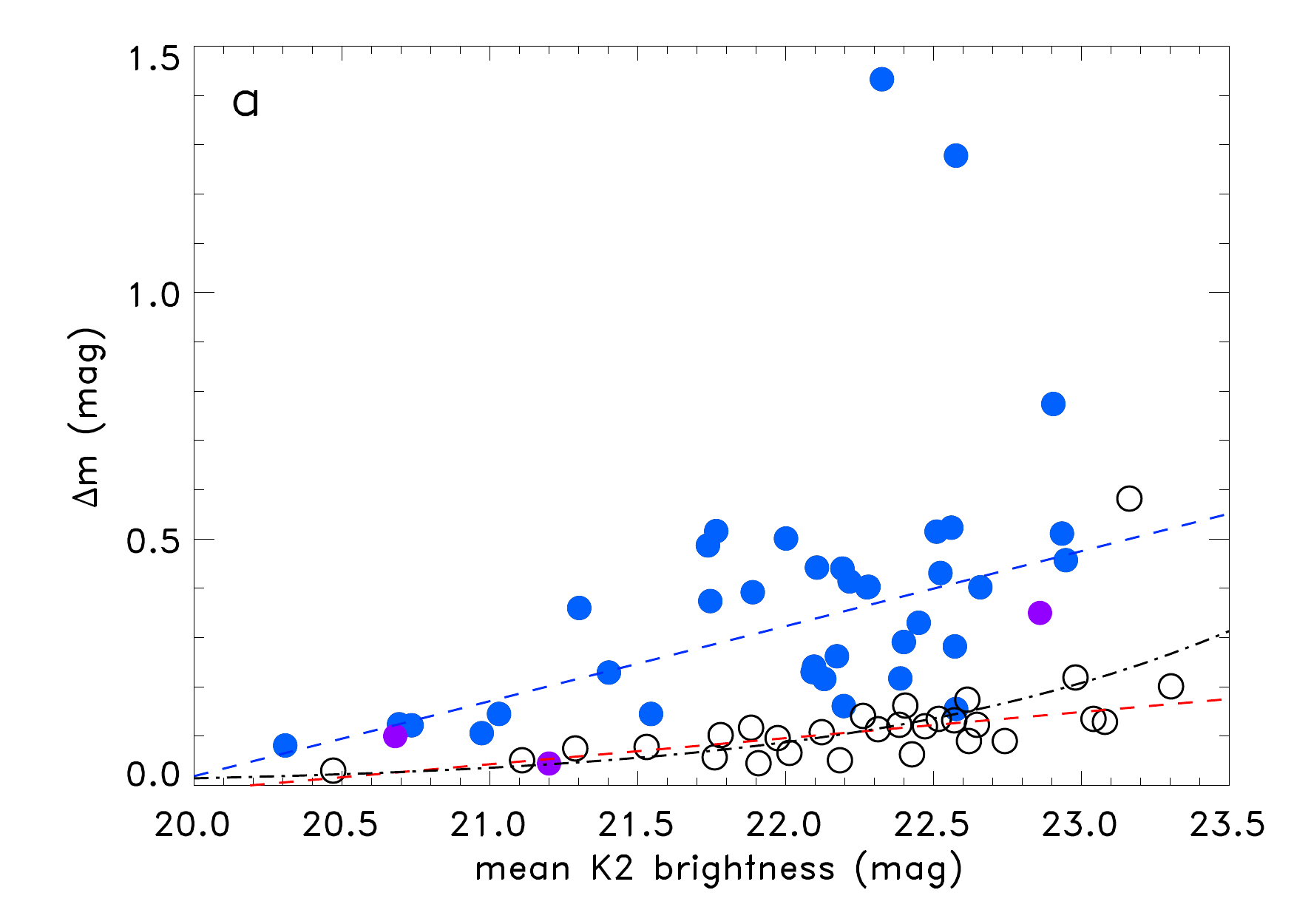}
\includegraphics[width=0.47\textwidth]{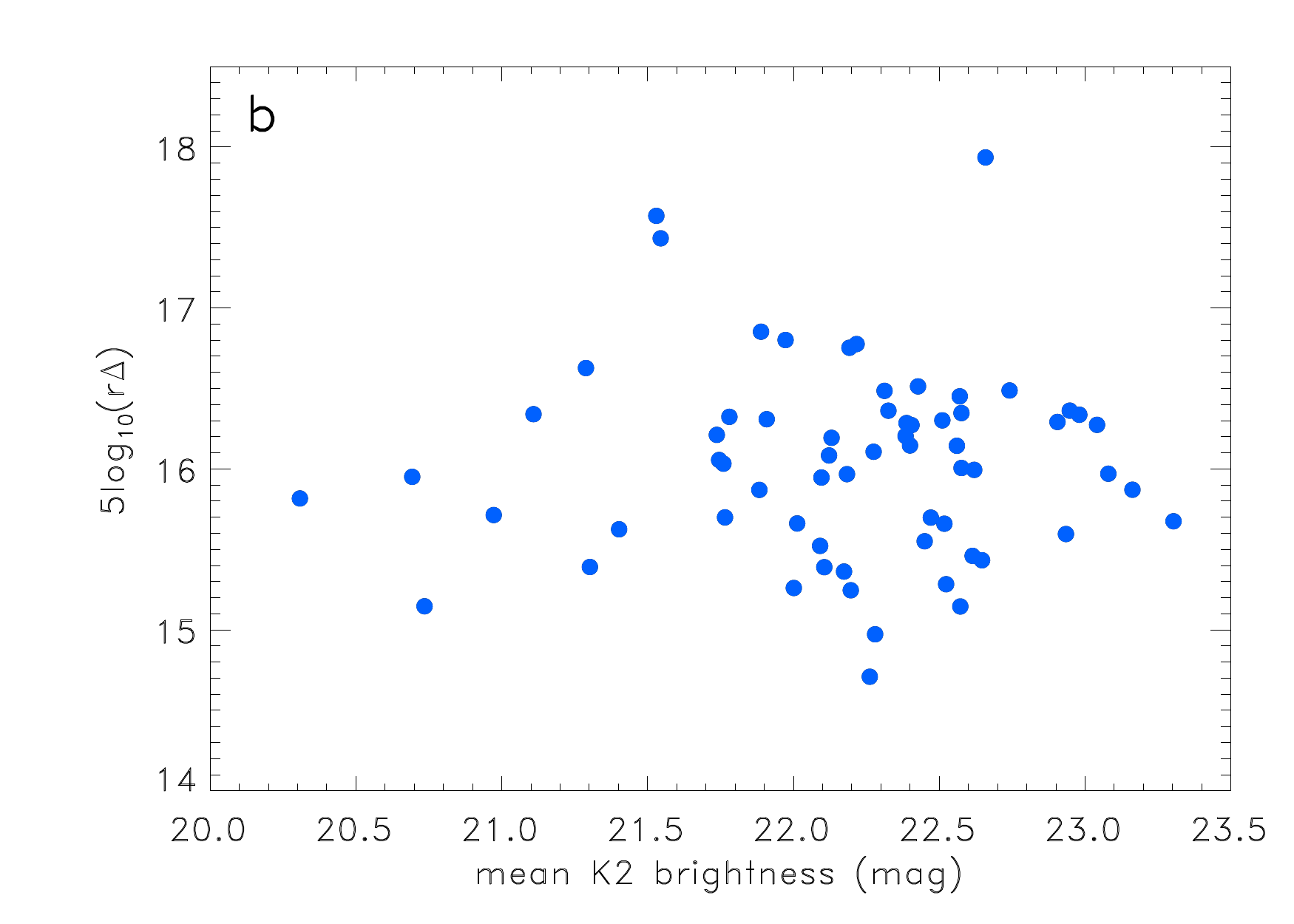}
\caption{(a) Light curve amplitude as a function of mean brightness. Blue and black filled circles represent the K2 TNOs with detected periods, and those with a tentative frequency, respectively. The data of three previous K2 TNO light curve detections 
\citep[purple symbols, Gonggong, \jj, \gv,][]{Pal2015,Pal2016} are also included.
Open circles shows the TNOs without a detected period, in these cases the amplitude upper limit is obtained as $\Delta m$\,=\,3$\sigma_f^1$. 
The blue and red dashed lines represent linear fits to the light curve amplitudes of the detections and the 3$\sigma_f^1$ upper limits of period non-detections, respectively. The black dash-dotted curve corresponds to a constant flux density detection limit fitted to the non-detection upper limits, with $\Delta m$\,=\,0.0133\,mag at $\langle m \rangle$\,=\,20\,mag.
(b) Distance modulus of our K2 targets versus their mean apparent brightness
} 
\label{fig:ampmean}
\end{center}
\end{figure}

\begin{table}[ht!]
    \centering
    \begin{tabular}{ccccc}
          \hline
          Correlation & slope & $\rho_{Sp}$ & S$_{Sp}$ & Fig. \\ 
          & (mag/mag) & & & \\ \hline
          $\Delta$m vs. $\overline{m}$, det. & 0.152$\pm$0.030 & 0.593 & 5.5e-5 & \ref{fig:ampmean} \\
          $\Delta$m vs. $\overline{m}$, non-det. & 0.053$\pm$0.001  & 0.728 & 7.54e-6 & \ref{fig:ampmean} \\ 
          $\Delta$m vs. H$_{V}$, K2+ & 0.072$\pm$0.022  & 0.433 & 0.005 & \ref{fig:amph}b \\ 
          $\Delta$m vs. H$_{V}$, K2 & 0.068$\pm$0.032  & 0.354 & 0.032 & \ref{fig:amph}b \\ 
          $\Delta$m vs. H$^{R}_{min}$, K2+ & 0.073$\pm$0.021  & 0.476 & 0.009 & \ref{fig:amph}c \\ 
          $\Delta$m vs. H$^{R}_{min}$, K2 & 0.073$\pm$0.031  & 0.339 & 0.040 & \ref{fig:amph}c \\ \hline
          \hline
    \end{tabular}
    \caption{Correlations between the light curve amplitude ($\Delta$m) and the mean apparent brightness ($\overline{m}$) or the \absmag\, and H$^{R}_{min}$ absolute magnitudes, listing the corresponding slope, the Spearman's rank correlation coefficient $\rho_{Sp}$ and the significance S$_{Sp}$. For the non-detections the 3$\sigma$ upper limits are considered.
    K2 refers to the targets in this paper, K2+ includes Gonggong, \jj\, and \gv\, in addition. The last column lists the numbers of the corresponding figures. }
    \label{table:slopes}
\end{table}

In \autoref{fig:ampmean} we present the light curve amplitudes of our targets versus their mean brightness. Detectable amplitudes show the expected trend and only light curves with amplitudes $\Delta m$\,$\geq$\,0.5\,mag are detected at mean brightness values of $\langle m \rangle$\,$\gtrsim$\,23\,mag. {Notably}, no large amplitude light curve was identified among the brighter targets at $\langle m \rangle$\,$\lesssim$\,21\,mag. Linear fits to the detected amplitudes and the upper limits versus the mean brightness of the targets are also shown in this figure, and the slopes and correlation coefficients are listed in Table~\ref{table:slopes}. Both the detected light curve amplitudes and the upper limits show a fairly strong correlation with the mean brightness.
We also fitted the upper limits (open circles) assuming a detection limit that is constant in flux density and independent of the brightness of the target, i.e. it is a higher fraction of the brightness for fainter targets (black dash-dotted curve). This provides a better fit than the linear correlation (dashed red line), with a $\sim$30\% lower total residual, and corresponds to a 3$\sigma$ detection limit of $\Delta$m\,=\,0.0133\,mag at $\langle m \rangle$\,=\,20\,mag and $\Delta$m\,=\,0.21\,mag at $\langle m \rangle$\,=\,23\,mag mean brightness values. 


\begin{figure*}
\begin{center}
\hbox{ 
\includegraphics[width=0.5\textwidth]{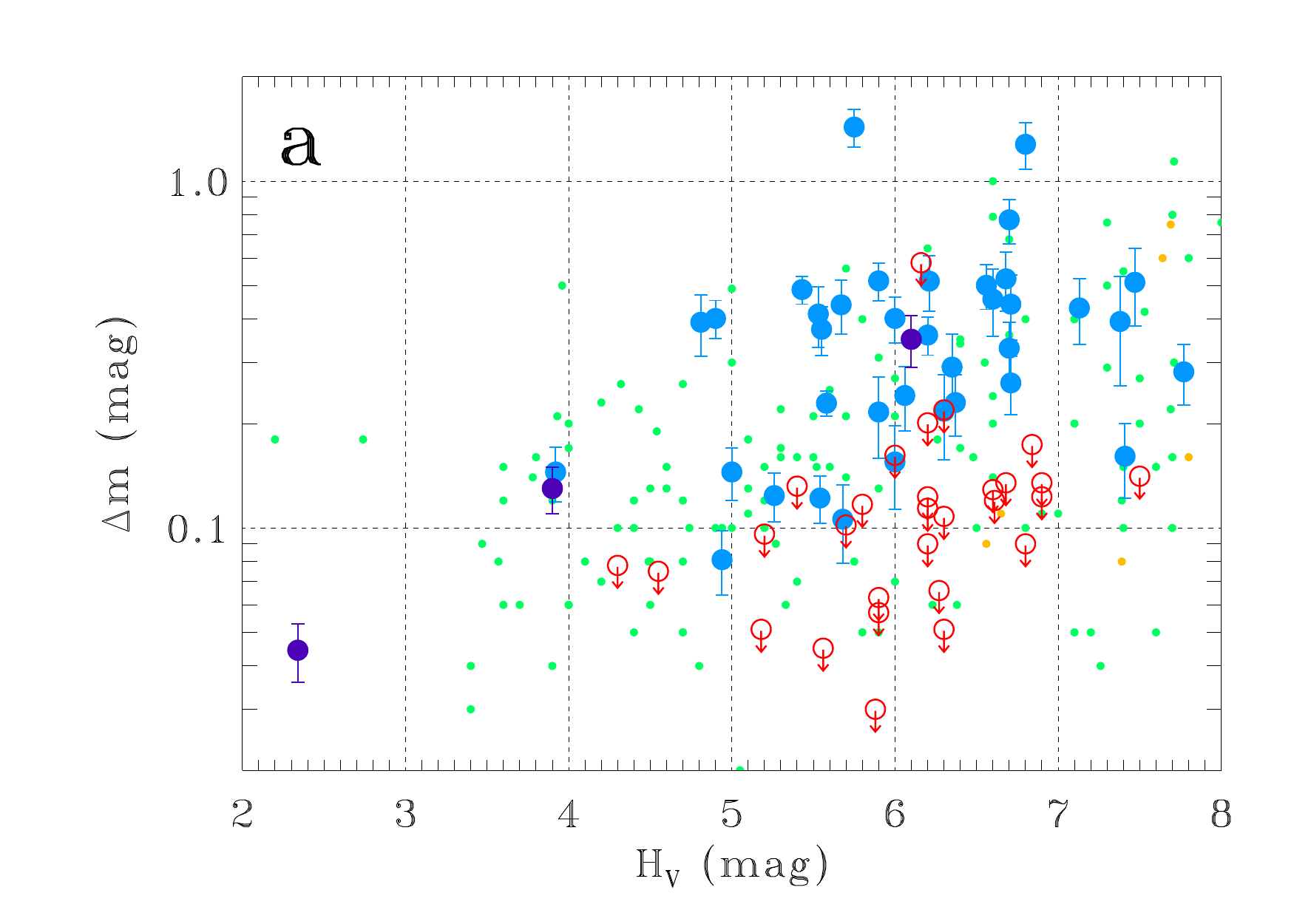}
\includegraphics[width=0.5\textwidth]{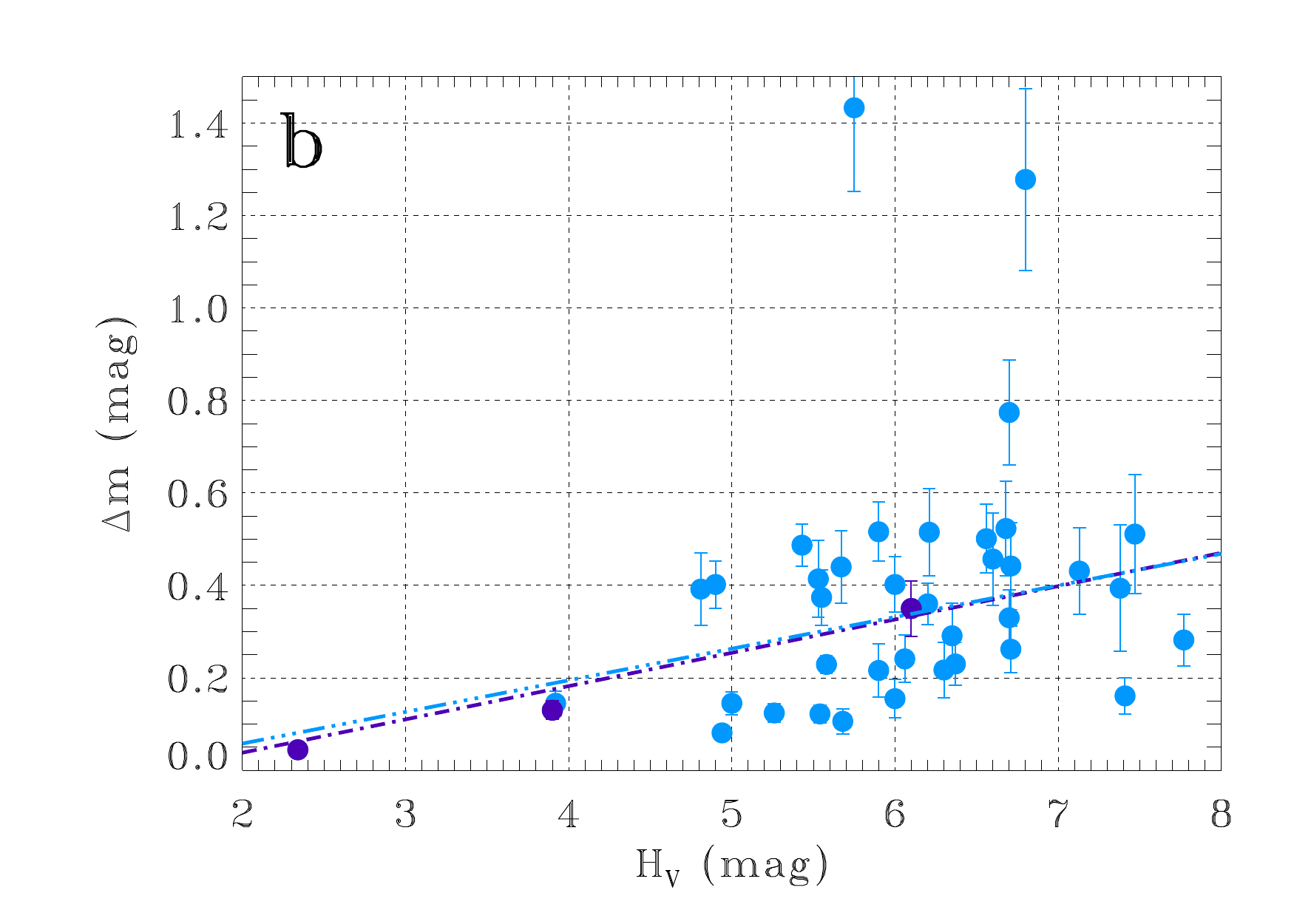}}
\hbox{
\includegraphics[width=0.5\textwidth]{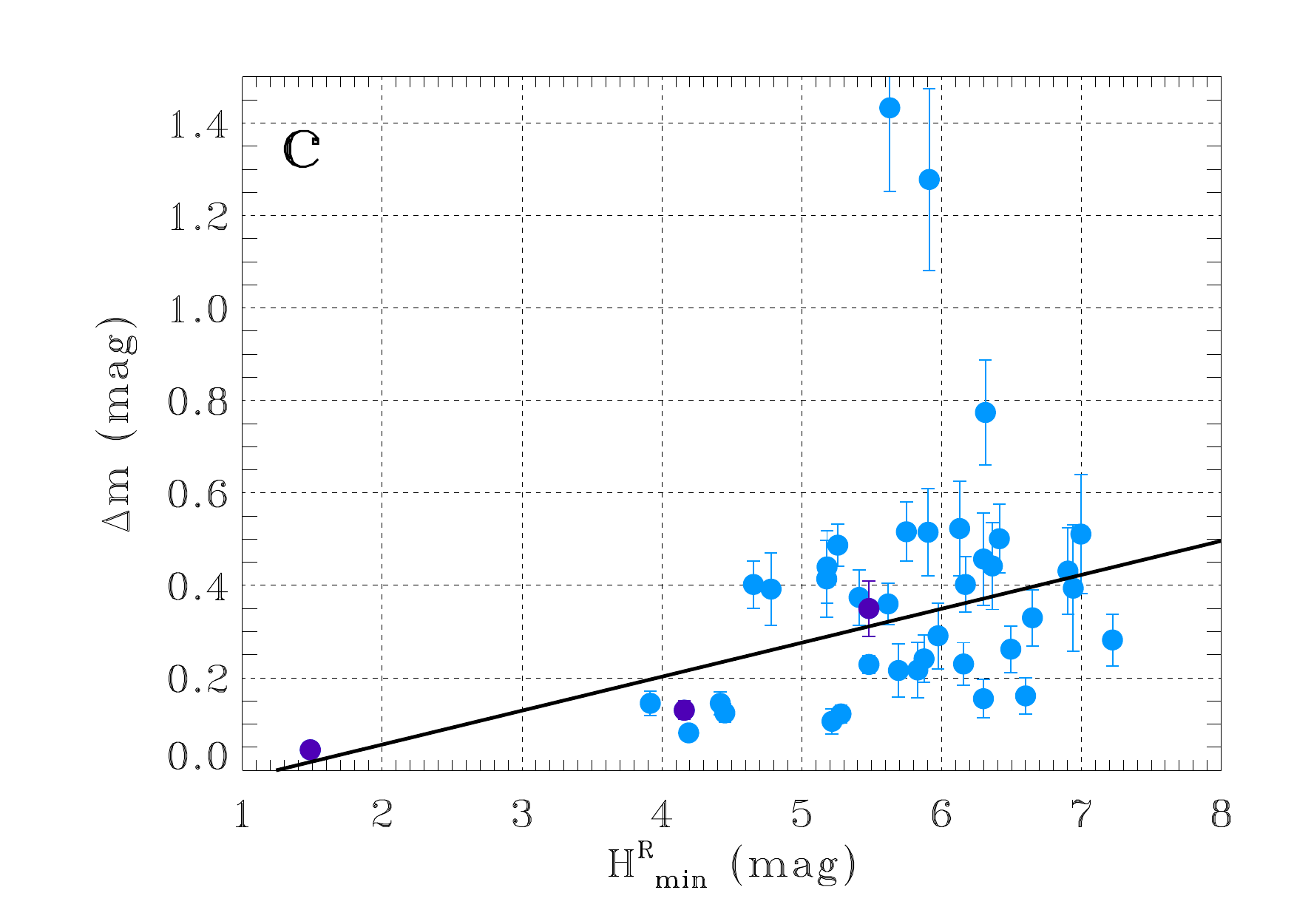}
\includegraphics[width=0.5\textwidth]{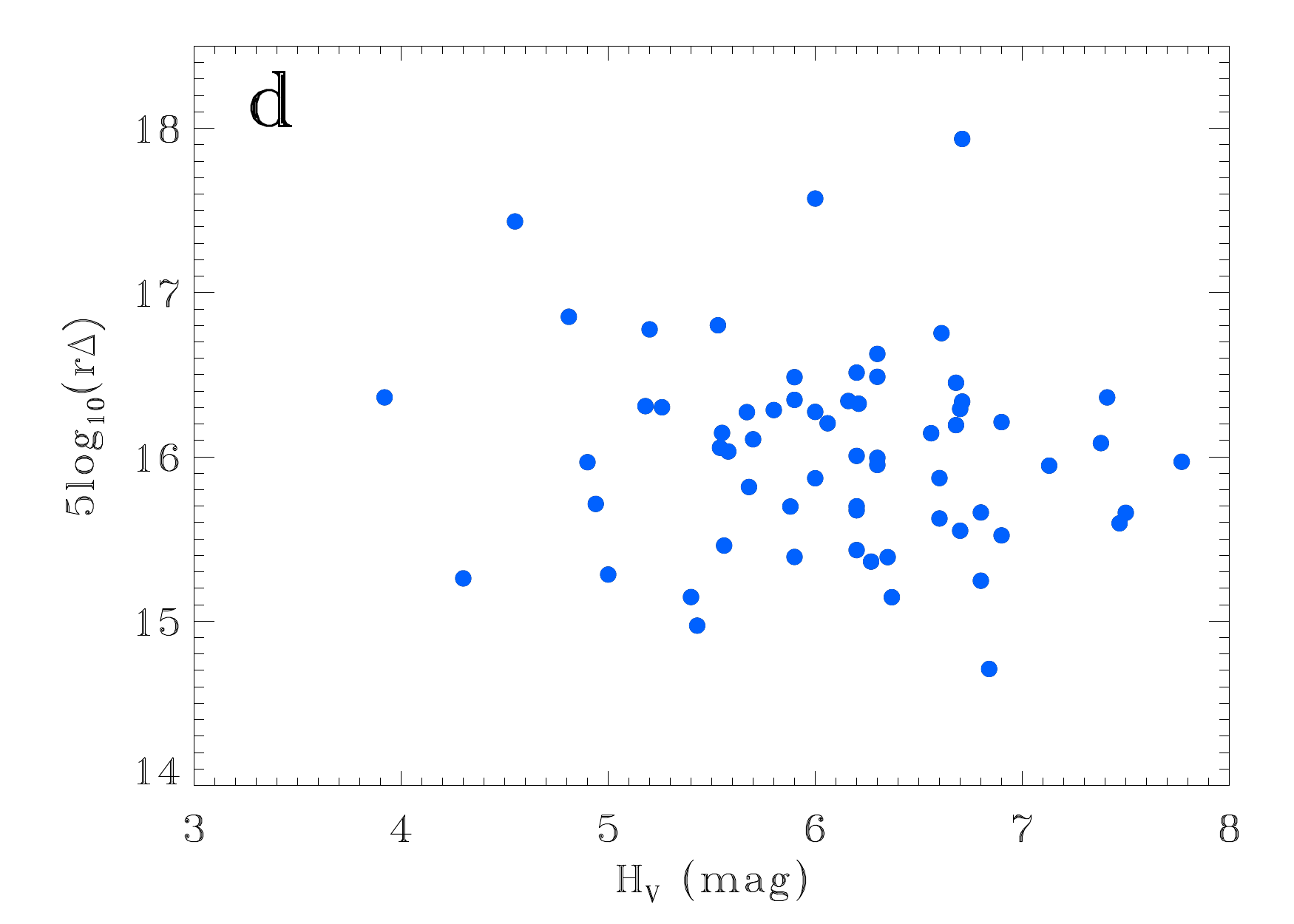}
}
\caption{a) Light curve amplitude ($\Delta m$) versus the \textit{V}-band absolute magnitude ($H_V$). 
{ Green} and orange symbols are TNOs and Centaurs from the LCDB (using the maximum amplitude value); Blue symbols correspond to TNOs with detected periods from our K2 survey (this paper); Red symbols are amplitude upper limits of TNOs without detected periods from our K2 survey (this paper, represented by 3$\sigma_f$ at f\,=\,1\,c/d). 
Purple symbols are K2 measurements of Gonggong (2007\,OR$_{10}$), \gv, and \jj\, \citep{Pal2015,Pal2016}. 
b) $\Delta$m versus \absmag\, relationship of K2 TNOs with a linear fit. The purple dash-dotted and blue dash-tripple-dotted lines correspond to fits with and without Gonggong, respectively. 
c) The same as b) but using $H_{min}^R = m_{110}^R - \Delta m/2$ instead of $H_V$. The linear fits to the data with and without Gonggong are so close to each other that they are represented by a single solid line. d) Distance modulus versus V-band absolute magnitude. }
\label{fig:amph}
\end{center}
\end{figure*}

We plot the amplitude of the light curves against the absolute magnitudes of our sample in \autoref{fig:amph}a; the detected light curve amplitudes versus \absmag, separately (\autoref{fig:amph}b); and also the light curve amplitudes versus the corrected 'Kepler band' (i.e. USNO B1.0 R) absolute magnitude, H$^R_{min}$\,=\,$m^R_{110}$--$\Delta$m/2 (\autoref{fig:amph}c), as defined in \citet{Showalter2021}. Targets with detected light curve period show a weak correlation between \absmag\, or H$^R_{min}$ and the light curve amplitude, at a similar level of significance as in \citet{Showalter2021}, and with slopes similar to the \citet{Showalter2021} 'non-OSSOS' values, considering the large error bars on the slopes due to the weak correlations. The high slope values obtained for their OSSOS sample is not seen among our targets. While the correlation between \absmag\, and $\Delta$m is often interpreted as a sign that smaller (higher \absmag) objects could in general have higher light curve amplitudes \citep[see e.g. ][]{Benecchi2013,Thirouin2019} an apparent-brightness-dependent light curve amplitude detection limit -- that we have seen above for our sample -- may easily lead to a weak \absmag--$\Delta$m correlation by selecting only the higher amplitude targets among the lowest-\absmag\, ones, concealing their true behaviour. However, as demonstrated in Fig.~\ref{fig:ampmean}b and \ref{fig:amph}d, we cannot observe notable correlation {either} between the apparent mean brightness and the distance modulus of the targets, nor between the absolute magnitude and the distance modulus which could explain a light curve amplitude -- \absmag\, correlation. 


\subsection{Individual objects (with previously "detected" light curve periods)}

We have cross-checked our light curves with previous light curve detection of the same targets. While in some cases we can confirm the earlier light curve periods, in many cases our data clearly rules out the previous detection, as the usually high light curve amplitudes claimed in those papers should have clearly been detected in our data at those frequencies. 

%
\paragraph{\de} An upper limit of $\Delta$m$<$0.1\,mag was obtained on the light curve amplitude and a lower limit of P\,$>$\,12\,h on the period by \citet{SJ02}; our new P\,=\,22.9\,h period and $\Delta$m\,=\,0.081\,mag amplitude are compatible with these previous values. 

\paragraph{\sn} This target has two well-defined frequencies in the residual spectrum which are the first and second harmonics of the same frequency. Our analysis provided the double-peak period, P\,=\,5.646\,h being the more likely one. This period is different from the periods obtained by \citet{P02} and \citet{LL06} (P\,=\,10.0\,h and 8.84\,h or 8.70\,h, respectively). 

\paragraph{Borasisi}
\citet{Kern2006} obtained a rotation period of P\,=\,6.4$\pm$0.1\,h, using data from one single night to determine the light curve period, and another night with some spare phase coverage almost a year later, when the fixed period from the first date was used to phase the data.  We could not recover this period from our data, but identified a single prominent period at f\,=\,1.208$\pm$0.002\,c/d or P\,=\,19.868\,$\pm$0.032\,h, with an amplitude of $\Delta m$\,=\,0.216$\pm$0.057\,mag. 

\paragraph{\cm} A light curve amplitude estimate of $\Delta$m\,$\approx$\,0.14\,mag was obtained by \citet{TS19} which is about the 3$\sigma$ limit of our data. We have not detected any period for this target though.  

\paragraph{\yh} Our \emph{single-peak} period, P\,=\,13.705\,h is roughly compatible with the 13.2\,h period obtained by \citet{T10}, however, we identified the double period, P\,=\,27.397\,h as the more likely one. 

\paragraph{\tb} While a light curve with a period of P\,=\,12.68\,h and amplitude of $\Delta$m\,=\,0.12\,mag was identified by \citet{T12} {we could only obtain} a 3$\sigma$ amplitude upper limit of $\Delta$m\,=\,0.075\,mag for this target. We could not identify any peak in the spectrum at the corresponding frequency. 

\paragraph{\ep} The P\,=\,7.48\,h period obtained by \citet{BS13} could not be recovered from our data. Instead, a very prominent period of P\,=\,12.422\,h with $\Delta$m\,=\,0.122\,mag was indentified from our K2 measurements.  

\paragraph{Praamzius}
A prominent minimum was identified at the low-frequency part of the residual spectum at f\,=\,0.352$\pm$0.004\,c/d, P\,=\,68.182$\pm$0.775\,h, with a large light curve ampltiude of $\Delta m$\,=\,1.433$\pm$0.181\,mag. 
There is another, prominent minimum at the high-frequency part at f\,=\,9.108$\pm$0.002\,c/d (P\,=2.635$\pm$0.001\,h, $\Delta m$\,=\,0.536$\pm$0.218\,mag) that technically meets the criterion of a detection. However, we consider this frequency as tentative and use f\,=\,0.352$\pm$0.004\,c/d as {the} primary frequency. 

\paragraph{\fe} While two rather different possible light curve periods were identified (P\,=\,7.727 and 41.203\,h) from our K2 data, no notable peak could be found at the previously found P\,=\,5.85\,h period \citep{Kern2006}.  

\paragraph{\et} Two prominent frequencies were obtained which correspond to the single/double peak light curves with the same base period (P\,=\,16.529 and 33.149\,h). We found no indication for the short period of P\,=\,3.94\,h found earlier by \citet{BS13}. 

\paragraph{\vt} A light curve amplitude estimate of $\Delta$m\,$\approx$\,0.21\,mag was obtained by \citet{T14} which is notably higher than our $\Delta$m\,=\,0.1\,mag 3$\sigma$ amplitude upper limit. Despite that we could not {detect} any light curve period for this target.  

\paragraph{\jq} A prominent contact binary signal was detected for \jq\, by \citet{TS18} (TS18), with a double-peaked light curve period of P\,=\,12.16\,h. Our analysis provided the period P\,=\,6.079\,h, (f\,=\,3.948 c/d) with the highest signal-to-noise over the r.m.s.\ noise in the residual spectrum. This is exactly the half of that found by \citet{TS18}, and agrees their single-peak period. While our associated light curve amplitude is definitely smaller than the \citet{TS18} one, this is expected considering the faintness of the target and the relative accuracy that could be achieved in K2 measurements due to the small telescope size and long integration times relative to the rotation period and compared with the width/depth of the minima of a contact binary light curve. Finding the previously identified period of \jq\, is a good example that our method can effectively select the 'right' periods from multiple possibilities. 

\paragraph{\ls} \citep{TS19} obtained a light curve period of P\,=\,5.52 or 11.04\,h, with a rather high amplitude of $\Delta$m\,=\,0.35\,mag. Although our 3$\sigma$ amplitude upper limit of $\sim$0.12\,mag is significantly lower that this amplitude, we could not find these periods in our data. 

\paragraph{\fa} A light curve amplitude estimate of $\Delta$m\,$\approx$\,0.1\,mag was obtained by \citet{TS19} which is close to our $\Delta$m\,=\,0.1\,mag 3$\sigma$ amplitude upper limit. We could not detect any light curve period for this target.

\subsection{Amplitude distribution \label{Sect:ampldist}}

\begin{figure*}[ht!]
\begin{center}
\includegraphics[width=0.98\textwidth]{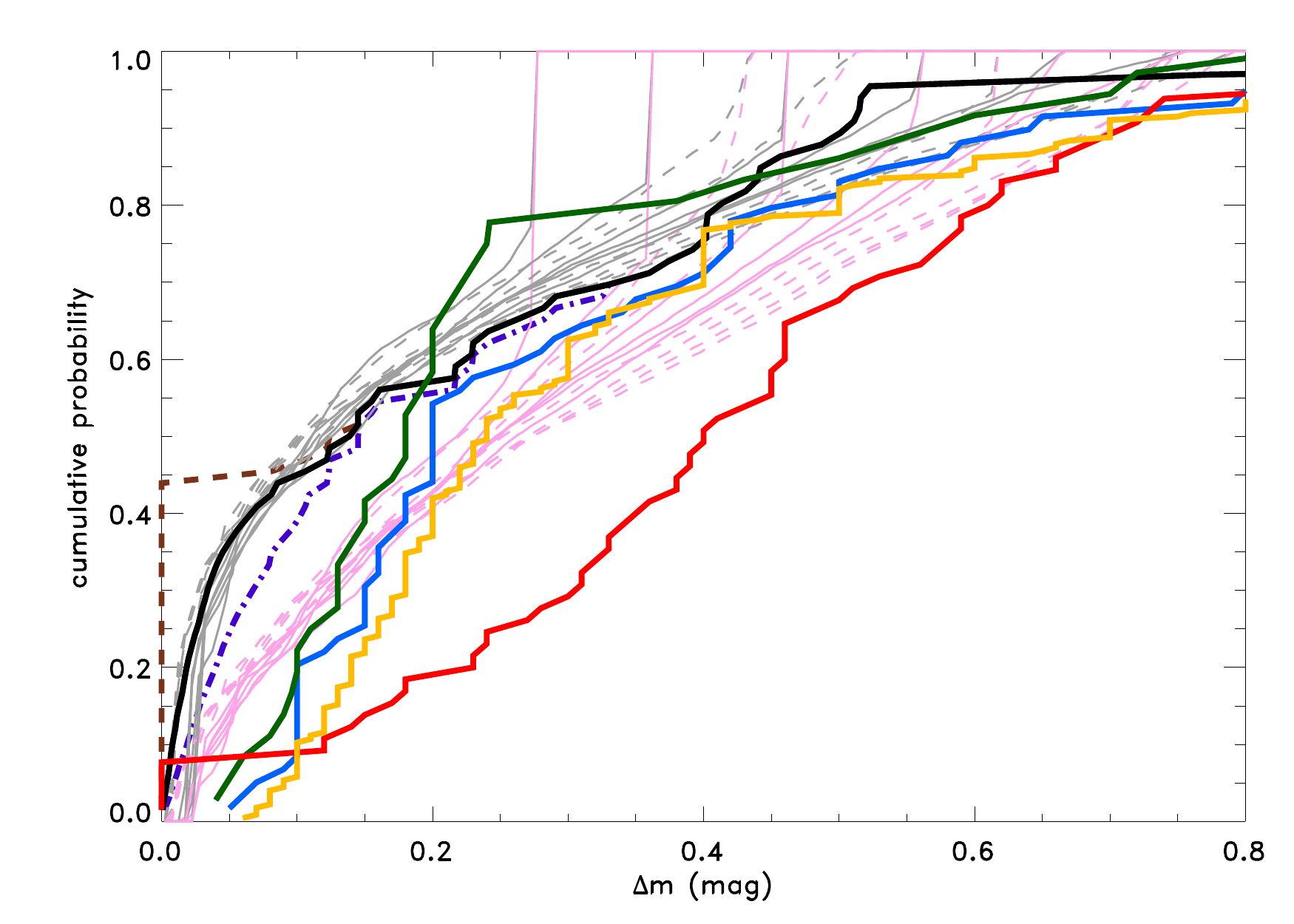}
\caption{Cumulative amplitude distribution of the K2 TNOs. The brown (dashed), black (solid) and purple (dash-dotted) curves represents our TNO sample, assuming different 
distributions for the non-detections, as discussed in Sect.~\ref{Sect:ampldist} (Cases (1), (2) and (3), respectively). The blue, green and red curves correspond to the 'Combined', 'Centaur' and 'OSSOS' curves in Fig.~11 in \citet{Showalter2021}. The pale gray and pink curves in the background show the cumulative amplitude distribution of contact binaries assuming spherical shapes, size ratios from 0.5 to 1.0, Lambert (gray) or Lommel-Seeliger (pink) scattering, and either uniform aspect angle (solid curves) or isotropic pole orientation (dashed curves) distributions for a set of simulated objects (see a more detailed discussion in \citet{Showalter2021}). The yellow curve is the cumulative amplitude distribution in the main belt as derived in \citet{Masiero2009}.}
\label{fig:cumulampl}
\end{center}
\end{figure*}

We compared our cumulative amplitude distribution with those discussed in \citet{Showalter2021}. To obtain the cumulative amplitude distribution curve of our K2 sample we considered both the targets with detected periods and well-determined amplitudes, and also those which only have amplitude upper limits (see Fig.~\ref{fig:cumulampl}). The amplitude upper limits in our sample were considered in three different ways when drawing the cumulative amplitude distributions: (1) using $\Delta$m\,=\,0 for all non-detections (brown dashed curve in Fig.~\ref{fig:cumulampl}); (2) assigning a random amplitude value assuming an underlying normal distribution and using the 1$\sigma$ detection limit as standard deviation around zero amplitude (black solid curve); and (3) using $\Delta$m\,=\,3$\sigma$ for all non-detections (purple dash-dotted curve). The light curve amplitudes of the targets with detected periods were considered with their actual value as an expectation value, using the amplitude uncertainty as standard deviation.
We generated a thousand curves in each cases, and their median curve is considered as our K2 cumulative amplitude distribution, presented in Fig.~\ref{fig:cumulampl}. Cases (1) and (3) represent the two most extreme cases for the underlying amplitude distribution of our non-detections and due to the preference of small amplitudes in Case (2) the true cumulative amplitude curve is very likely between the curves of Cases (2) and (3). Above $\Delta$m\,$\approx$\,0.1\,mag there is no real difference between the curves of (1), (2) and (3). 

In the same figure we also plot the 'Combined', 'Centaur' and 'OSSOS' curves from \citet{Showalter2021} which mostly use ground-based data. We also plot the original cumulative amplitude curve derived by the Thousand Asteroid Light Curve Survey \citep{Masiero2009} to cover the main belt asteroids. There is a considerable difference between the ground-based and the K2 cumulative amplitude distribution curves of transneptunian objects: a notably larger number of objects appear at low amplitudes compared with any of the curves derived by \citet{Showalter2021}. As mentioned in \citet{Showalter2021}, there may be multiple bias factors in their KBO amplitude statistics, e.g., high light curve amplitude targets are more favourably detected and reported than low amplitude targets from ground based observations. This alone can explain the difference between any of their distribution curves and our K2 one; the consideration of non-detections naturally increases the cumulative probability at low amplitudes. While our K2 sample cannot be considered to be non-biased, it is from a sample in which targets were selected merely by their celestial location (to be in actual K2 fields), and no other criteria (brightness, dynamic class, etc.) were applied. As biases in the ground-based data dominated sample are hard to be corrected for, we believe that our K2 cumulative amplitude curve is currently the best representation of the KBO light curve amplitude distribution. 

\citet{Showalter2021} compared their cumulative amplitude distributions to a group of models assuming single triaxial bodies or contact binaries with various configurations. As any specific curve could be fitted with a weighted combination of such curves from various shape and scattering models, we do not aim to try to find a `best model population' for our K2 curve. However, it is obvious from the \citet{Showalter2021} models that the current shape of the K2 curve (with a `bulge' at low amplitudes) is expected to be best fitted with `contact binary'-type curves. To demonstrate this we also calculated model cumulative amplitude curves of contact binaries assuming spherical shapes, size ratios from 0.5 to 1.0, Lambert or Lommel-Seeliger scattering, and either uniform aspect angle or isotropic pole orientation distributions for a set of simulated objects (pale gray and pink curves in Fig.~\ref{fig:cumulampl}). In this very simple approach the best match is provided by models with Lambert scattering and uniform aspect angle distribution, but our relatively poor statistics does not allow for a more profound analysis of the underlying shape distribution, especially if light curve of binaries and elongated objects have to be combined. 

Our K2 TNO cumulative amplitude curve is also different from the K2 Hilda and Jovian Trojan curves which have a notably lower number of objects at small amplitudes -- in this sense the Hilda and Jovian Trojan curves are more similar to the main belt cumulative amplitude distribution \citep{Masiero2009} where the light curve amplitude distribution is explained by elongated objects with a large fraction of asteroids with axis ratios b/a\,$\sim$\,0.8, and a smaller distinct group at b/a\,$\sim$\,0.3. A $\sim$20-25 {$\%$} binary fraction among Hildas and Jovian Trojans indicated by the K2 data \citep{Ryan2017,Szabo2017,Szabo2020,Kalup2021} would still mean that the majority of these light curve can be explained by elongated bodies. 


\subsection{Light curve period distribution \label{sect:perioddist}}

\begin{figure}[h!]
\begin{center}
\includegraphics[width=0.5\textwidth]{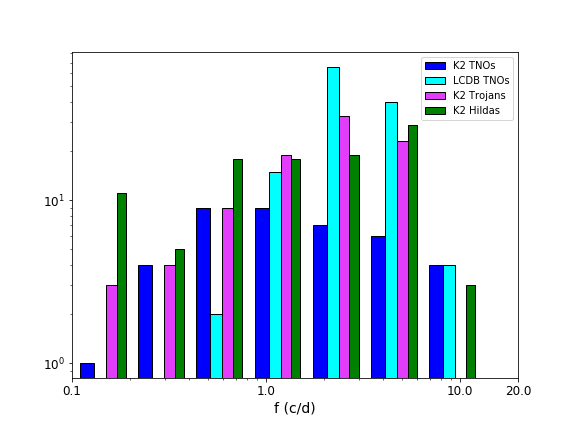}
\includegraphics[width=0.5\textwidth]{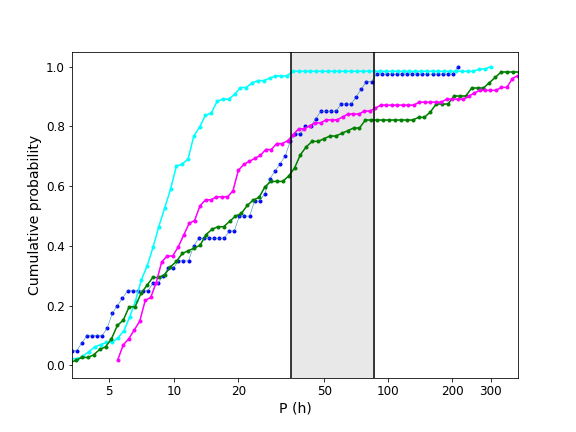}
\caption{Top panel: Frequency distribution of asteroids. The cyan, magenta, green and blue colours represent the TNOs in the LCDB and Jovian trojans, Hildas and TNOs from K2, respectively. 
Bottom panel: Cumulative distribution of the same data sets as in the top panel. The gray regime between the vertical black lines shows where TNOs overcome Hilda asteroids.}
\label{fig:histfreq}
\end{center}
\end{figure}

\begin{figure}[h!]
\begin{center}
\includegraphics[width=0.5\textwidth]{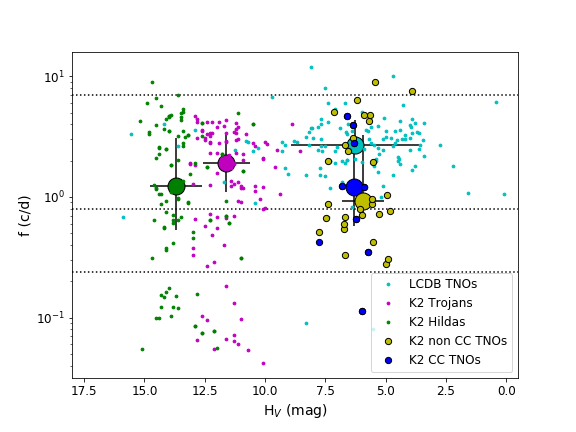}
\caption{Frequency as a function of absolute magnitude. Big circles with error bars mark the median values with standard deviations for the different samples. The actual values are listed in \autoref{table:fratio}. The boundary of fast, slow and very slow rotators are also added on the figure (black dotted lines). }
\label{fig:fh}
\end{center}
\end{figure}

The light curve period distribution obtained from K2 mission data was found to be substantially different from that of ground based measurements, with a notably higher fraction of slowly rotating objects found in K2 measurements, throughout all Solar system small body populations \citep{Szabo2017,Szabo2020,Kalup2021,Molnar2018}. The same difference was observed for main belt asteroids using about ten thousand light curve obtained from TESS data \citep{Pal2020}. Here we compare our period distribution (using our detected light curve frequencies/periods) with that obtained from the LCDB, and also with the K2 Hilda and Jovian Trojan distributions in Table~\ref{table:fratio} and Figs.~\ref{fig:histfreq} and \ref{fig:fh}. 

As demonstrated in Table~\ref{table:fratio} and Fig.~\ref{fig:histfreq} K2 TNOs show a similar fraction of $\sim$38\% of slow rotators \citep[f\,$\leq$\,0.8\,d$^{-1}$, see][]{PravecHarris2000} than that found among K2 Jovian Trojans and Hildas, and this fraction is much higher than that in the LCDB TNO data ($\sim$3\%). While the fraction of very slow rotators (f\,$\leq$\,0.24\,d$^{-1}$) is relatively high among Jovian Trojans and Hildas (13 and 18\%) these are essentially missing from the K2 TNO sample (2.5 \%). The median light curve frequency is $\sim$1.00\, c/d  ($\sim$24.0\,h) among the K2 TNOs while it is $\sim$2.7\, c/d ($\sim$8.9\,h) among LCDB TNOs. This $\sim$1\,d K2 TNO median light curve period is even longer that those of K2 Jovian Trojans and Hildas. Note that while the absolute magnitude (and therefore the size) range we explore is similar or at least somewhat overlapping for Jovian Trojans and Hildas (see Fig.~\ref{fig:fh}), it is very different for the K2, as well as the LCDB TNOs. It means that {we are looking} at objects about an order of magnitude \emph{larger} among TNOs, in the D\,$\geq$\,100\,km diameter range. Some possible consequences are discussed in Sect.~\ref{sect:conclusions}. 

\citet{TS19} compared the mean light curve periods of cold classical TNOs and TNOs in other dynamical classes and found that cold classicals rotate somewhat slower (mean rotation periods of $\overline{P}(CC)$\,=\,9.47$\pm$1.53\,h and $\overline{P}$(non-CC)\,=\,8.45$\pm$0.58\,h). These values are significantly different from our median rotation periods (20.00\,h and 26.37\,h), and in our sample cold classicals rotate slightly faster, however, the small difference between the two groups is {likely} due to statistical errors than real physical differences.  


\begin{table*}
\begin{center} \footnotesize
\begin{tabular}{c|rrrrrr}
\hline
\hline
& TNOs & K2 TNOs & K2 TNOs & K2 TNOs & K2    & K2     \\
& LCDB & full    & non-CC  & CC      & JTs  & Hildas \\              
\hline
\hline
N             & 129     & 40 & 30 & 10     & 101        & 112       \\
f$_{m}$ (c/d) & $2.71_{-0.34}^{+1.63}$    &  1.00$_{-0.33}^{+2.10}$   & $1.00_{-0.28}^{+2.30}$  &    $1.01_{-0.51}^{+1.53}$    & $1.90_{-0.79}^{+1.34}$      & $1.22_{-0.69}^{+1.88}$      \\
H$_{V}$ (mag) &   $6.26\pm2.66$  &  $5.90\pm$0.97   & $5.79\pm{1.12}$  &   $6.25\pm{0.55}$     &      $11.6\pm{0.99}$      &    $13.7\pm{1.07}$        \\
P$_{m}$ (h)   & 8.85   &  23.94   & 24.00  &  23.76   & 12.63      & 19.67     \\
\hline
N$_{f}$       & 3      &  2   & 2       &    0   & 0           & 2         \\
r$_{f}$ (\%)  & 2.3    &  5.0   &  6.66 &    0   & 0           & 1.8       \\
\hline
N$_{s}$       & 4      &  15   & 12  &   3    & 26          & 43        \\
r$_{s}$ (\%)  & 3.1    &  37.5   & 40.0  &   30.0    & 25.7        & 38.9      \\
\hline
N$_{vs}$      & 2     &  1  &  0 &  1     & 13          & 20        \\
r$_{vs}$ (\%) & 1.5   & 2.5 &  0 &  10.0   & 12.9        & 17.85 \\
\hline
\end{tabular}
\caption{Summary table containing some basic statistics, including the number of asteroids (N), median rotation rates (f$_{m}$) and the corresponding rotation period (P$_{m}$), and the fraction of targets in the fast, slow, and very slow rotator groups of the LCDB TNOs, K2 TNOs (this work), K2 Jovian trojans \citep{Szabo2017,Kalup2021} and K2 Hildas  \citep{Szabo2020}. Fast rotators (subscript 'f') are defined as f $\geq$ 7 d$^{-1}$, slow rotators ('s') as f $\leq$ 0.8 d$^{-1}$ and very slow rotators ('vs') as f $\leq$ 0.24 d$^{-1}$. \label{table:fratio}
 }
\end{center}
\end{table*}

\subsection{Light curve amplitude versus frequency \label{sect:fdm}}

\begin{figure*}
\begin{center}
\hbox{
\includegraphics[width=0.5\textwidth]{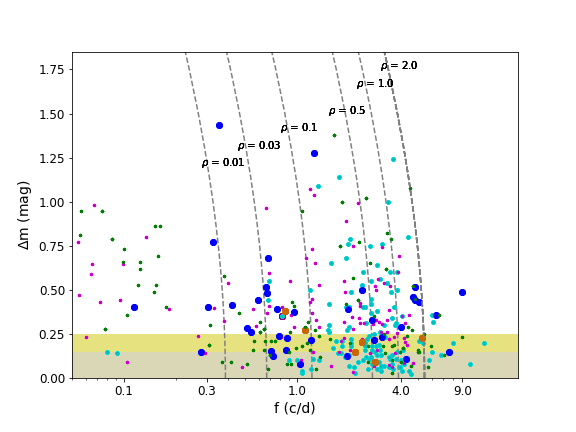}
\includegraphics[width=0.5\textwidth]{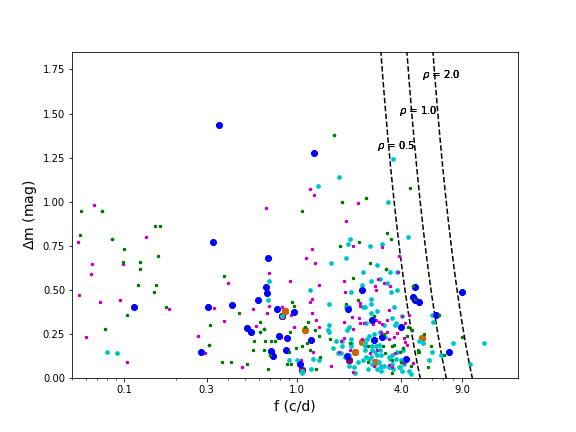}
}
\caption{Left panel: Light curve amplitude as a function of the light curve (rotational) frequency. 
Colour coding is the same as in the previous figures; brown filled circles show the Centaurs from \citet{Marton2020}. 
Each dashed curve represent a specific density (indicated by the numbers in units of $g\cdot cm^{-3}$) that corresponds to the rotation of a strengthless body with the shape of a Jacobi ellipsoid. 
Right panel: Here the same data are presented as on the left panel, but instead of the constant density Jacobi ellipsoid curves here we show the critical density curves calculated as $\Delta m = (\rho_c/[1\,g\,cm^{-3}]) (f/[7.27\,c/d])^{-2} - 1 $ \citep{PravecHarris2000}.}
\label{fig:freqamp}
\end{center}
\end{figure*}

The light curve amplitude versus rotational frequency plane (see Fig.~\ref{fig:freqamp}) is usually divided into three main zones \citep{Leone1984,Sheppard2004,T10,Benecchi2013}.  Light curve variations of objects with small amplitudes ($\Delta m$\,$\leq$\,0.25\,mag or 0.15\,mag) can either be caused by albedo and shape features or can as well be binaries (colored areas in Fig.~\ref{fig:freqamp}). If the rotational equlibrium of a strengthless body is considered and approximated by a Jacobi ellipsoid, constant density curves can be drawn (blue dash-dotted curves in Fig.~\ref{fig:freqamp}) \citet[see eqs. 1 \& 2 in][and references therein]{Lacerda2007}, assuming $\vartheta$\,=\,$\pi/2$ aspect angle, i.e. equator-on viewing geometry and maximum light curve amplitude. Objects to the right of a curve of a constant density are likely rotating single bodies, if their rotational speed is below the breakup limit \citet[see][]{PravecHarris2000}, and the left panel of Fig~\ref{fig:freqamp}. 
{ Rotations of the objects to the left of a constant density Jacoby ellipsoid curve are too slow to cause elongation and their corresponding rotational light curve.} For these objects the light curves are sometimes explained by binarity \citep[see][for a discussion]{Marton2020}. 

The typical density of Kuiper belt objects in the size range of our study -- 100--1000\,km, see Table~\ref{table:orbital} -- is 0.5--1.0\,\gcc\, \citep{Grundy2019} and density estimates from binary orbits of Kuiper belt objects range from  $\rho$\,=\,0.44$^{+0.44}_{-0.17}$\,g\,cm$^{-3}$ \citep[Typhon-Echinda,][]{Grundy2008} to 
$\rho$\,=\,1.37$^{+0.66}_{-0.32}$\,g\,cm$^{-3}$\citep[Ceto-Phorcys,][]{Grundy2007}. Smaller objects with sizes of a few kilometers may have similar or somewhat lower densities and high porosities. E.g. the density of Arrokoth was estimated to be 0.235\,\gcc\, \citep{Keane2022}, while the bulk density of 67P is 0.533\,\gcc\, \citep{Patzold2016}. 

Based on their location on the amplitude-frequency diagram the density {of} Hildas and Jovian Trojans were estimated to be $\lesssim$\,0.5\,\gcc\, using a $\sim$5\,h spin period limit \citep{Ryan2017,Szabo2017,Szabo2020,Kalup2021} for diameters D\,$\gtrsim$\,10\,km. The study by \citet{FOSSIL1} included smaller objects down to D\,$\sim$\,2\,km, and estimated a bulk density of 0.9\,\gcc\, for Jovian Trojans, and \citet{FOSSIL2} obtained a bulk density of 1.5\,\gcc\, for Hildas from the light curves of small objects of \lele{1}{D}{3}\,km. 


Majority of the TNOs both in our K2 sample and in the LCDB can be found at {frequency-amplitudes} characterised by low Jacobi ellipsoid and breakup limit densities, typically below $\rho$\,=\,0.5-1.0\,\gcc. There are a few objects in our K2 sample, however, which have relatively high spin frequencies (e.g. \jf, \ez\, and \yg, see Table~\ref{table:big} and Fig.~\ref{fig:big0}). These targets would require high, $\rho$\,$\geq$\,2.0\,\gcc\, density and/or considerable internal strength to maintain their rotational state with rotation periods $\sim$3\,h, or faster. For most of these targets a single peak light curve is associated with the primary frequency. Despite that our analysis found these light curve frequencies to be the most likely ones, a double peak light curve, and a corresponding double period is similarly likely which would also reduce the density/internal strength limits for these objects. However, even in this case the rotation periods are shorter than the fastest ones in the Jovian Trojan K2 sample. This suggests that moderate density objects ($\rho$\,=\,1.0-2.0\,\gcc) may exist in the transneptunian region at this few hundred kilometer size range, as previously obtained from the studies of binaries. 

As in the previous cases of Jovian Trojans, Hildas and Centaurs observed by K2 \citep{Szabo2017,Szabo2020,Marton2020,Kalup2021} a considerable fraction of our targets falls into the low frequency - high amplitude regime where their rotation is traditionally explained by binarity. Simply considering the $\rho$\,=\,0.5\,\gcc\, Jacobi ellipsoid curve and an amplitude limit of $\Delta$m\,=\,0.25\,mag 21\% of the K2 targets would fall into the 'binary' category which is very similar to the values obtained in the K2 samples of other populations \citep{Ryan2017,Szabo2017,Szabo2020,Marton2020,Kalup2021}. 


\section{Conclusions \label{sect:conclusions}}

Little is known about the true shape of transneptunian objects. Accurate shapes are available for those visited by a spacecraft, {and at the large end of the size range} Pluto and Charon were found to be very round \citep{Nimmo2017}. On the small end of the size range, however, Arrokoth was most probably formed by a merger of two bodies, and has a bilobate shape \citep[see e.g.][]{Keane2022}.  
In recent years combination of light curve and occultation measurements {have} revealed the shape and size of some Centaurs and transneptunian objects, including 2002\,GZ32 \citep{SS21}, 2003\,AZ$_{84}$ \citep{Dias2017}, Huya \citep{SS22}, Chariklo \citep{Morgado2021}, 2002\,VS$_2$ \citep[][]{VL22} and Haumea \citep{Ortiz2017} . While the very distorted shape of Haumea is certainly caused by its fast rotation, the other targets also show triaxial ellipsoid shapes as discussed in the respective papers. These objects are either small, below D\,$\approx$\,300\,km, or moderately deformed at larger sizes.

\begin{figure}[ht!]
\begin{center}
\includegraphics[width=0.5\textwidth]{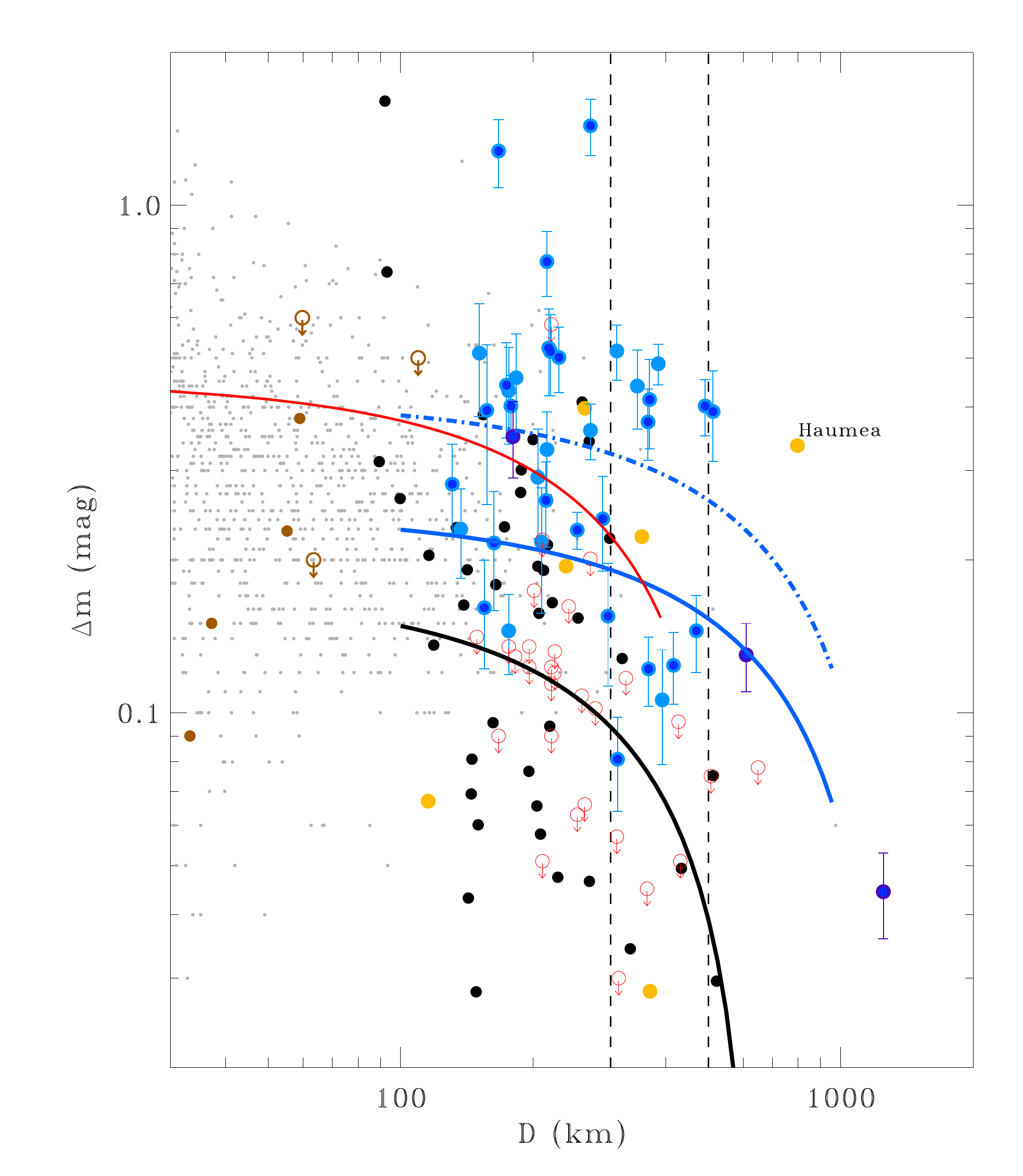}
\caption{Light curve amplitude versus the estimated size of the targets in our sample. Symbols/colors are the same as in Fig.~\ref{fig:amph}; blue symbols with dark blue 'cores' mark those targets for which the actual rotation period would indicate a density $\rho$\,$\leq$\,0.5\,\gcc\, assuming a Jacobi ellipsoid. The region between the vertical dashed lines mark the irregular-to-spherical transition size range in the main belt. Black symbols represent the 'deformed shape' light curve amplitudes of main belt asteroids studied by \citet{Vernazza2021}, and the solid black curve is the 'mean' light curve amplitude curve obtained from this sample (see the main text for a detailed explanation). The blue curves are the mean K2 TNO amplitudes considering the detections only (dash-dotted) and also considering non-detections (solid curve). Orange symbols correspond to the Centaurs and TNOs whose shape and size are obtained from the combination of occultation and light curve measurements; the dwarf planet Haumea is marked. Brown symbols represent the Centaurs from \citet{Marton2020}. The solid red curve is the mean amplitude curve from \citet{Showalter2021}. }
\label{fig:dmvsdiam}
\end{center}
\end{figure}

Our work samples the largest asteroids in the transneptunian region (in most cases D\,$\geq$\,200\,km, see Table~\ref{table:orbital}) which are, as discussed above, believed to be primordial {with} their rotational characteristics have likely not been considerably modified since their formation, assuming that they are single bodies. It is generally accepted that periodic changes in asteroid light curves are caused by asteroid rotation due to either deformed shape or albedo variegations on the surface of a spherical (or rotationally flattened) body {\citep[see e.g.][]{lac}}. In the last decades most works {have} tried to explain the light curves by deformed shape, and light curve inversion techniques {have been} very successful in this matter \citep[see e.g.][]{Durech2015}. 

The vast majority of the main belt asteroids studied this way have diameters D\,$\leq$\,100\,km and therefore their shapes are not expected to be close that of hydrostatic equlibrium, due to their insufficient mass and gravity. Recent works using high resolution imaging \citep[e.g. the VLT/SPHERE instrument][]{Hanus2020,Vernazza2020,Vernazza2021} {show} that asphericity drops sharply ($1-\varphi$\,$\leq$0.01) for objects D\,$\geq$\,400\,km in the main belt and the transitional size from `irregular' to `spherical' objects may be as low as D\,$\approx$\,300\,km. This transitional size is expected to be even lower for the icy objects in the transneptunian region due to their lower compressive strength \citep{PotatoRadius}. {Notably} TNOs in the D\,$\approx$\,500\,km size range have densities $\rho$\,$\leq$\,1\,\gcc{} and a considerable level of porosity { \citep[e.g.][]{BN19,Grundy2019}}.  
Due to their lower densities TNOs are more susceptible to be deformed due to rotation, but as discussed above, most of the K2 TNO targets have spin rates too low to be deformed by rotation with any plausible bulk density. 
{ We note that the targets discussed in this paper have diameters $D$\,$<$\,1000\,km with the exception of Gonggong. These TNOs have  low densities and high porosities, also suggesting that these smaller objects may have formed later, several million years after the time of CAI formation. Due to the insufficient heat from radiogenic decay they likely did not have enough heat to reach even partial differentiation, and may remain  undifferentiated for their whole existence \citep{BN19}}.

In this sense it is interesting to plot the light curve amplitudes of our targets versus their estimated size (Fig.~\ref{fig:dmvsdiam}). While there are only three objects with D\,$\geq$\,500\,km in our sample, there are a number of objects -- both with and without detected light curve periods -- that fall in the \lele{300}{D}{500}\,km transitional zone where asphericity -- hence light curve amplitude -- is expected to drop assuming a single rotating body and main belt composition. Main belt asteroids are already almost extinct in this size range, and so are Centaurs -- for these bodies irregular shapes are expected in most cases. The light curve amplitudes of large main belt asteroids presented in Fig.~\ref{fig:dmvsdiam} are not their actual light curve amplitudes{,} but they are derived from the \citet{Vernazza2021} triaxial ellipsoid shape models, calculated from the ratios of the semi-axes $b/a$ and assuming homogeneous albedo distribution on the surface. The mean main belt curve is obtained assuming random spin axis orientations. Similar mean curves are derived from the K2 TNO data (blue curves in Fig.~\ref{fig:dmvsdiam}), considering the detections only, and including the non-detections by considering the upper limits assigning random {and} amplitudes to each non-detection from a normal distribution with a standard deviation corresponding to the 1$\sigma$ upper limit around zero amplitude.

While the general trend is the same among K2 TNOs {as} among large main belt asteroids, the deformed shape light curve amplitude is notably, a factor of 2-3 higher among TNOs than in the main belt, in contrast to the expectation. This remains true even if the lower TNO densities are considered: the majority of the K2 TNO targets would require very low ($\rho$\,$\ll$\,0.5\,\gcc) densities to be deformed by rotation with their actual low spin rate. \citep{Showalter2021} found a similar trend (see the red curve in Fig.~\ref{fig:dmvsdiam}), however, the sample they consider in their paper is restricted to targets with \absmag\,$\gtrsim$\,5.5\,mag (D\,$\lesssim$\,400\,km) and cannot be readily extrapolated to larger sizes. 

The contradiction of large amplitudes at large sizes compared with the main belt could be resolved if TNOs had higher-than-expected compressive strength and become spherical for sizes larger than their main belt counterparts, and remain 'irregular' in the \lele{300}{D}{500}\,km range. However, their general low density and high porosity point against this scenario. A notable fraction of contact or semi-contact binary systems in which the members themselves are in hydrostatical equlibrium could produce a population of high-amplitude light curves in this size range \citep{Lacerda2006,Lacerda2014}. As some authors {have} pointed out, contact binaries may be very frequent, especially in the plutino population \citep{Thirouin2018,Thirouin2019}. However, a binary fraction higher than the currently deduced $\sim$20\% would be needed to explain the large number of large amplitudes among TNOs. The long term stability of such systems against their tidal evolution should also be investigated to answer the reliability of this assumption. 
Spherical (rotationally flattened spheroidal) bodies with large albedo variegations could also explain the observed large amplitudes. The large number of such objects would be a major difference compared to the main belt where most asteroids have relatively homogeneous albedo distributions on their surface. 


\section*{Acknowledgement}
\begin{acknowledgements}

The research leading to these results has received funding from the K-115709, K-138962, PD-116175, KKP-137523 and GINOP-2.3.2-15-2016-00003 grants of the National Research, Development and Innovation Office (NKFIH, Hungary); and from the LP2012-31 and LP2018-7/2021 Lend\"ulet grants of the Hungarian Academy of Sciences.
The research leading to these results have been supported by the  \'UNKP-19-2, \'UNKP-20-2 and \'UNKP-21-2 New National Excellence Programs of the Ministry of Innovation and Technology from the source of the National Research, Development and Innovation Fund. 
Funding for the \textit{Kepler} and K2 missions are provided by the NASA Science Mission Directorate. The data presented in this paper were obtained from the Mikulski Archive for Space Telescopes (MAST). STScI is operated by the Association of Universities for Research in Astronomy, Inc., under NASA contract NAS5-26555. Support for MAST for non-HST data is provided by the NASA Office of Space Science via grant NNX09AF08G and by other grants and contracts. This research has made use of data and services provided by the International Astronomical Union's Minor Planet Center. The authors thank the hospitality of the Veszpr\'em Regional Centre of the Hungarian Academy of Sciences (MTA~VEAB), where part of this project was carried out. { We are also thankful to our reviewers for their fair and balanced reports.} 
\end{acknowledgements}

\vspace{5mm}
\facilities{Kepler/K2 \citep{K2}}


\software{astropy \citep{astropy:2013,astropy:2018},  
          Period04 \citep{P04}, {fitsh \citep{Pal2012}}
          }


\newpage



\setlength{\bibsep}{0pt}

\bibliography{tno}
\bibliographystyle{aasjournal}




%


\appendix
\onecolumngrid

\section{Basic tables of K2 TNO observations}

Table~\ref{table:big} summarises the main circumstances of the K2 observations of our targets, including start and end dates, length, duty, heliocentric and observer range, and phase angle.

\startlongtable
\begin{deluxetable*}{lcccrrrrrr}
\scriptsize
\tabletypesize{\scriptsize}
\tablecaption{Transneptunian objects observed by K2, studied in this work, ordered by asteroid number and provisional designation. The columns are: (1) number, name and/or designation of the target; (2) K2 campaign; (3-4) start and end date in Julian Date; (5) the length of observations (day); (6) the number of frames considered in the analysis; (7) the duty cycle of the observations (ratio of useful cadences and all cadences over the time span of the observations); (8) $r_h$: range of heliocentric distance during the observations (au); (9) $\Delta$: observer -- target distance (au); (10) $\alpha$: phase angle (deg);
}
\tablehead{
\colhead{Name}  &  \colhead{Cam.}  &  \colhead{Start}  &  \colhead{End}  &  \colhead{Length}  &  \colhead{\#frame}  &  \colhead{Duty}  &  \colhead{$r$} & \colhead{$\Delta$} & \colhead{$\alpha$ } 
}
\colnumbers
\startdata
    (026375)         \de &  C10 &  7584.098 &  7651.713 & 67.6148 & 2549 & 0.770 &  38.426...38.490 &  37.903...39.017 & 1.331...1.563 \\  
    (035671)         \sn &  C08 &  7418.913 &  7448.930 & 30.0170 & 1424 & 0.969 &  37.479...37.483 &  37.060...37.572 & 1.351...1.495 \\ 
    (066652)    Borasisi &  C12 &  7738.372 &  7816.653 & 78.2811 & 1222 & 0.319 &  42.013...42.030 &  41.235...42.498 & 0.847...1.332 \\
    (080806)         \cm &  C14 &  7905.723 &  7984.903 & 79.1802 & 1469 & 0.379 &  42.795...42.808 &  42.014...43.209 & 0.888...1.401 \\
    (119878)         \cy &  C14 &  7942.708 &  7946.999 &  4.2910 &  161 & 0.766 &  38.779...38.782 &  38.508...38.577 & 1.489...1.514 \\
    (126154)         \yh &  C05 &  7139.606 &  7214.434 & 74.8278 & 2233 & 0.609 &  36.858...36.871 &  36.179...37.344 & 1.171...1.610 \\
    (127871)         \fc &  C10 &  7582.607 &  7630.196 & 47.5899 & 1452 & 0.623 &  33.763...33.774 &  33.172...33.926 & 1.474...1.780 \\
    (135182)         \qt &  C08 &  7415.092 &  7443.168 & 28.0758 & 1099 & 0.799 &  37.144...37.145 &  36.733...37.214 & 1.372...1.508 \\
    (138537)         \ok &  C12 &  7767.061 &  7798.038 & 30.9773 & 1195 & 0.788 &  40.087...40.088 &  39.646...40.174 & 1.256...1.395 \\
    (145480)         \tb &  C12 &  7767.551 &  7796.403 & 28.8522 &  473 & 0.335 &  46.197...46.198 &  45.789...46.284 & 1.108...1.211 \\
    (149348)         \vs &  C13 &  7848.918 &  7879.711 & 30.7934 &  951 & 0.631 &  42.706...42.715 &  42.316...42.837 & 1.224...1.346 \\
    (160147)         \kn &  C15 &  7989.440 &  8077.406 & 87.9667 & 2091 & 0.485 &  39.958...39.964 &  39.125...40.432 & 0.910...1.479 \\
    (182934)         \gj &  C15 &  8022.910 &  8039.461 & 16.5512 &  678 & 0.836 &  44.365...44.370 &  43.975...44.243 & 1.251...1.328 \\
    (307463)         \vu &  C13 &  7852.698 &  7870.802 & 18.1042 &  824 & 0.929 &  40.299...40.309 &  39.927...40.216 & 1.306...1.421 \\
    (308379)         \rs &  C08 &  7426.841 &  7457.206 & 30.3643 & 1386 & 0.932 &  43.256...43.270 &  42.852...43.382 & 1.180...1.298 \\
    \qwlong             &  C08 &  7415.072 &  7444.823 & 29.7513 & 1349 & 0.926 &  43.560...43.566 &  43.153...43.657 & 1.167...1.286 \\
    (312645)         \ep &  C17 &  8179.574 &  8246.412 & 66.8383 & 2438 & 0.745 &  33.103...33.109 &  32.319...33.369 & 1.093...1.699 \\
    (385266)         \qb &  C12 &  7760.931 &  7786.003 & 25.0720 &  843 & 0.686 &  39.975...39.980 &  39.540...39.972 & 1.269...1.400 \\
    (385437)         \gh &  C15 &  8011.120 &  8050.740 & 39.6208 & 1372 & 0.707 &  41.051...41.055 &  40.554...41.197 & 1.289...1.444 \\
    (408832)         \qj &  C12 &  7761.768 &  7783.203 & 21.4348 &  651 & 0.620 &  44.708...44.710 &  44.328...44.691 & 1.163...1.252 \\
    (420356) Praamzius &  C16 &  8095.469 &  8100.394 &  4.9245 &  175 & 0.726 &  43.026...43.026 &  43.516...43.585 & 1.100...1.153 \\
    (469420)         \xp &  C16 &  8095.469 &  8173.710 & 78.2403 & 2465 & 0.643 &  33.068...33.072 &  32.336...33.574 & 1.129...1.711 \\
    (469420)         \xp &  C18 &  8251.541 &  8300.541 & 48.9998 & 1667 & 0.695 &  33.077...33.080 &  32.293...32.995 & 1.105...1.789 \\
    (469421)         \xd &  C16 &  8114.738 &  8139.749 & 25.0107 &  356 & 0.291 &  38.794...38.802 &  38.479...38.914 & 1.381...1.461 \\
    (469505)         \fe &  C15 &  7989.440 &  8077.447 & 88.0075 & 2723 & 0.632 &  36.081...36.095 &  35.243...36.569 & 1.010...1.638 \\
    (469704)         \ez &  C17 &  8179.533 &  8246.596 & 67.0631 & 2299 & 0.700 &  34.143...34.157 &  33.377...34.458 & 1.109...1.647 \\
    (470523)         \cs &  C14 &  7936.251 &  7973.890 & 37.6387 & 1299 & 0.705 &  37.049...37.061 &  36.600...37.211 & 1.456...1.620 \\
    (471137)         \et &  C10 &  7583.301 &  7651.120 & 67.8191 & 2173 & 0.654 &  39.661...39.666 &  39.069...40.127 & 1.251...1.515 \\
    (471150)    2010FC49 &  C10 &  7582.586 &  7651.713 & 69.1269 & 2557 & 0.755 &  39.675...39.684 &  39.104...40.164 & 1.263...1.515 \\
    (471318)         \jf &  C15 &  7990.032 &  8070.827 & 80.7945 & 2675 & 0.676 &  42.199...42.227 &  41.401...42.648 & 0.930...1.403 \\
    (472235) \gefourteen &  C17 &  8202.930 &  8213.841 & 10.9115 &  413 & 0.773 &  33.709...33.714 &  33.444...33.636 & 1.625...1.668 \\
    (508869)         \vt &  C13 &  7853.720 &  7883.144 & 29.4244 &  447 & 0.310 &  43.071...43.073 &  42.712...43.207 & 1.234...1.337 \\
    (523658)         \dw &  C17 &  8179.533 &  8246.535 & 67.0018 & 2193 & 0.668 &  39.725...39.726 &  38.928...39.980 & 0.895...1.415 \\
    (523687)         \df &  C17 &  8179.533 &  8219.665 & 40.1316 & 1579 & 0.803 &  42.937...42.942 &  42.555...43.237 & 1.224...1.310 \\
    (523692) \ezfourteen &  C15 &  7989.440 &  8077.059 & 87.6193 & 3044 & 0.709 &  55.777...55.819 &  54.970...56.226 & 0.634...1.059 \\
    (523698) \gdfourteen &  C17 &  8179.533 &  8246.474 & 66.9405 & 2696 & 0.822 &  34.763...34.776 &  33.992...35.033 & 1.052...1.618 \\
    (523706)         \hf &  C15 &  8028.182 &  8050.781 & 22.5996 &  769 & 0.695 &  37.327...37.339 &  36.963...37.316 & 1.494...1.581 \\
    (523769)         \ws &  C16 &  8095.469 &  8173.710 & 78.2403 & 2215 & 0.578 &  40.646...40.711 &  40.004...41.189 & 0.962...1.390 \\
    (523769)         \ws &  C18 &  8257.222 &  8295.269 & 38.0474 & 1361 & 0.730 &  40.781...40.812 &  40.031...40.598 & 0.957...1.421 \\
    (525462)         \eo &  C17 &  8179.840 &  8246.596 & 66.7566 & 1372 & 0.420 &  43.055...43.062 &  42.301...43.378 & 0.896...1.307 \\
    (533207) \djfourteen &  C15 &  8017.066 &  8037.479 & 20.4132 &  915 & 0.915 &  37.366...37.367 &  36.901...37.218 & 1.440...1.575 \\
    (533562)         \jq &  C15 &  8017.597 &  8048.043 & 30.4461 &  700 & 0.469 &  31.673...31.683 &  31.177...31.672 & 1.673...1.865 \\
    (533676)         \ls &  C15 &  7989.440 &  8007.380 & 17.9407 &  728 & 0.828 &  42.137...42.140 &  41.327...41.544 & 0.907...1.177 \\
    (535018)     \wanine &  C14 &  7916.553 &  7923.521 &  6.9679 &  298 & 0.873 &  42.744...42.745 &  42.033...42.117 & 1.002...1.104 \\
    (535023)         \wo &  C14 &  7905.723 &  7984.535 & 78.8124 & 2159 & 0.559 &  39.933...39.934 &  39.100...40.248 & 0.857...1.503 \\
    (535028)      \waten &  C14 &  7905.723 &  7984.985 & 79.2619 & 1986 & 0.512 &  45.224...45.232 &  44.382...45.540 & 0.739...1.327 \\
    (535030)         \wj &  C14 &  7929.855 &  7970.314 & 40.4585 & 1511 & 0.762 &  34.845...34.848 &  34.337...34.981 & 1.494...1.722 \\
    (535228)         \ye &  C18 &  8267.970 &  8297.599 & 29.6287 & 1169 & 0.806 &  47.853...47.880 &  47.346...47.842 & 1.058...1.236 \\
    (535231)         \yj &  C16 &  8111.428 &  8134.273 & 22.8448 &  449 & 0.401 &  35.254...35.265 &  35.057...35.432 & 1.569...1.608 \\
         \am             &  C16 &  8095.490 &  8173.710 & 78.2198 & 2426 & 0.633 &  48.222...48.249 &  47.529...48.813 & 0.829...1.171 \\
 \apthirteen             &  C16 &  8121.665 &  8163.677 & 42.0115 & 1272 & 0.618 &  47.635...47.672 &  47.063...47.709 & 0.929...1.188 \\
         \at             &  C16 &  8095.551 &  8173.812 & 78.2607 & 2177 & 0.568 &  62.520...62.591 &  61.813...62.971 & 0.549...0.906 \\
         \at             &  C18 &  8251.541 &  8289.486 & 37.9452 & 1389 & 0.747 &  62.662...62.696 &  61.920...62.496 & 0.623...0.923 \\
\bbfifteen               &  C18 &  8261.922 &  8301.277 & 39.3551 & 1454 & 0.754 &  42.844...42.881 &  42.312...42.976 & 1.158...1.380 \\
         \bc             &  C16 &  8114.902 &  8134.395 & 19.4937 &  734 & 0.769 &  36.020...36.035 &  35.774...36.121 & 1.523...1.575 \\
         \bc             &  C18 &  8262.269 &  8288.056 & 25.7872 &  895 & 0.709 &  35.895...35.916 &  35.344...35.713 & 1.343...1.619 \\
         \bz             &  C15 &  7989.685 &  8077.427 & 87.7419 & 1204 & 0.280 &  48.805...48.843 &  48.091...49.385 & 0.861...1.215 \\
         \da             &  C18 &  8265.048 &  8299.519 & 34.4715 &  950 & 0.563 &  36.545...36.560 &  35.998...36.518 & 1.332...1.622 \\
         \fa             &  C17 &  8179.533 &  8246.596 & 67.0631 & 2061 & 0.627 &  44.919...44.926 &  44.162...45.239 & 0.854...1.253 \\
 \gjfourteen             &  C15 &  8014.144 &  8045.182 & 31.0387 &  720 & 0.474 &  29.815...29.829 &  29.332...29.812 & 1.779...1.986 \\
         \hz             &  C15 &  7989.440 &  8077.447 & 88.0075 & 2500 & 0.580 &  43.692...43.698 &  42.857...44.180 & 0.837...1.353 \\
         \jv             &  C15 &  7989.440 &  8077.427 & 87.9871 & 2160 & 0.501 &  41.570...41.576 &  40.776...42.095 & 0.940...1.424 \\
         \qa             &  C12 &  7757.968 &  7786.105 & 28.1371 &  708 & 0.514 &  37.031...37.035 &  36.591...37.063 & 1.365...1.512 \\
         \qx             &  C12 &  7760.134 &  7786.105 & 25.9711 & 1029 & 0.809 &  43.238...43.240 &  42.803...43.243 & 1.171...1.295 \\
         \tf             &  C13 &  7842.951 &  7871.538 & 28.5866 &  809 & 0.578 &  40.791...40.805 &  40.382...40.869 & 1.266...1.405 \\
         \tv             &  C13 &  7856.682 &  7872.294 & 15.6113 &  613 & 0.802 &  35.017...35.021 &  34.853...35.113 & 1.606...1.636 \\
         \xr             &  C13 &  7856.315 &  7885.473 & 29.1588 &  796 & 0.557 &  57.336...57.343 &  57.025...57.503 & 0.942...1.005 \\
         \xu             &  C16 &  8095.469 &  8173.812 & 78.3424 & 1778 & 0.463 &  41.552...41.566 &  40.756...41.995 & 0.794...1.365 \\
         \yg             &  C13 &  7847.508 &  7856.192 &  8.6843 &  412 & 0.969 &  34.810...34.815 &  34.390...34.529 & 1.482...1.571 \\
         \ys             &  C16 &  8097.881 &  8169.092 & 71.2111 &  542 & 0.155 &  43.490...43.495 &  42.761...43.900 & 0.869...1.304 \\

\enddata
\label{table:big}
\end{deluxetable*}
\clearpage
\onecolumngrid

\smallskip

\noindent Table~\ref{table:orbital} lists the main orbital elements obtained from the JPL Horizons service \citep{JPLHorizons}, as well as the dynamical group or mean motion resonance of the target, the R-band (Kepler) absolute magnitudes obtained in this work (see Sect.~\ref{sect:absmag}), and the estimated diameters of our targets. These latter are radiometric size estimates, typically obtained from Herschel and Spitzer thermal infrared emission measurements, whenever they are available \citep{Lellouch2013,Vilenius2014,FT20}. When no radiometric size estimate was found we used the \absmag\, absolute magnitude of the object (also listed in Table~\ref{table:orbital}), assuming a geometric albedo of \geomalb\,=\,0.12 for cold classicals which are predominantly red objects with relatively high albedos, while we used \geomalb\,=\,0.08 for all other dynamical groups \citep{Lacerda2014,FT20}. The diameter is calculated as: D(km)\,=\,1329\,$p_V^{-1/2}$\,10$^{-0.2H_V}$. 

\startlongtable
\begin{deluxetable*}{lccccc|ccc|c}
\tabletypesize{\scriptsize}
\tablecaption{Summary table of the main orbital characteristics and dynamical classification of our targets. 
The columns are: (1) name and/or provisional designation; (2) V-band absolute magnitude (mag); (3) semi-major axis (au);
(4) eccentricity; (5) inclination (deg); (6) dynamical class or mean-motion resonance with Neptune. 
Cold and hot classicals are separated by the inclination i\,=\,4\fdg5 \citep[see e.g.][]{Gladman2008}; 
(7) phase-angle-uncorrected mean absolute brightness (Kepler\,$\approx$\,R-band, mag); (8) mean phase angle (deg);
(9) phase-angle-corrected mean absolute brightness (Kepler\,$\approx$\,R-band, mag). 
(10) estimated diameter (km), superscript letters mark targets with estimates based on radiometric data: 
a -- \citet{Lellouch2013}, 
b -- \citet{Vilenius2014} 
c -- \citet{FT20}.
Diameters of unmarked targets are estimated using their \absmag\, absolute magnitudes (see the text for details). 
} 
\tablehead{\colhead{Name}  & \colhead{H$_{V}$}  & \colhead{a}  &  \colhead{e}  &  \colhead{i}  
&  \colhead{Dynamical group} & \colhead{$\langle m_{11}^R(\alpha) \rangle$} & \colhead{$\langle \alpha \rangle$} & 
\colhead{$m_{110}^R(\alpha)$} & 
\colhead{D}}
\label{table:orbital}
\colnumbers
\startdata
\sn & 5.68 & 37.8239563 & 0.0399826 & 4.61245 & hot classical  &    5.24   &    1.45  &   5.23$\pm$0.02 & 393$^b$\\ 
\de & 4.94 & 55.5040117 & 0.4168371 & 7.62938 & 2:5 resonance  &    4.45   &    1.49  &   4.20$\pm$0.06 & 311$^a$\\  
\cm & 6.68 & 42.1255750 & 0.0640919 & 3.75516 & cold classical &    6.75   &    1.18  &   6.54$\pm$0.16 & 176 \\
\ok & 6.0 & 46.5813354 & 0.1403453 & 4.88465 & hot classical   &    6.56   &    1.35  &   6.32$\pm$0.18 & 296 \\
\hz & 6.6 & 43.0180466 & 0.0323816 & 2.93273 & cold classical  &    6.55   &    1.20  &   6.35$\pm$0.16 & 183\\
\kn & 6.6 & 44.0637410 & 0.0956953 & 2.64496 & cold classical  &    7.08   &    1.30  &   6.85$\pm$0.17 & 182\\
\qb & 6.8 & 42.7192221 & 0.0995969 & 1.79485 & cold classical  &    6.62   &    1.35  &   6.38$\pm$0.18 & 167\\
\qj & 6.2 & 44.1753883 & 0.0365290 & 2.15477 & cold classical  &    5.82   &    1.22  &   5.61$\pm$0.16 & 220\\
\qt & 6.2 & 37.0258514 & 0.0220311 & 1.84662 & inner classical &    7.62   &    1.47  &   7.37$\pm$0.20 & 270\\
\qx & 6.3 & 44.2115741 & 0.0270365 & 0.90745 & cold classical  &    6.66   &    1.25  &   6.45$\pm$0.17 & 210\\
\xd & 5.8 & 39.2882985 & 0.1118632 & 18.14109 & plutino        &    5.99    &   1.44  &   5.74$\pm$0.19 & 325\\
\xp & 7.77 & 42.0465415 & 0.2142201 & 2.61611 & 3:5 resonance  &    7.51    &   1.52  &   7.25$\pm$0.20 & 131\\
\xu & 6.68 & 43.2829795 & 0.0780917 & 6.52230 & hot classical  &    6.38    &   1.18  &   6.18$\pm$0.16 & 217\\
\yh & 5.58 & 42.1828001 & 0.1369339 & 11.09996 & 3:5 resonance  &   5.75    &   1.50  &   5.49$\pm$0.20 & 252$^c$ \\
\cy & 6.16 & 53.7570643 & 0.3432009 & 15.76009 & 5:12 resonance &   7.27    &   1.50  &   7.01$\pm$0.22 & $<$220$^c$\\
\gj & 5.4 & 44.5446573 & 0.1073526 & 11.57225 & hot classical   &   6.11    &   1.30  &   5.89$\pm$0.17 & 224$^a$\\
\vs & 6.3 & 44.8124374 & 0.1201659 & 3.00131 & cold classical   &   6.09    &   1.31  &   5.86$\pm$0.17 & 209\\
\vt & 5.7 & 42.2145036 & 0.0356054 & 1.16426 & cold classical   &   5.43    &   1.30  &   5.21$\pm$0.17 & 277\\
\vu & 5.9 & 38.9541796 & 0.2103117 & 1.37856 & plutino          &   5.72    &   1.38  &   5.48$\pm$0.18 & 310\\
\fc & 7.41 & 35.0390366 & 0.0839447 & 2.37160 & 4:5 resonance   &   6.92    &   1.71  &   6.62$\pm$0.23 & 155\\
\fe & 6.37 & 48.4332947 & 0.2592514 & 3.38301 & 1:2 resonance   &   6.53    &   1.43  &   6.18$\pm$0.46 & 137$^c$\\
\gh & 6.0 & 44.4898956 & 0.0844735 & 1.10267 & cold classical   &   6.16    &   1.40  &   6.20$\pm$0.64 & 178$^b$\\
\qa & 6.9 & 38.2803717 & 0.0583336 & 3.42715 & inner classical  &   6.80    &   1.48  &   6.54$\pm$0.20 & 196\\
\qw  & 5.0 & 43.6941429 & 0.0734856 & 10.36787 & hot classical  &   4.65    &   1.26  &   4.43$\pm$0.17 & 470\\
\ys & 6.8 & 43.4354691 & 0.0266317 & 3.73349 & cold classical   &   6.19    &   1.05  &   6.01$\pm$0.14 & 167\\
\tf & 6.3 & 79.8663748 & 0.5072393 & 23.23663 & scattered disk  &   6.02    &   1.37  &   5.78$\pm$0.18 & 258\\
\tv & 6.9 & 47.3800039 & 0.2722973 & 9.78559 & 1:2 resonance    &   7.21    &   1.63  &   6.93$\pm$0.22 & 196\\
\xr & 4.3 & 57.2561486 & 0.1073786 & 46.79343 & scattered disk  &   3.96    &   0.98  &   3.79$\pm$0.13 & 649\\
\eo & 6.21 & 45.8983194 & 0.0739658 & 3.41227 & cold classical  &   6.16    &   1.23  &   5.95$\pm$0.16 & 219\\
\ez & 7.13 & 39.6766516 & 0.1557282 & 1.77550 & plutino         &   7.21    &   1.47  &   6.95$\pm$0.19 & 176\\
\rs & 5.18 & 47.7306105 & 0.2003342 & 10.02598 & 1:2 resonance  &   4.76    &   1.27  &   4.54$\pm$0.17 & 432\\
\tb & 4.55 & 75.9529877 & 0.3917797 & 26.49004 & scattered disk &   4.63    &   1.17  &   4.63$\pm$0.13 & 507$^a$\\
\cs & 6.27 & 42.2933803 & 0.1579361 & 15.94423 & 3:5 resonance  &   6.33    &   1.58  &   6.06$\pm$0.21 & 262\\
\yg & 6.2 & 55.2226347 & 0.4040541 & 5.16230 & 2:5 resonance    &   5.91    &   1.53  &   5.64$\pm$0.20 & 270\\
\ep & 5.54 & 47.8794282 & 0.3094197  & 18.86373 & 1:2 resonance &   5.55    &   1.55  &   5.29$\pm$0.20 & 366\\
\et & 5.26 & 62.5104891 & 0.3659258 & 30.58345 & scattered disk &   4.71    &   1.44  &   4.46$\pm$0.19 & 417\\
\fcten  & 5.88 & 39.0790779 & 0.0535418 & 39.73235 & plutino    &   4.49    &   1.45  &   4.24$\pm$0.19 & 313\\
\jf & 5.43 & 41.6753240 & 0.1307786 & 27.64613 & hot classical         &  5.50  &  1.26 &  5.28$\pm$0.17 & 385\\
\dw & 6.06 & 41.2840227 & 0.0379671 & 18.99296 & hot classical         &  6.12  &  1.28 &  5.90$\pm$0.17 & 288\\
\apthirteen & 5.67 & 57.5543033 & 0.3668924 & 3.33318 & 3:8 resonance  &  5.41  &  1.10 &  5.22$\pm$0.15 & 345\\
\at & 4.9 & 61.6173083 & 0.4217387 & 28.12567 & scattered disk         &. 4.82. &  0.81 &  4.68$\pm$0.11 & 492\\
\fa & 6.2 & 44.4091197 & 0.0426125 & 1.53768 & cold classical          &  6.22  &  1.14 &  6.02$\pm$0.15 & 220\\
\jv & 6.35 & 42.9996513 & 0.0424042 & 3.21441 & cold classical         &  6.23  &  1.29 &  6.01$\pm$0.17 & 205\\
\am & 5.2 & 47.0230084 & 0.1423797 & 7.18355 & hot classical           &  5.14  &  1.06 &  4.94$\pm$0.14 & 428\\
\df & 5.56 & 42.8307595 & 0.0456454 & 23.68086 & hot classical         &  5.58  &  1.28 &  5.36$\pm$0.17 & 363\\
\djfourteen & 6.61 & 38.1807753 & 0.0232355 & 6.93181 & 4:13 resonance &  6.76  &  1.52 &  6.50$\pm$0.20 & 224\\
\ezfourteen & 3.92 & 52.5242987 & 0.2263985 & 10.25719 & scattered disk&  4.09  &  0.94 &  3.93$\pm$0.12 & 176\\ 
\gdfourteen & 6.71 & 90.1672371 & 0.6157590 & 4.74239 & scattered disk &  6.78  &  1.47 &  6.52$\pm$0.19 & 214\\
\gefourteen & 6.56 & 56.0228149 & 0.4158196 & 0.80808 & 2:5 resonance  &  6.73  &  1.65 &  6.45$\pm$0.22 & 229\\
\gjfourteen & 7.5 & 39.8224386 & 0.2801272 & 17.22611 & plutino        &  7.55  &  1.90 &  7.22$\pm$0.25 & 149\\
\hf & 5.9 & 61.7601933 & 0.4255242 & 9.71007 & scattered disk          &  6.05  &  1.55 &  5.78$\pm$0.20 & 310\\
\jq & 7.38 & 39.7549394 & 0.2233159 & 7.97258 & plutino                &  7.32  &  1.78 &  7.01$\pm$0.24 & 157\\
\ls & 6.2 & 43.6029549 & 0.0681892 & 3.82390 & cold classical          &  6.17  &  1.04 &  5.99$\pm$0.14 & 220\\
\wanine & 6.0 & 44.0261031 & 0.0531885 & 3.00216 & cold classical      &  6.13  &  1.06 &  5.95$\pm$0.14 & 241\\
\waten & 5.9 & 45.4170544 & 0.0367855  & 2.15623 & cold classical      &  5.89  &  1.08 &  5.71$\pm$0.14 & 252\\
\wj & 6.71 & 68.4462110 & 0.4912992 & 24.36183 & scattered disk        &  6.69  &  1.67 &  6.41$\pm$0.22 & 174\\
\wo & 6.3 & 43.9992014  & 0.0923767  & 3.75719 & cold classical        &  6.19  &  1.28 &  5.96$\pm$0.17 & 210\\
\ws & 5.55 & 54.9409609 & 0.3708612 & 8.90970 & 2:5 resonance          &  5.65  &  1.25 &  5.44$\pm$0.16 & 365\\
\ye & 5.53 & 59.1067258 & 0.3757100 & 26.91854 & scattered disk        &  5.42  &  1.17 &  5.22$\pm$0.16 & 368\\
\yj & 6.84 & 39.3040558 & 0.1951859 & 7.28212 & plutino                &  7.13  &  1.60 &  6.86$\pm$0.21 & 201\\
\bbfifteen & 6.7 & 63.1405090  & 0.4398171 & 27.62710 & scattered disk &  6.59  &  1.32 &  6.37$\pm$0.17 & 215\\
\bc & 6.7 & 55.2299967  & 0.4166677 & 1.72294 & 2:5 resonance          &  6.94  &  1.52 &  6.68$\pm$0.20 & 215\\
\bz & 4.81 & 47.2750427 & 0.1850024 & 11.35989 & other TNO             &  5.00  &  1.03 &  4.82$\pm$0.14 & 513\\
\da & 7.47  & 72.9126980 & 0.5110819 & 27.66698 & scattered disk       &  7.32  &  1.51 &  7.06$\pm$0.20 & 151\\
Borasisi & 5.9 & 43.8343021 & 0.0906165 & 0.56342 & cold classical     &  5.90  &  1.08 &  5.72$\pm$0.14 & 163$^b$\\
Praamzius & 5.75 & 42.5834453 & 0.0104114 & 1.09950 & cold classical   &  5.93  &  1.24 &  5.72$\pm$0.17 & 270\\ \hline
\enddata
\end{deluxetable*}
\normalsize
\onecolumngrid
\clearpage

\section{Individual objects with detected light curve periods \label{sect:detect}}


\onecolumngrid
\begin{figure*}[ht!]
\hbox{
\includegraphics[width=0.33\textwidth]{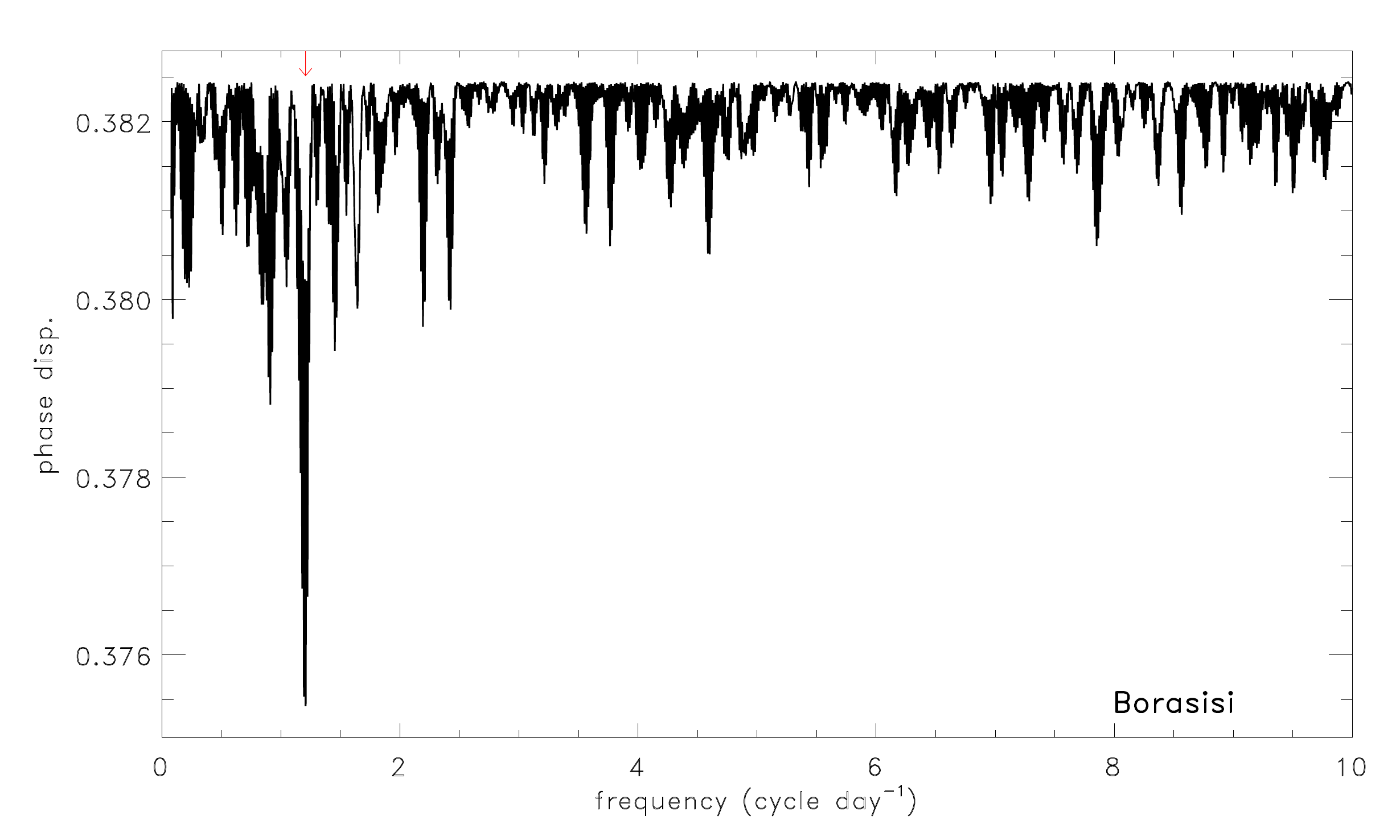}
\includegraphics[width=0.33\textwidth]{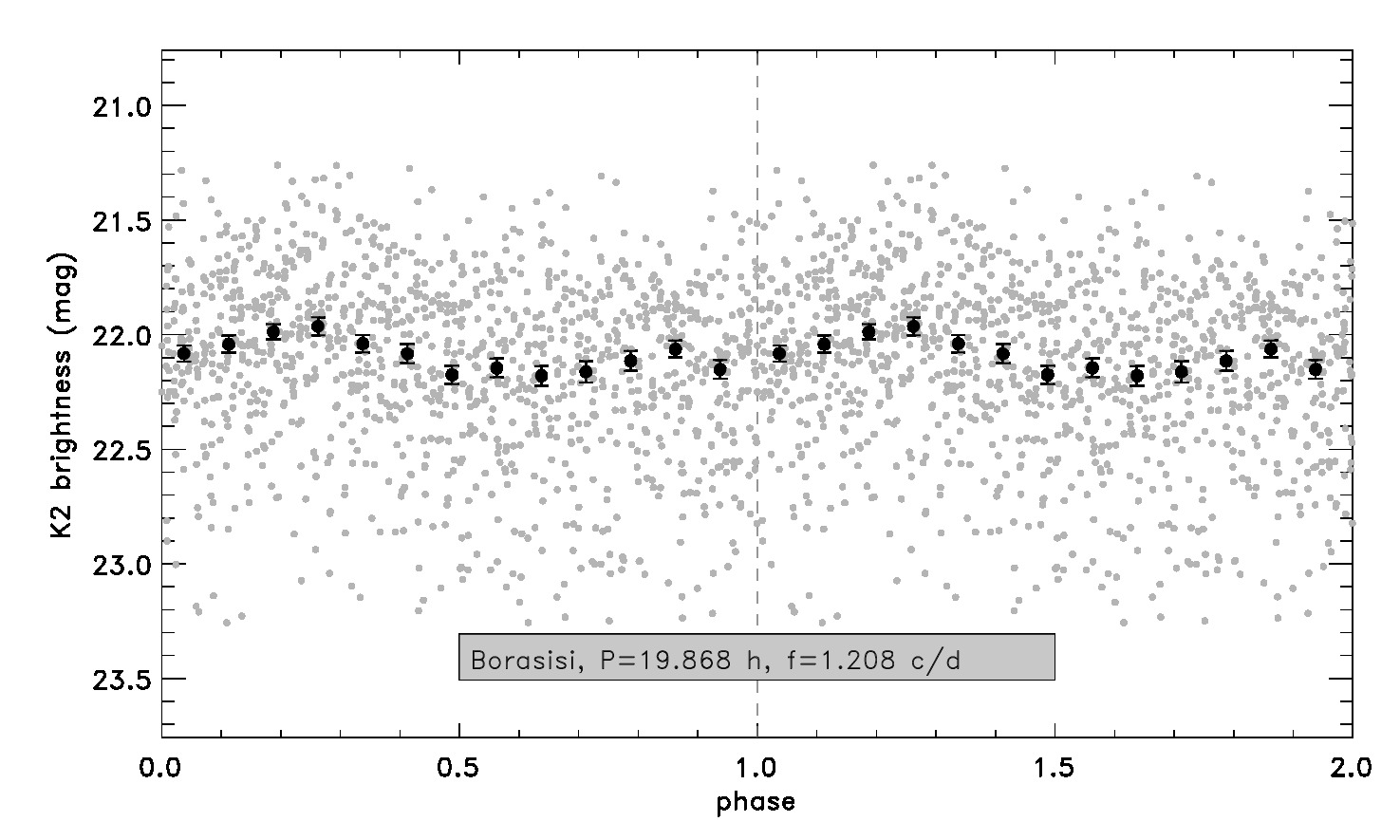}
}
\hbox{
\includegraphics[width=0.33\textwidth]{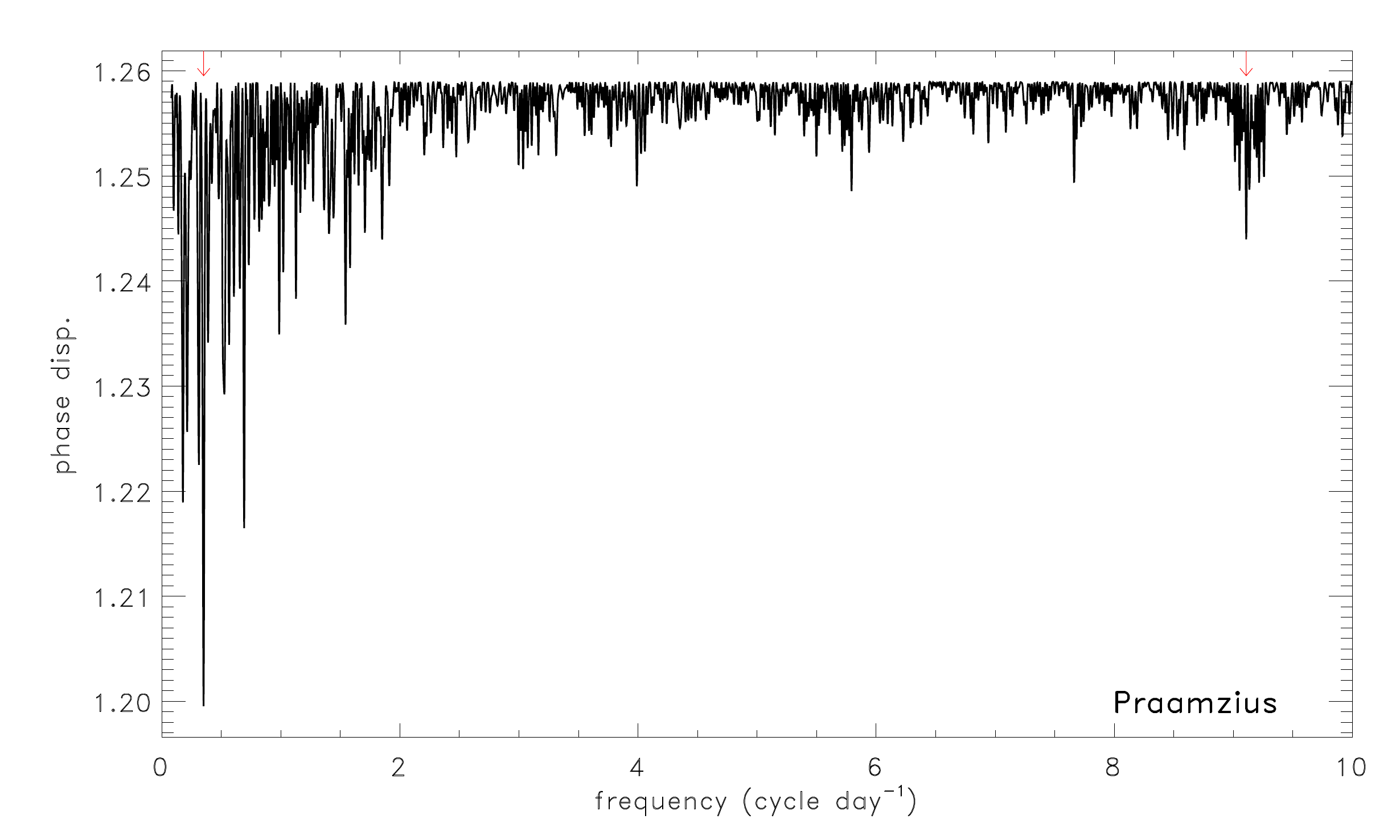}
\includegraphics[width=0.33\textwidth]{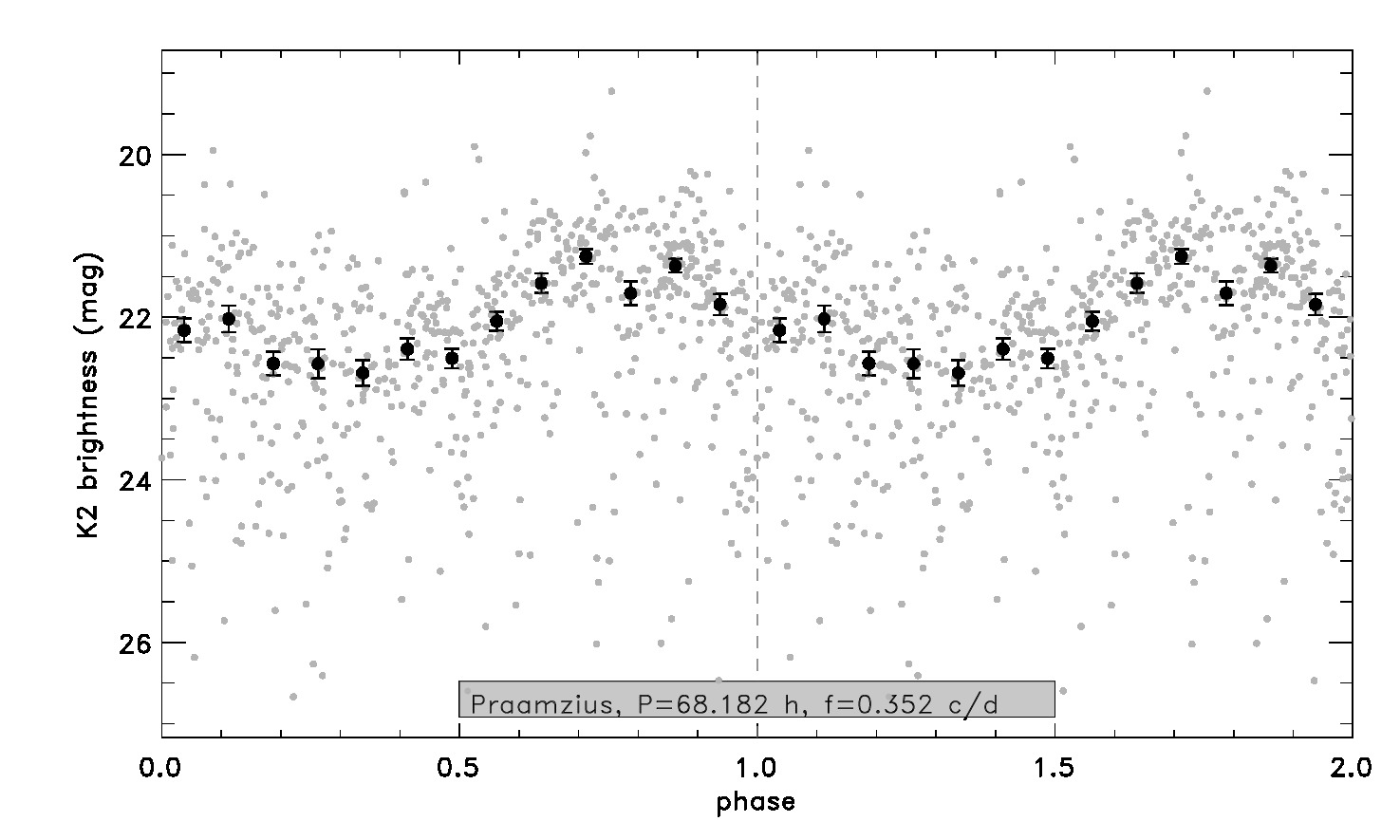}
\includegraphics[width=0.33\textwidth]{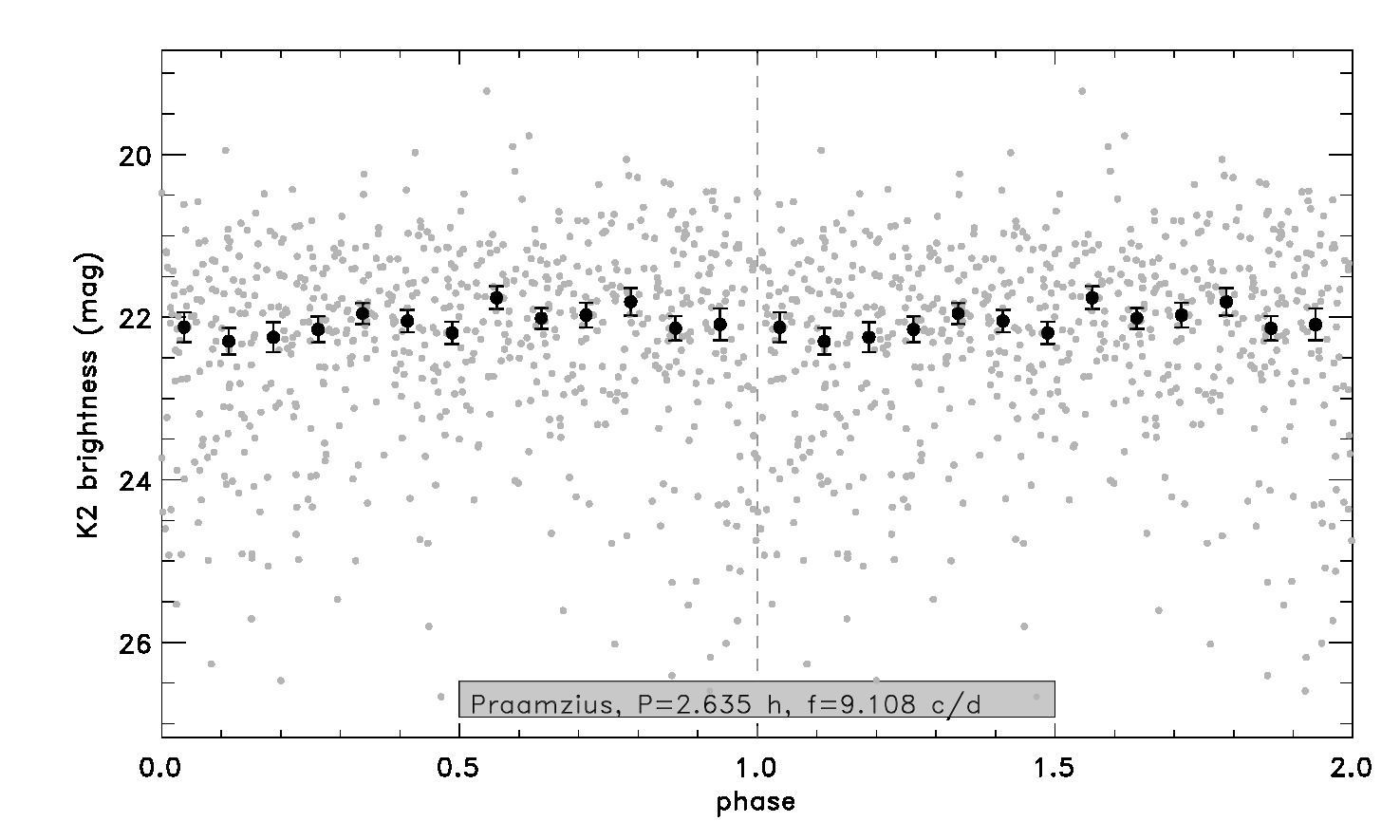}
}
\hbox{
\includegraphics[width=0.33\textwidth]{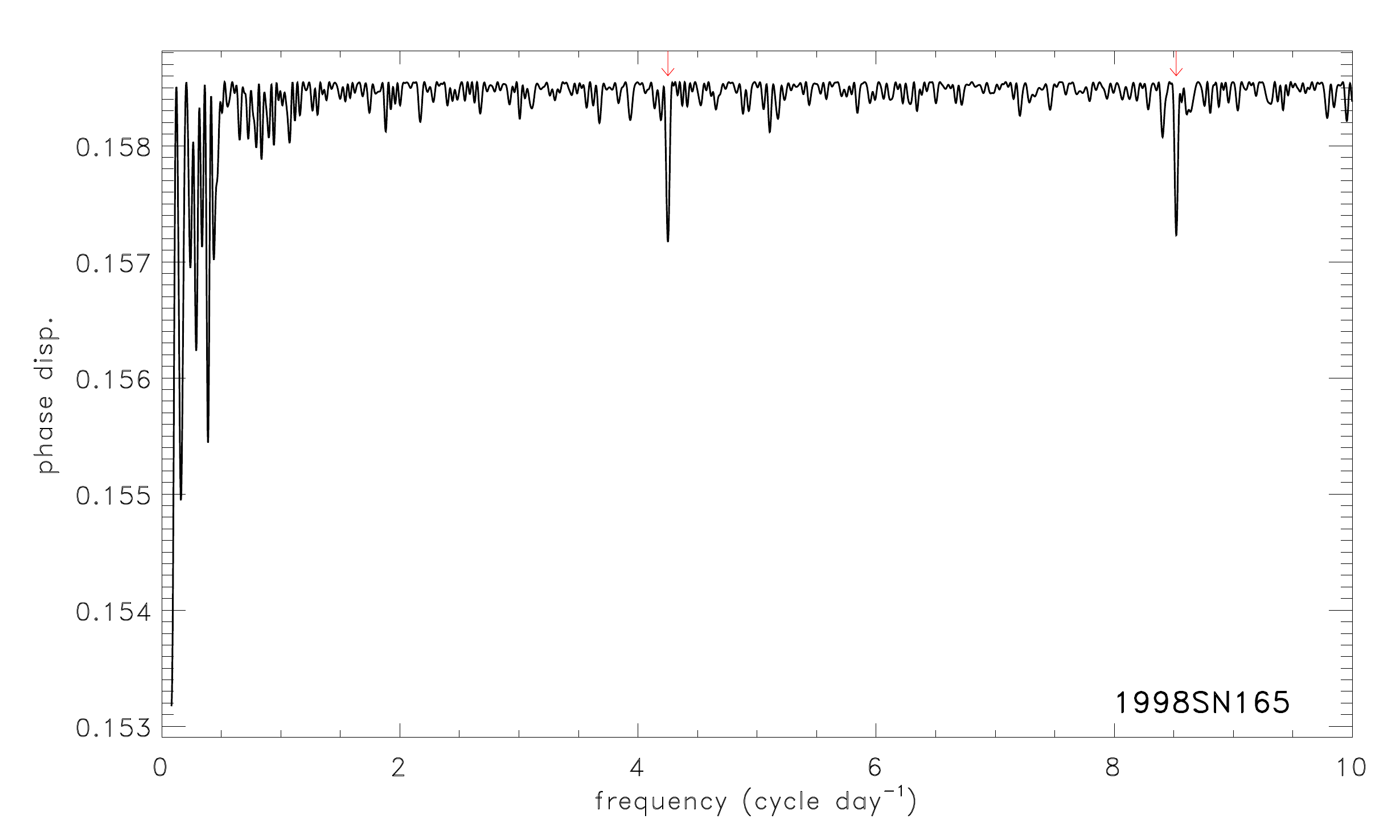}
\includegraphics[width=0.33\textwidth]{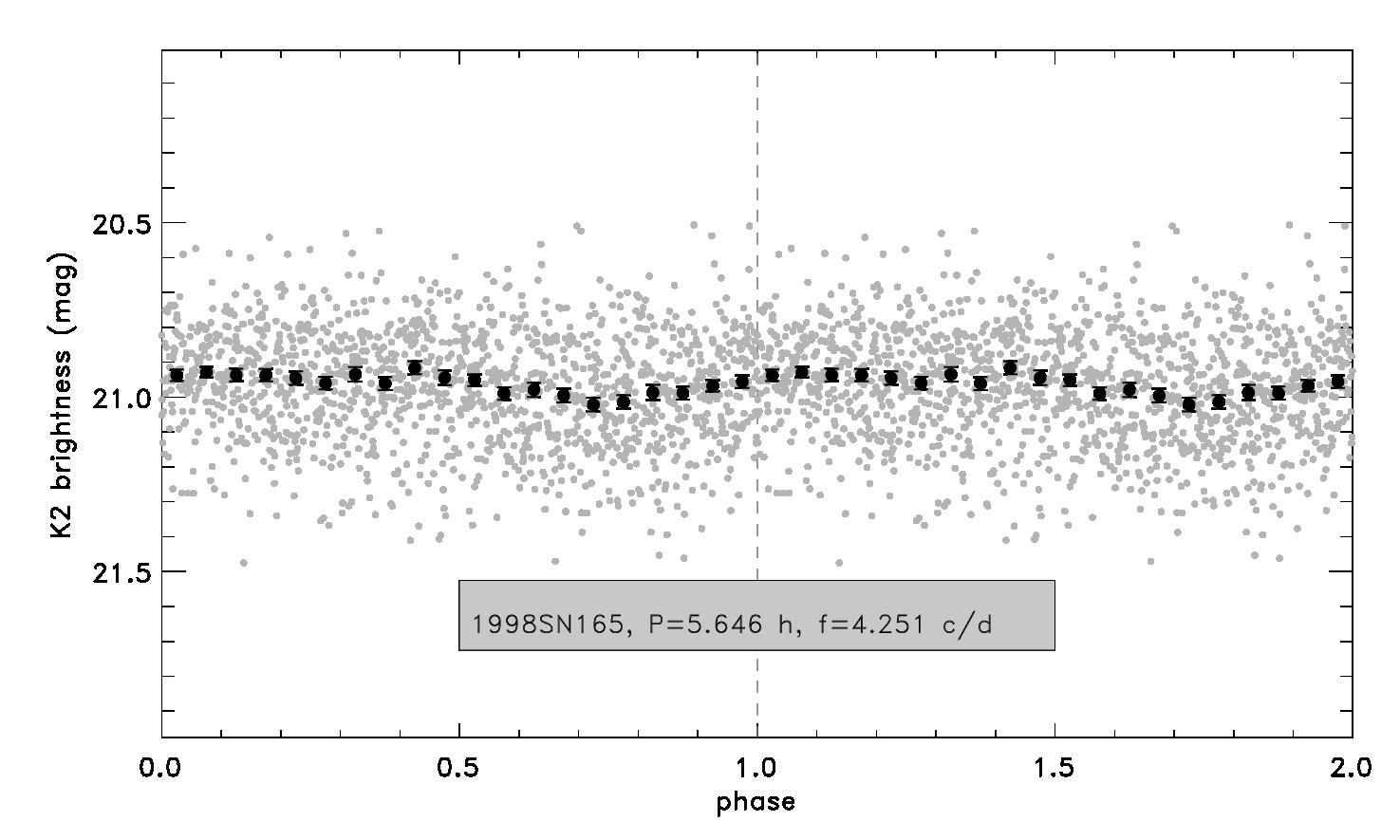}
\includegraphics[width=0.33\textwidth]{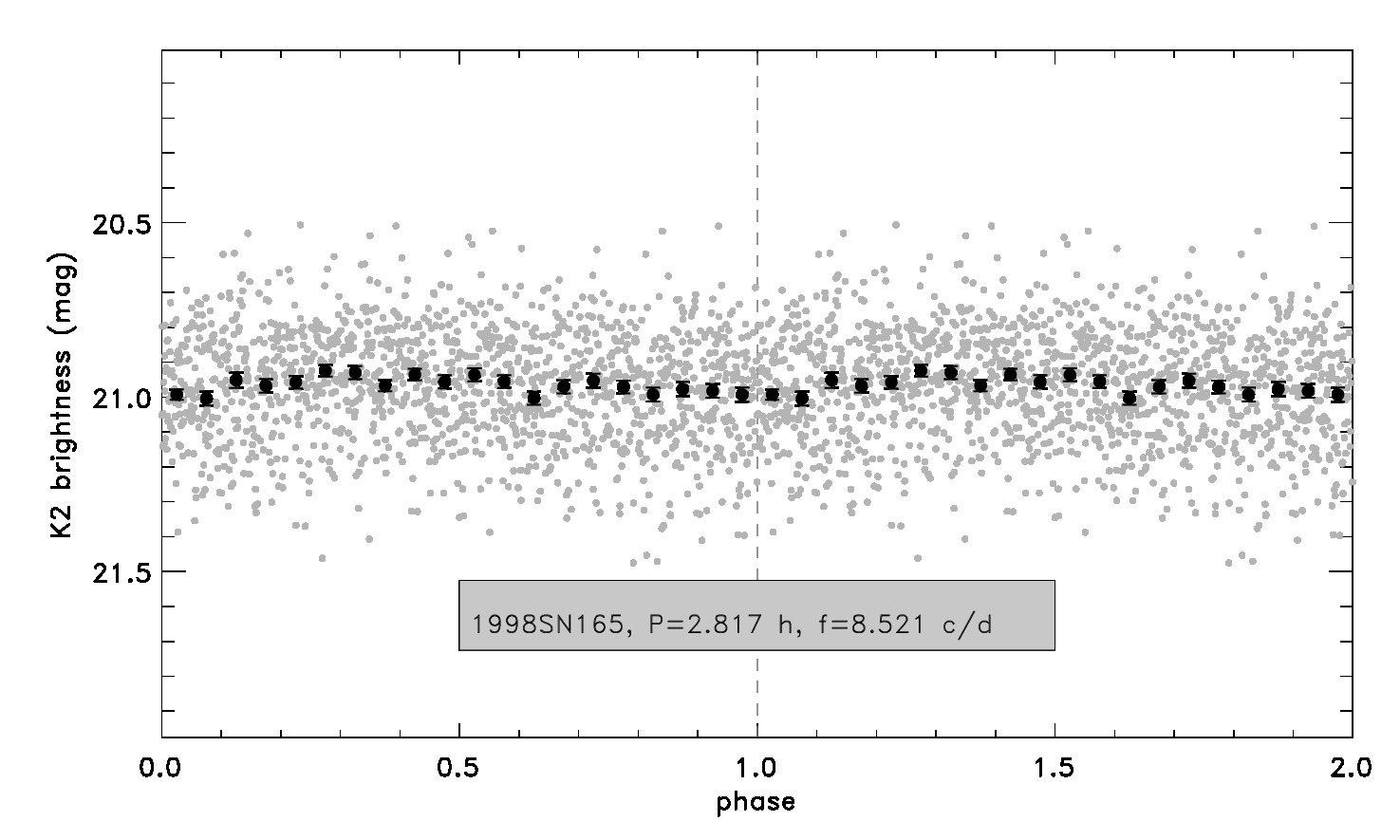}
}
\hbox{
\includegraphics[width=0.33\textwidth]{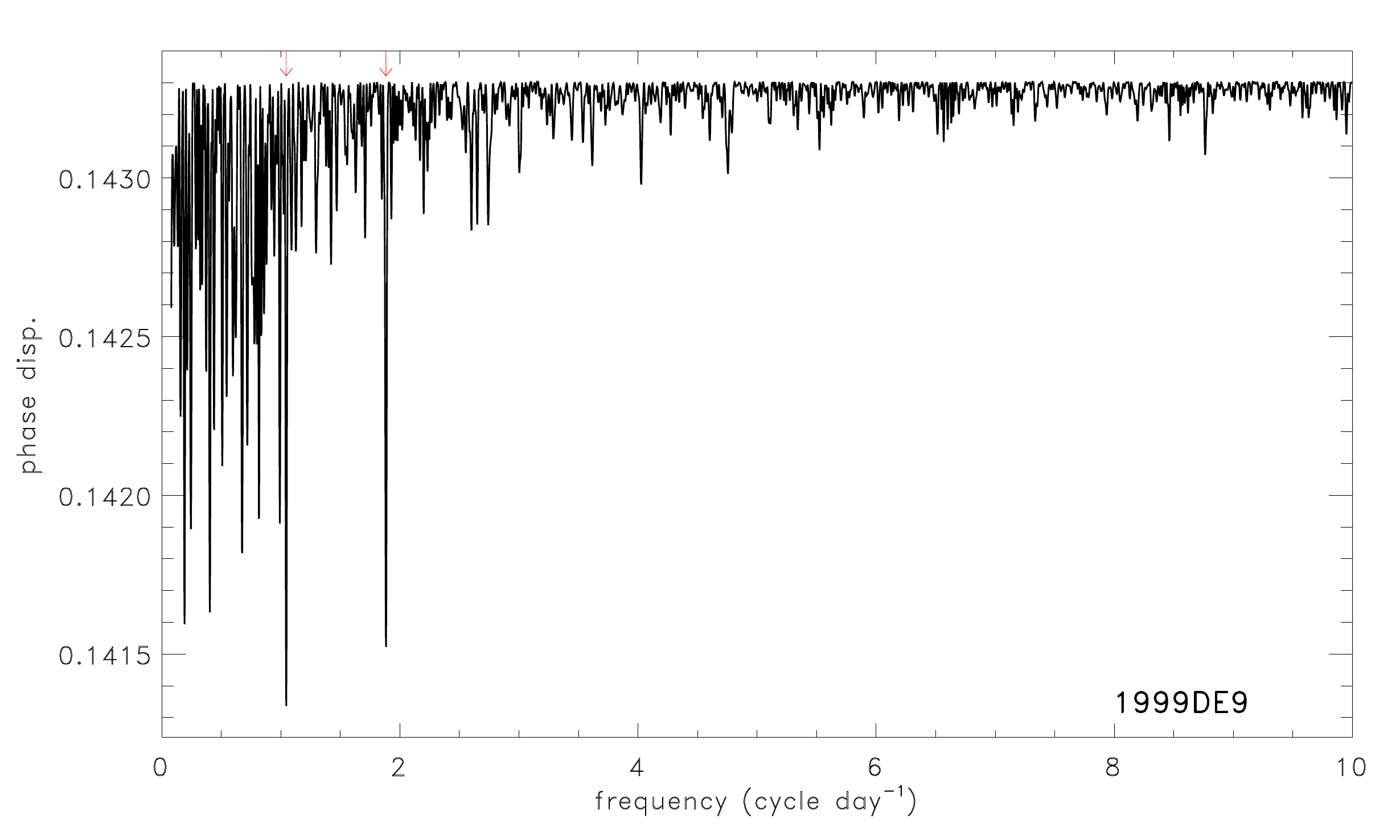}
\includegraphics[width=0.33\textwidth]{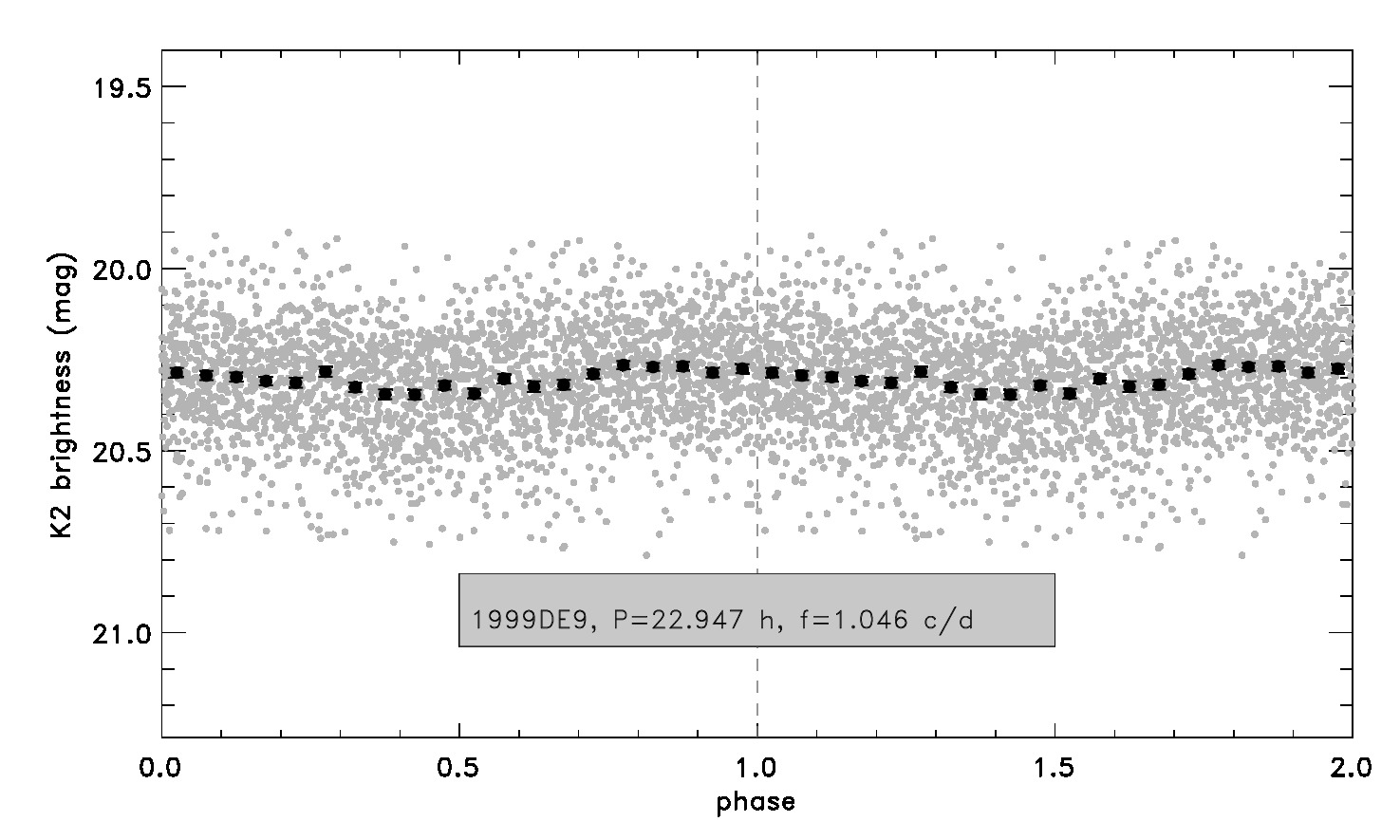}
\includegraphics[width=0.33\textwidth]{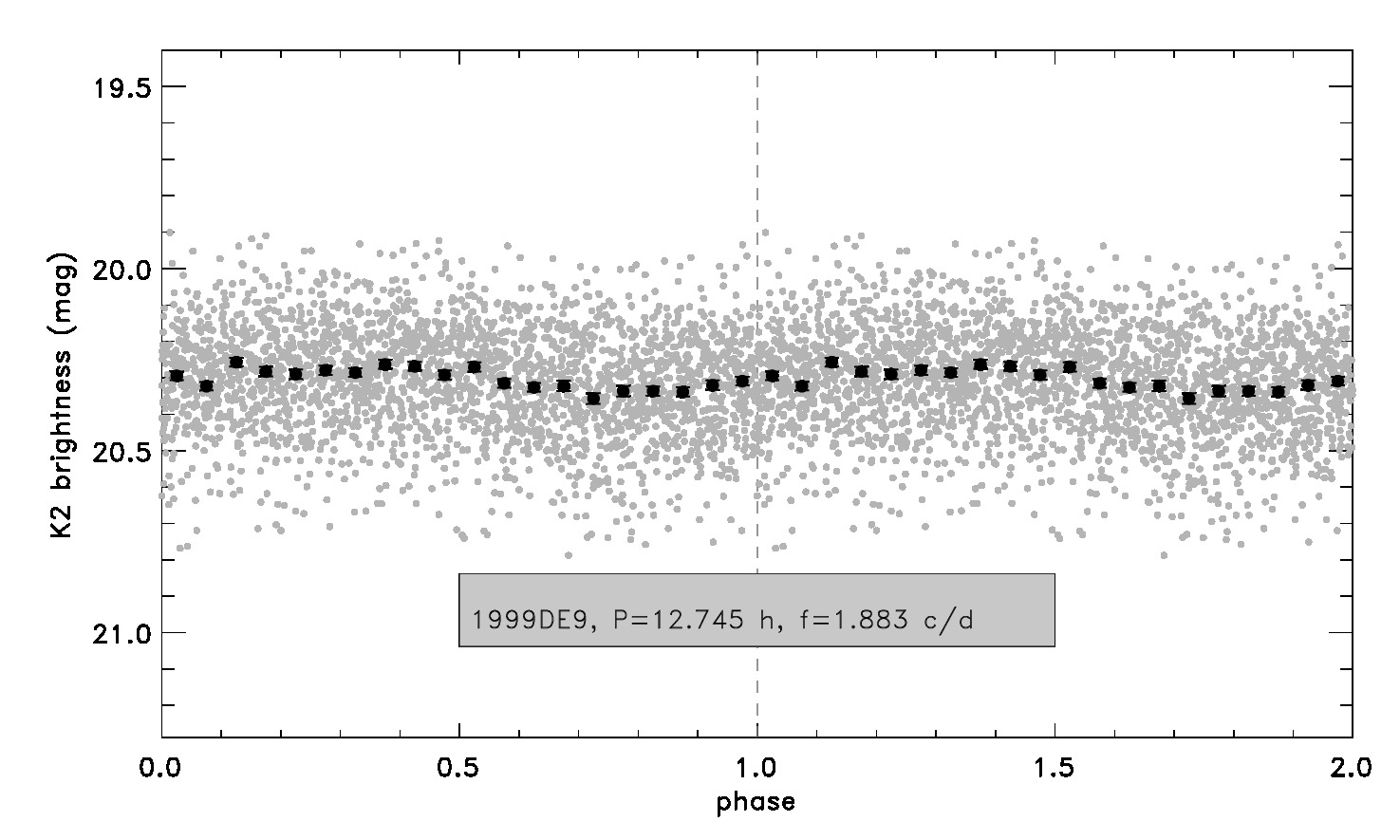}
}
\hbox{
\includegraphics[width=0.33\textwidth]{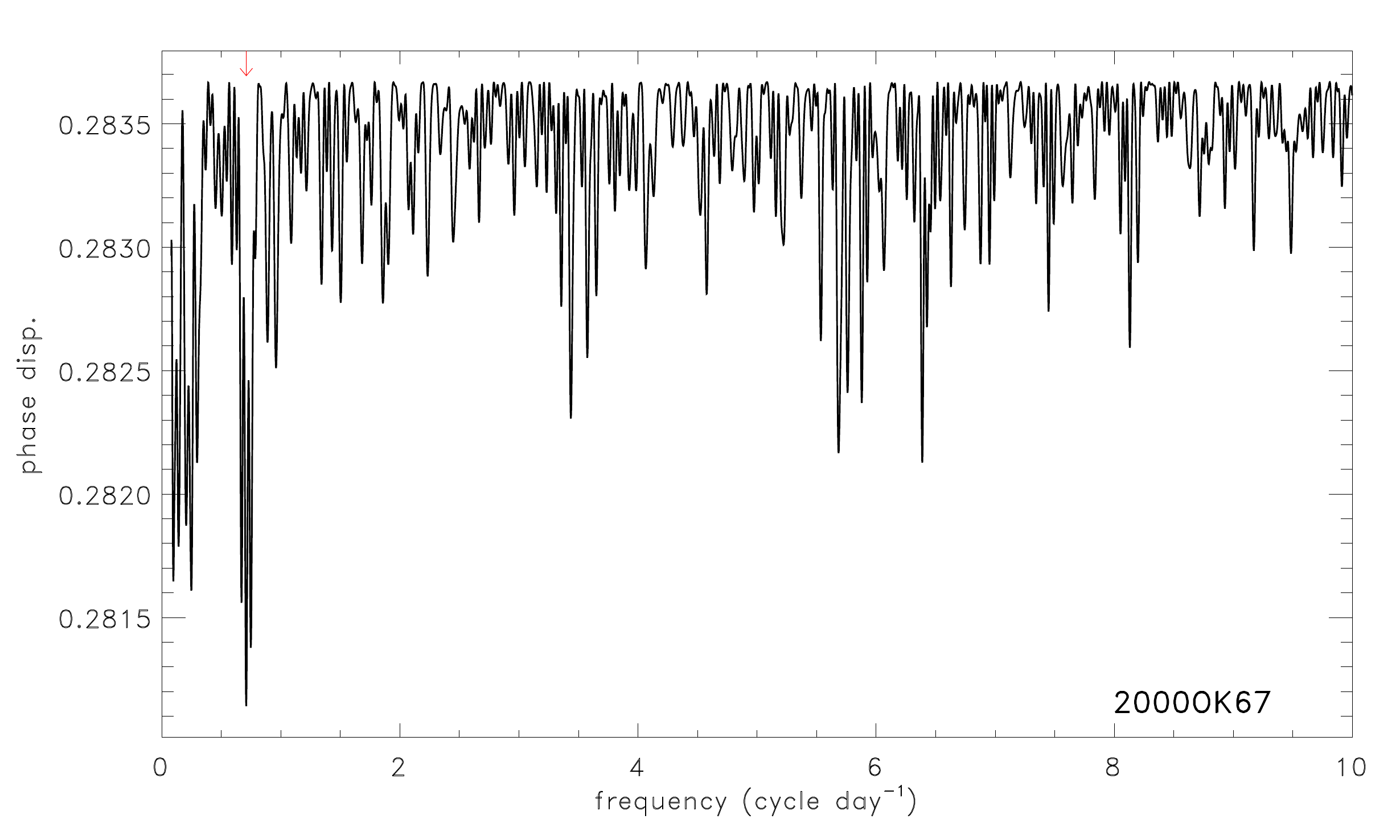}
\includegraphics[width=0.33\textwidth]{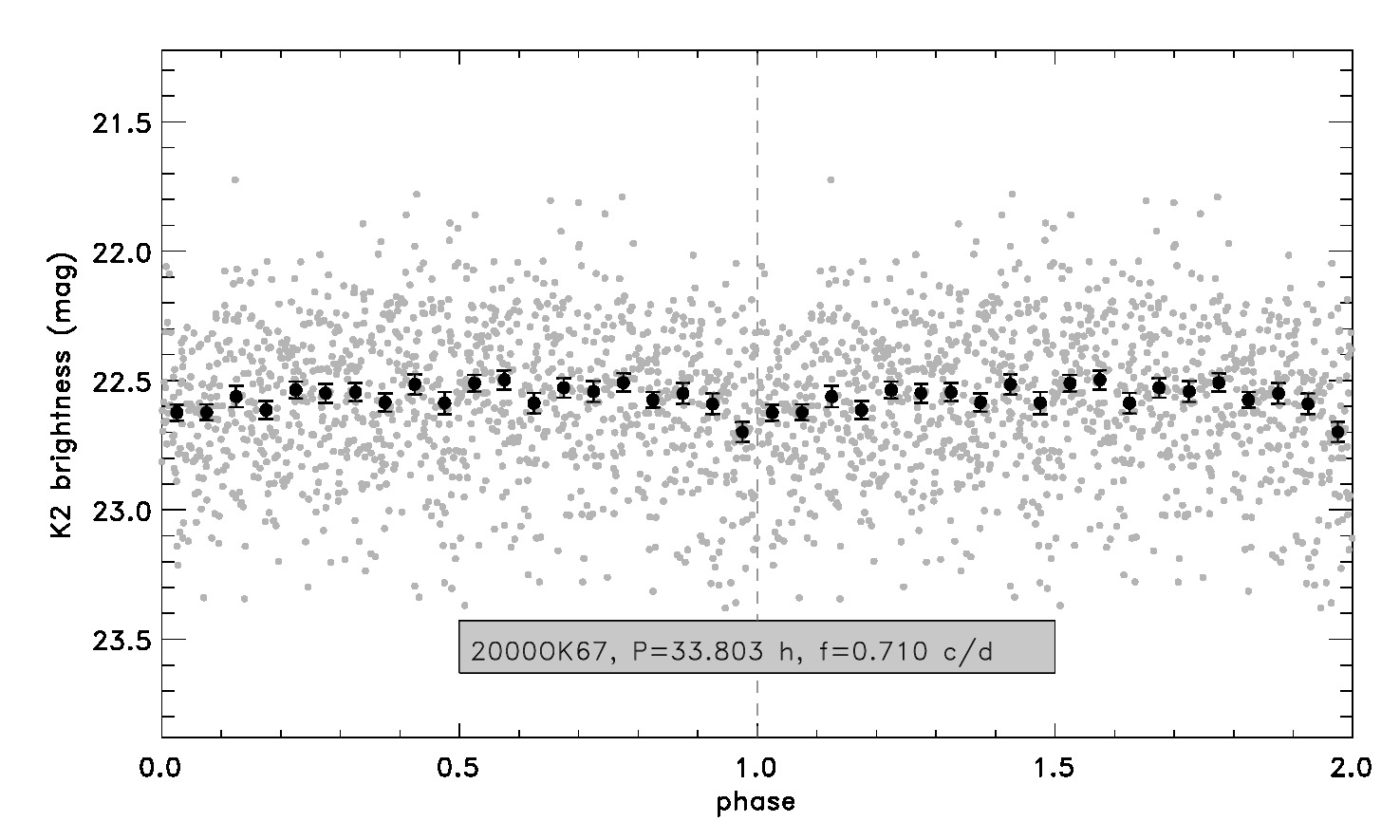}
}
\caption{For each target (one target a row) we present the phase dispersion versus frequency on the left column of the figure. Red arrows mark the light curve frequencies detected by our analysis. On the panels to the right we show the light curves folded with the periods/frequencies identified, as listed in Table~\ref{table:freqs}. The name of the target and the folding periods/frequencies are indicated. \label{fig:big0}}
\end{figure*}

\begin{figure*}[ht!]
\ContinuedFloat
\hbox{
\includegraphics[width=0.33\textwidth]{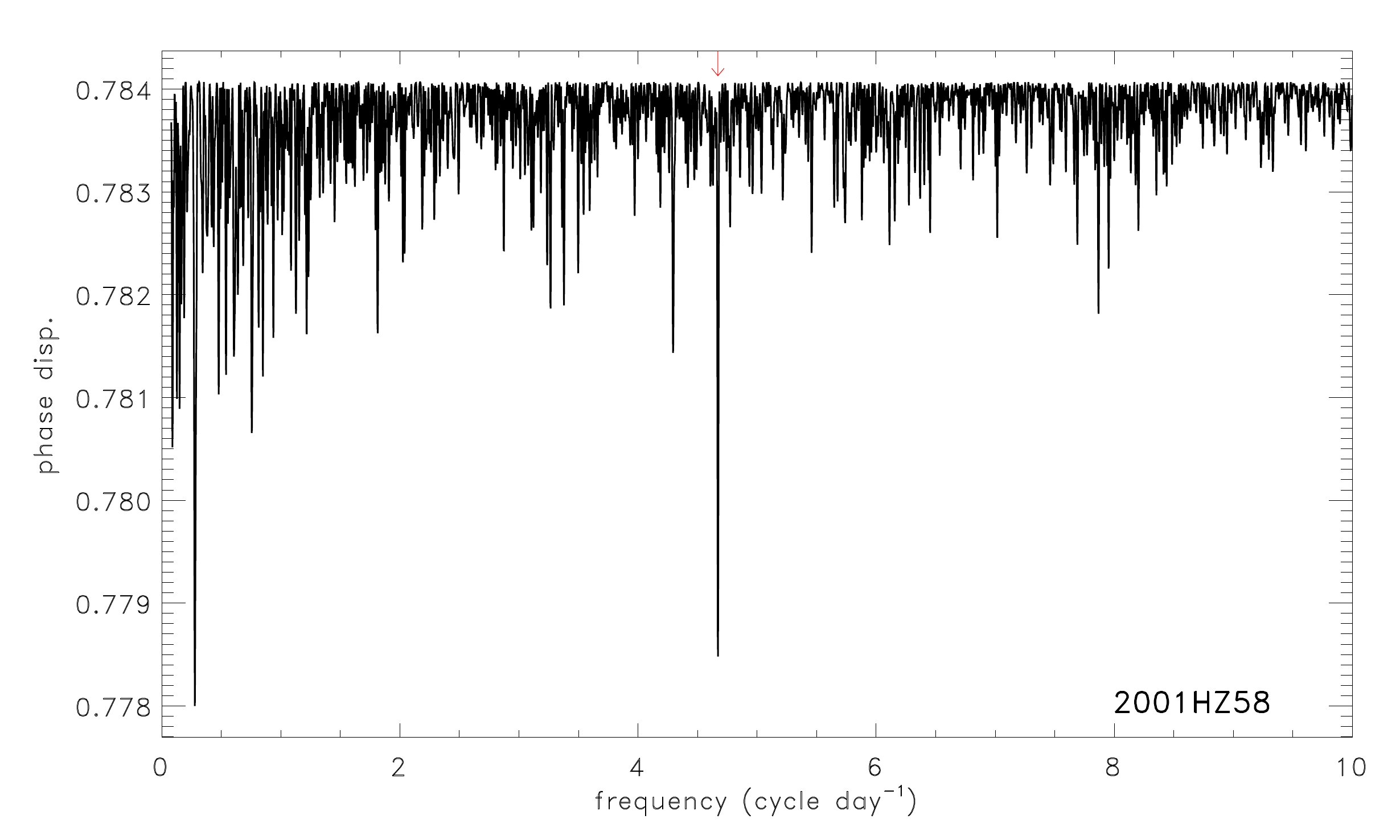}
\includegraphics[width=0.33\textwidth]{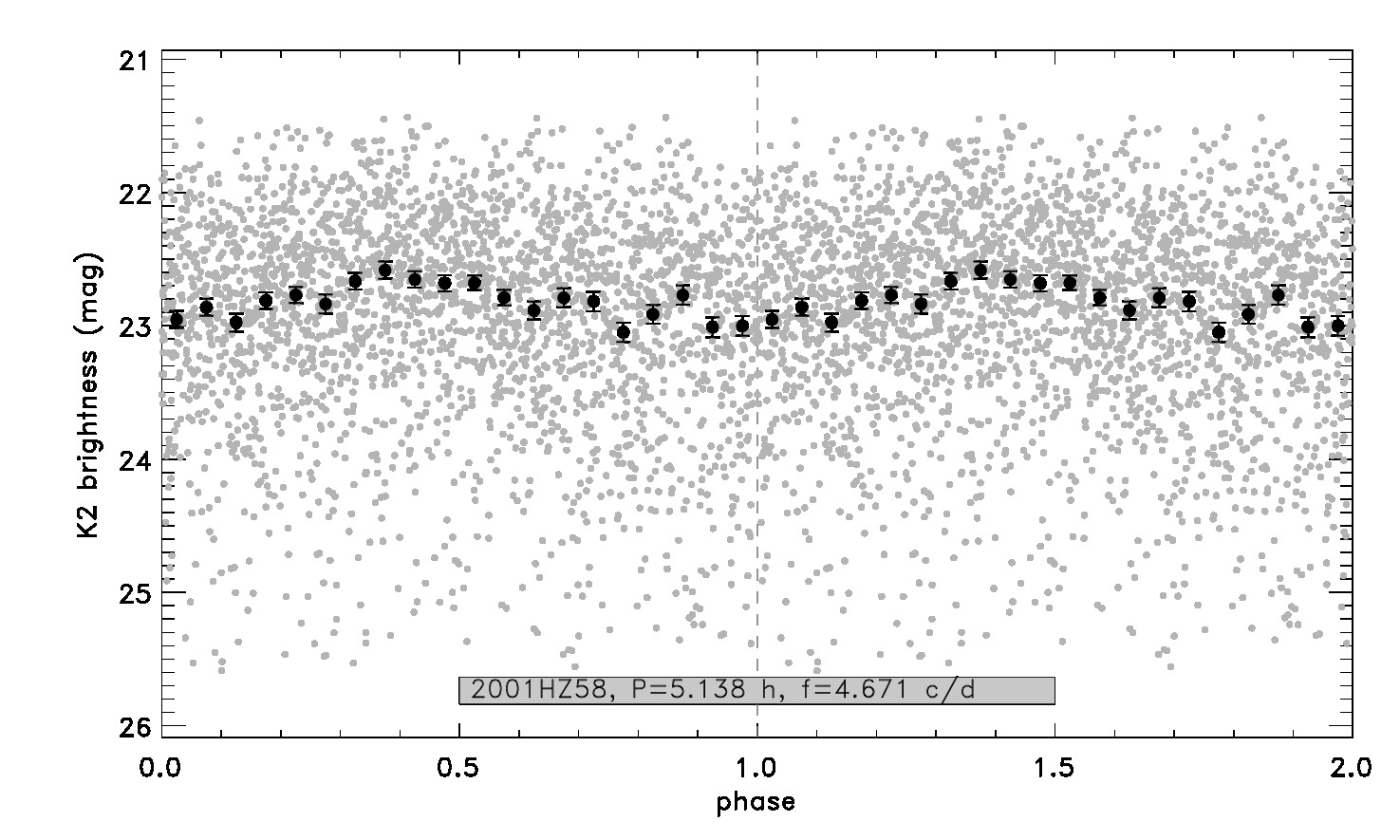}
}
\hbox{
\includegraphics[width=0.33\textwidth]{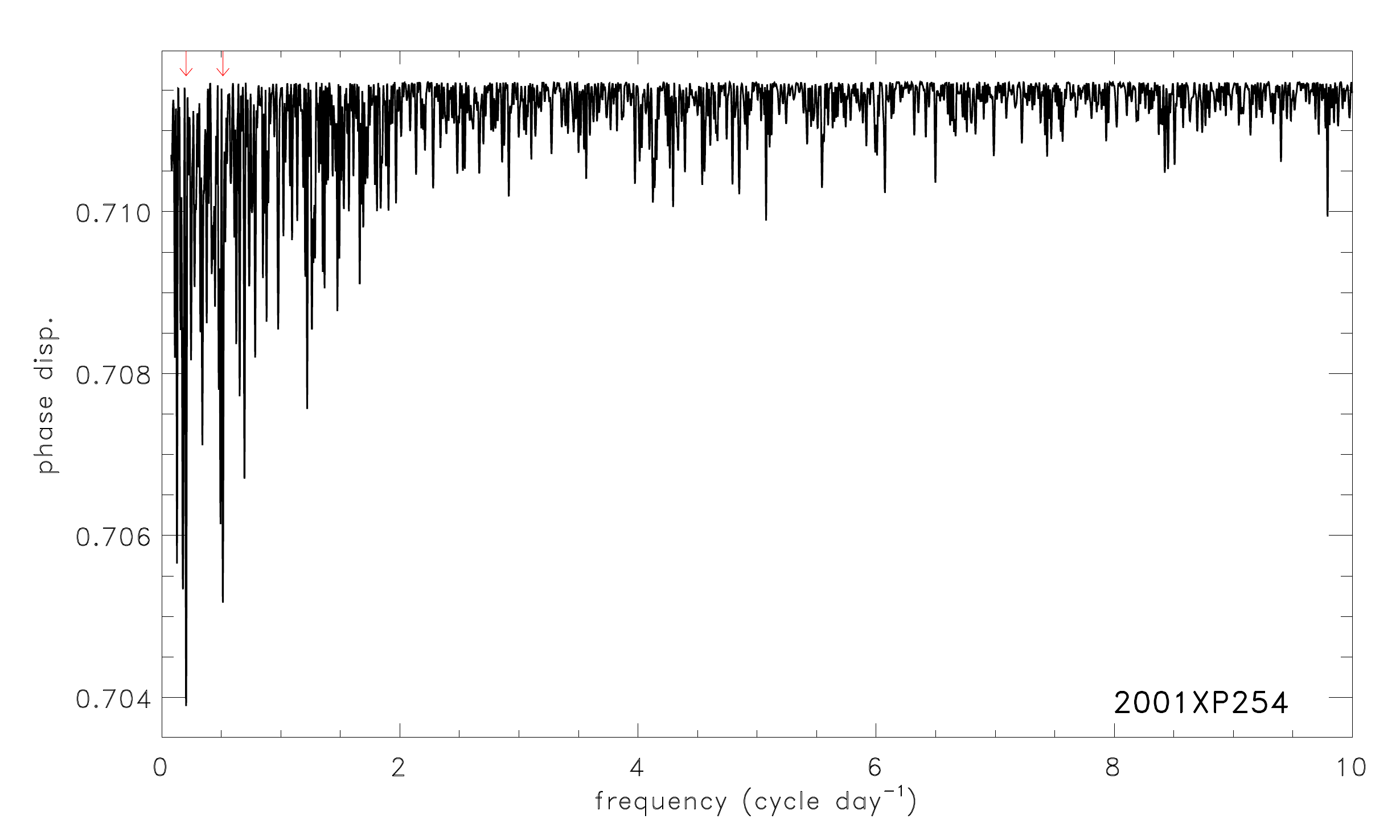}
\includegraphics[width=0.33\textwidth]{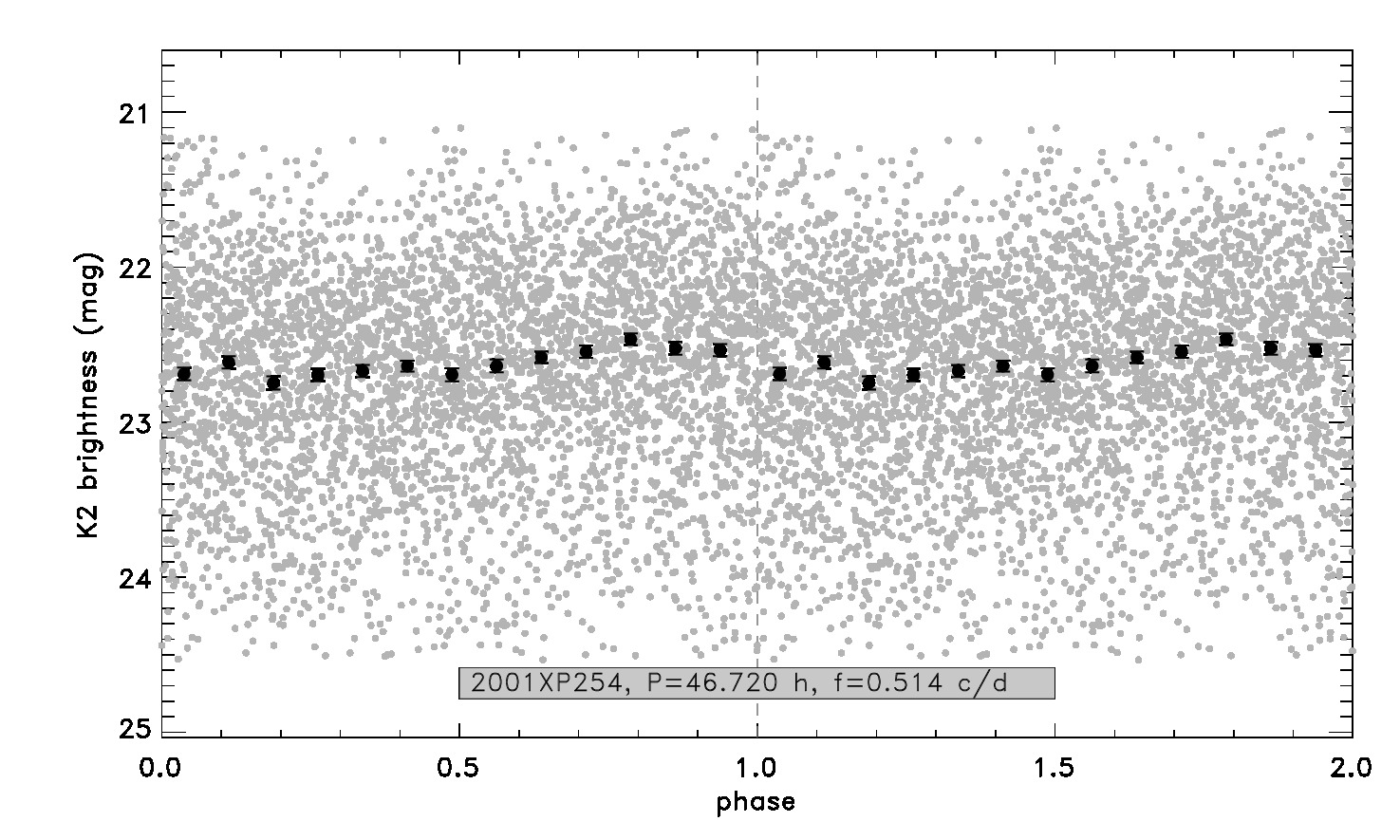}
\includegraphics[width=0.33\textwidth]{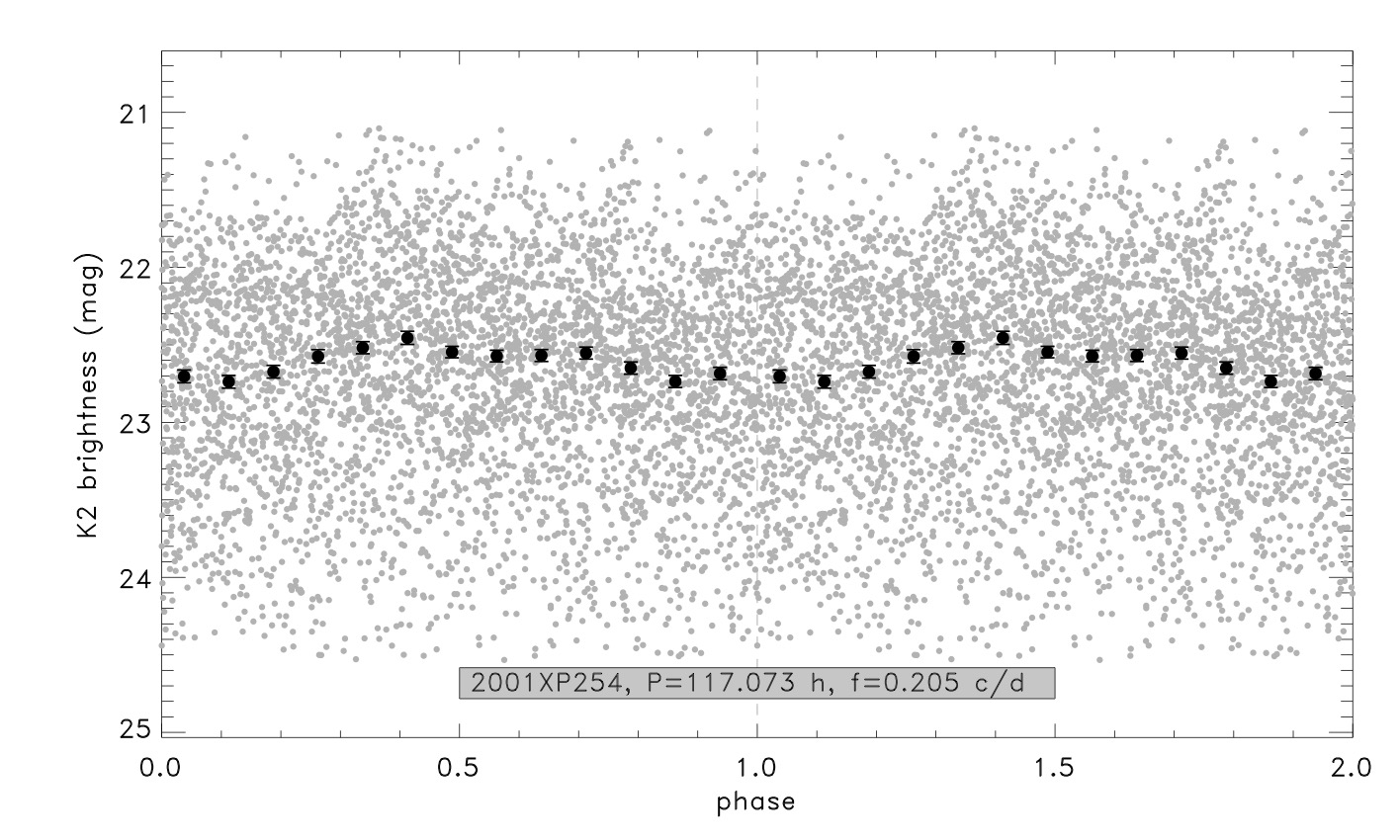}
}
\hbox{
\includegraphics[width=0.33\textwidth]{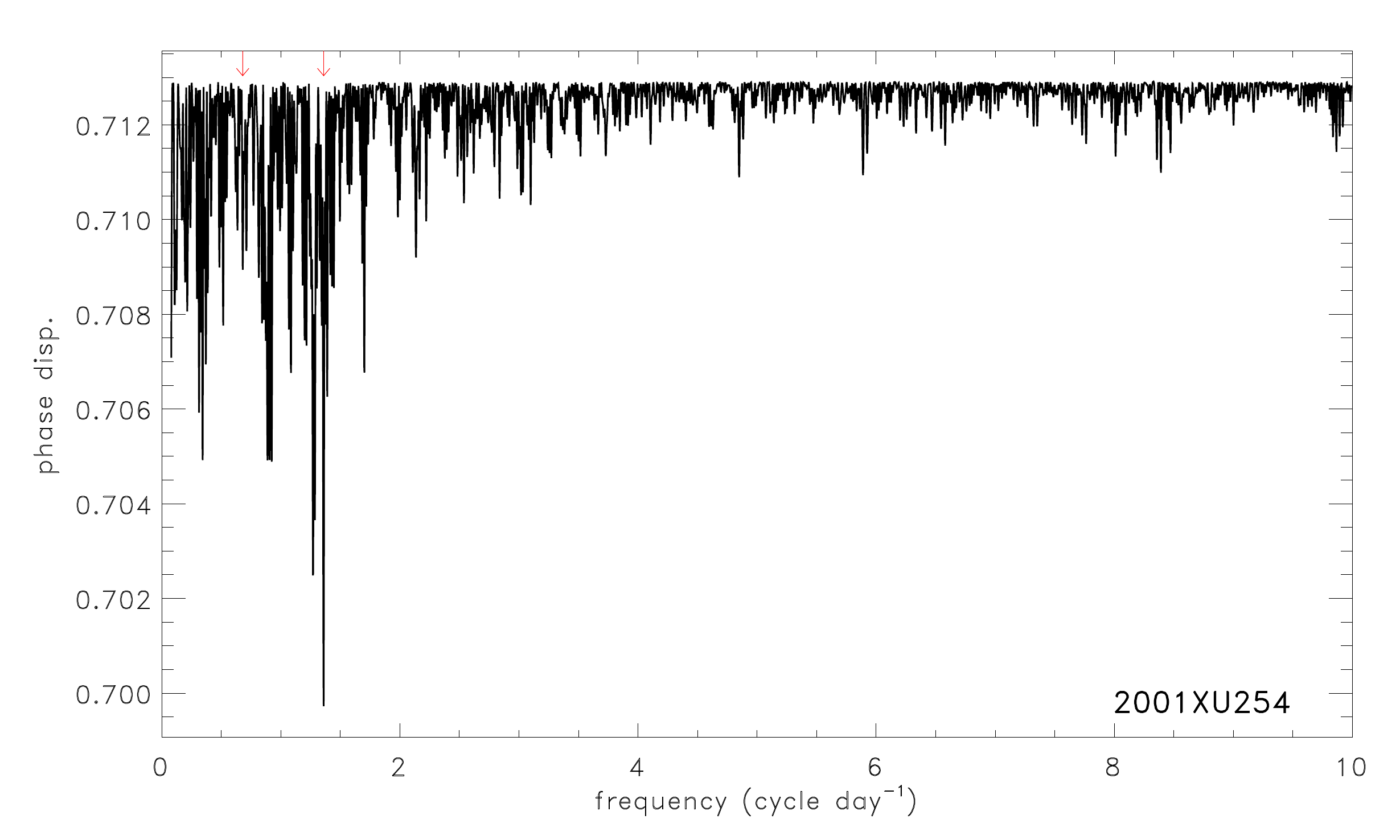}
\includegraphics[width=0.33\textwidth]{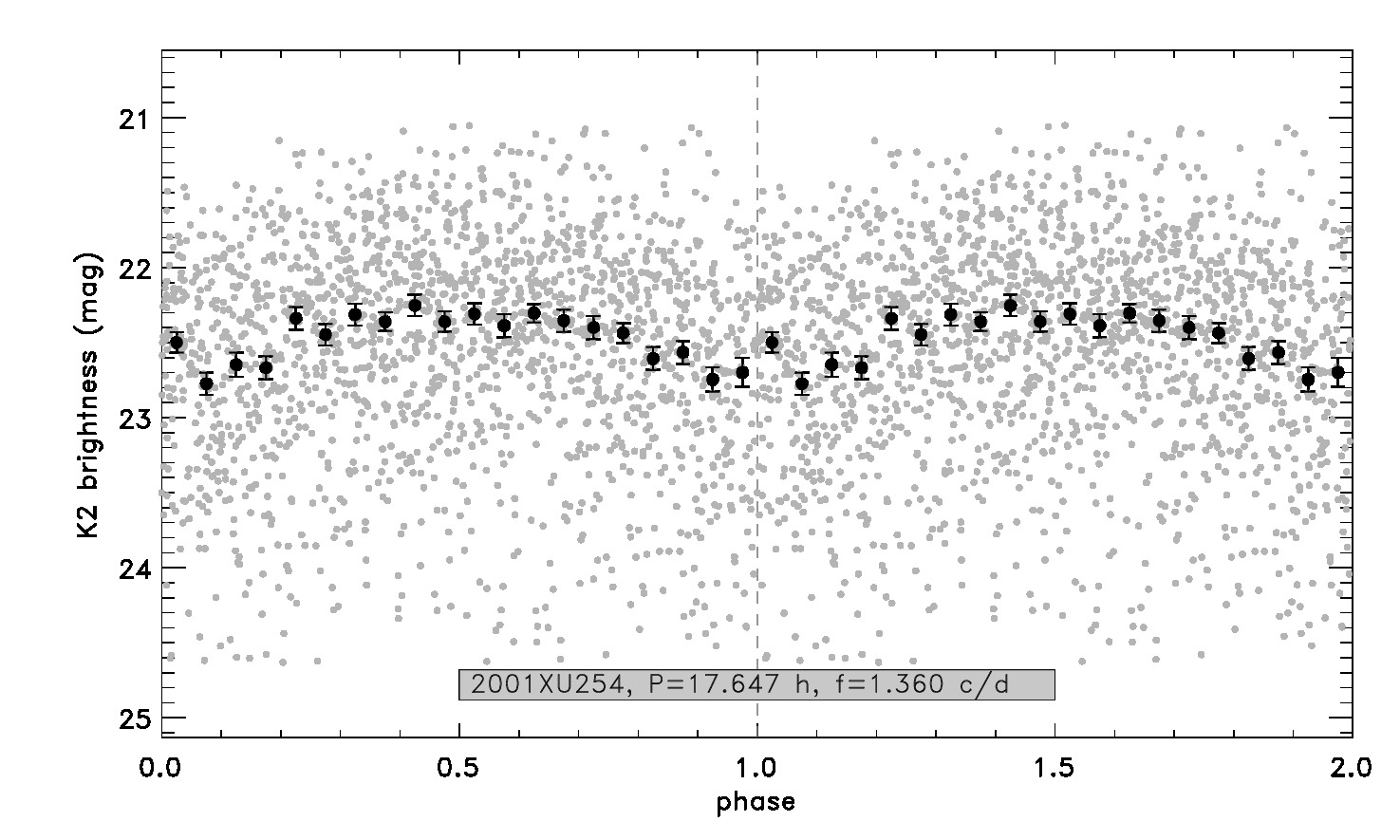}
\includegraphics[width=0.33\textwidth]{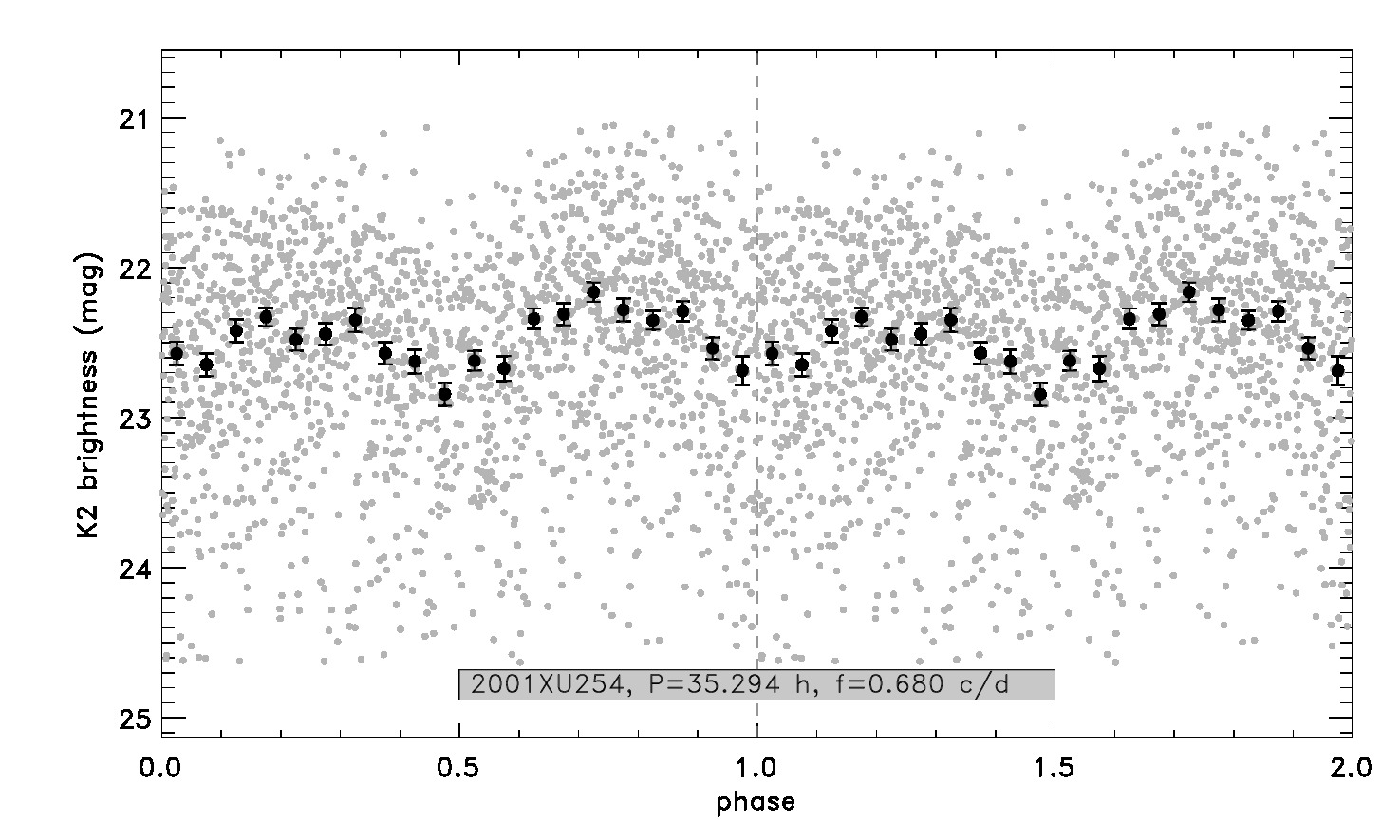}
}
\hbox{
\includegraphics[width=0.33\textwidth]{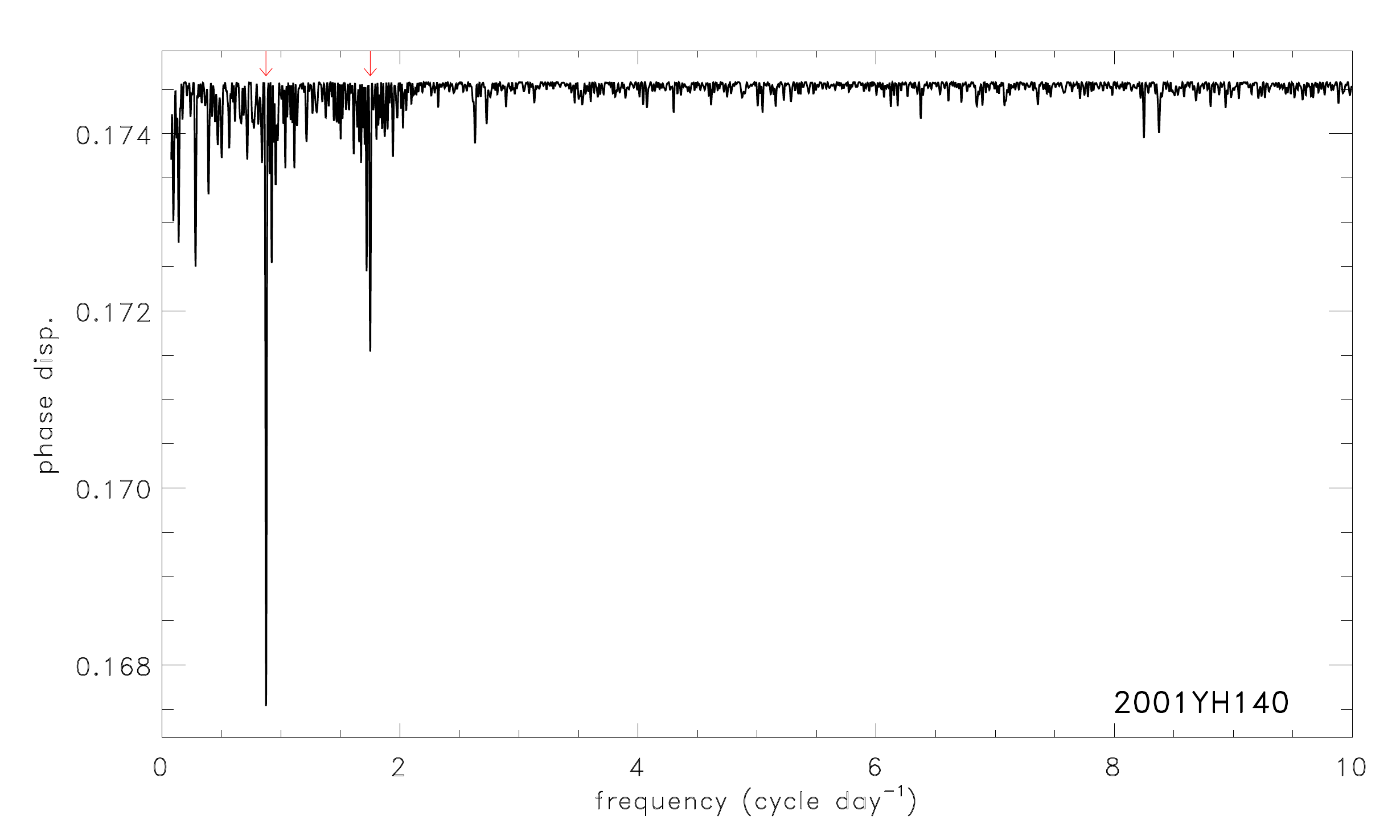}
\includegraphics[width=0.33\textwidth]{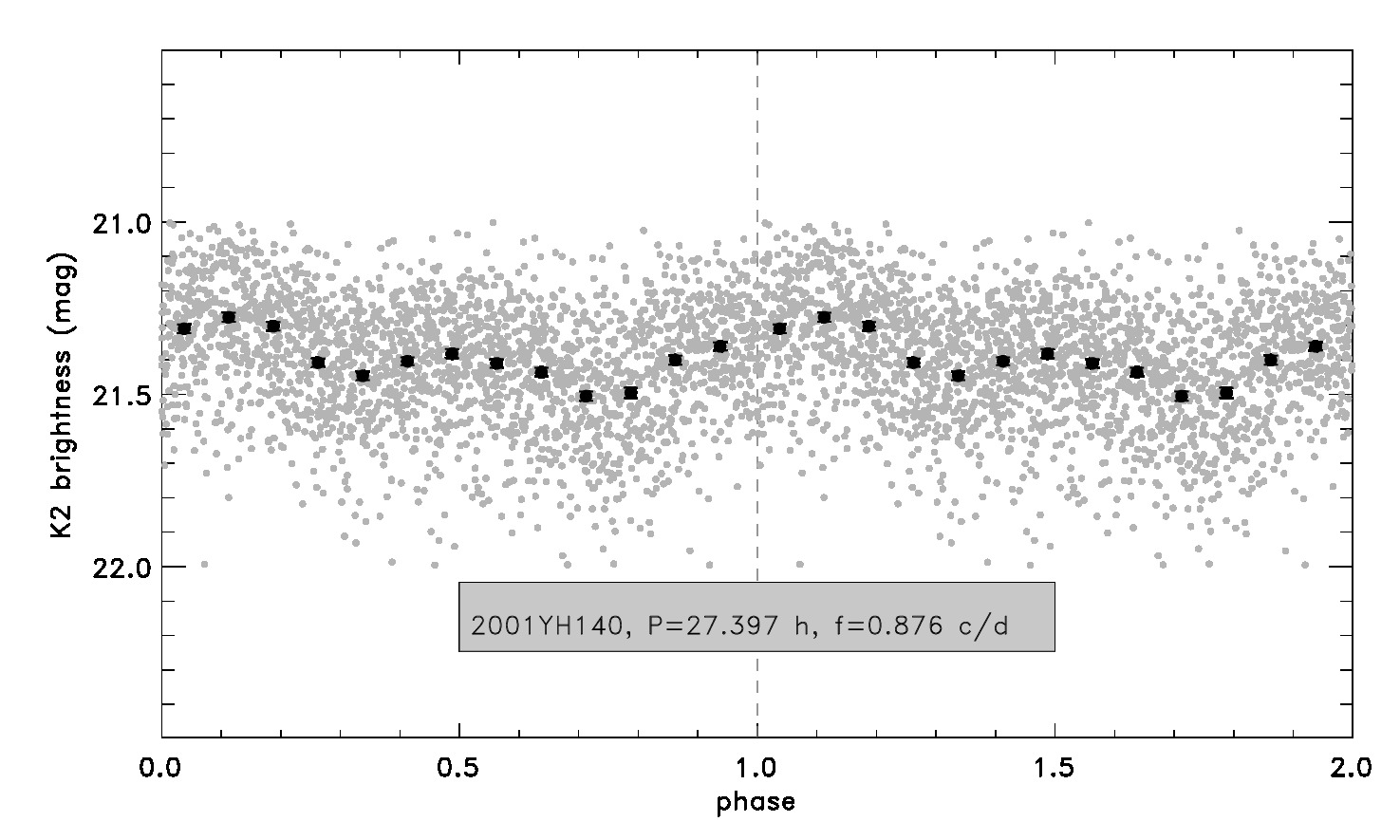}
\includegraphics[width=0.33\textwidth]{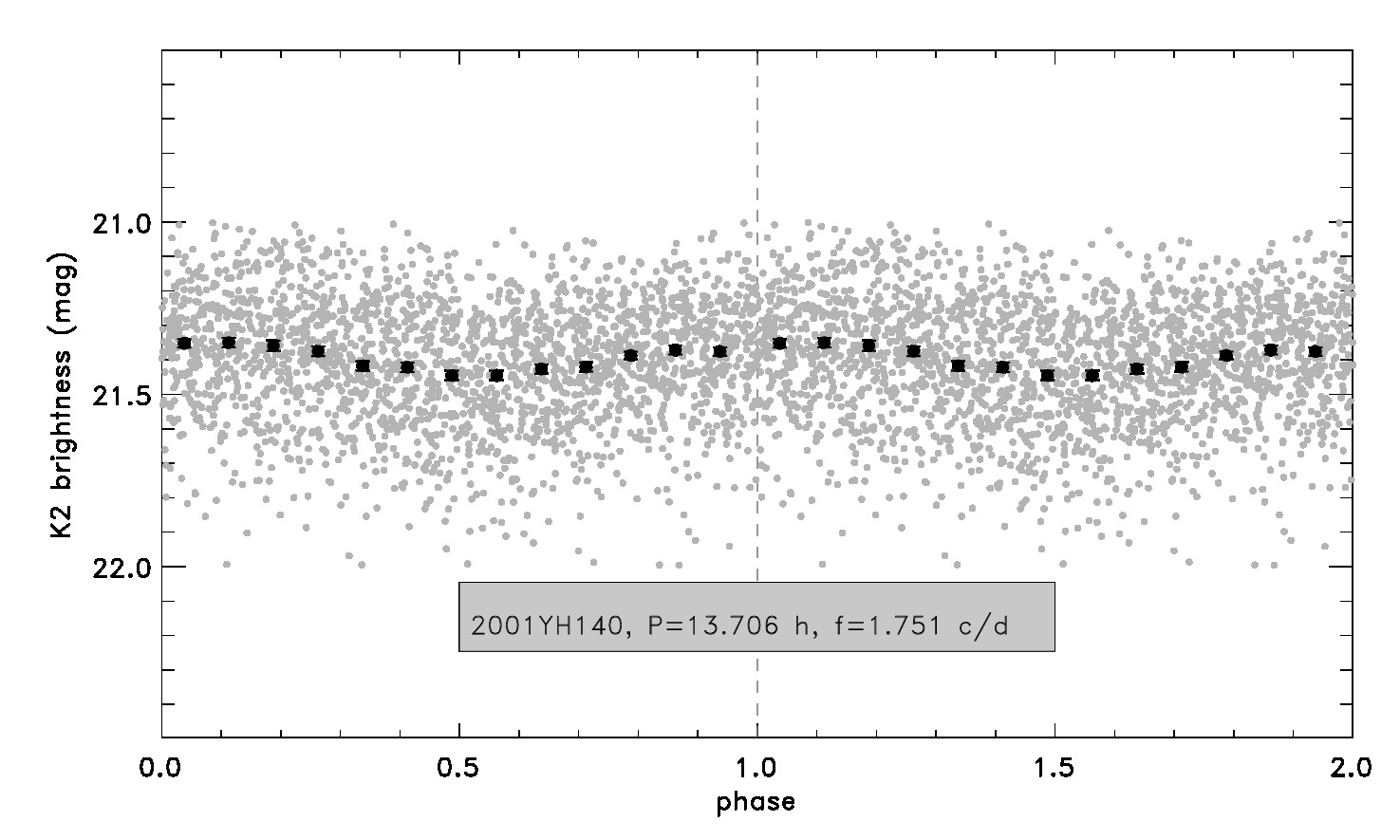}
}
\hbox{
\includegraphics[width=0.33\textwidth]{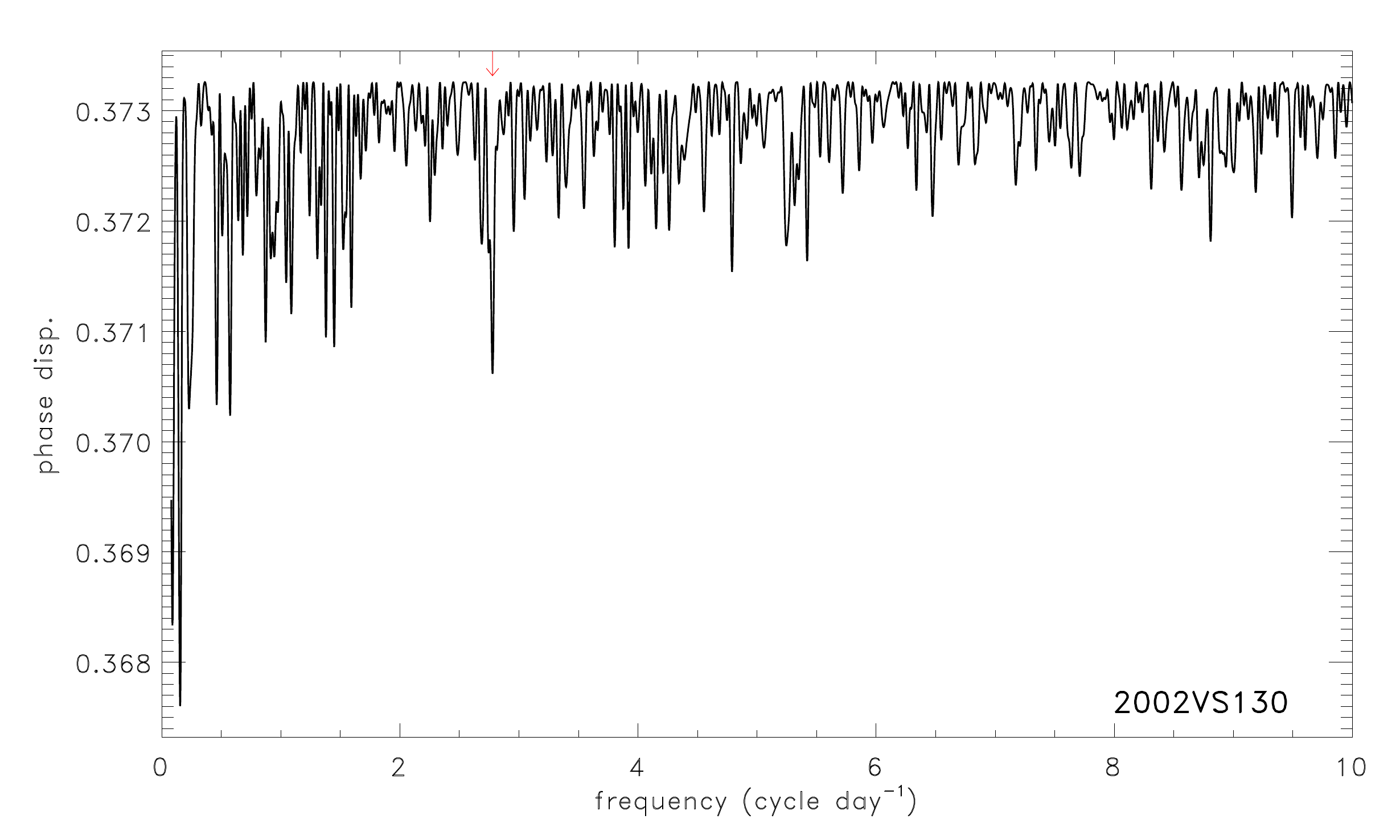}
\includegraphics[width=0.33\textwidth]{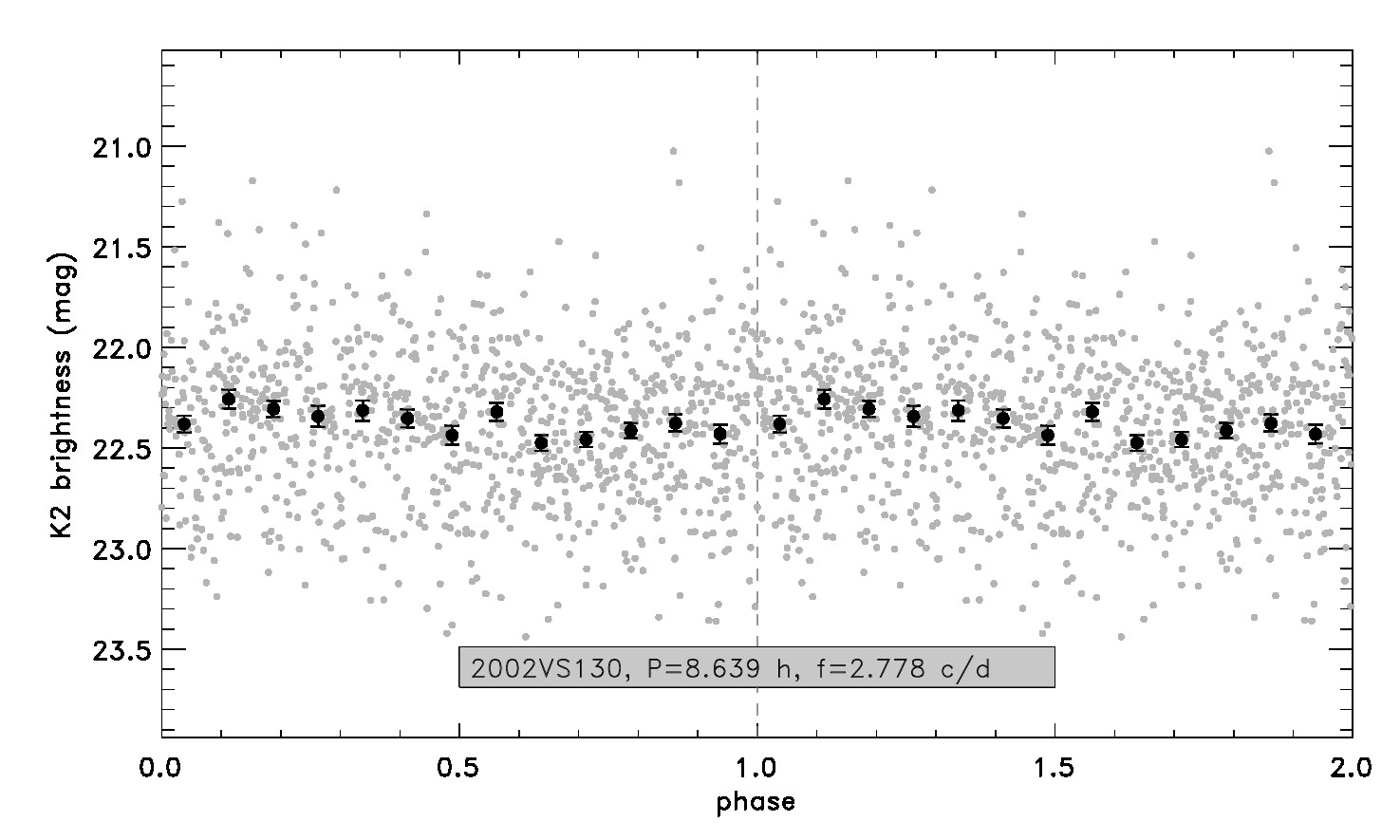}
}
\hbox{
\includegraphics[width=0.33\textwidth]{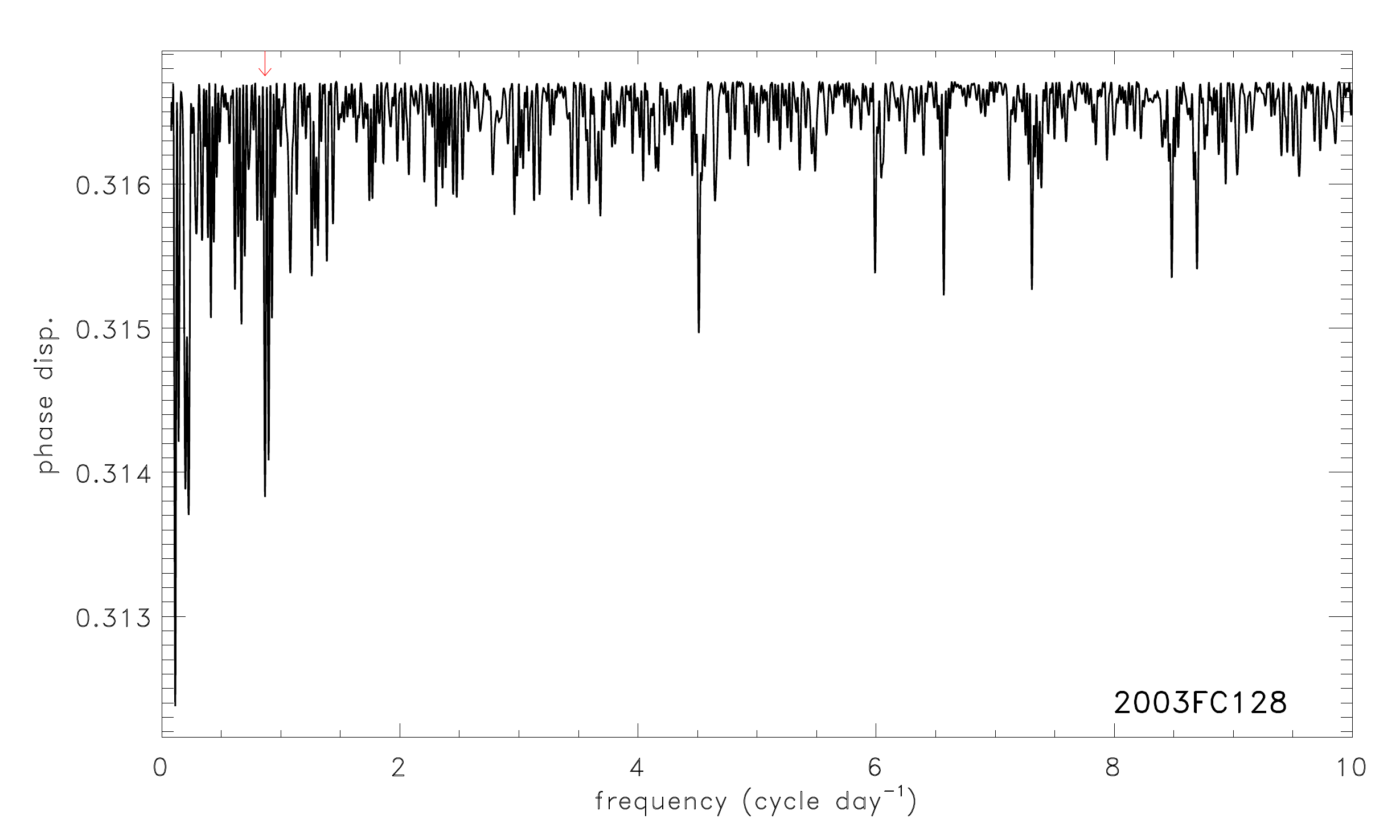}
\includegraphics[width=0.33\textwidth]{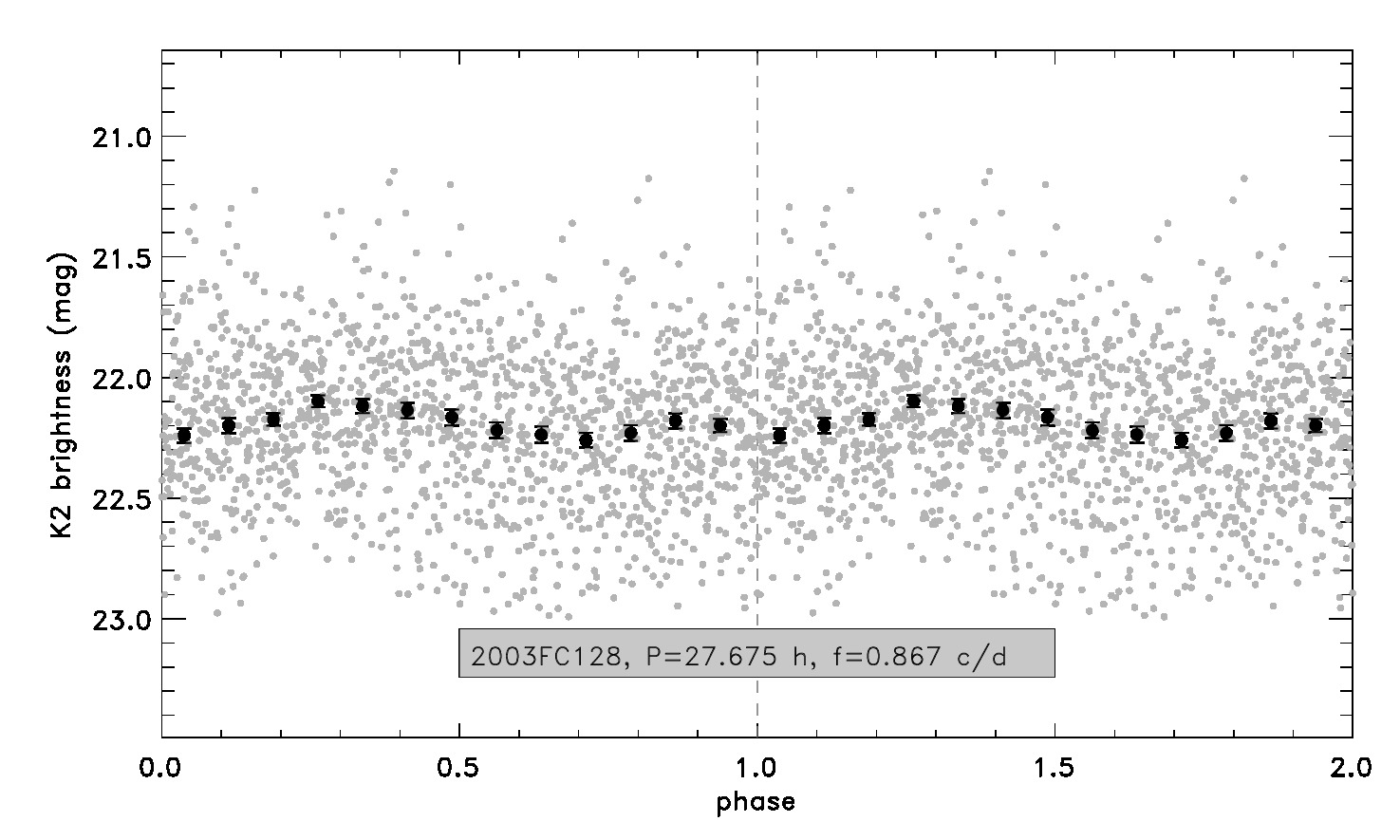}
}
\caption{
}
\end{figure*}

\begin{figure*}[ht!]
\ContinuedFloat
\hbox{
\includegraphics[width=0.33\textwidth]{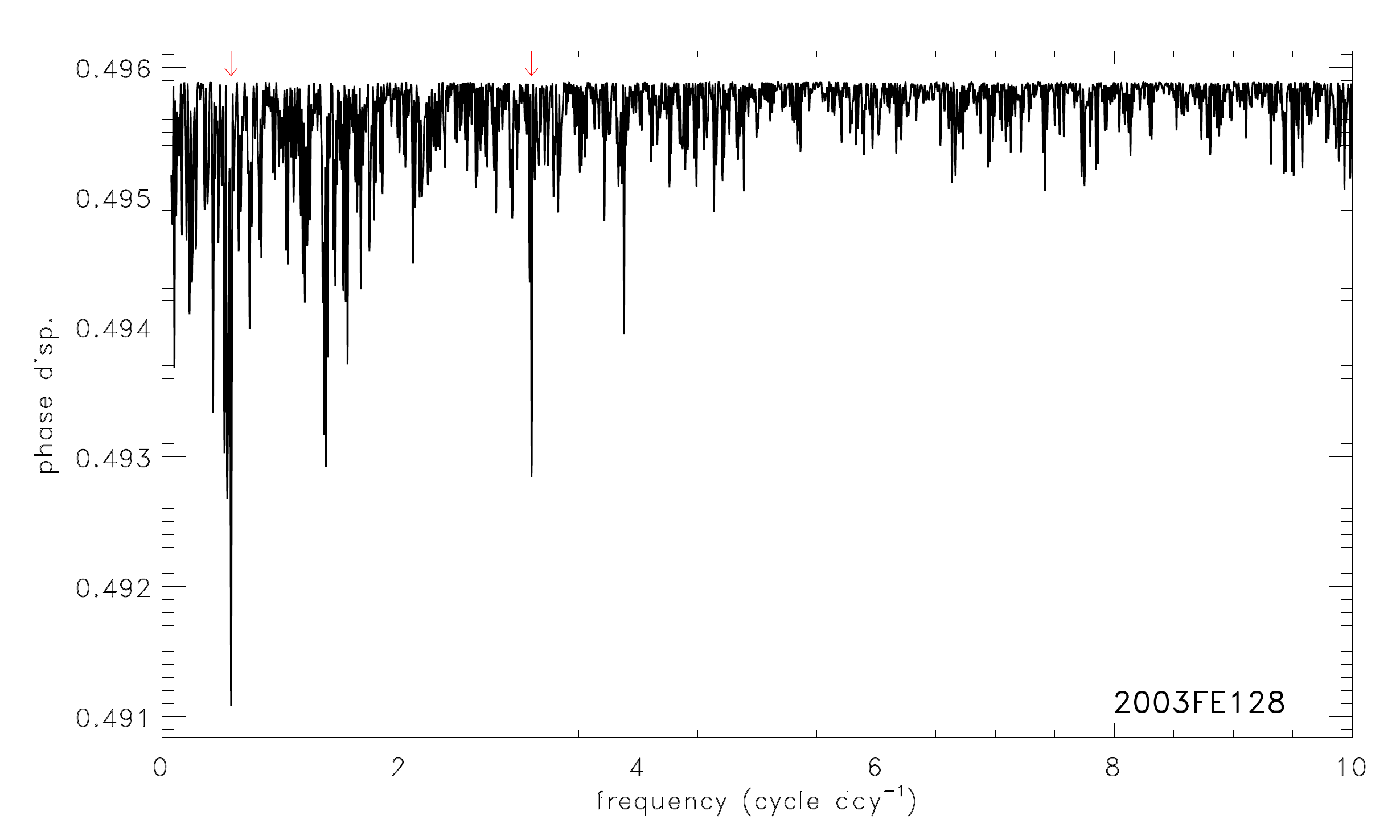}
\includegraphics[width=0.33\textwidth]{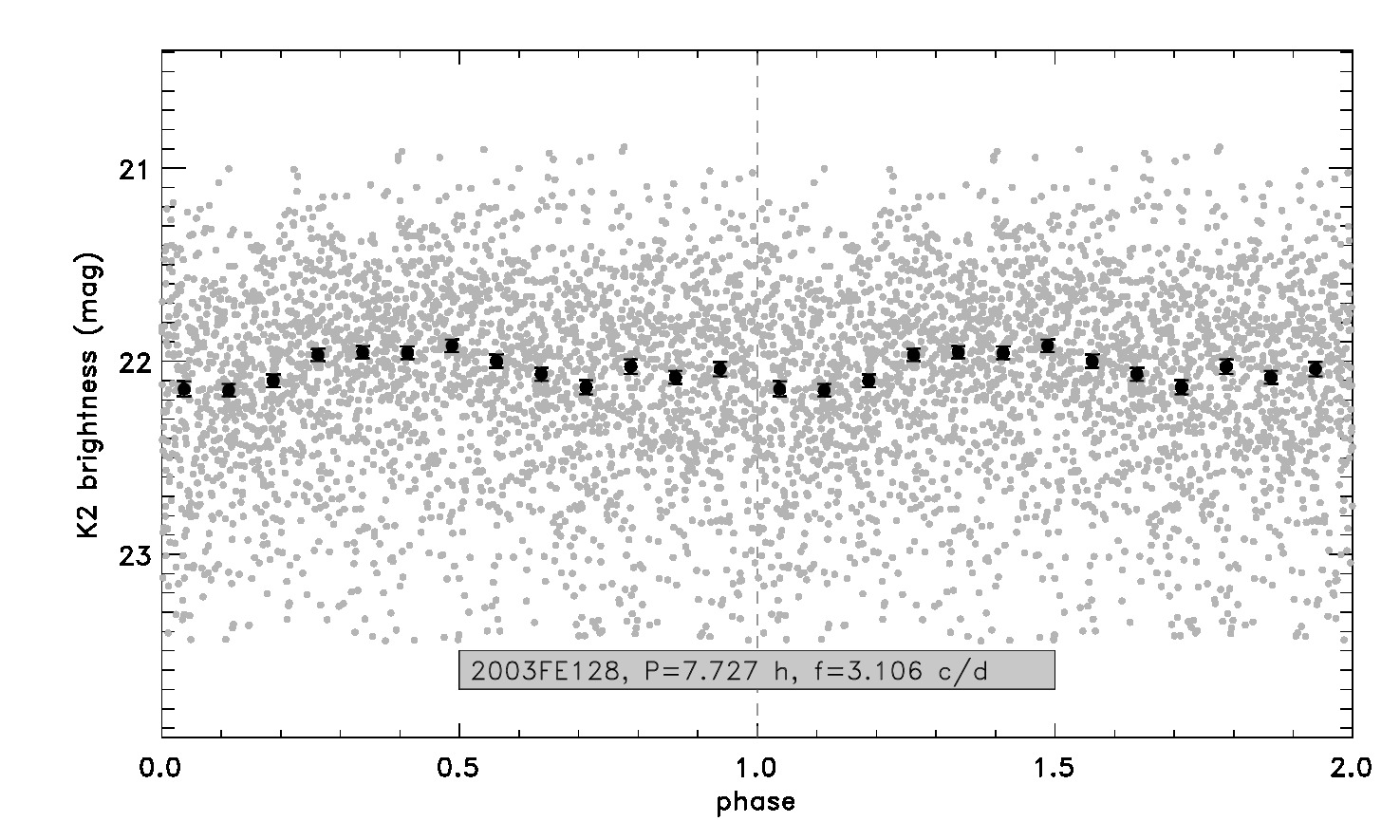}
\includegraphics[width=0.33\textwidth]{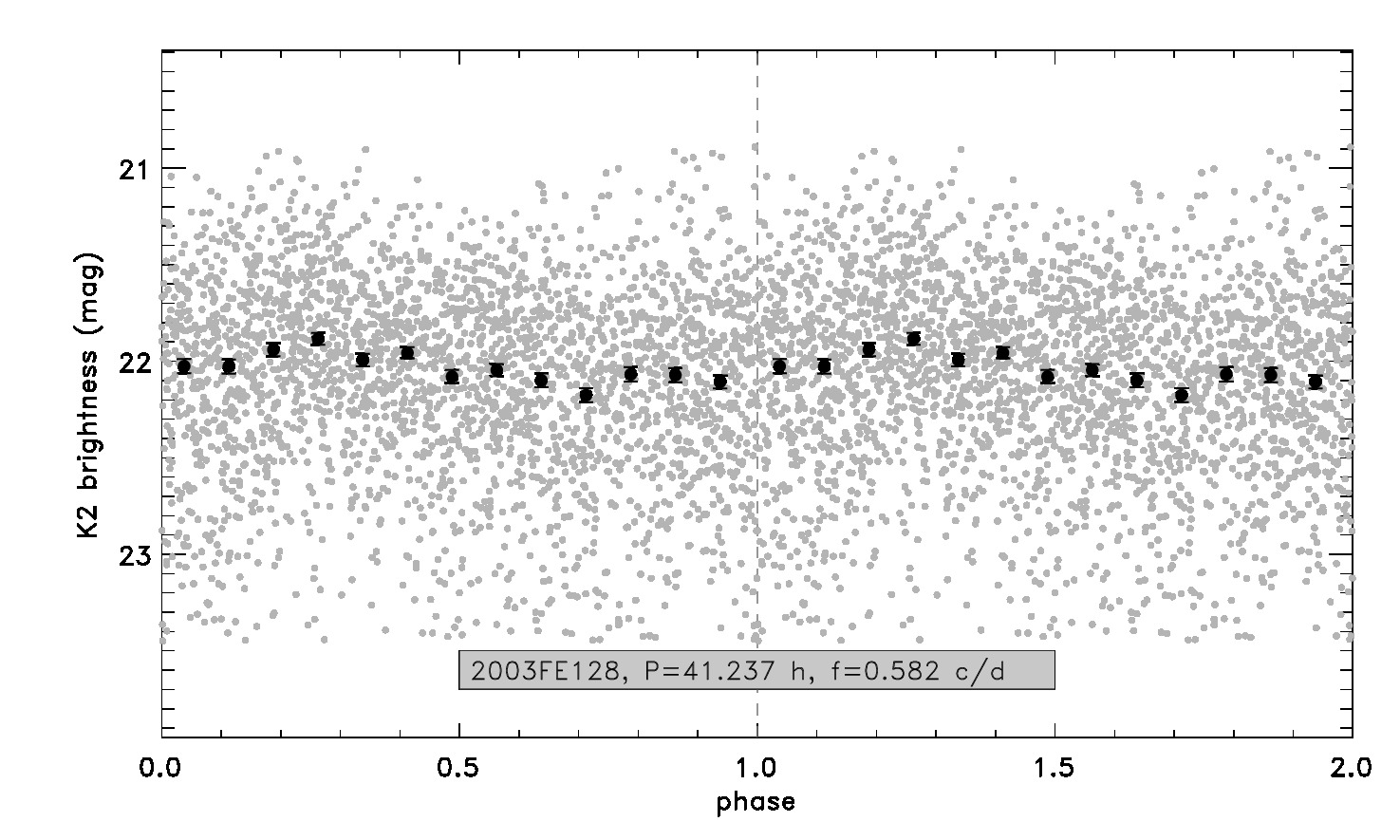}
}
\hbox{
\includegraphics[width=0.33\textwidth]{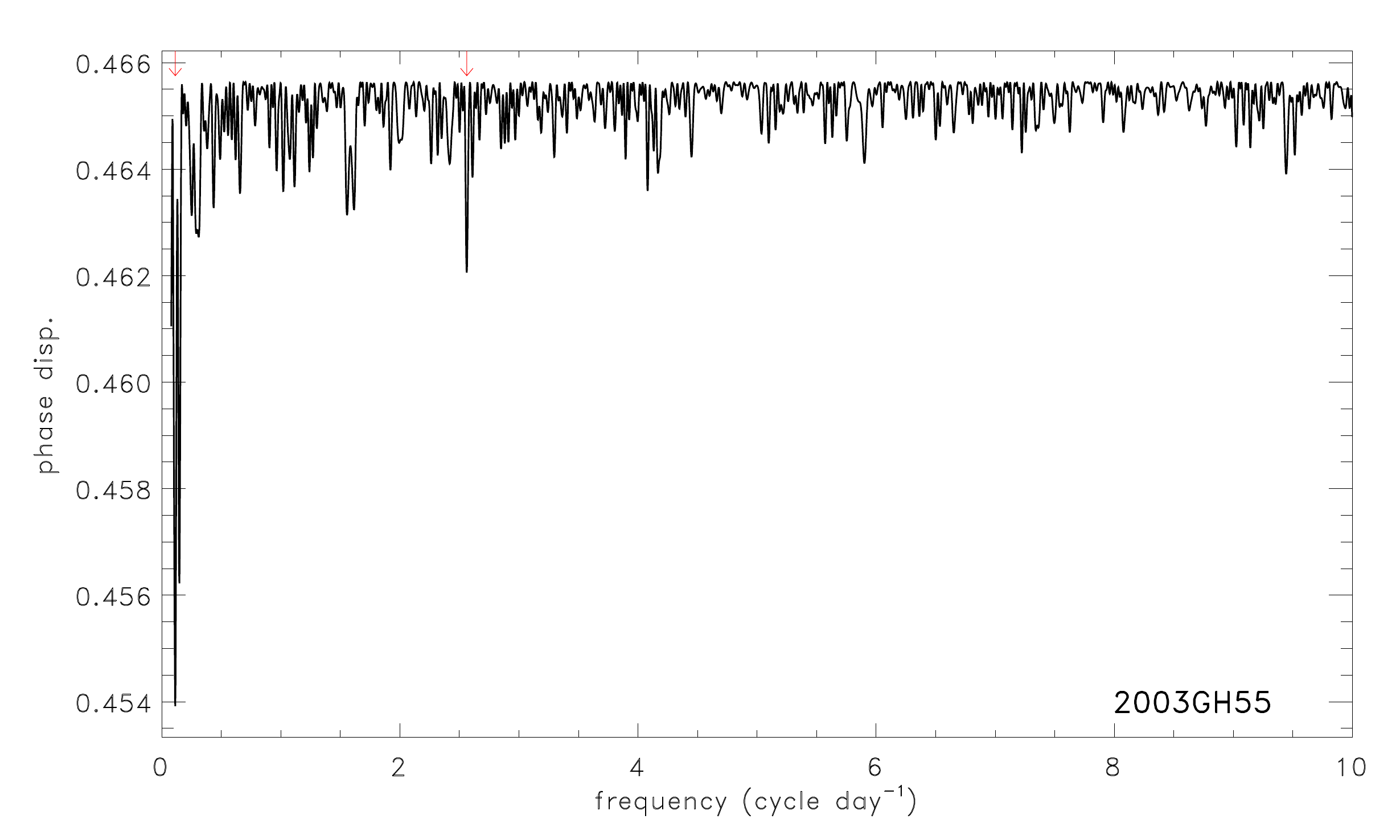}
\includegraphics[width=0.33\textwidth]{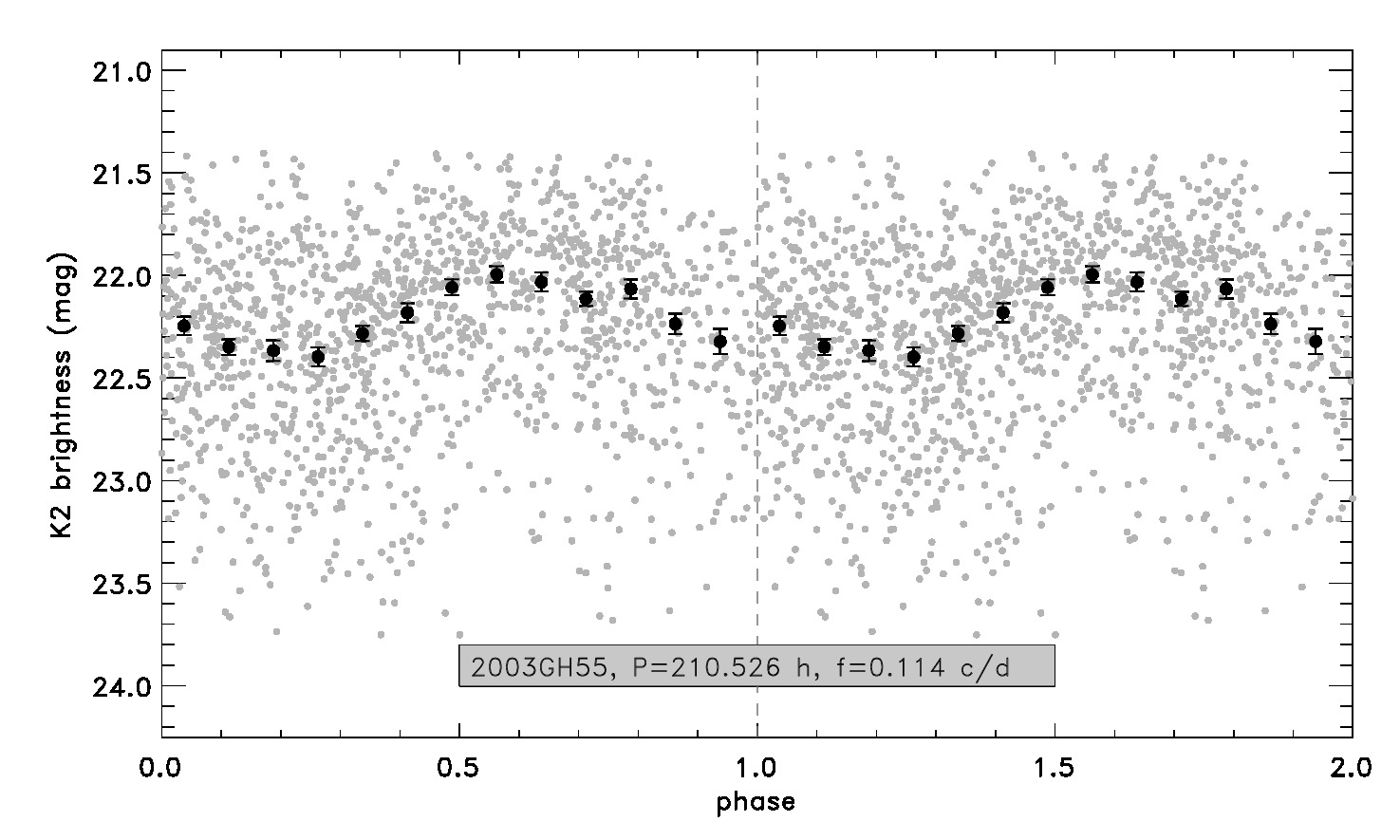}
\includegraphics[width=0.33\textwidth]{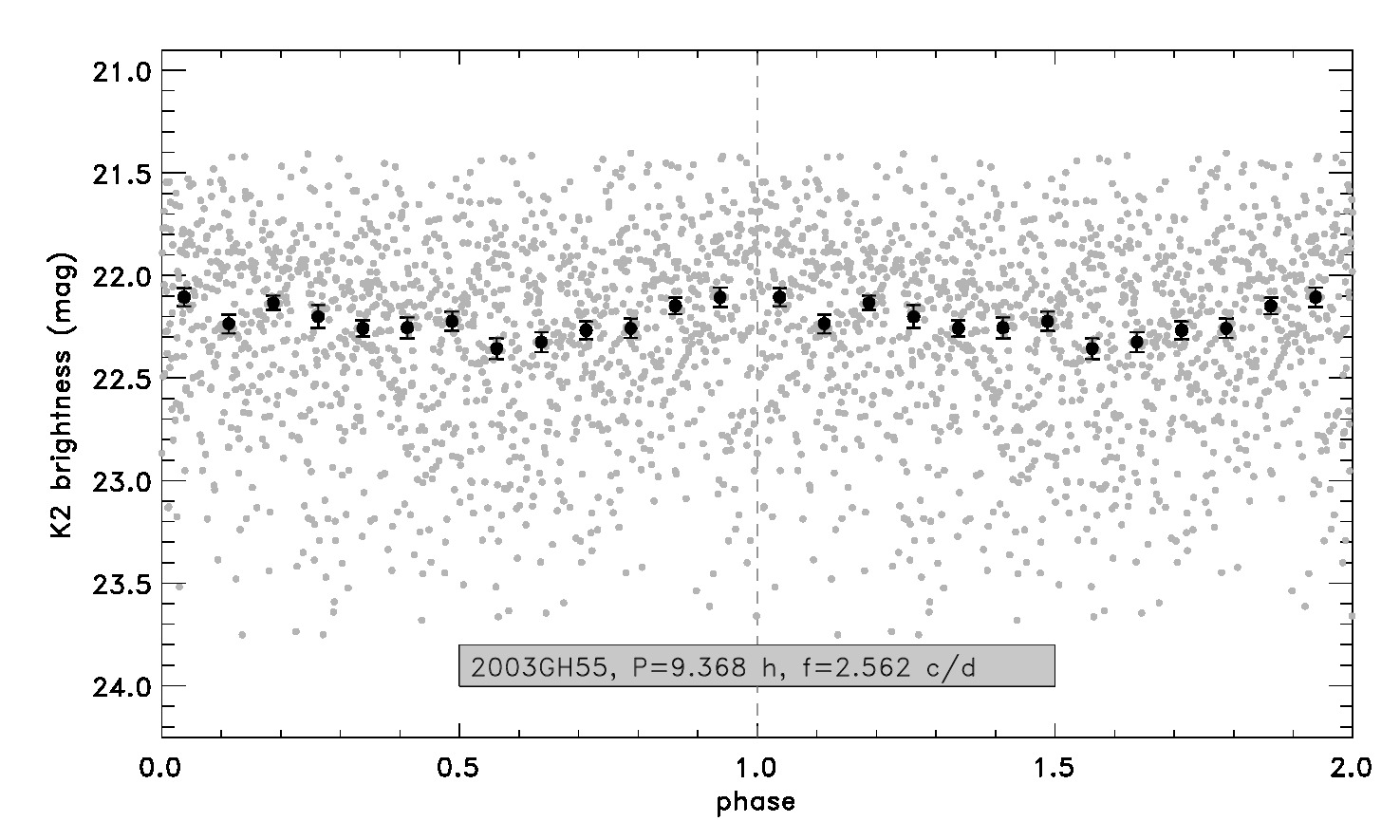}
}
\hbox{
\includegraphics[width=0.33\textwidth]{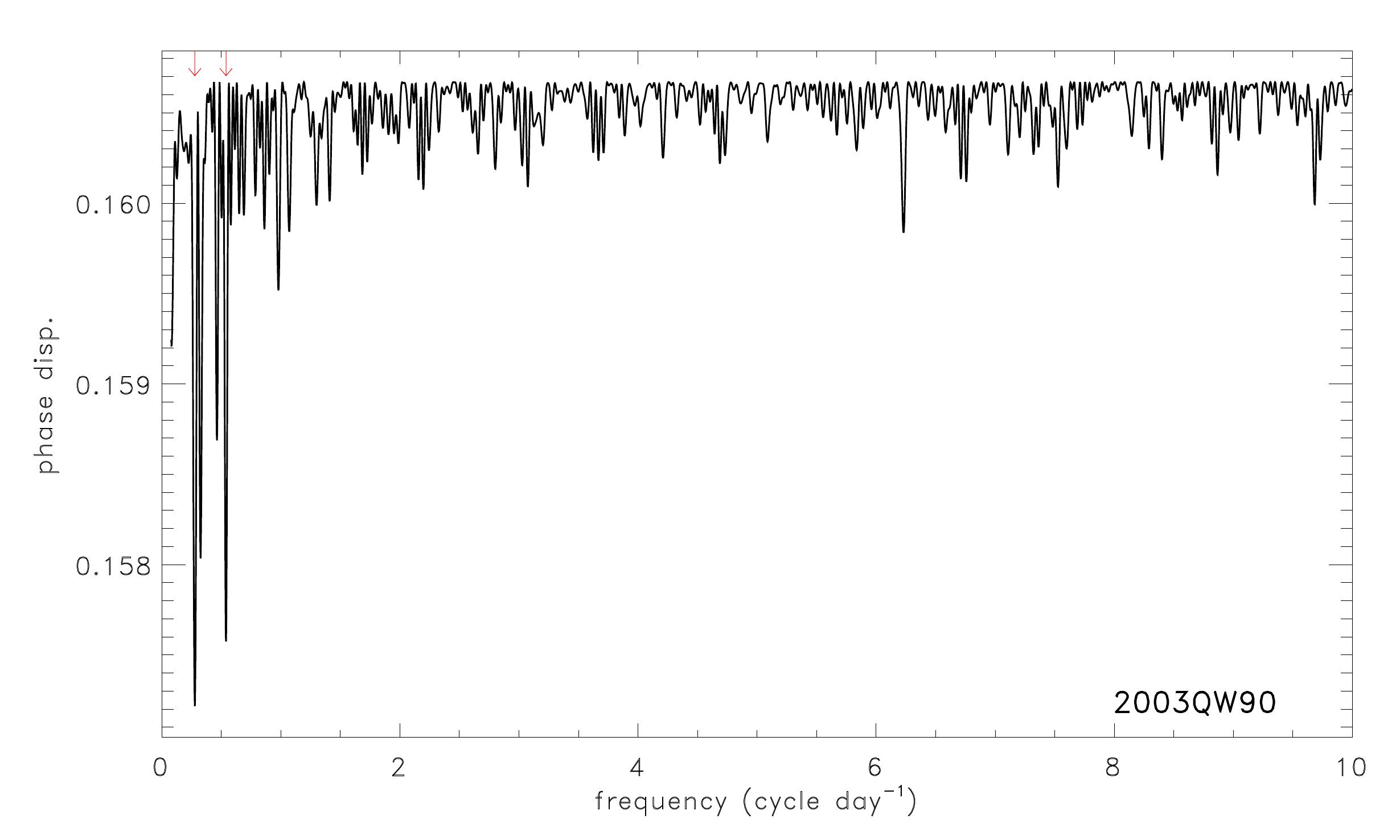}
\includegraphics[width=0.33\textwidth]{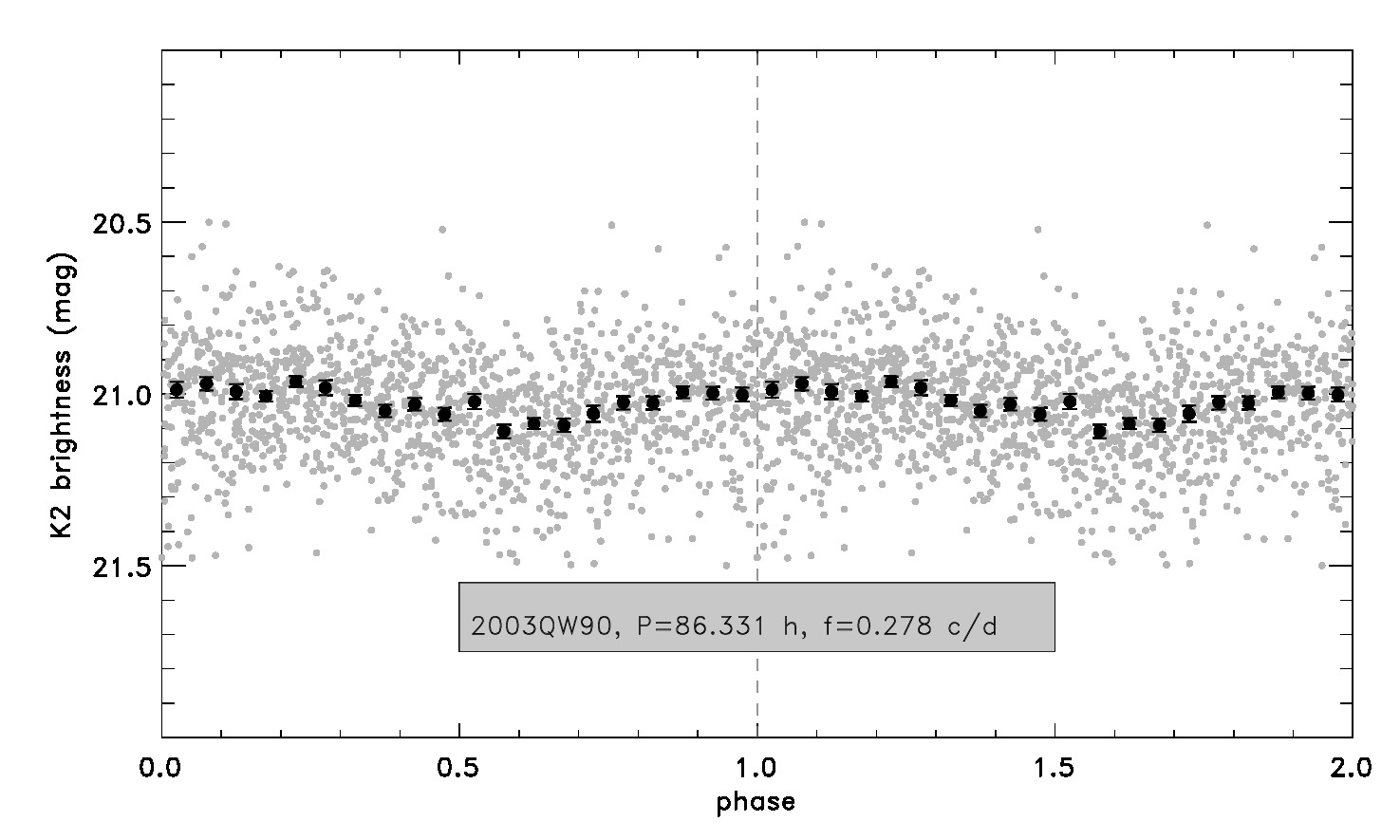}
\includegraphics[width=0.33\textwidth]{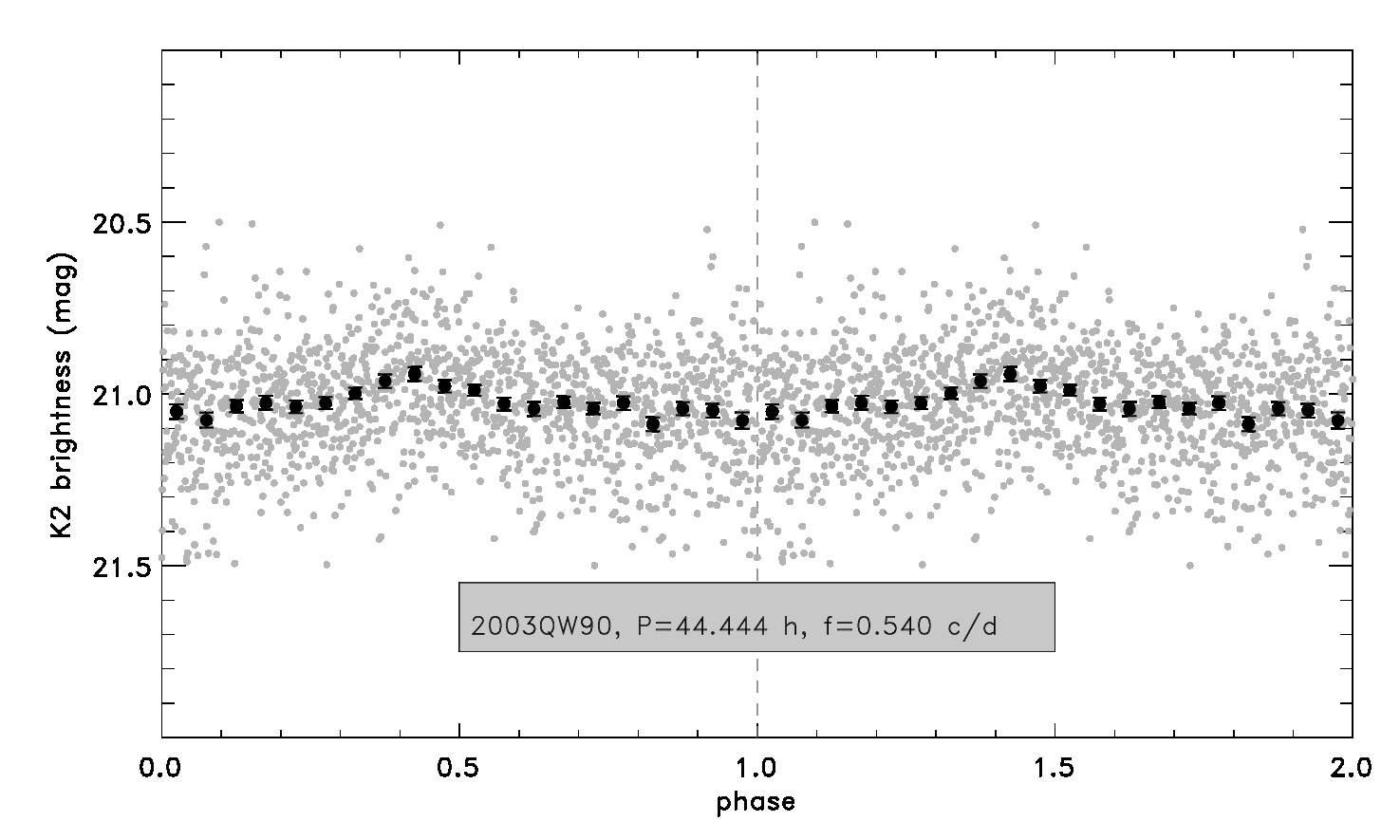}
}
\hbox{
\includegraphics[width=0.33\textwidth]{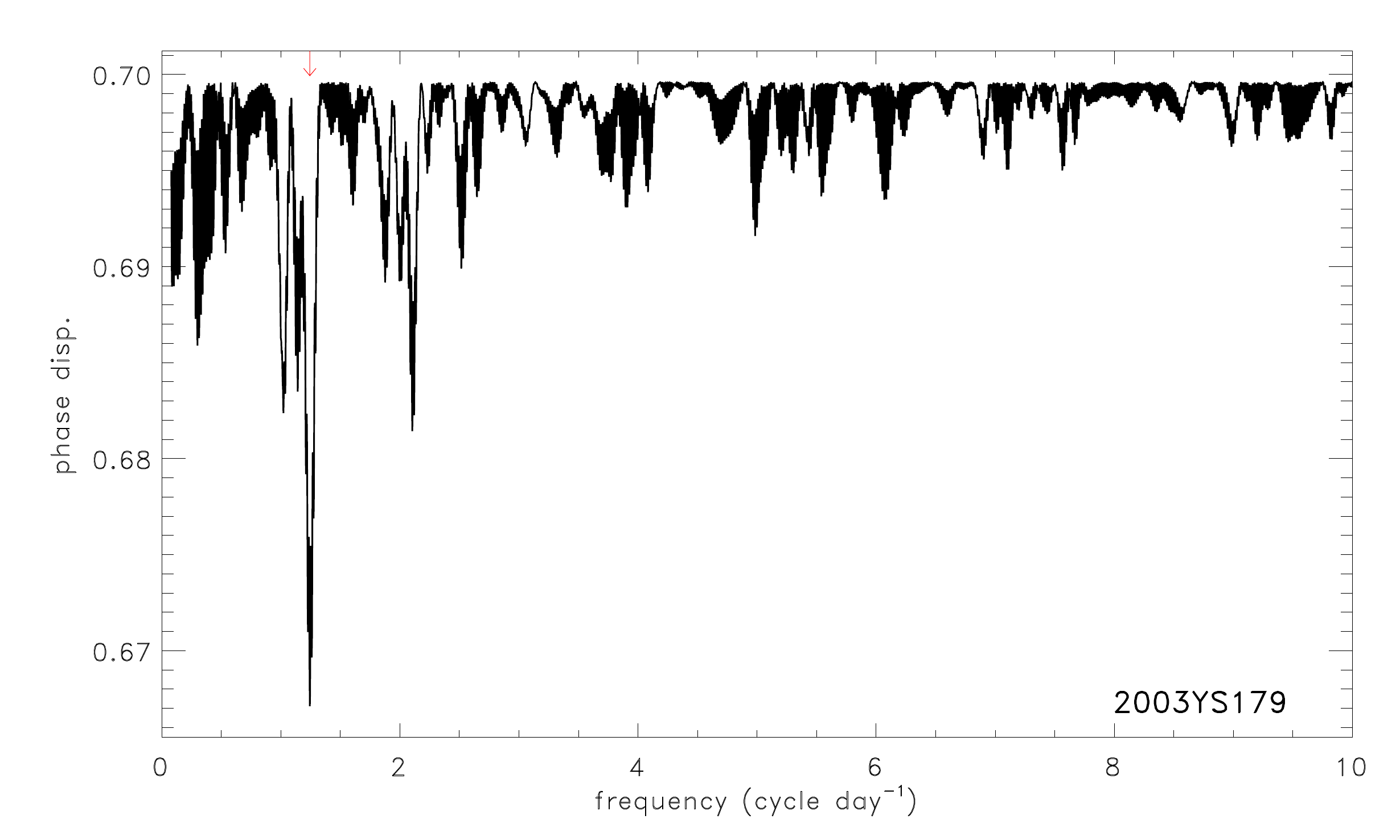}
\includegraphics[width=0.33\textwidth]{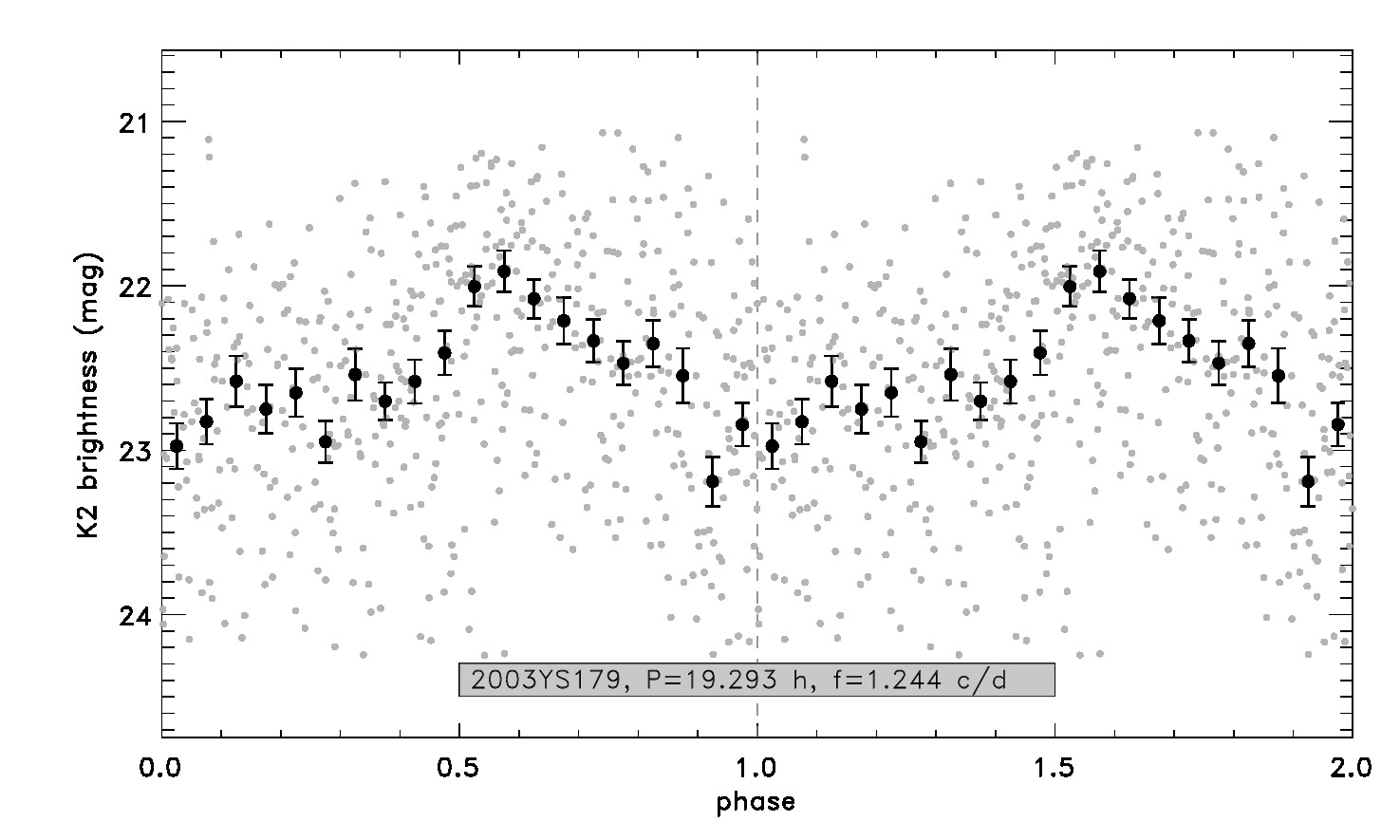}
}
\hbox{
\includegraphics[width=0.33\textwidth]{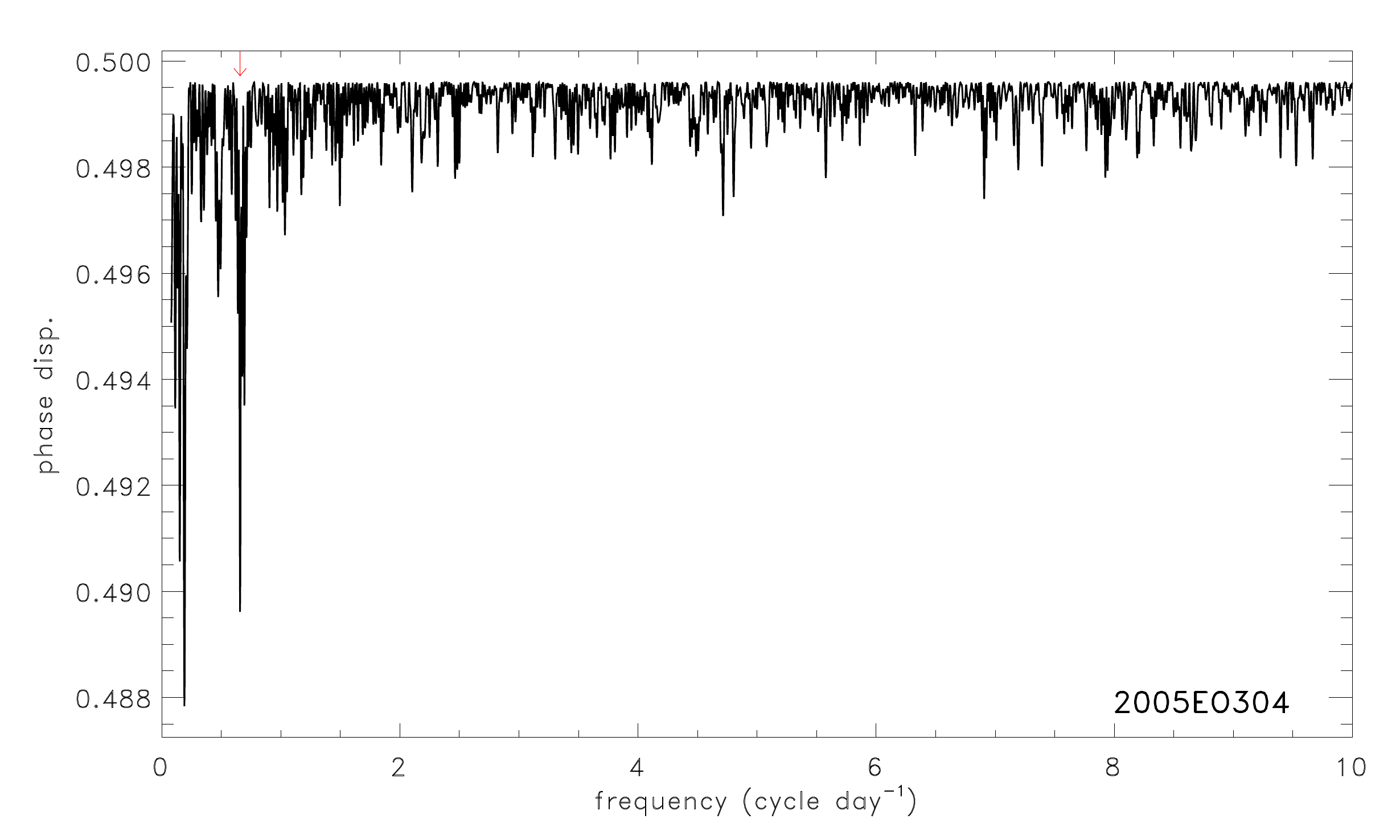}
\includegraphics[width=0.33\textwidth]{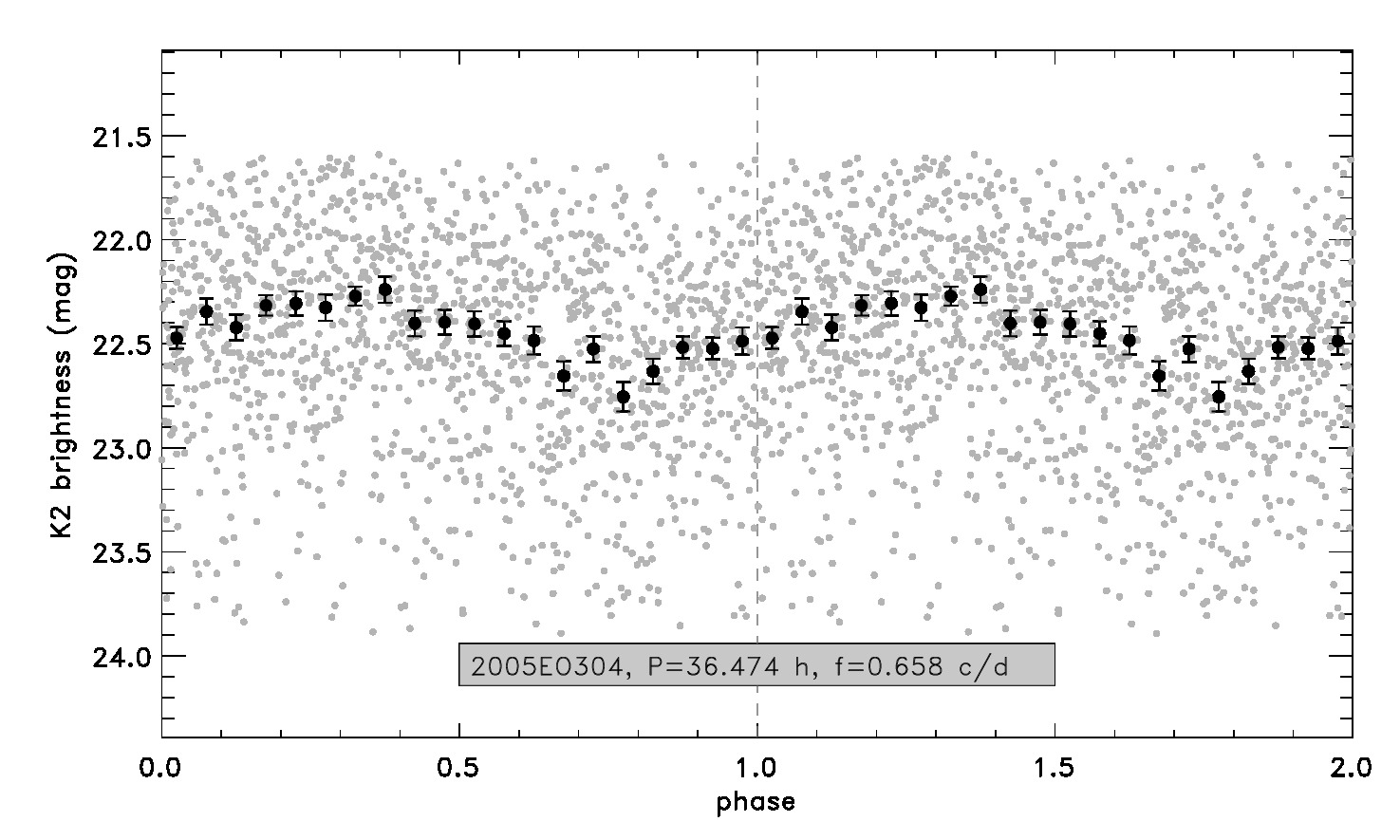}
}
\hbox{
\includegraphics[width=0.33\textwidth]{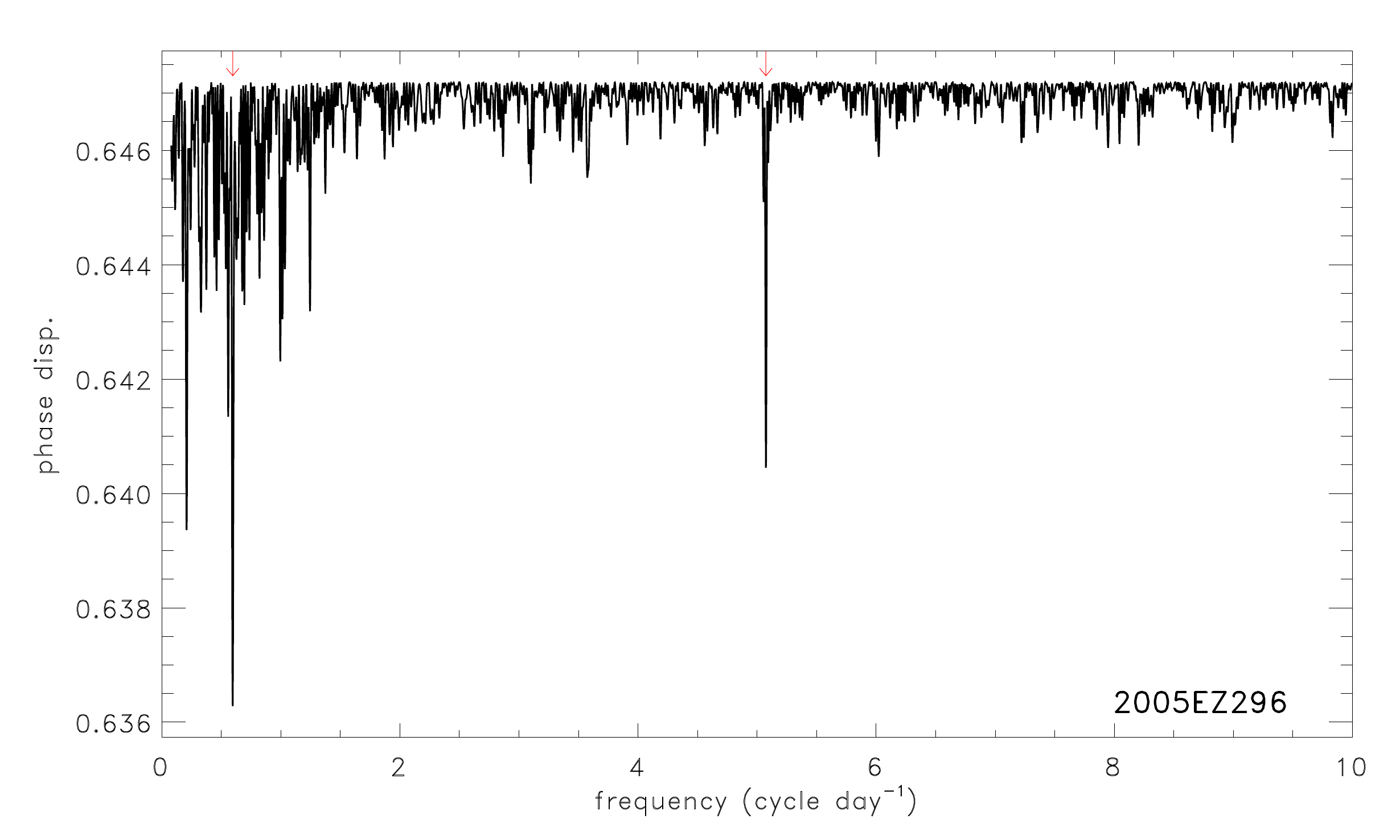}
\includegraphics[width=0.33\textwidth]{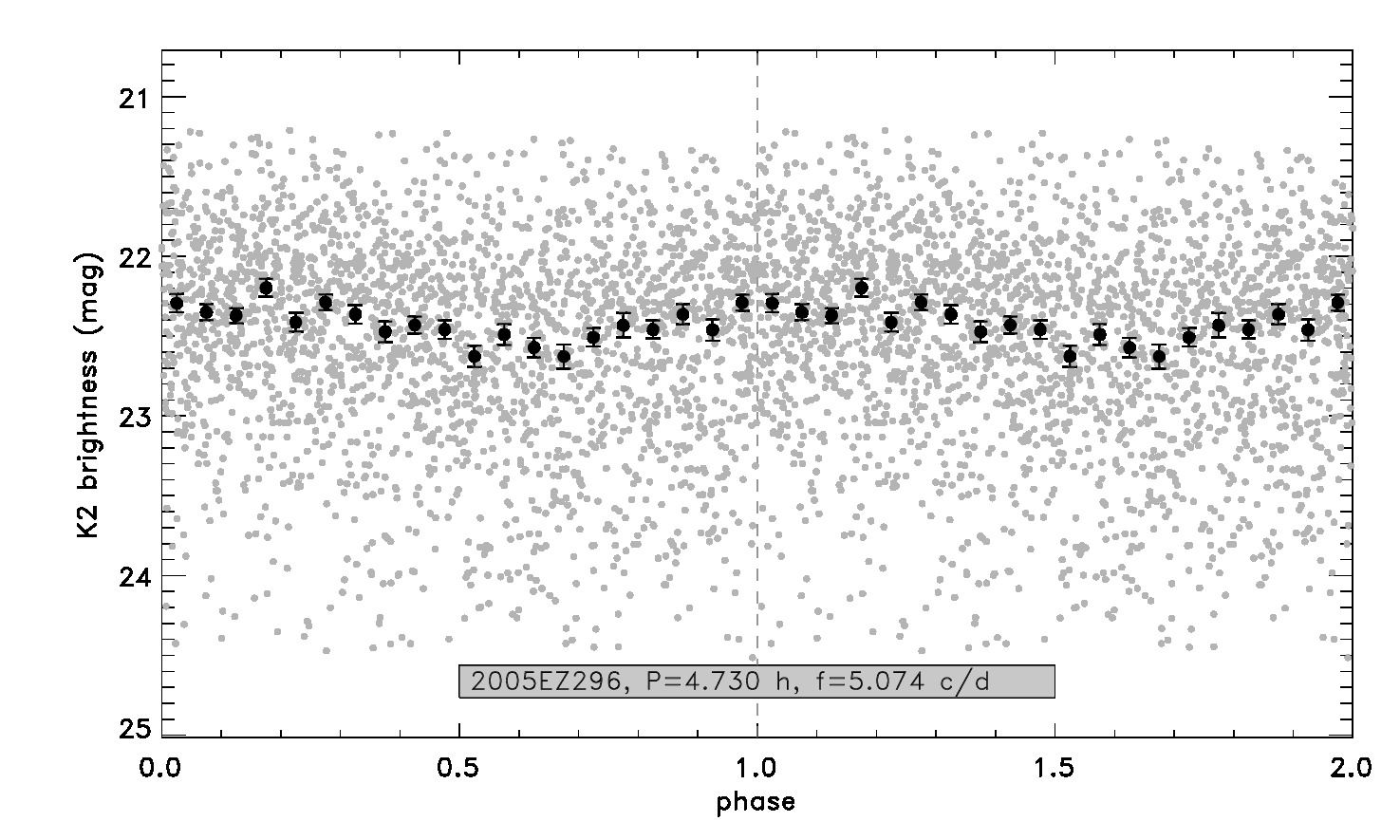}
\includegraphics[width=0.33\textwidth]{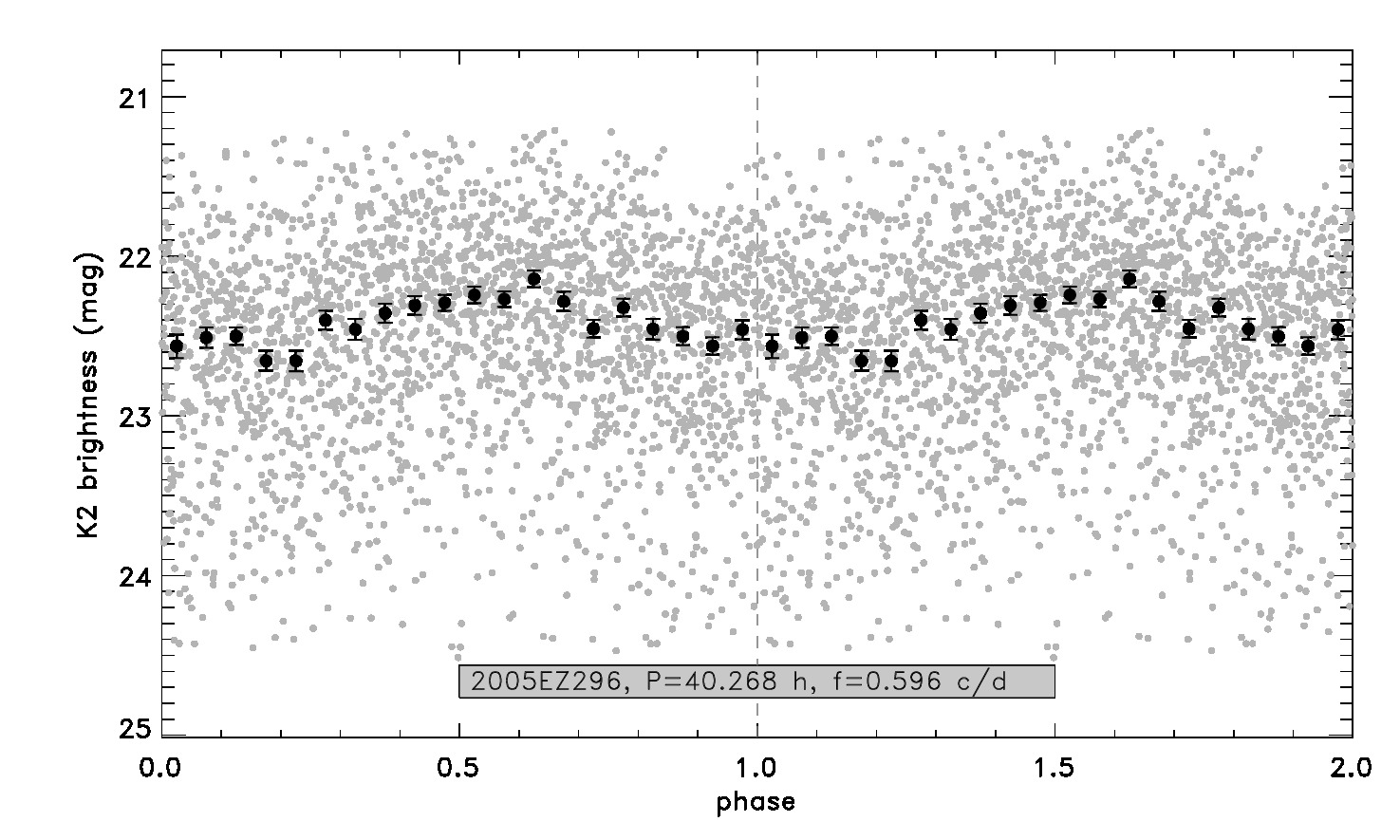}
}
\caption{
}
\label{fig:big2}
\end{figure*}

\begin{figure*}[ht!]
\ContinuedFloat
\hbox{
\includegraphics[width=0.33\textwidth]{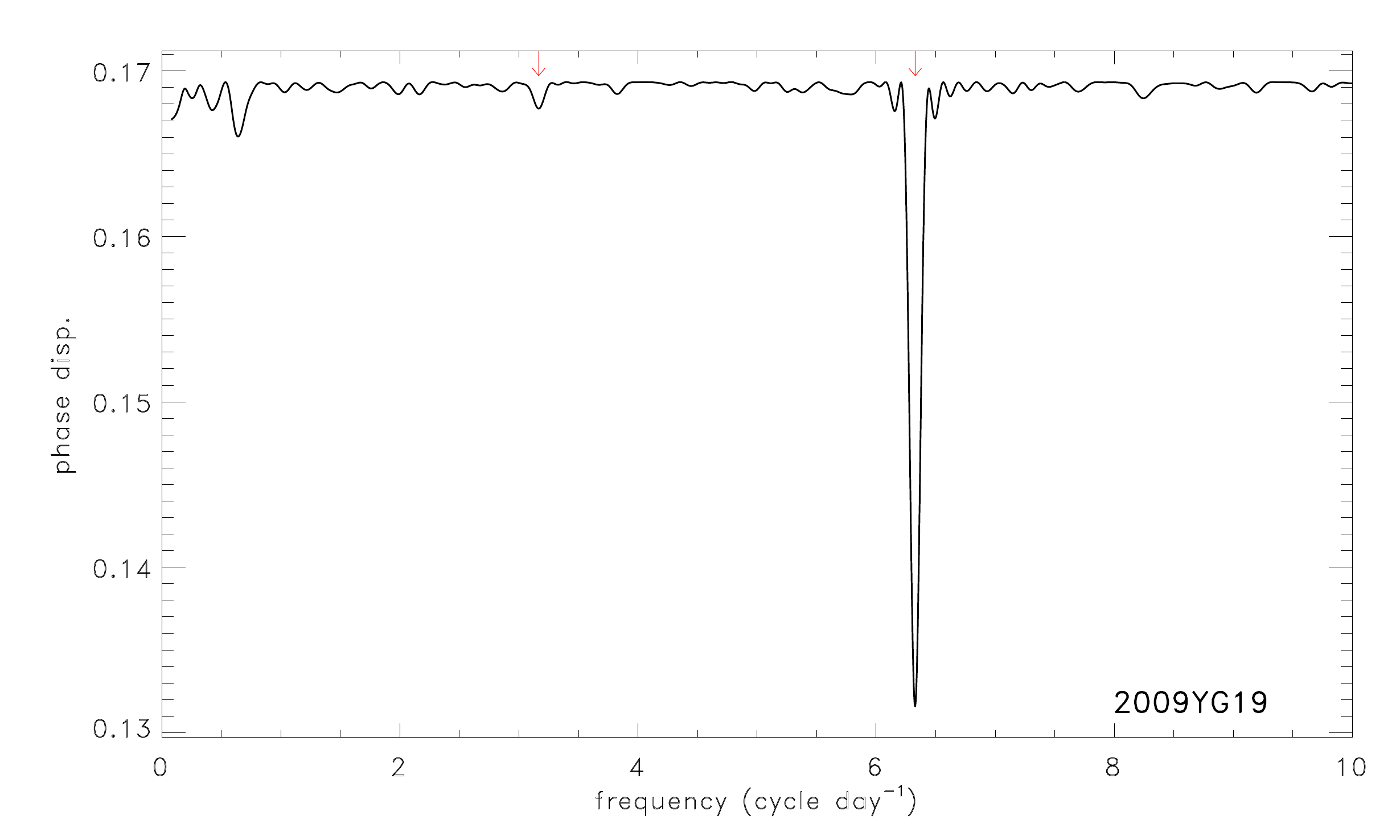}
\includegraphics[width=0.33\textwidth]{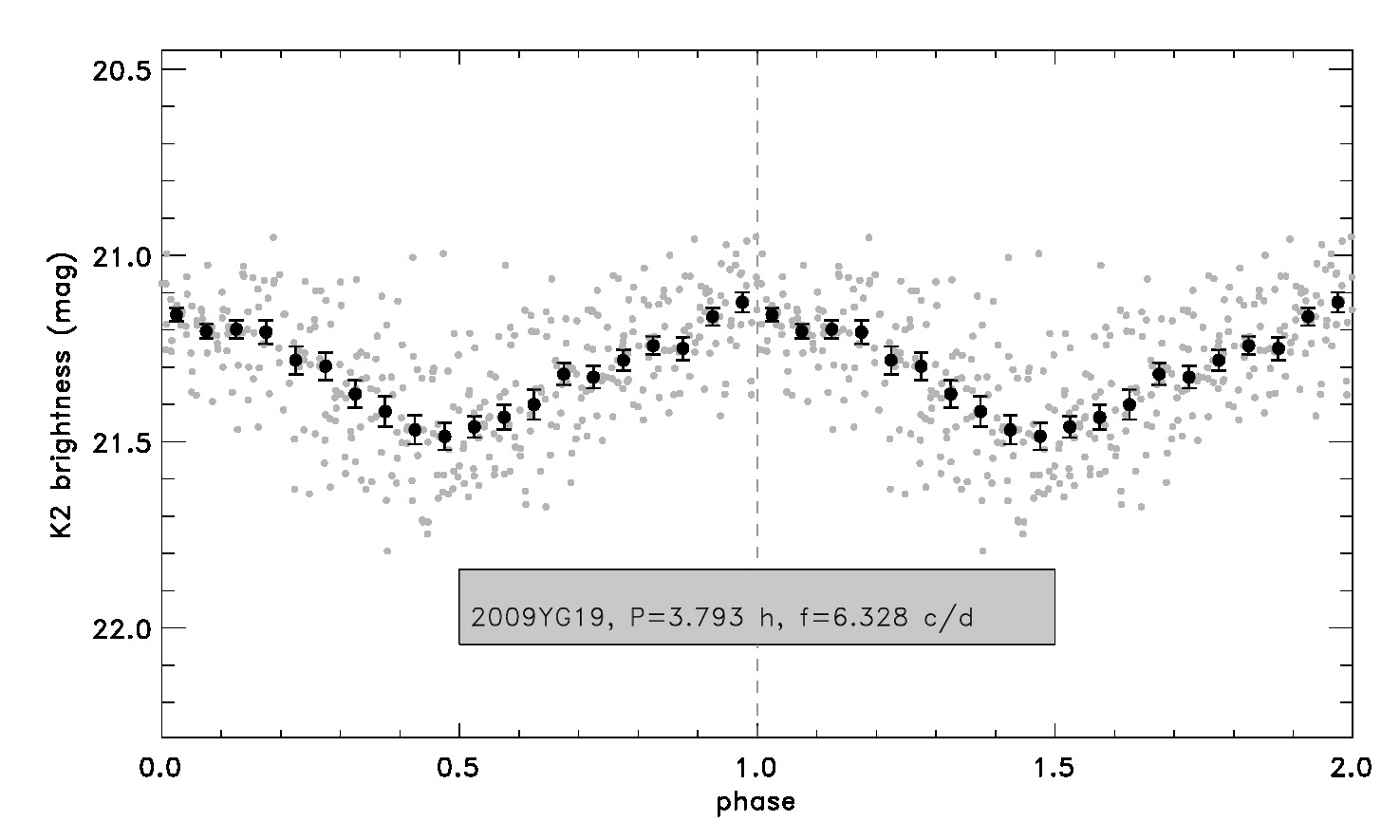}
\includegraphics[width=0.33\textwidth]{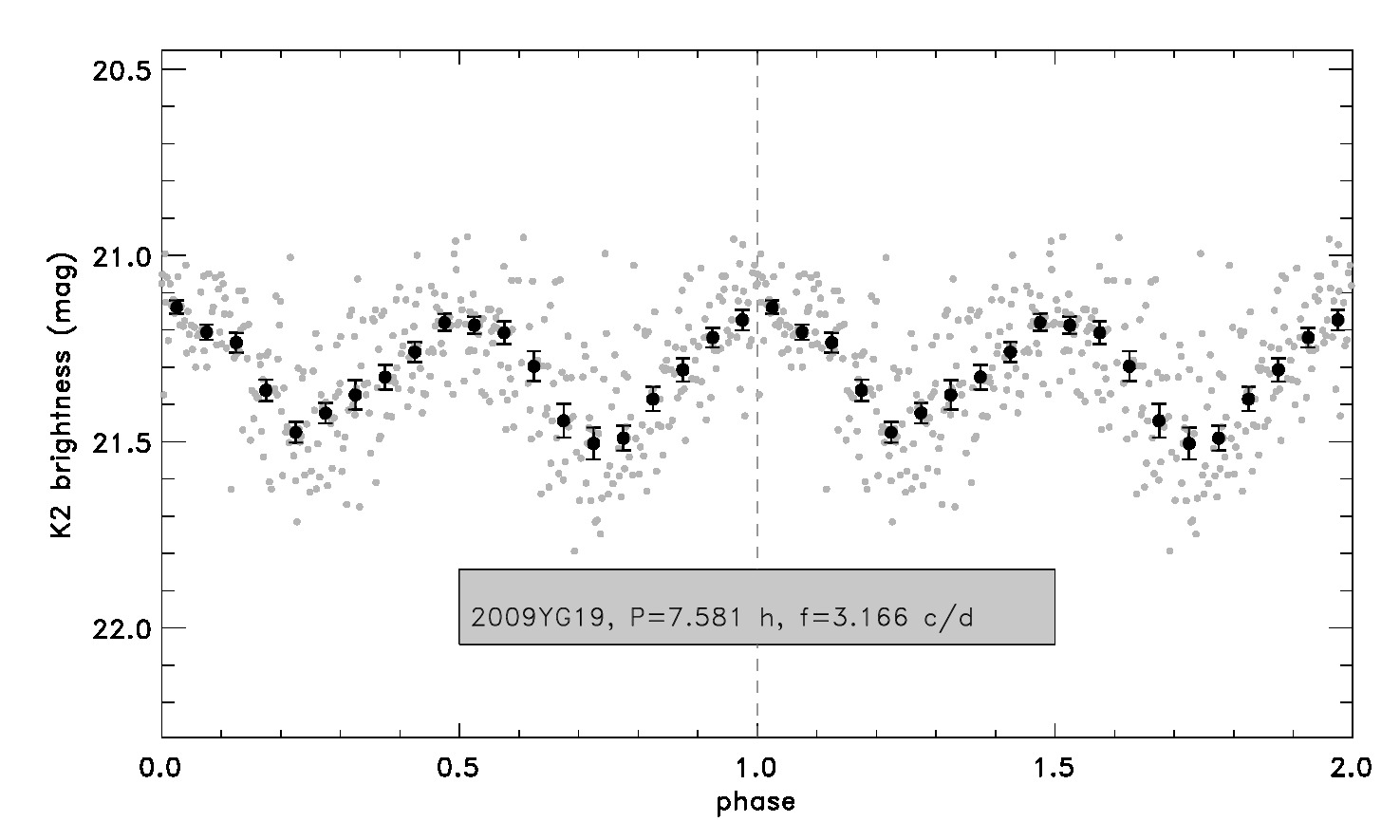}
}
\hbox{
\includegraphics[width=0.33\textwidth]{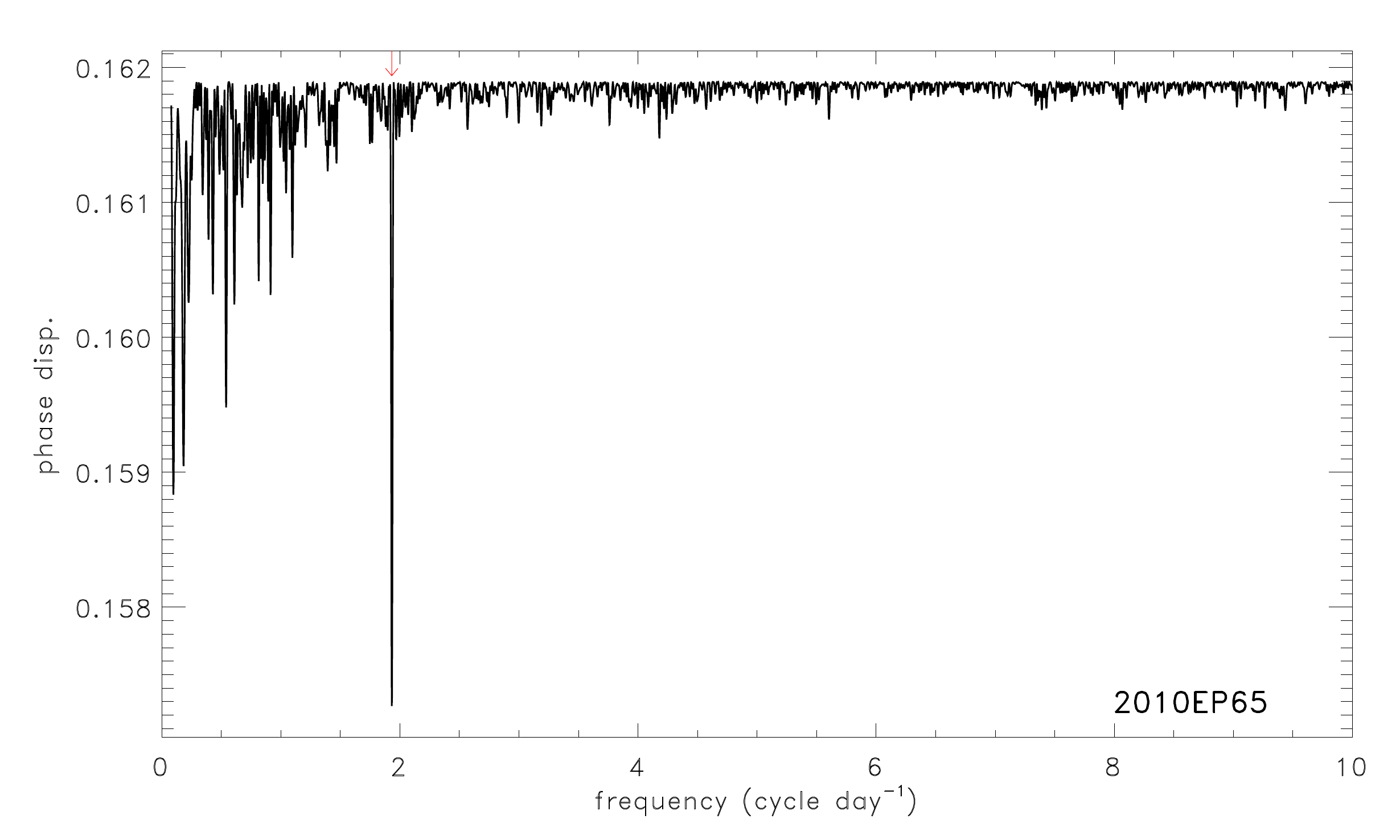}
\includegraphics[width=0.33\textwidth]{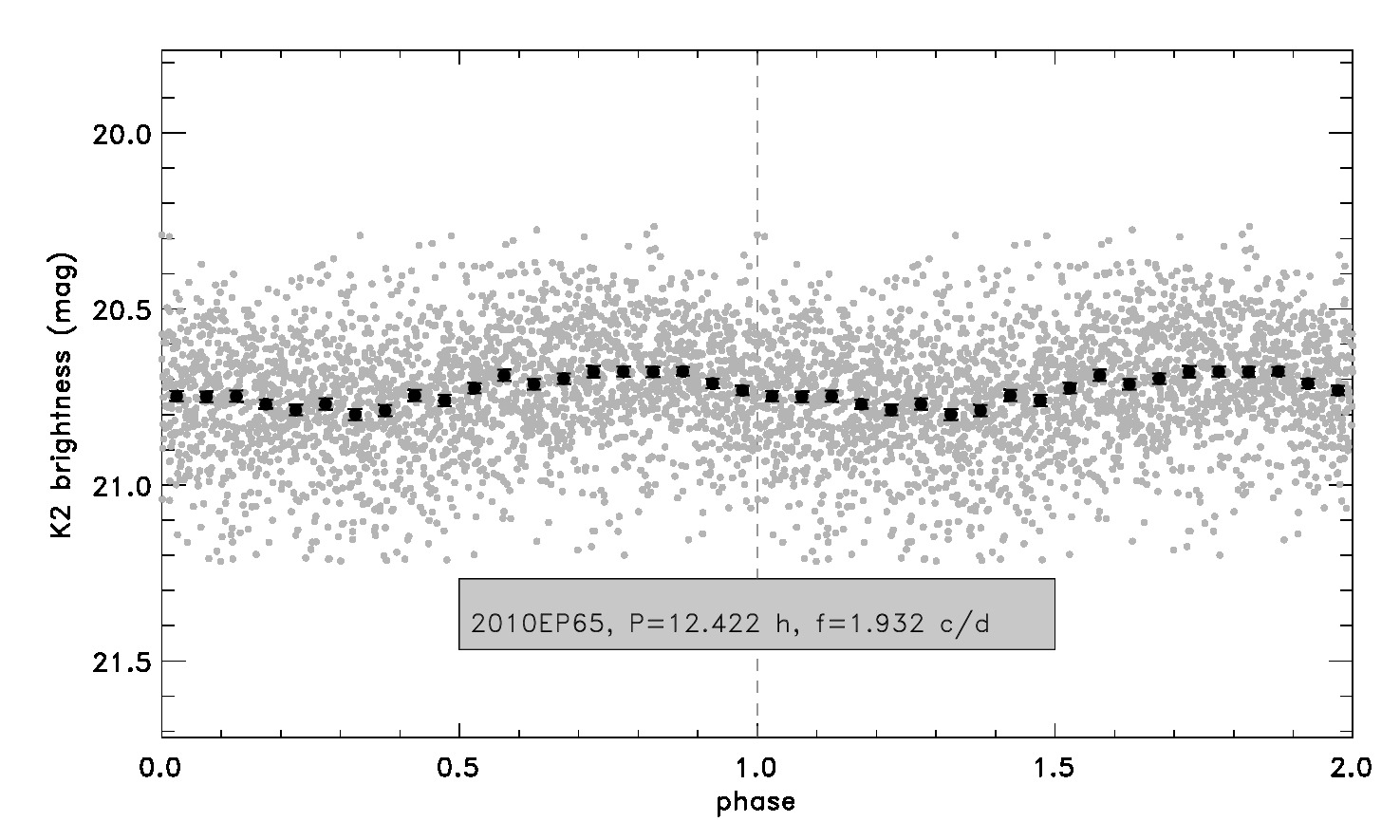}
}
\hbox{
\includegraphics[width=0.33\textwidth]{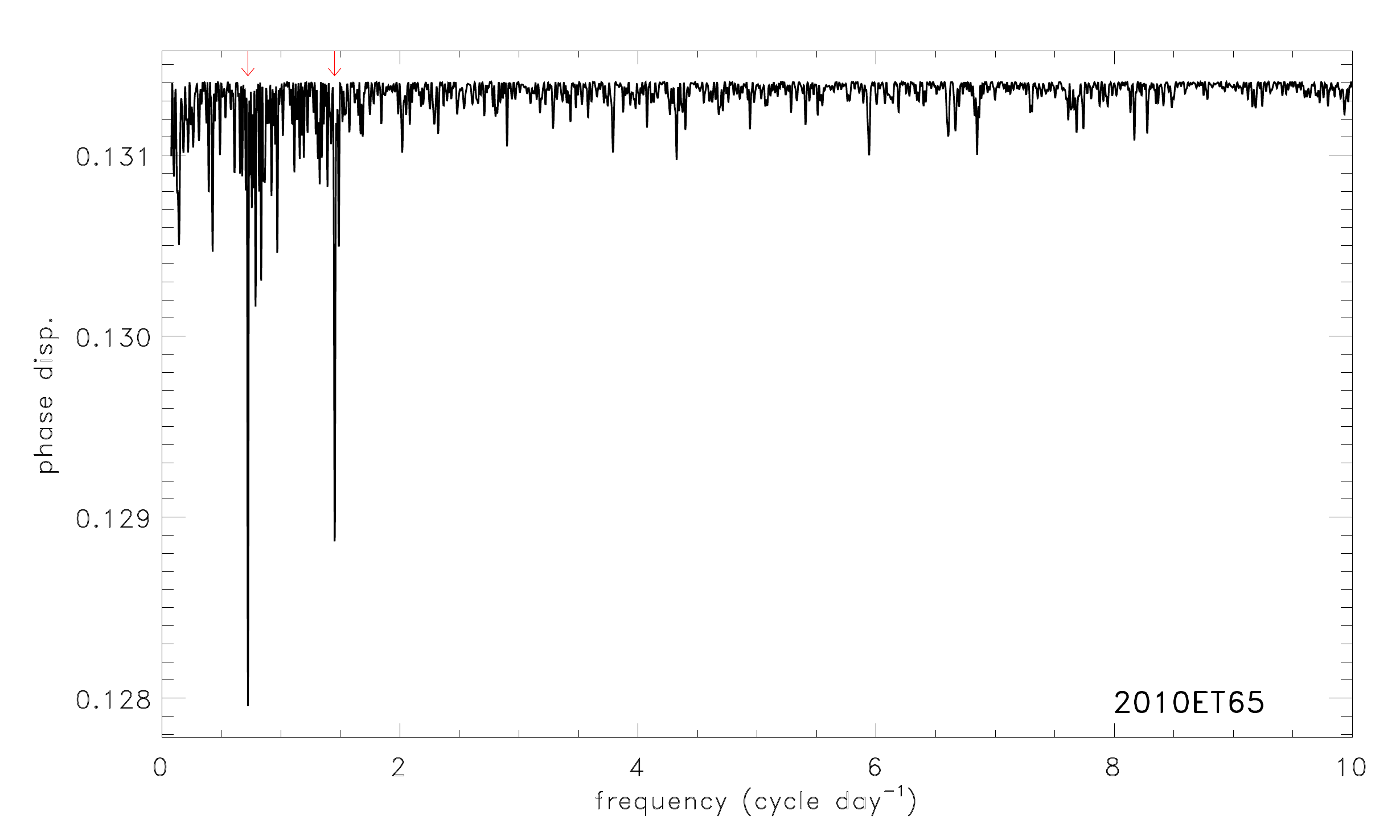}
\includegraphics[width=0.33\textwidth]{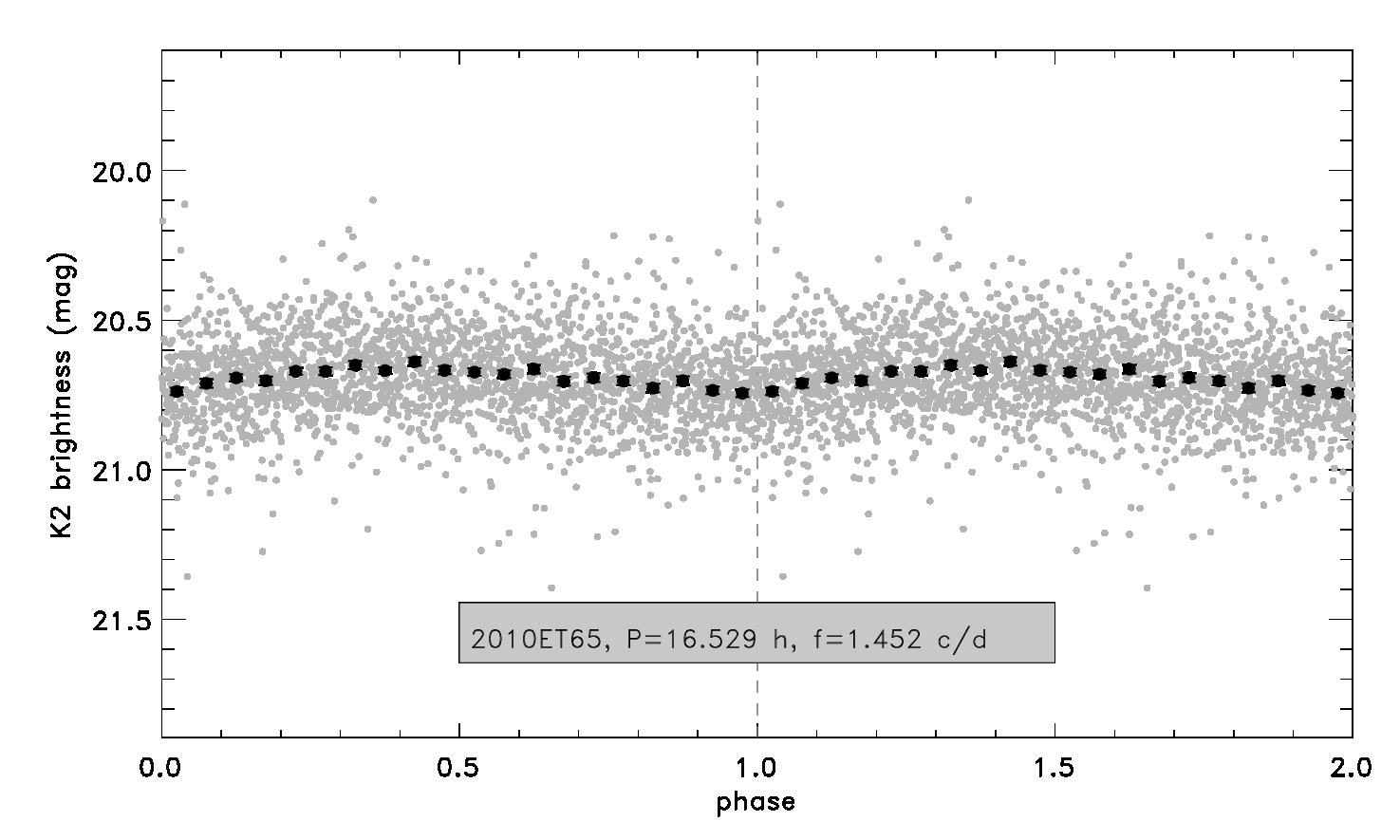}
\includegraphics[width=0.33\textwidth]{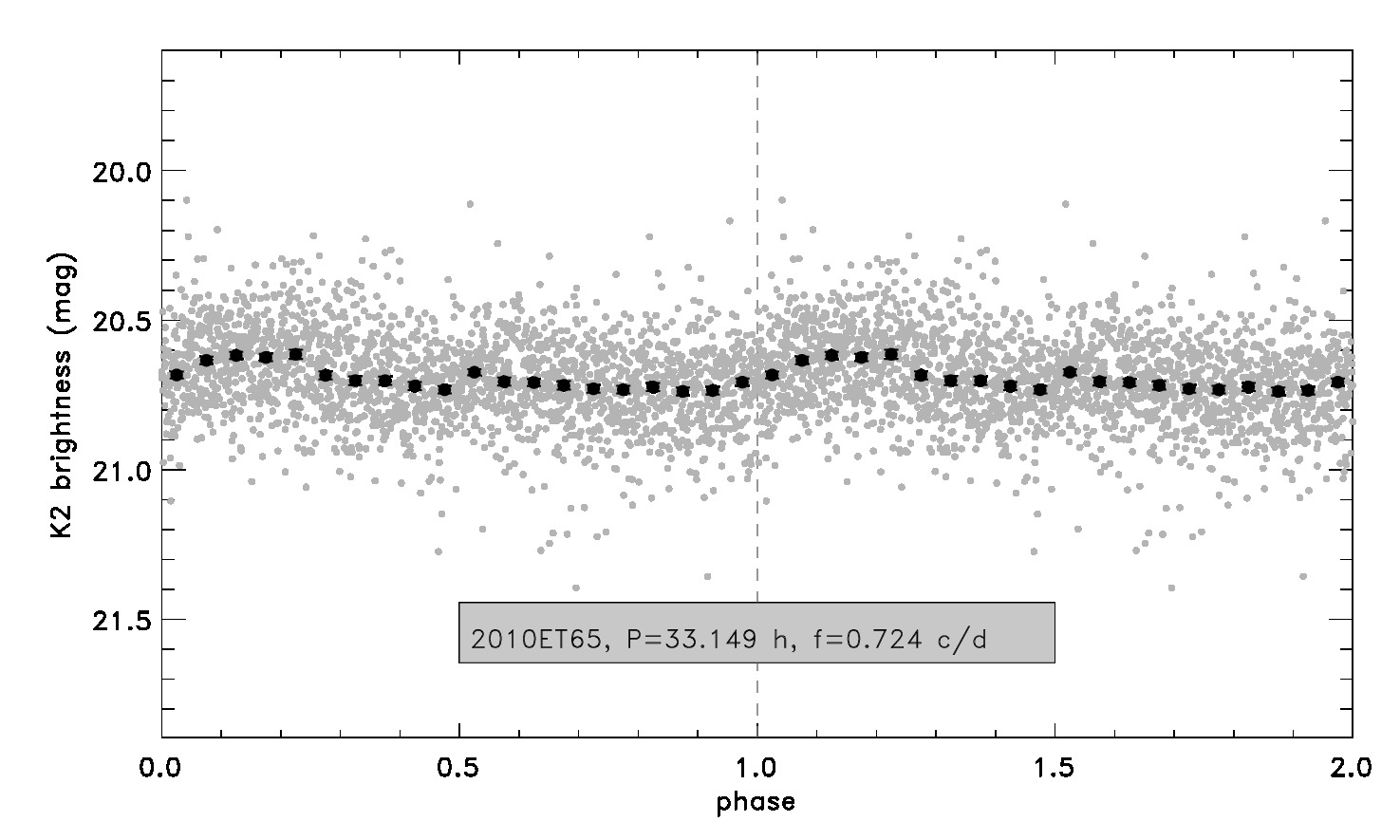}
}
\hbox{
\includegraphics[width=0.33\textwidth]{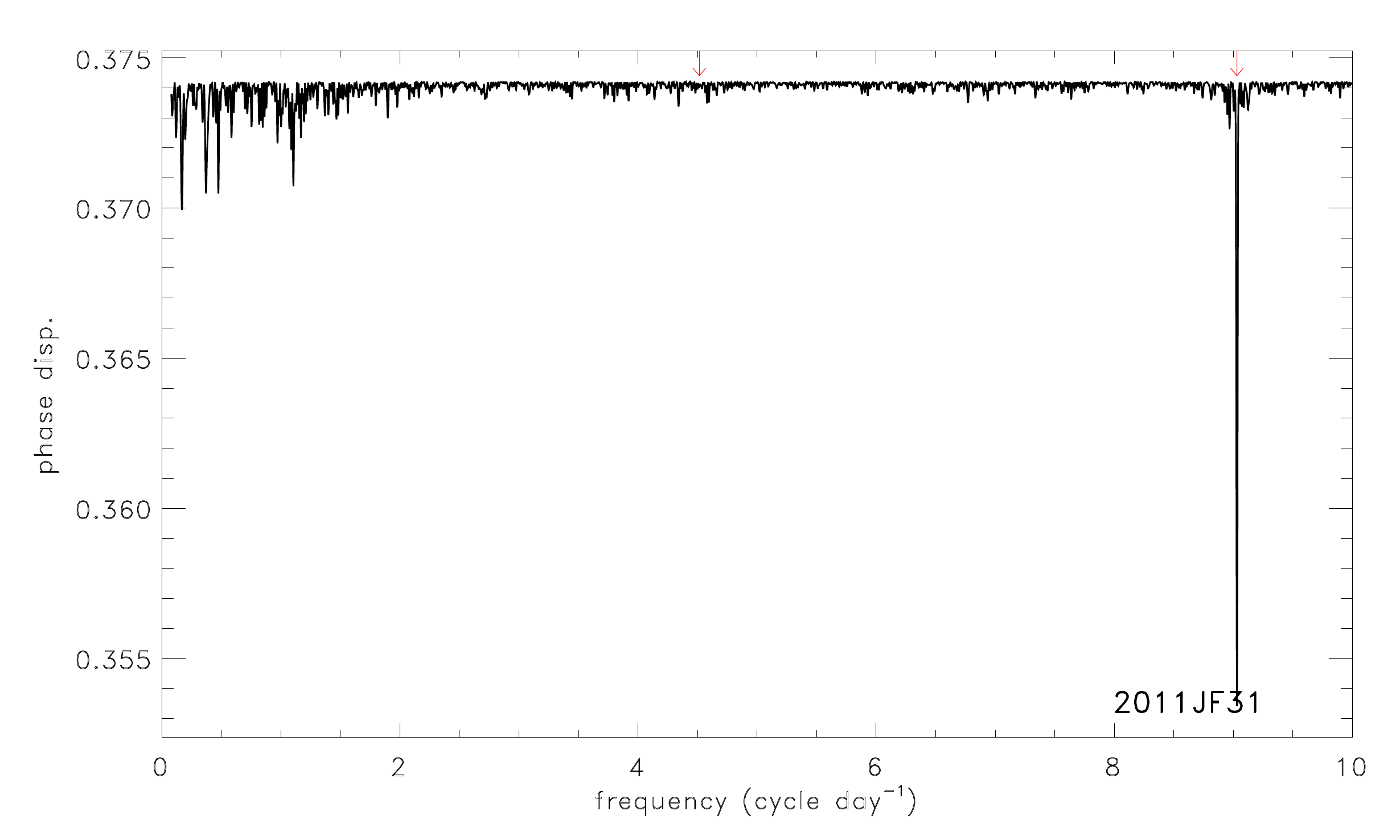}
\includegraphics[width=0.33\textwidth]{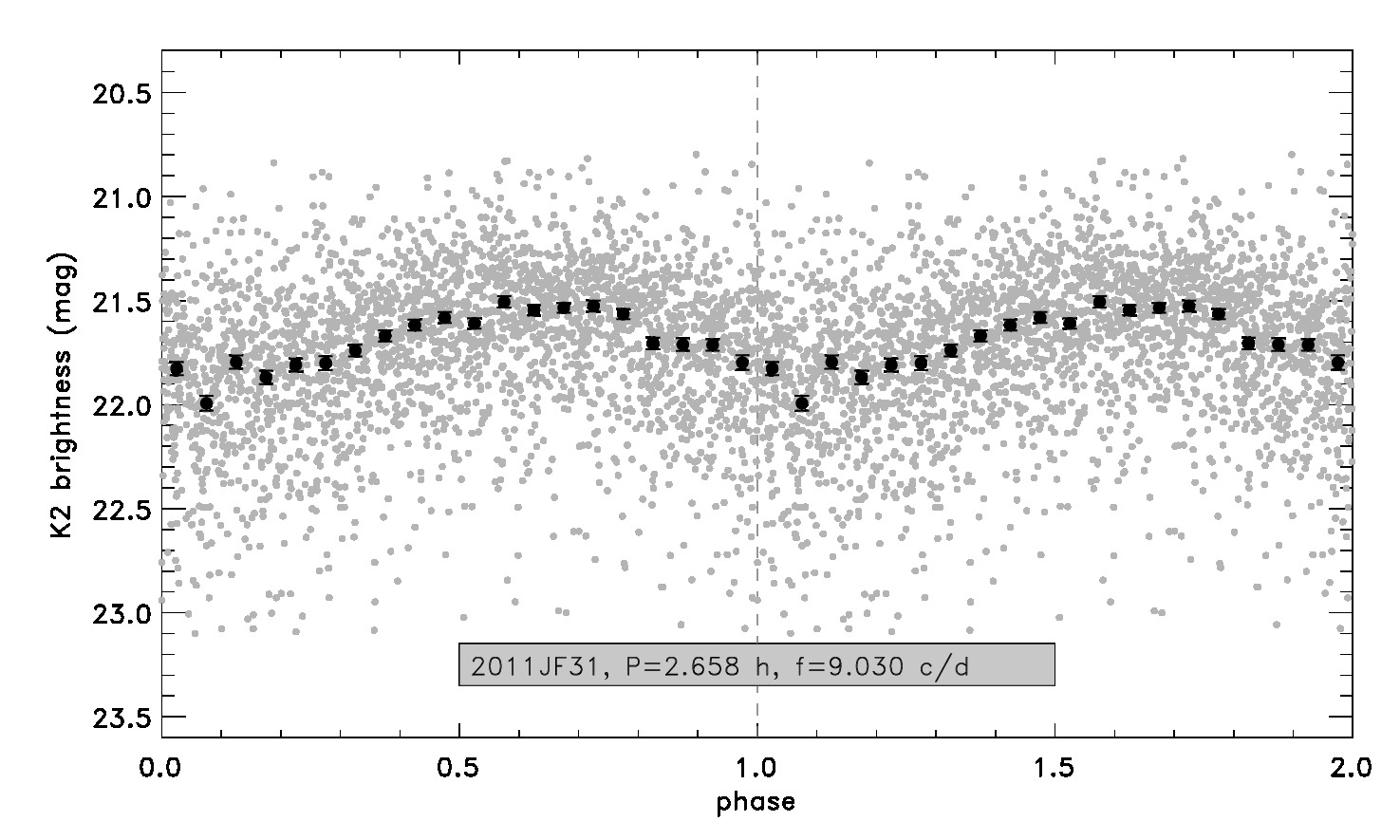}
\includegraphics[width=0.33\textwidth]{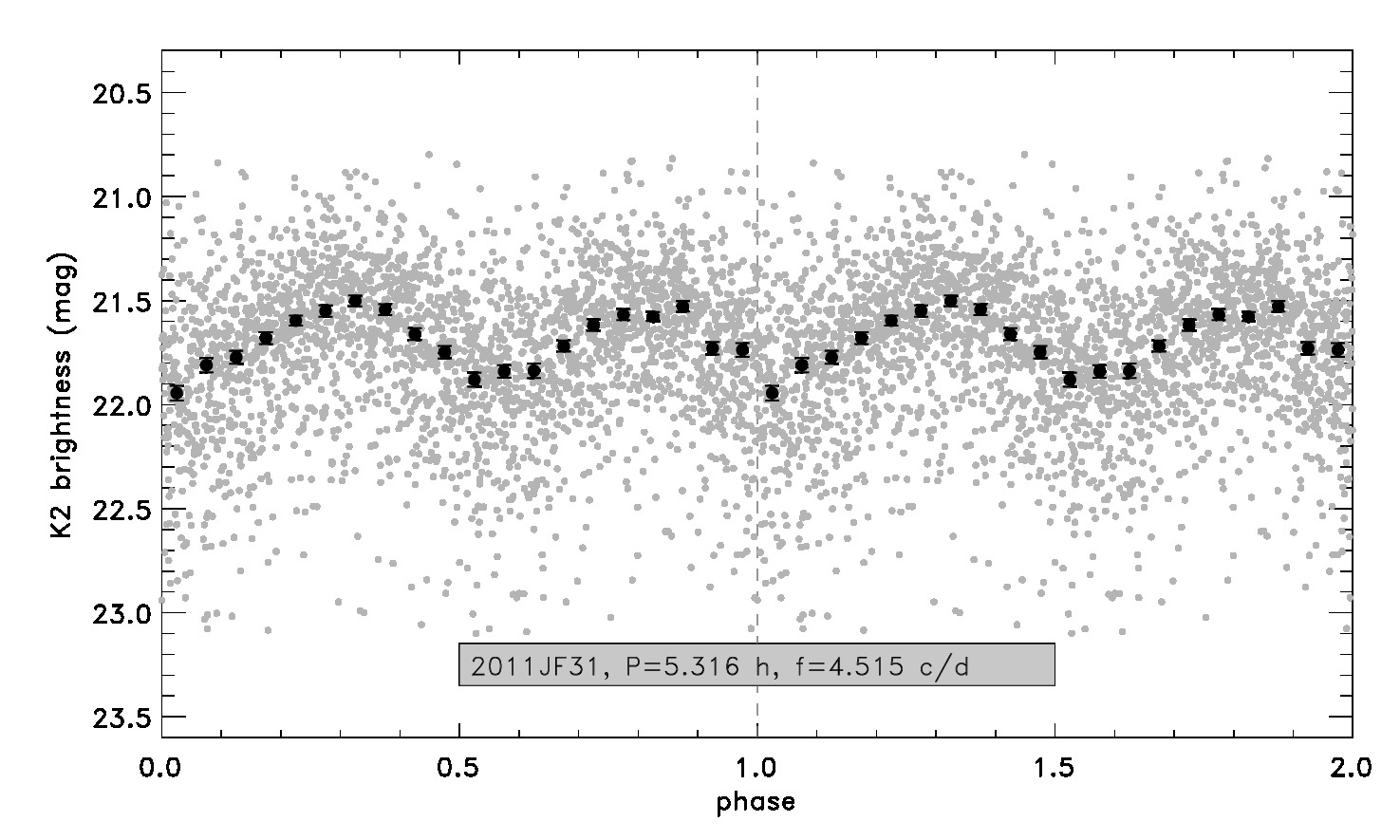}
}
\hbox{
\includegraphics[width=0.33\textwidth]{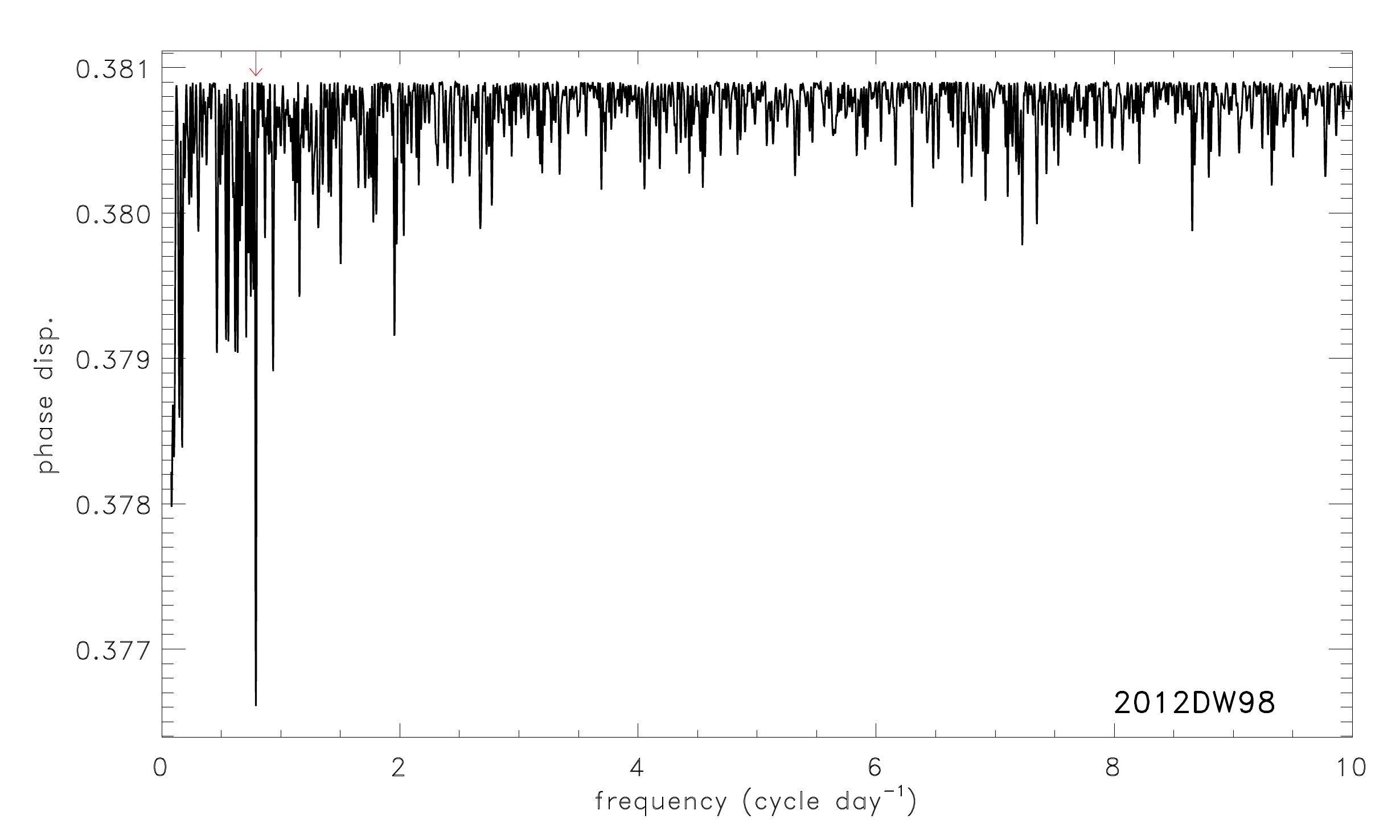}
\includegraphics[width=0.33\textwidth]{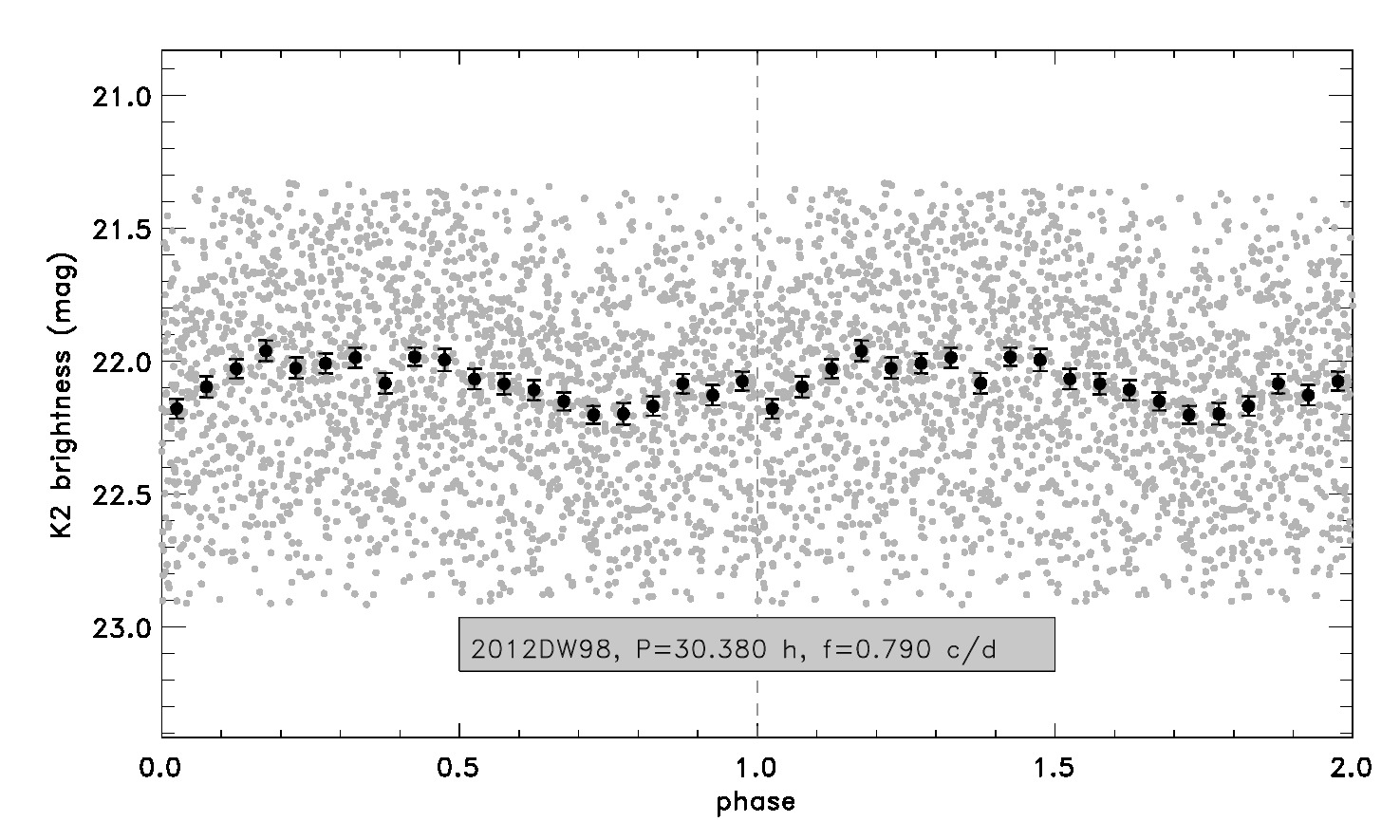}
}
\hbox{
\includegraphics[width=0.33\textwidth]{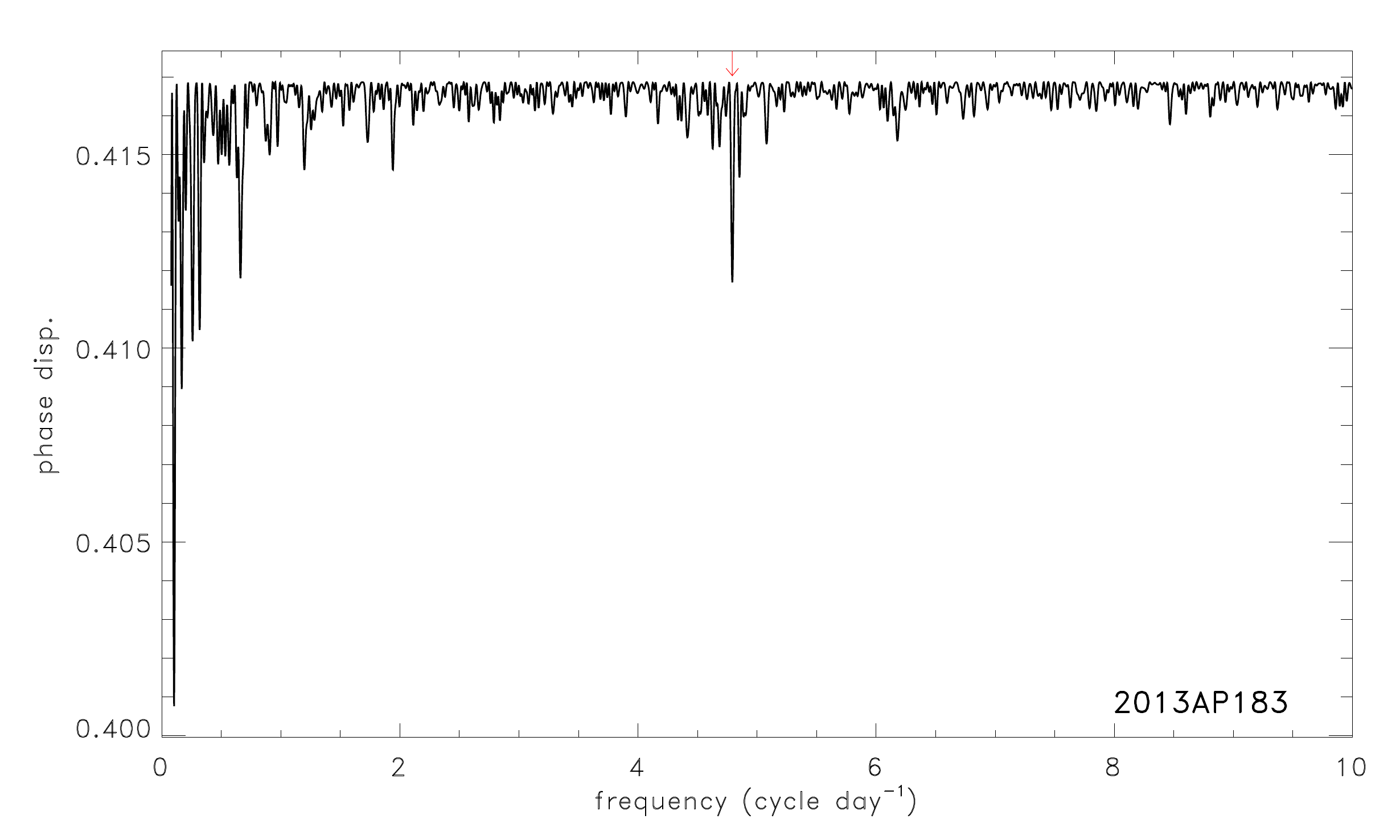}
\includegraphics[width=0.33\textwidth]{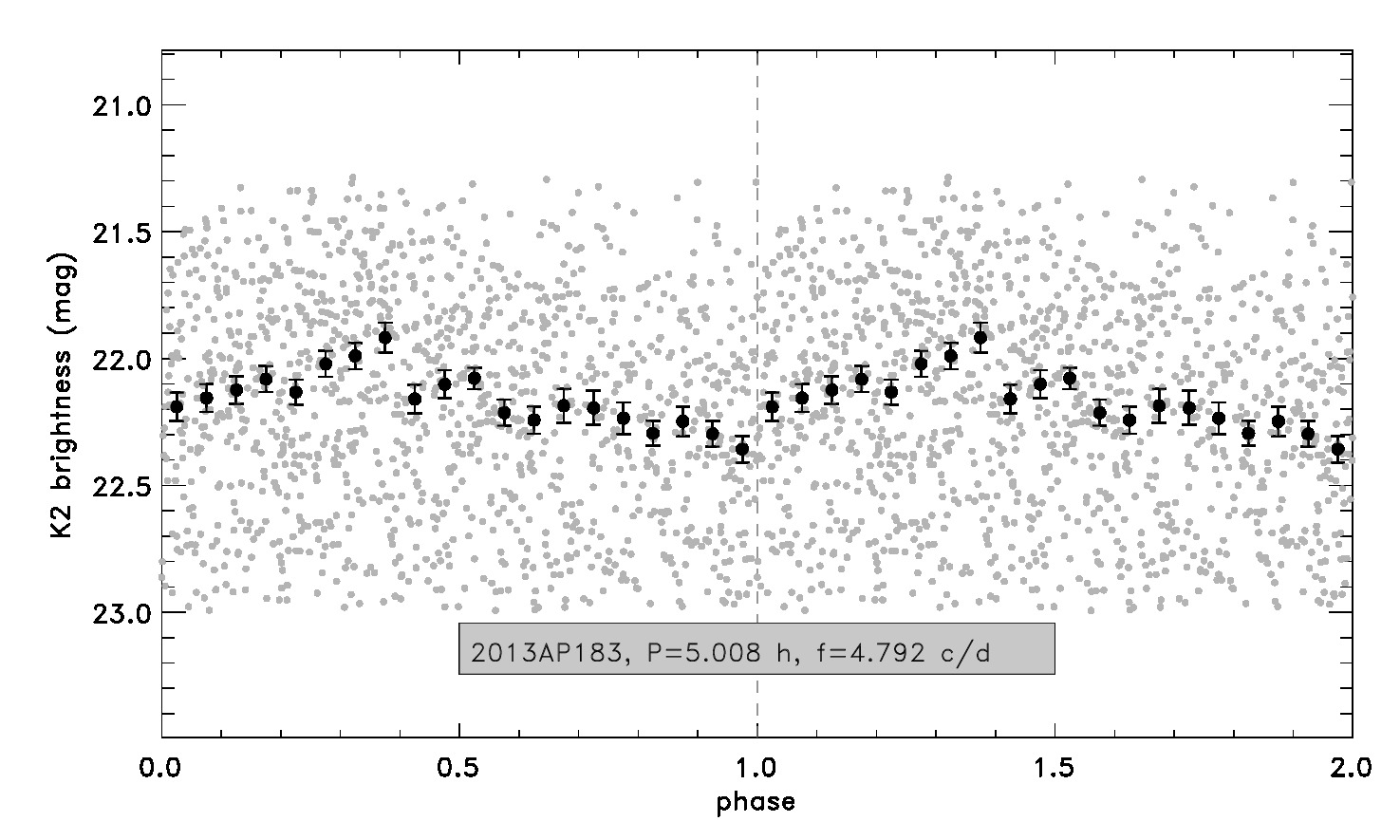}
}
\caption{
}
\label{fig:big3}
\end{figure*}

\begin{figure*}[ht!]
\ContinuedFloat
\hbox{
\includegraphics[width=0.33\textwidth]{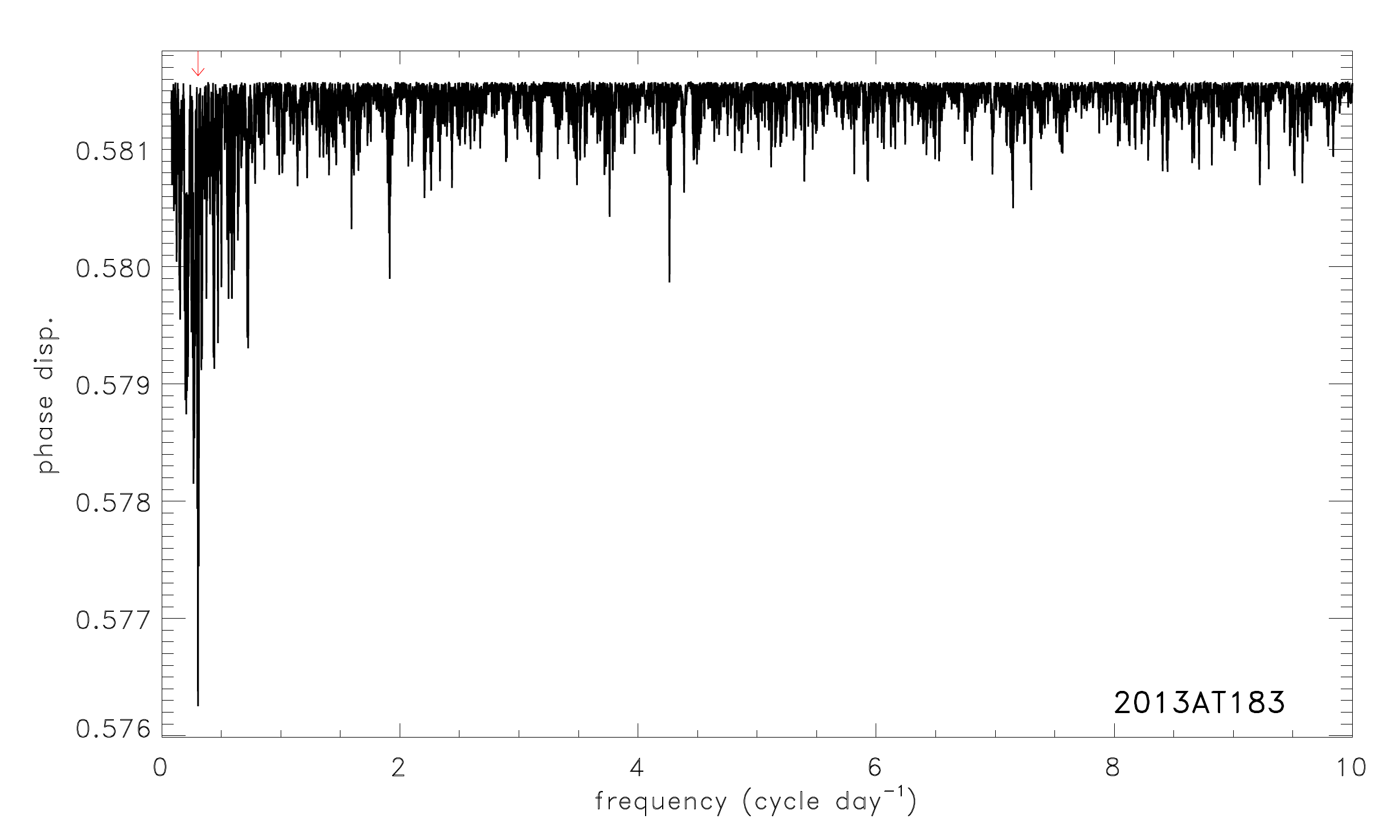}
\includegraphics[width=0.33\textwidth]{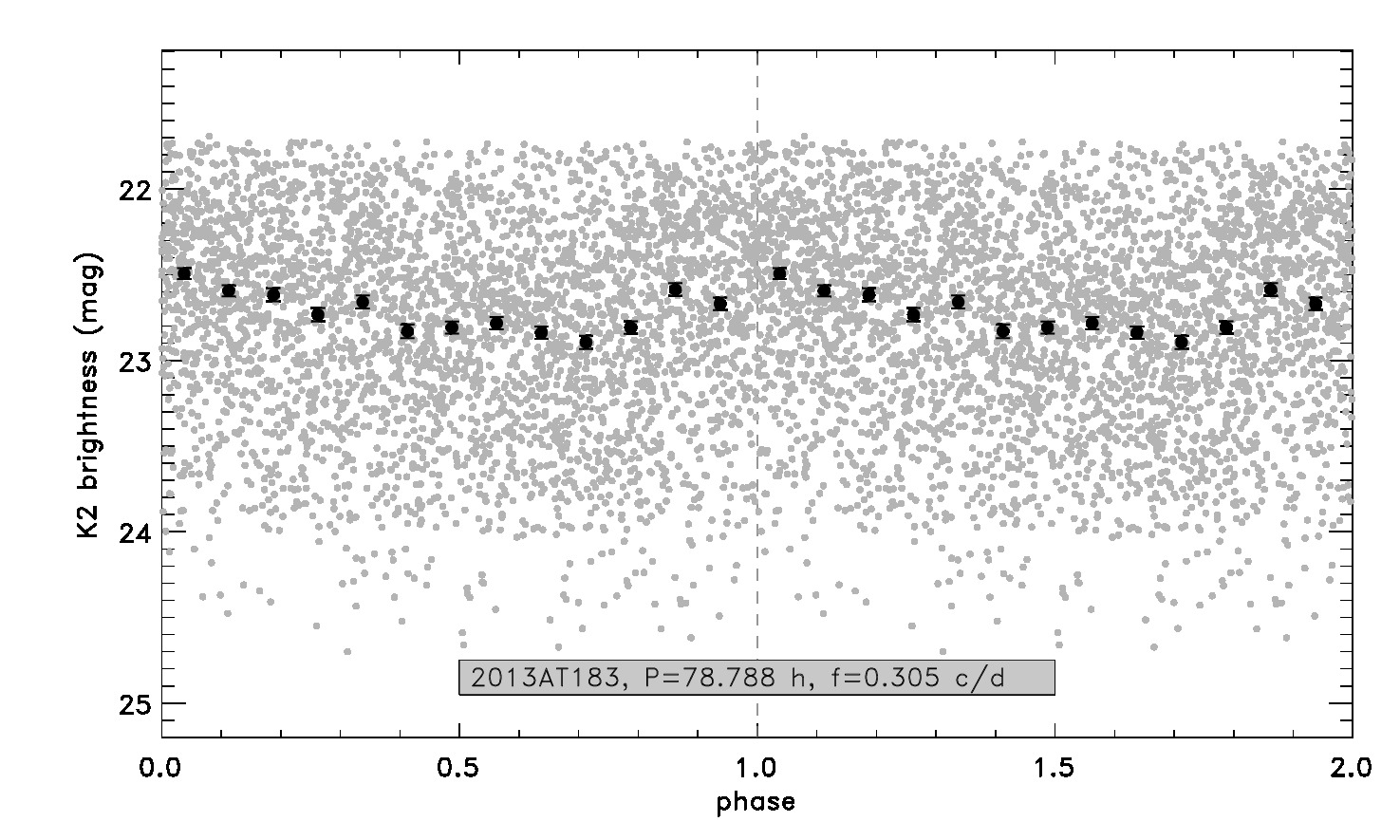}
}
\hbox{
\includegraphics[width=0.33\textwidth]{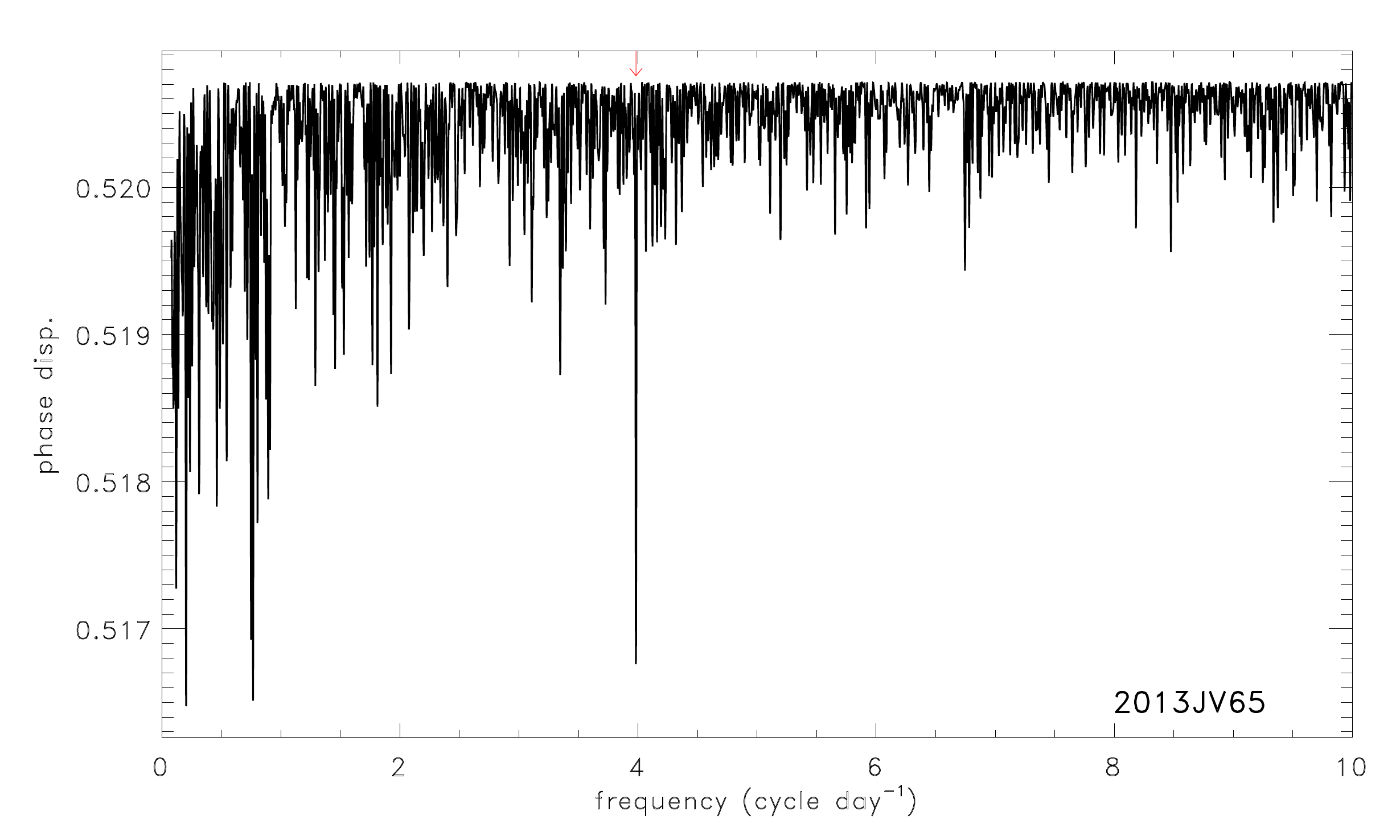}
\includegraphics[width=0.33\textwidth]{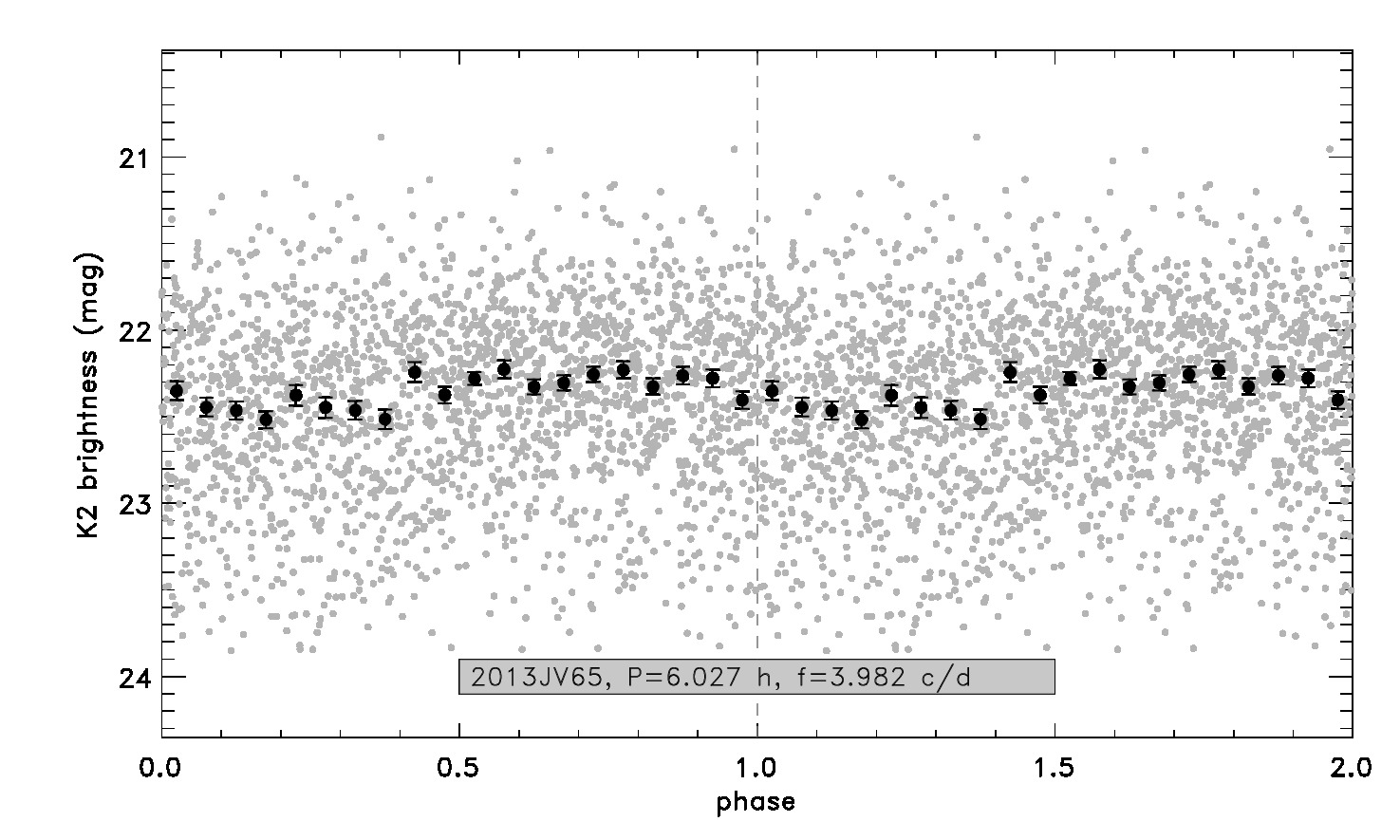}
}
\hbox{
\includegraphics[width=0.33\textwidth]{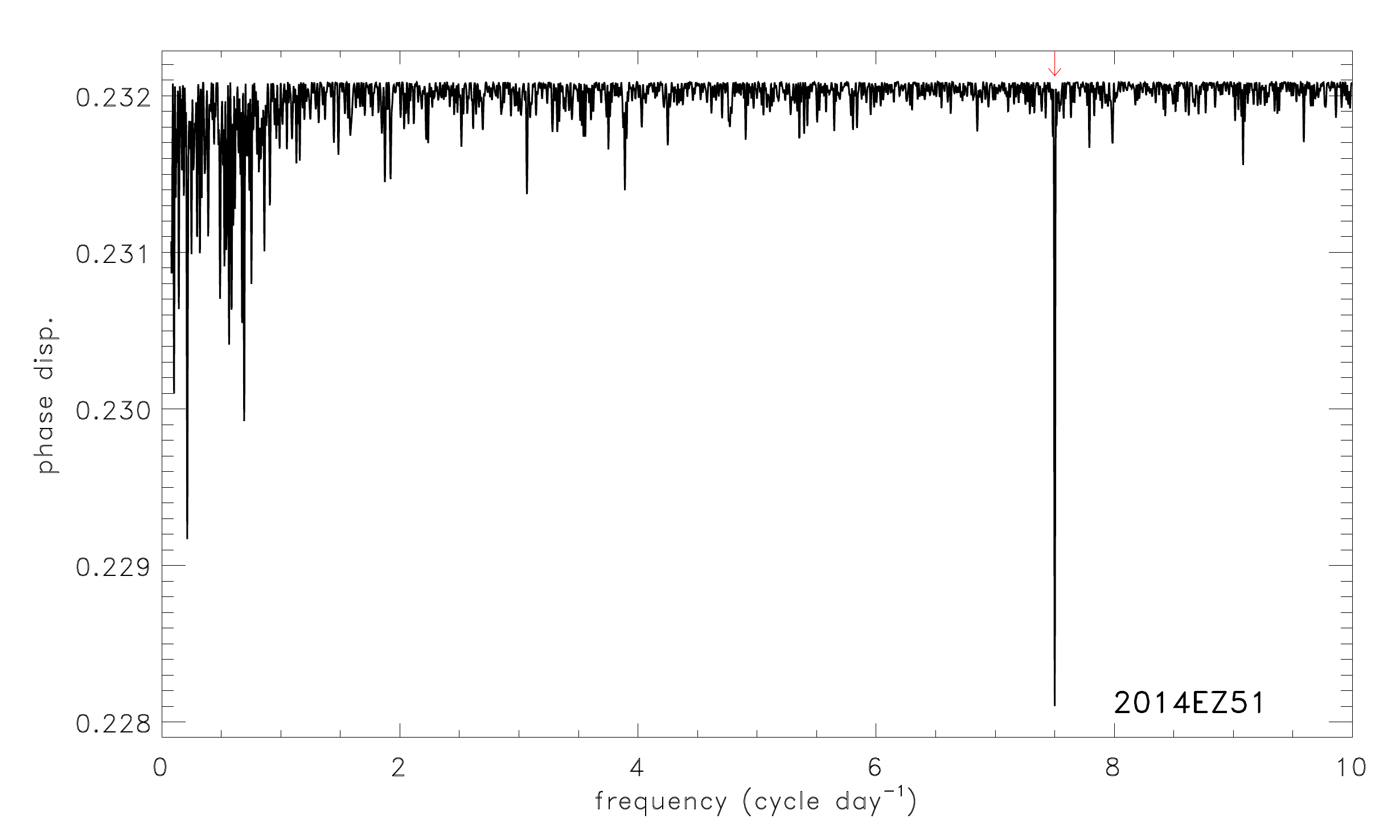}
\includegraphics[width=0.33\textwidth]{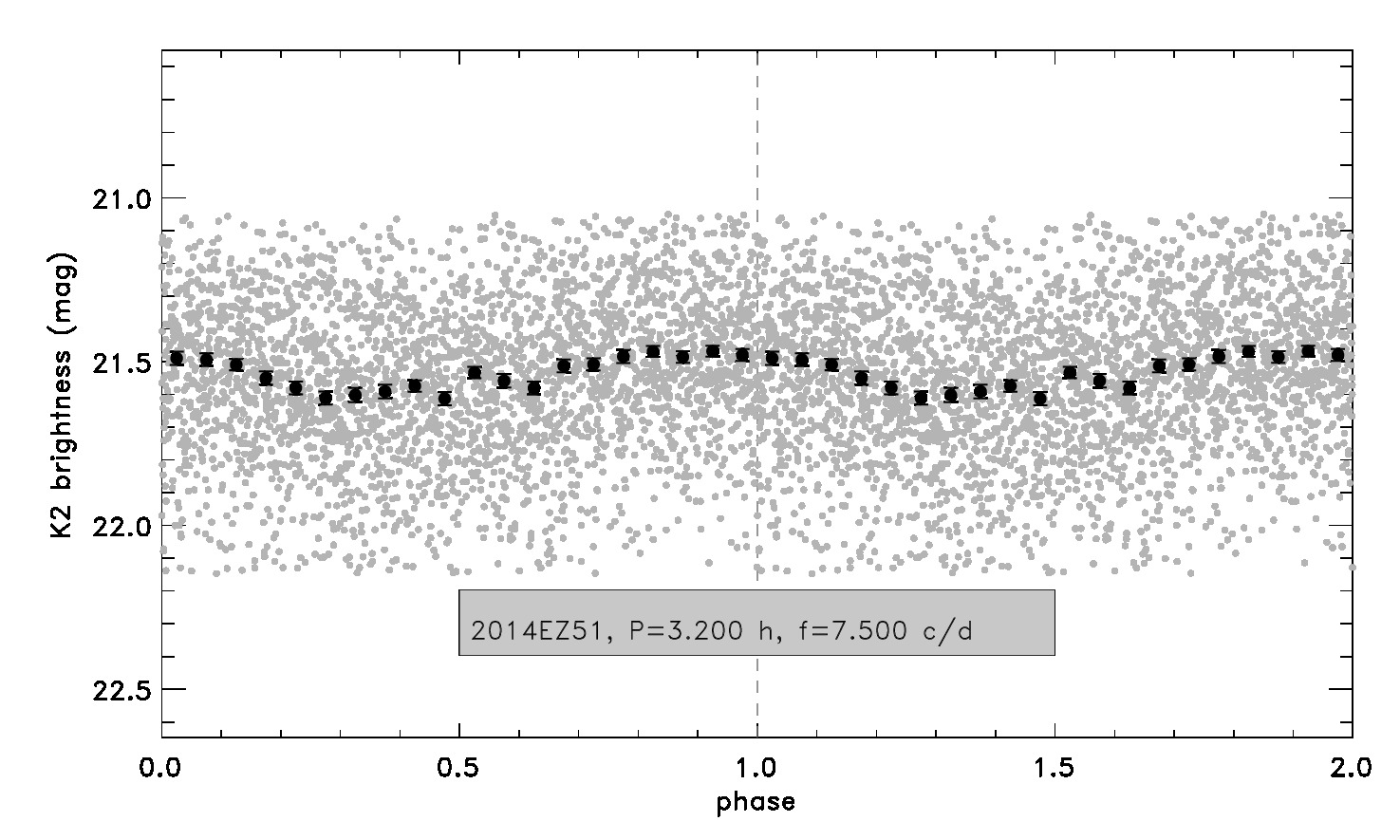}
}
\hbox{
\includegraphics[width=0.33\textwidth]{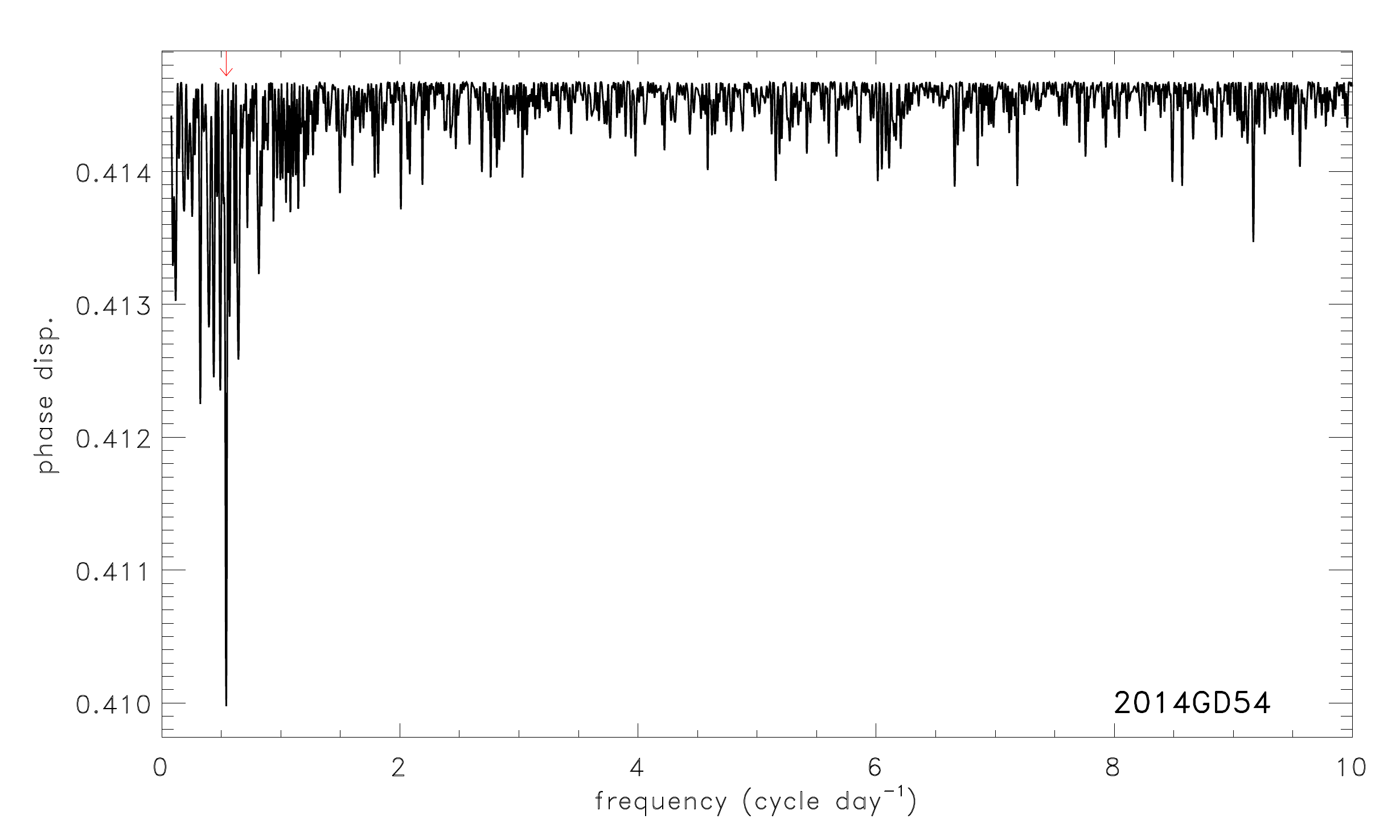}
\includegraphics[width=0.33\textwidth]{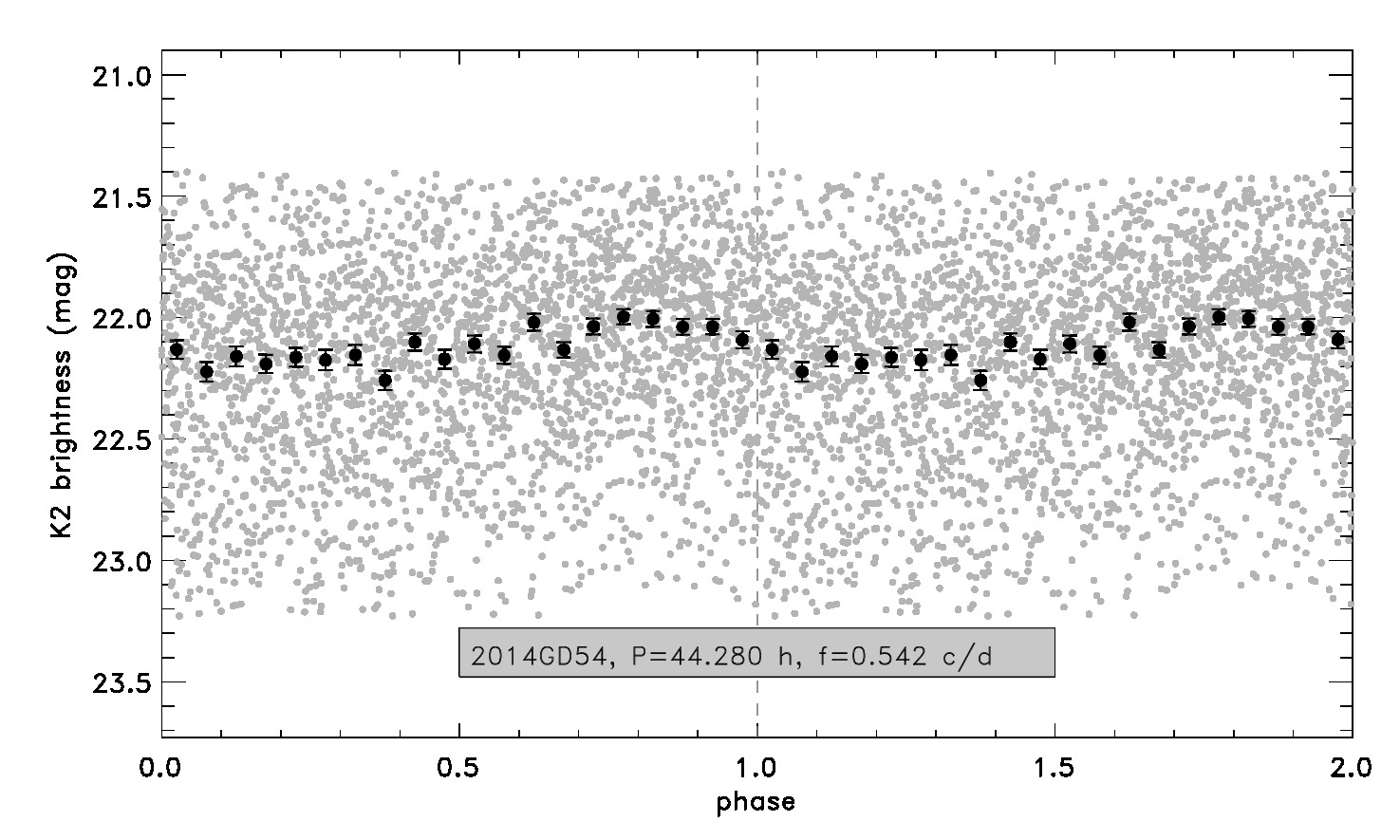}
}
\hbox{
\includegraphics[width=0.33\textwidth]{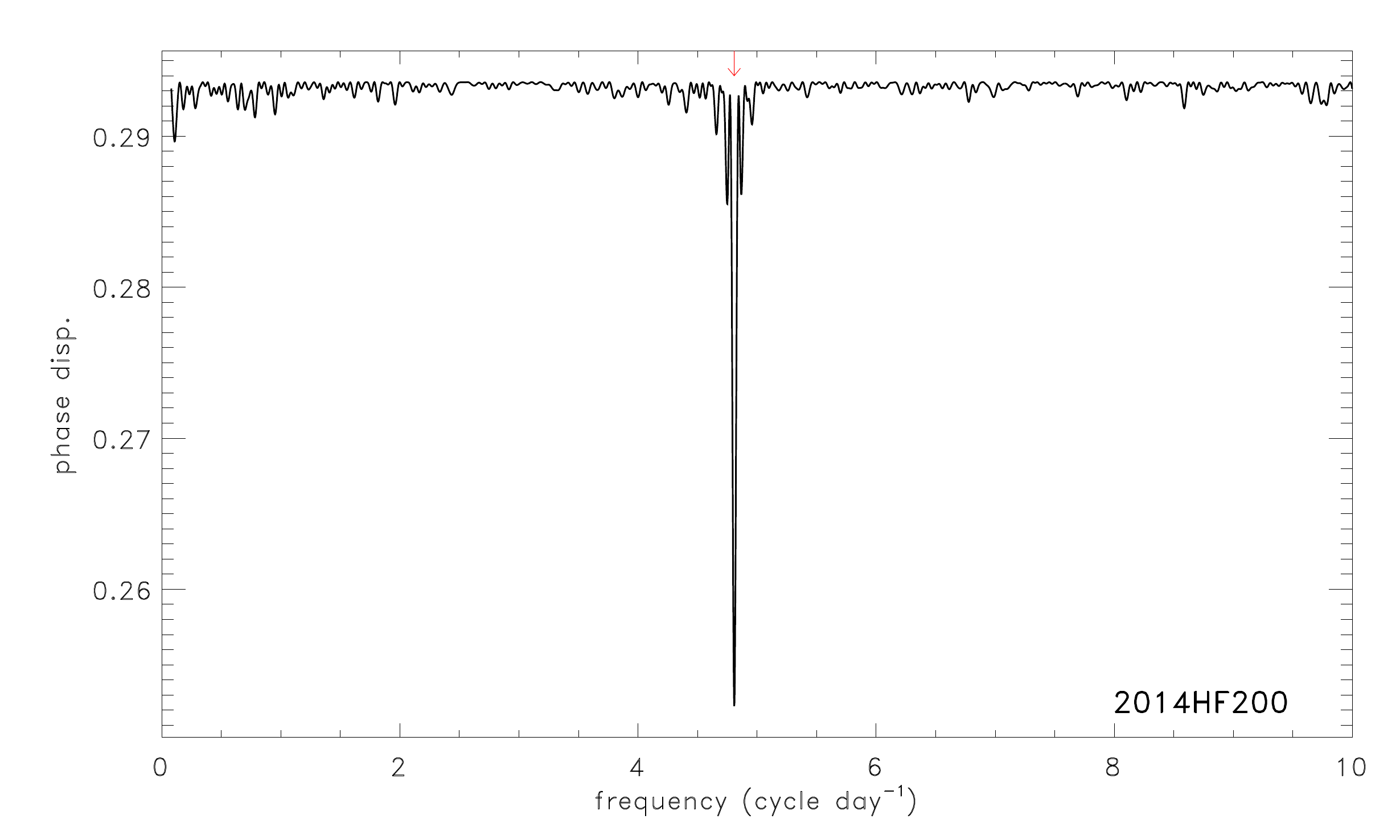}
\includegraphics[width=0.33\textwidth]{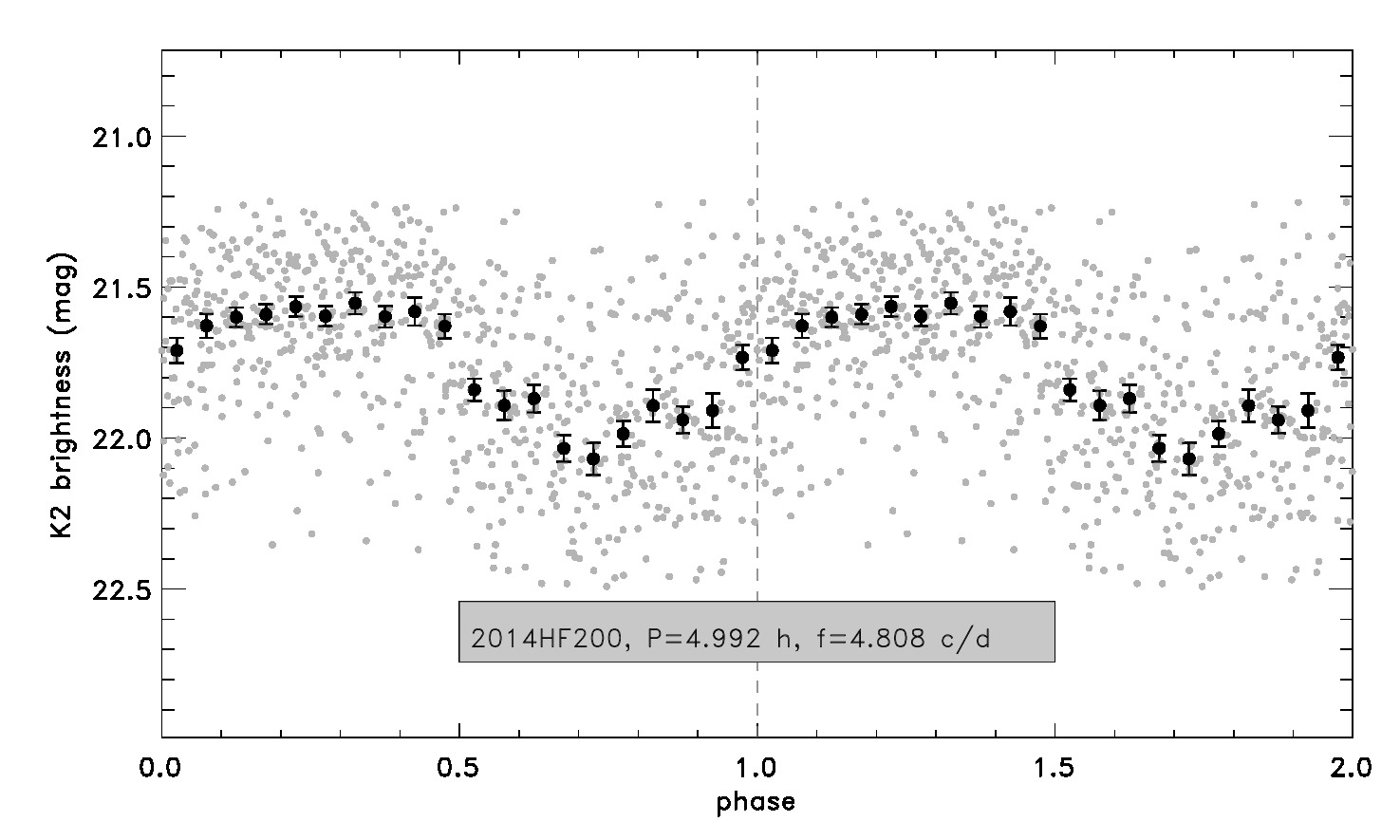}
}
\hbox{
\includegraphics[width=0.33\textwidth]{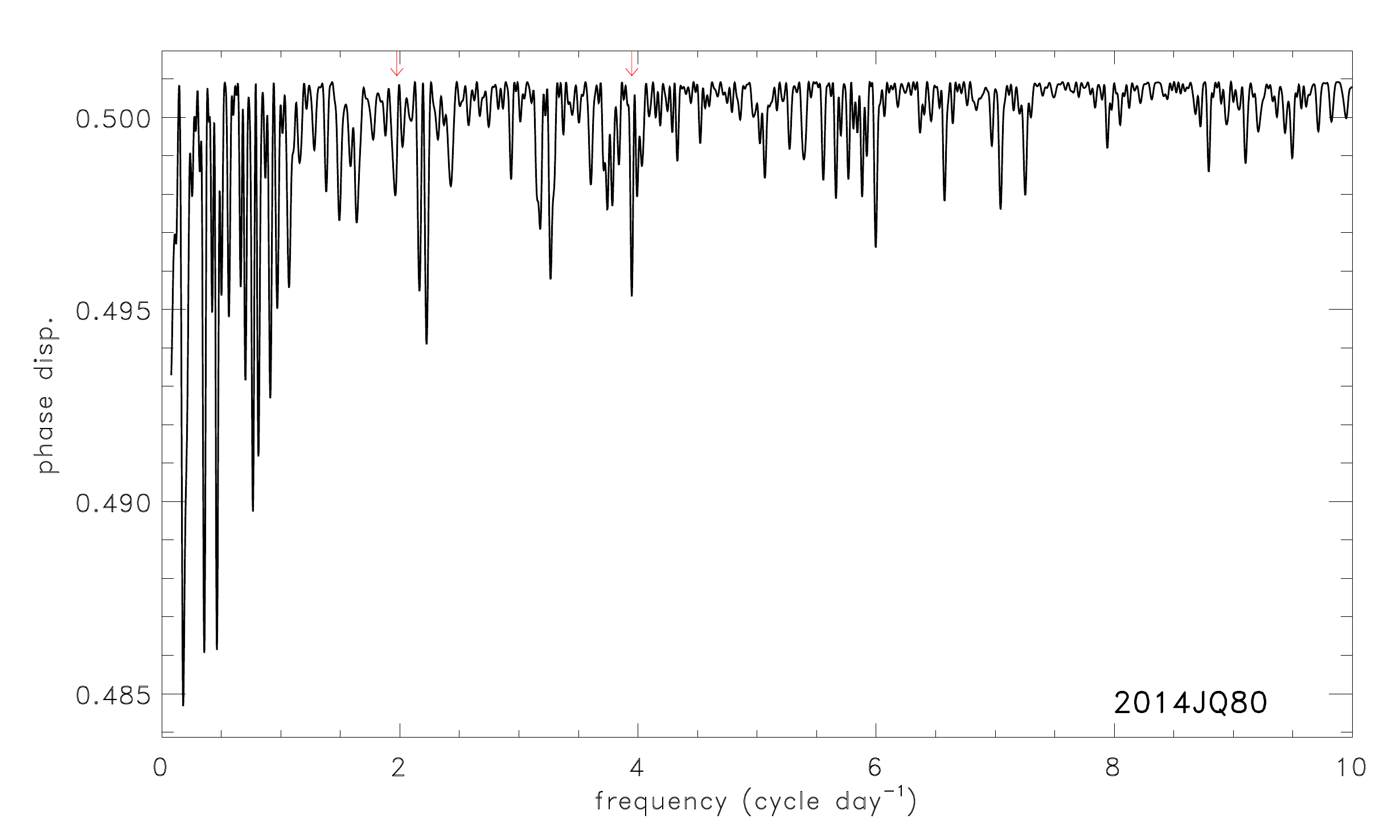}
\includegraphics[width=0.33\textwidth]{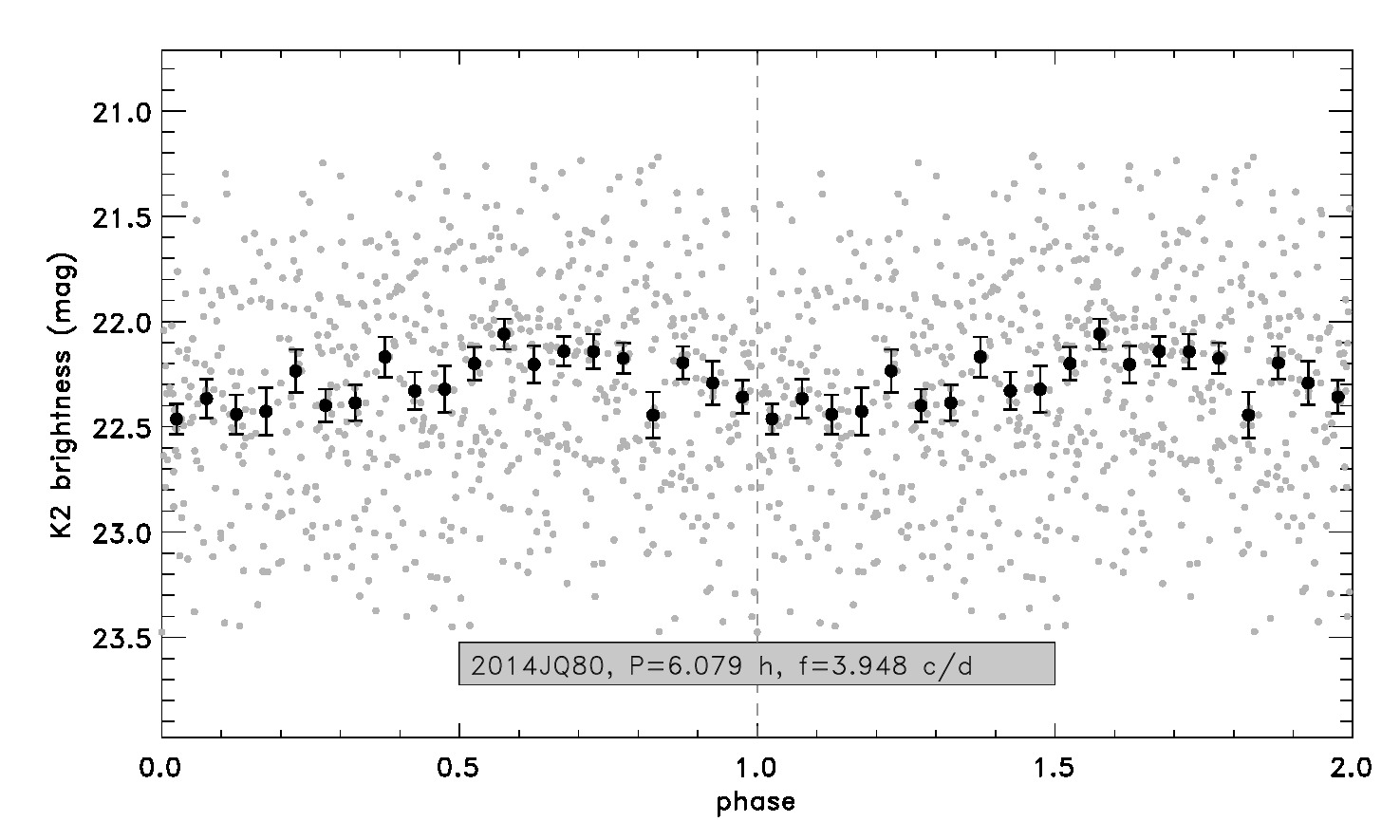}
\includegraphics[width=0.33\textwidth]{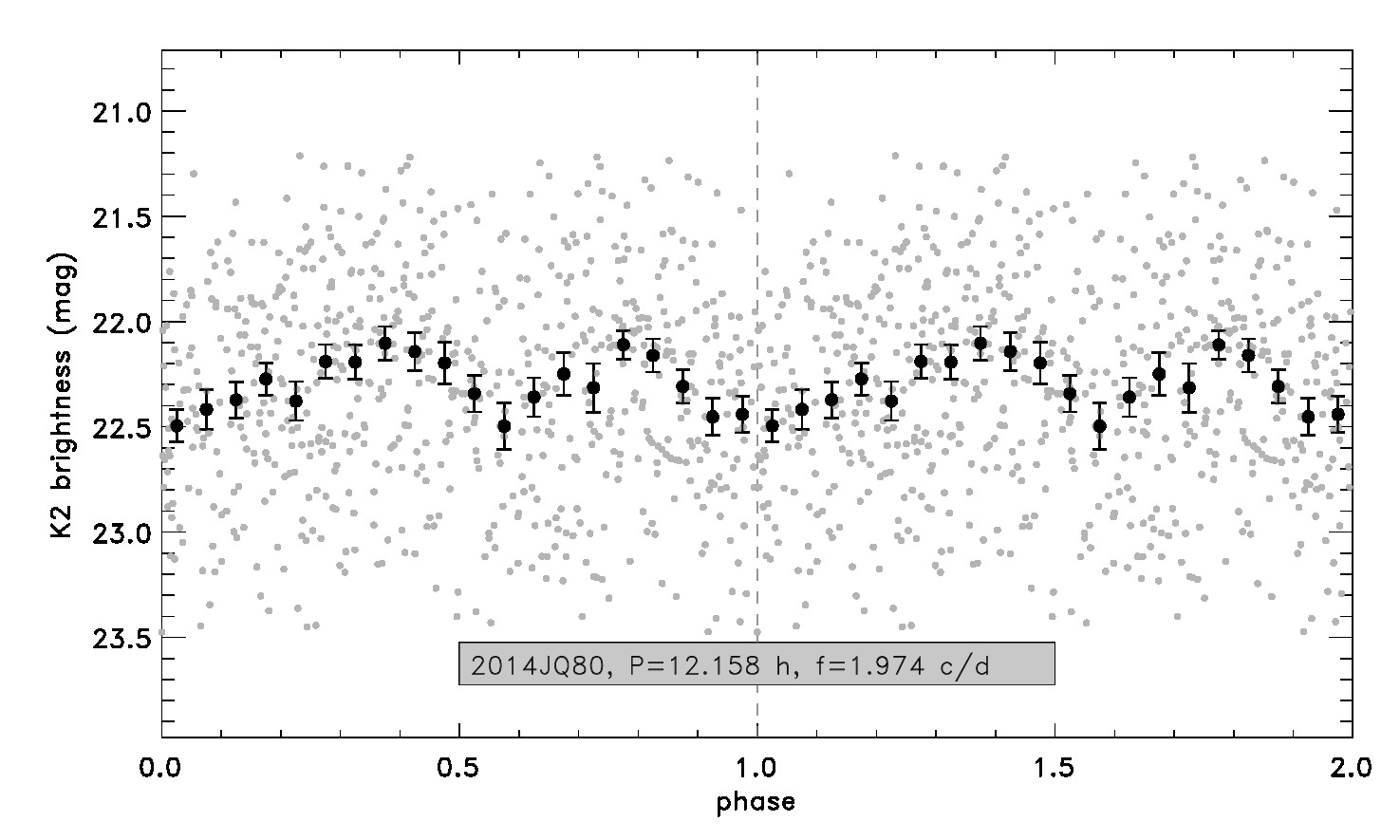}
}
\caption{
}
\end{figure*}

\begin{figure}[ht!]
\ContinuedFloat
\hbox{
\includegraphics[width=0.33\textwidth]{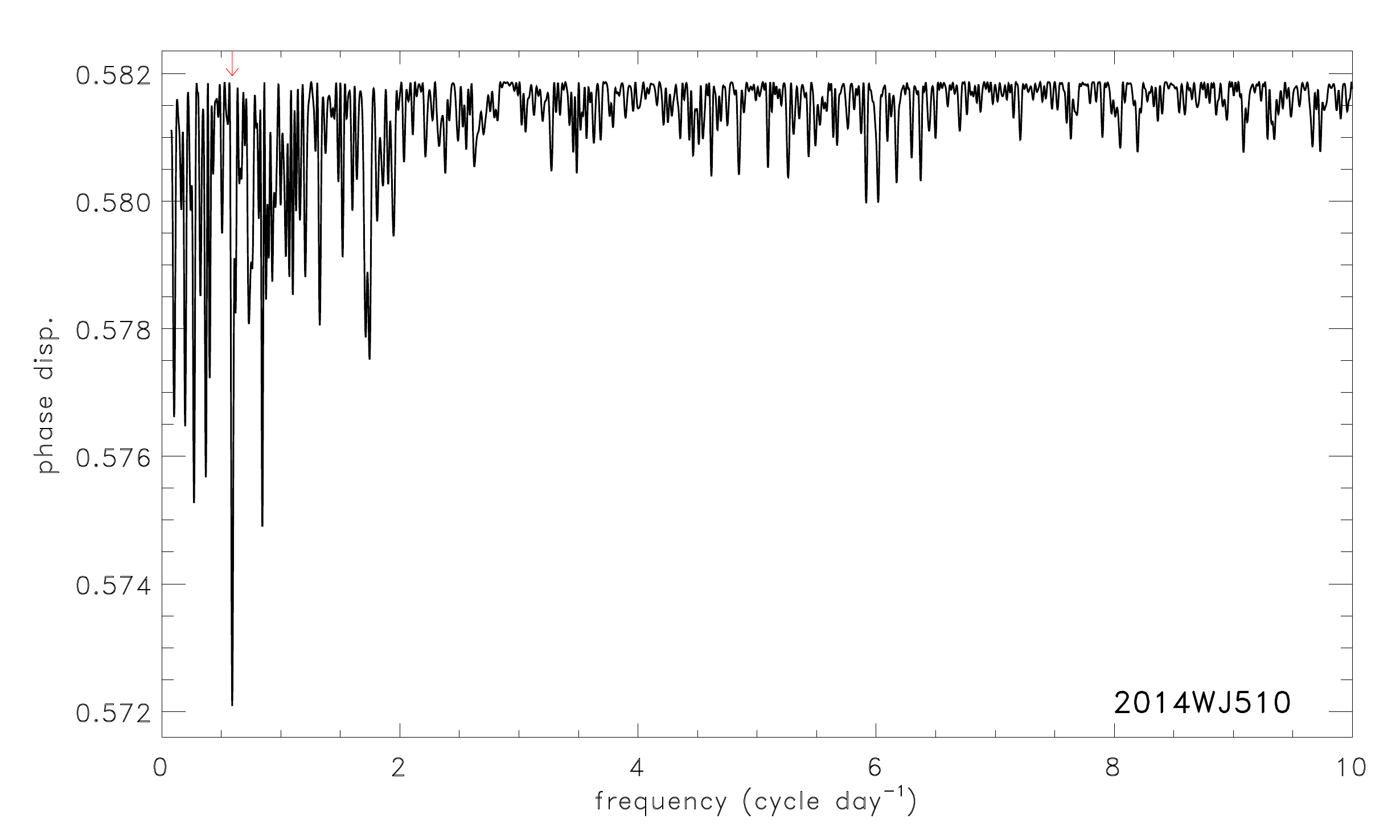}
\includegraphics[width=0.33\textwidth]{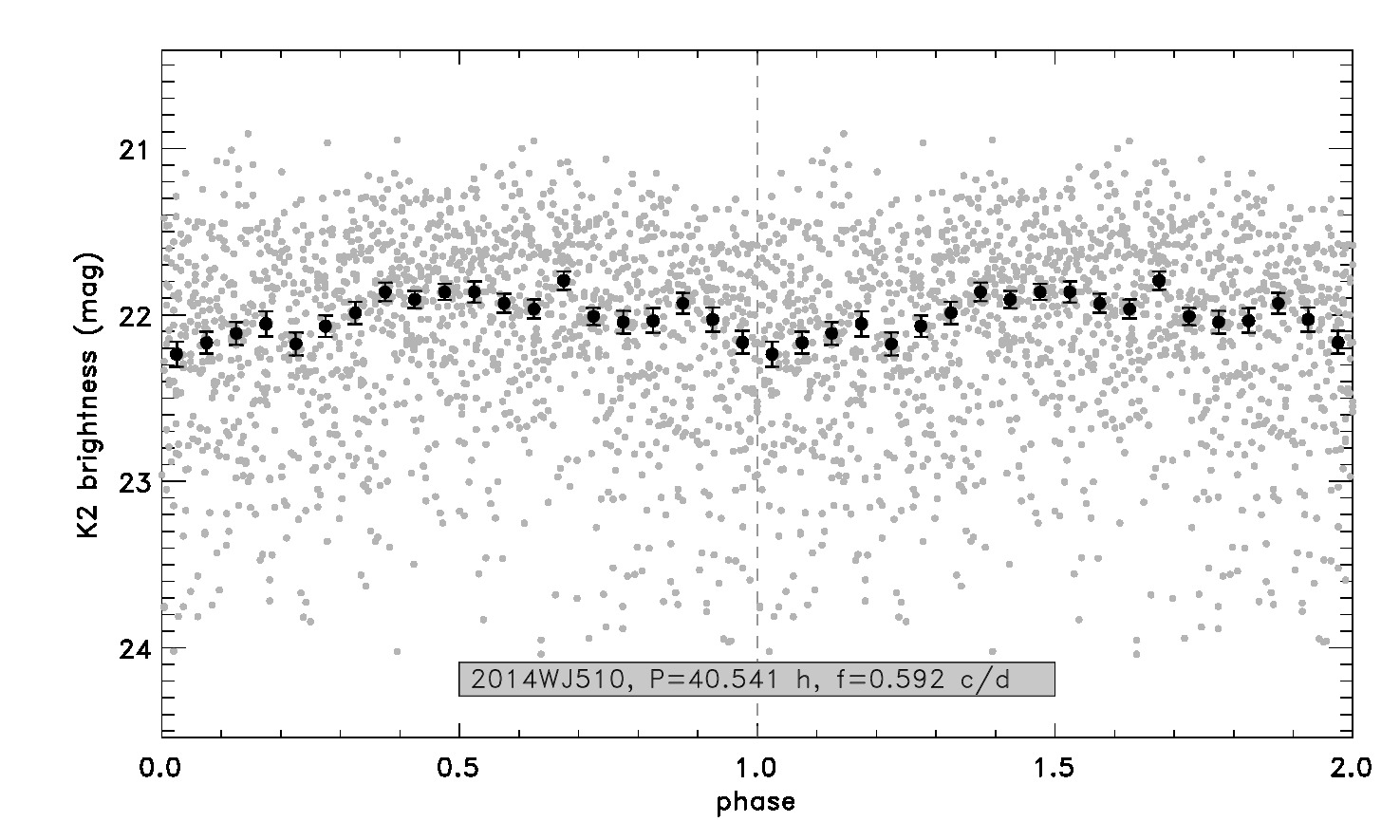}
}
\hbox{
\includegraphics[width=0.33\textwidth]{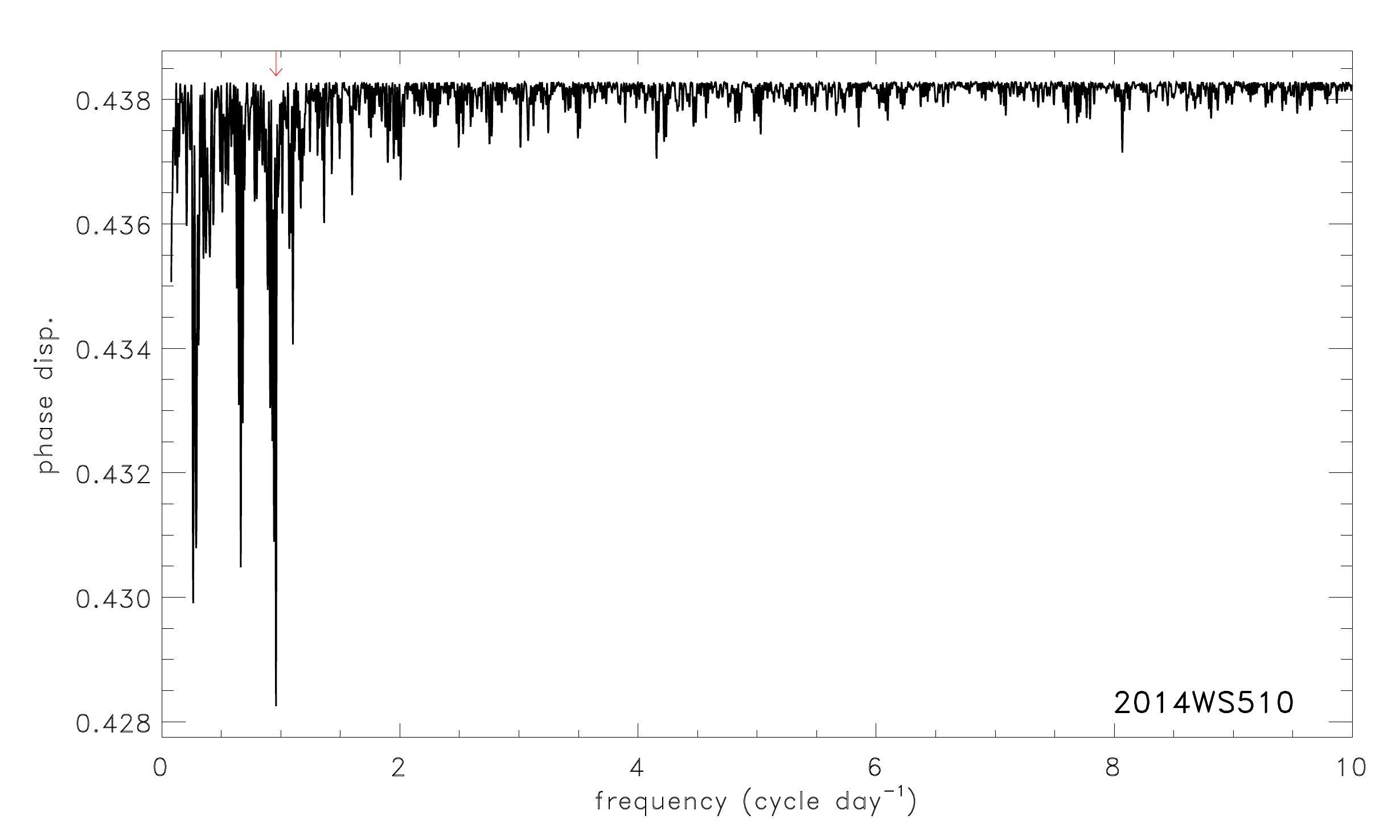}
\includegraphics[width=0.33\textwidth]{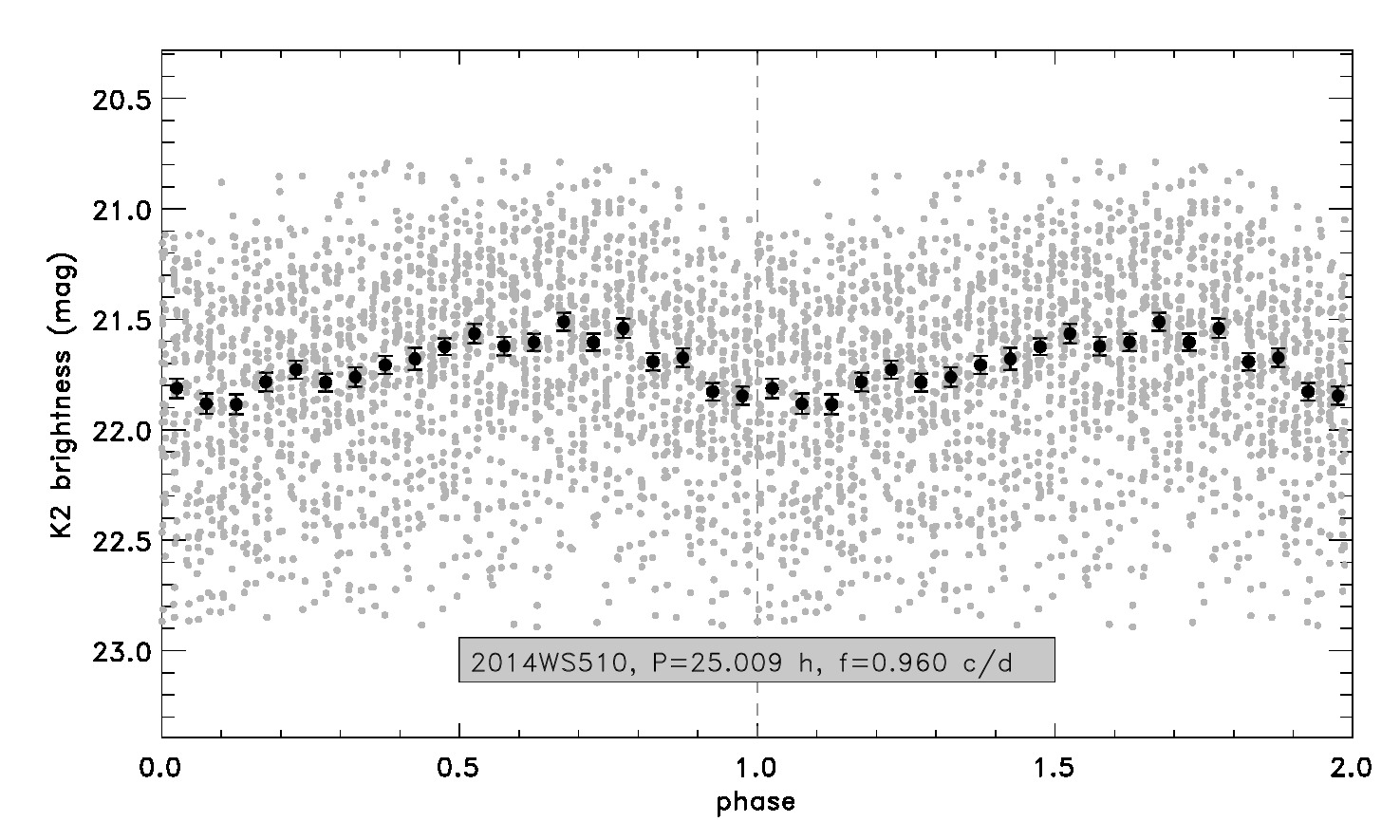}
}
\hbox{
\includegraphics[width=0.33\textwidth]{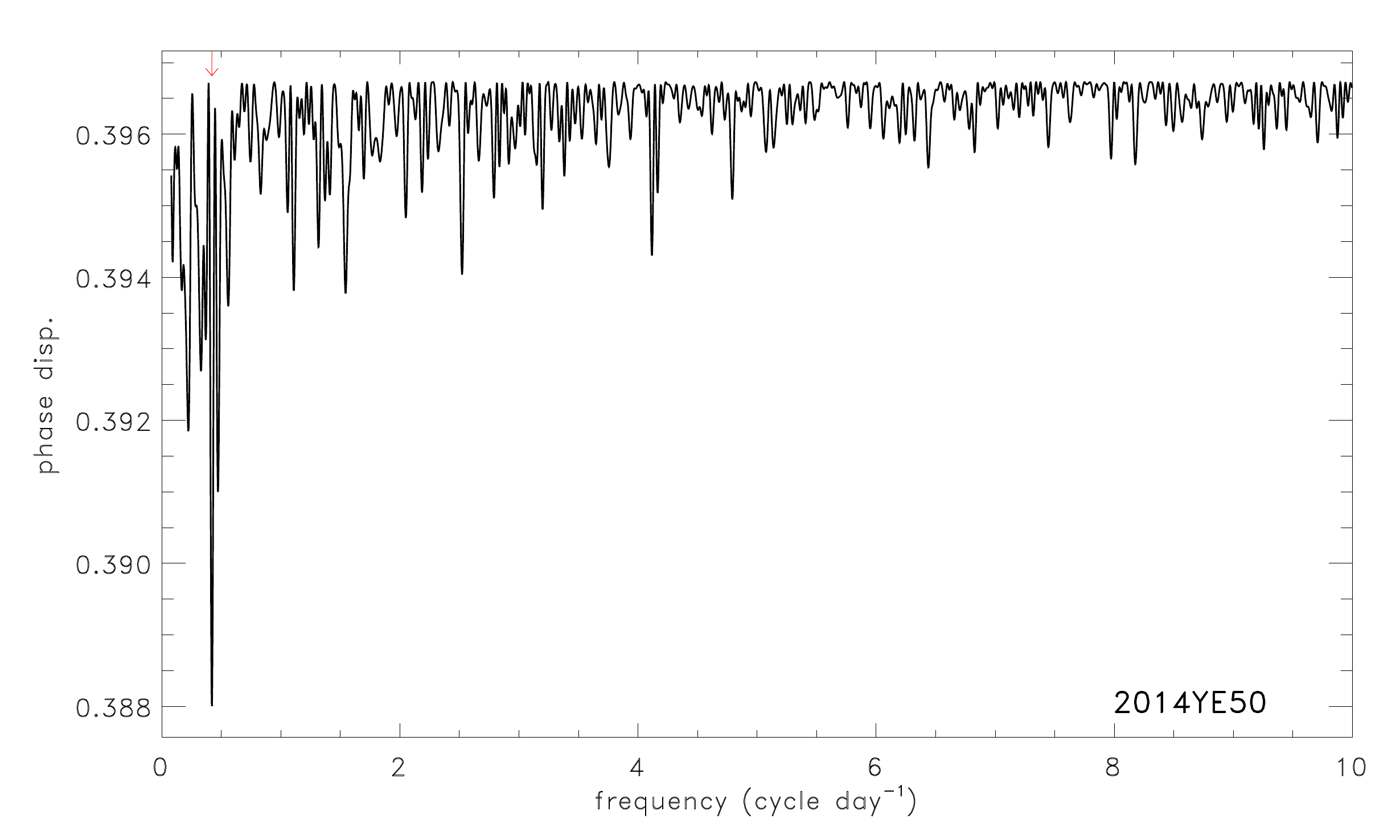}
\includegraphics[width=0.33\textwidth]{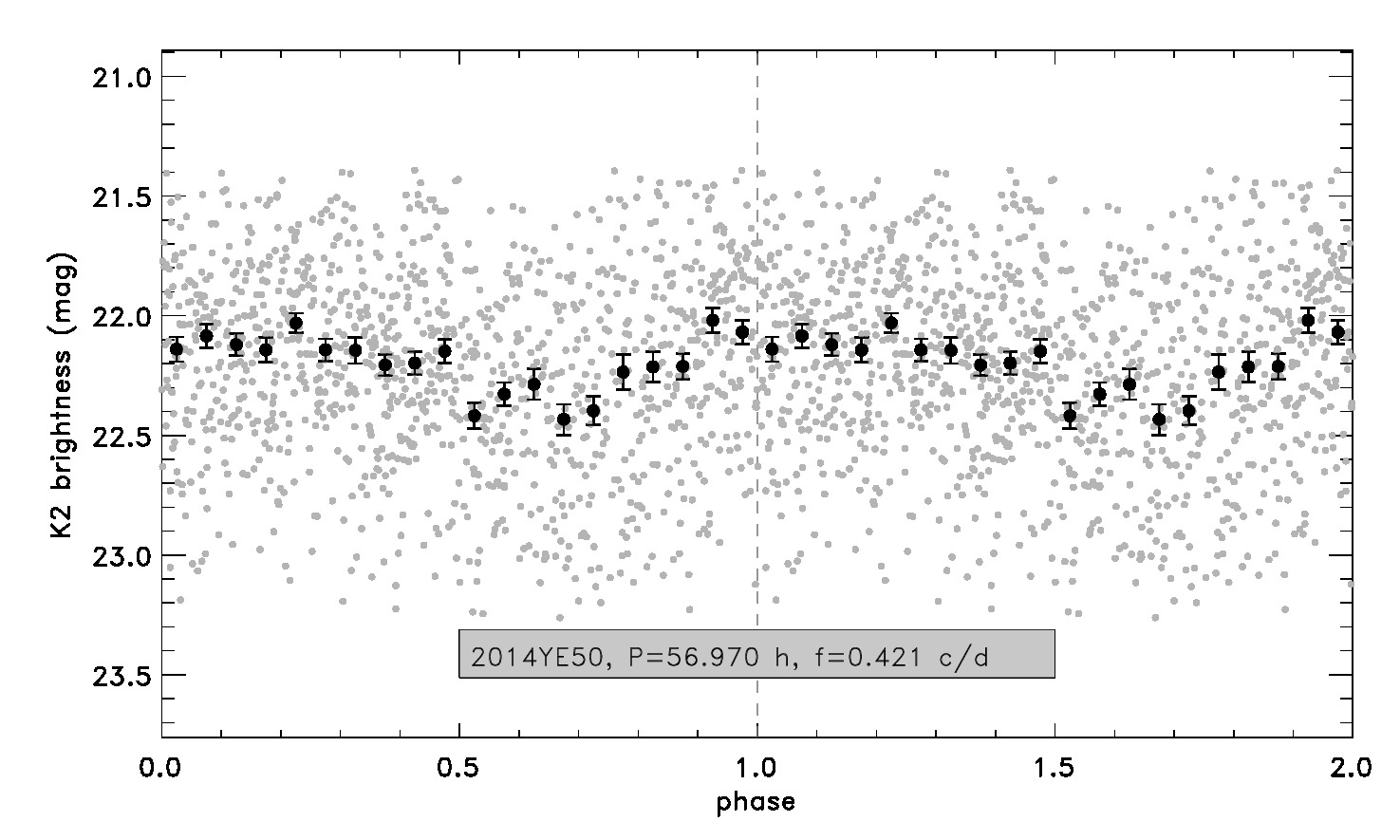}
}
\hbox{
\includegraphics[width=0.33\textwidth]{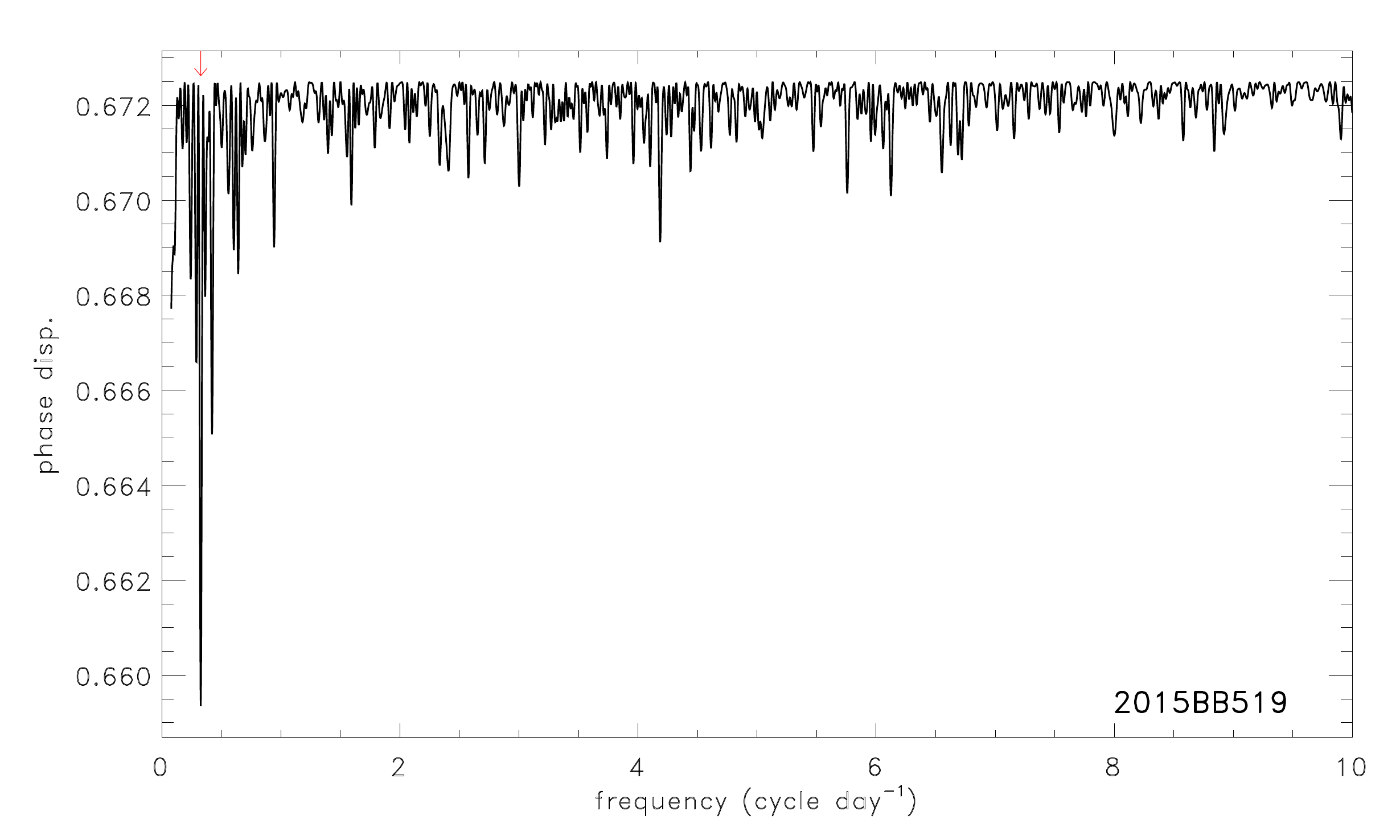}
\includegraphics[width=0.33\textwidth]{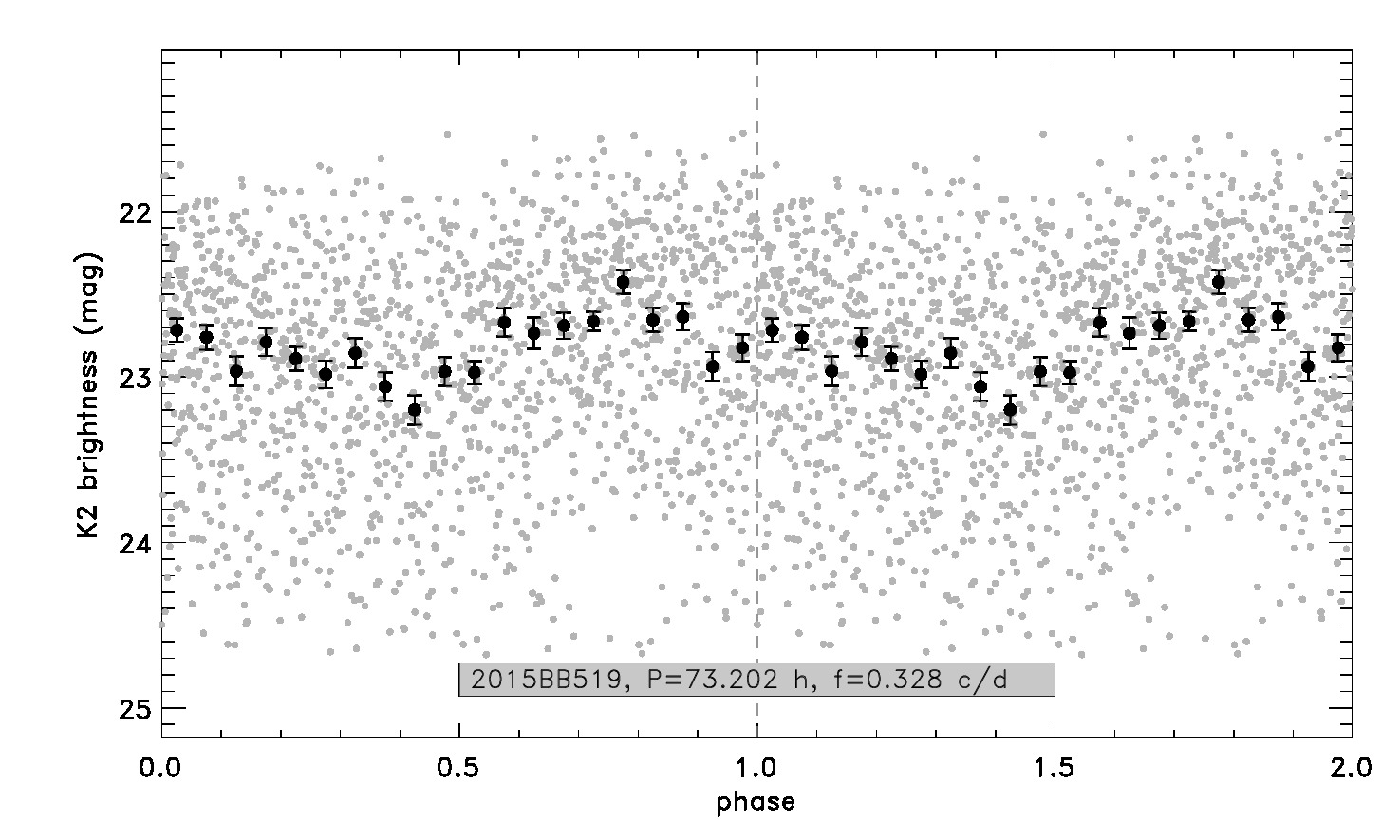}
}
\hbox{
\includegraphics[width=0.33\textwidth]{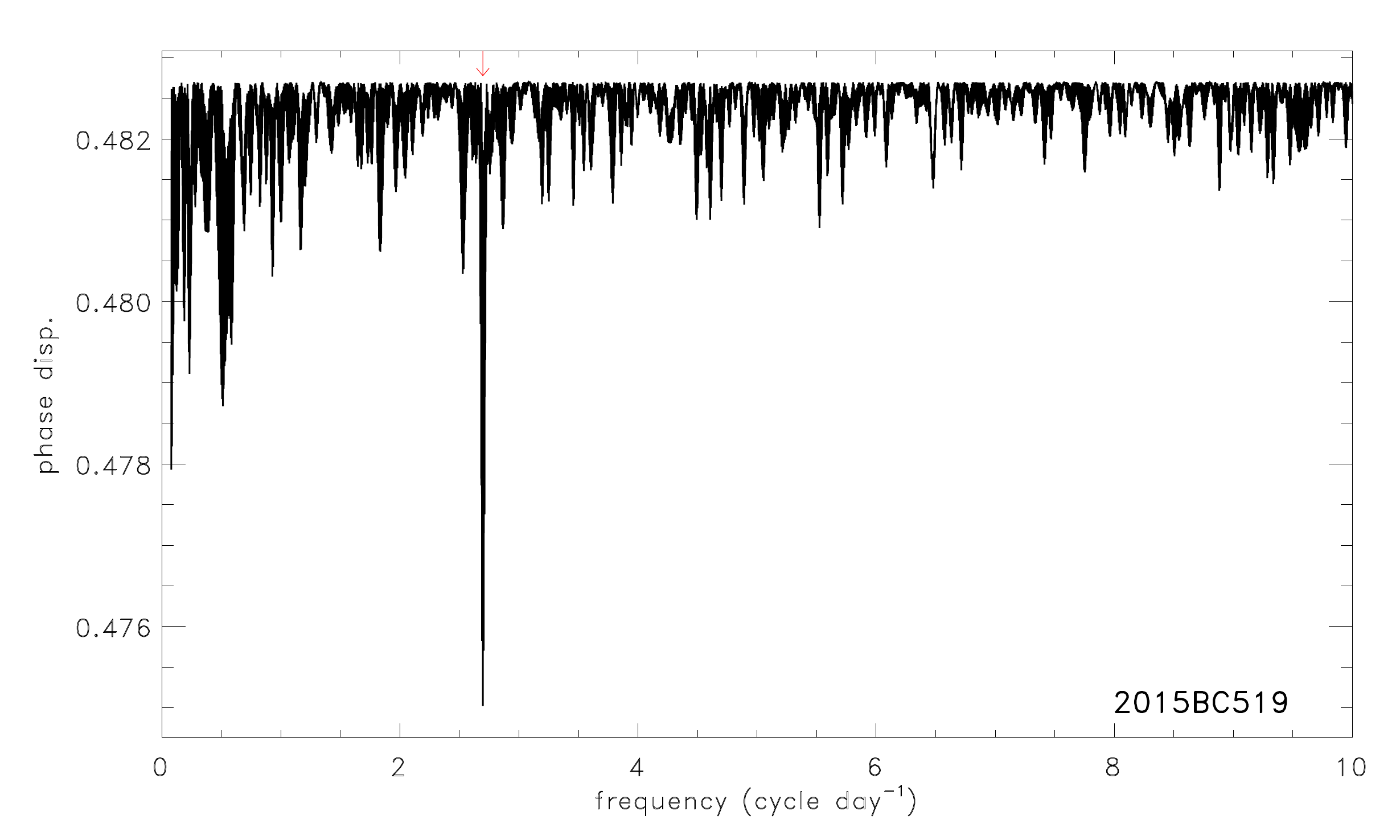}
\includegraphics[width=0.33\textwidth]{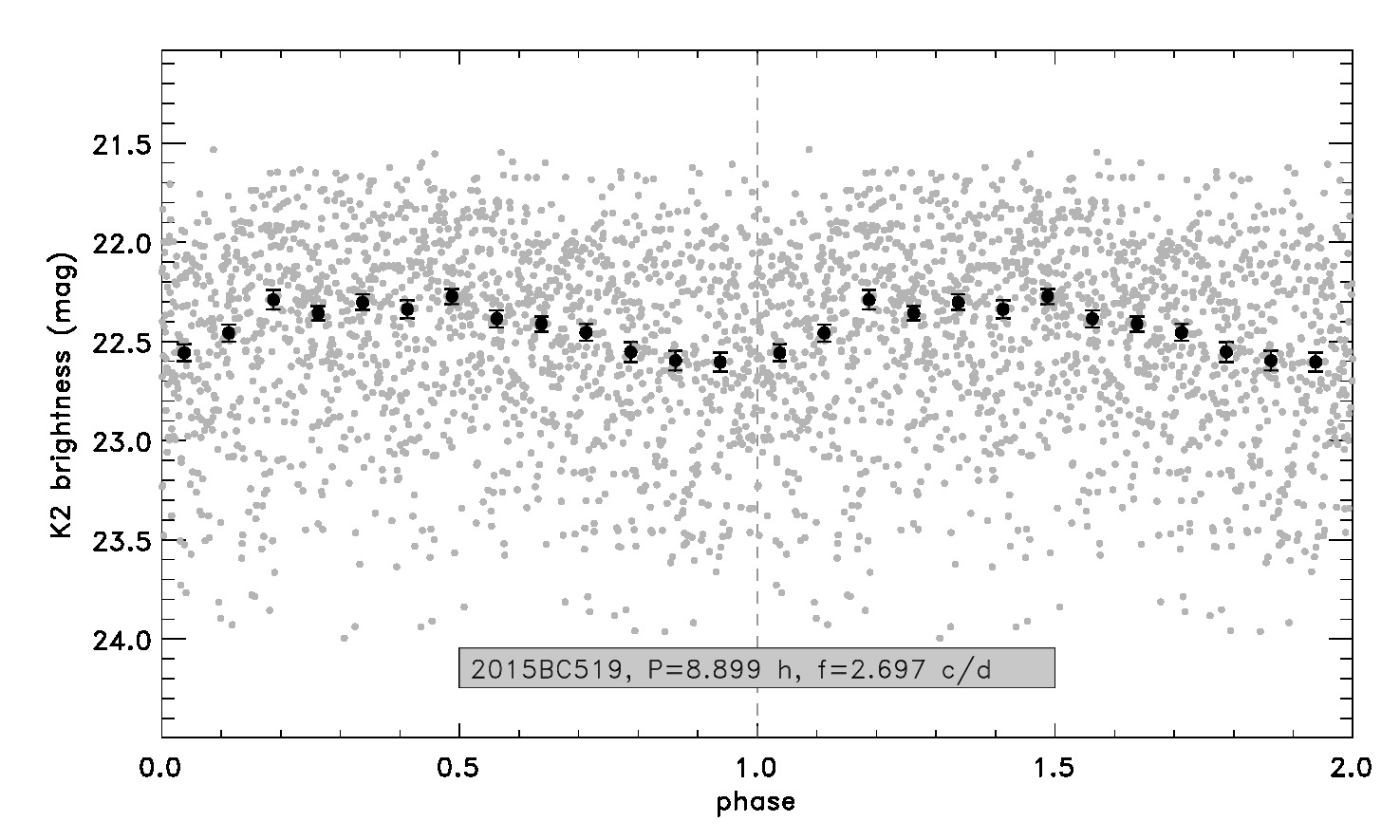}
}
\hbox{
\includegraphics[width=0.33\textwidth]{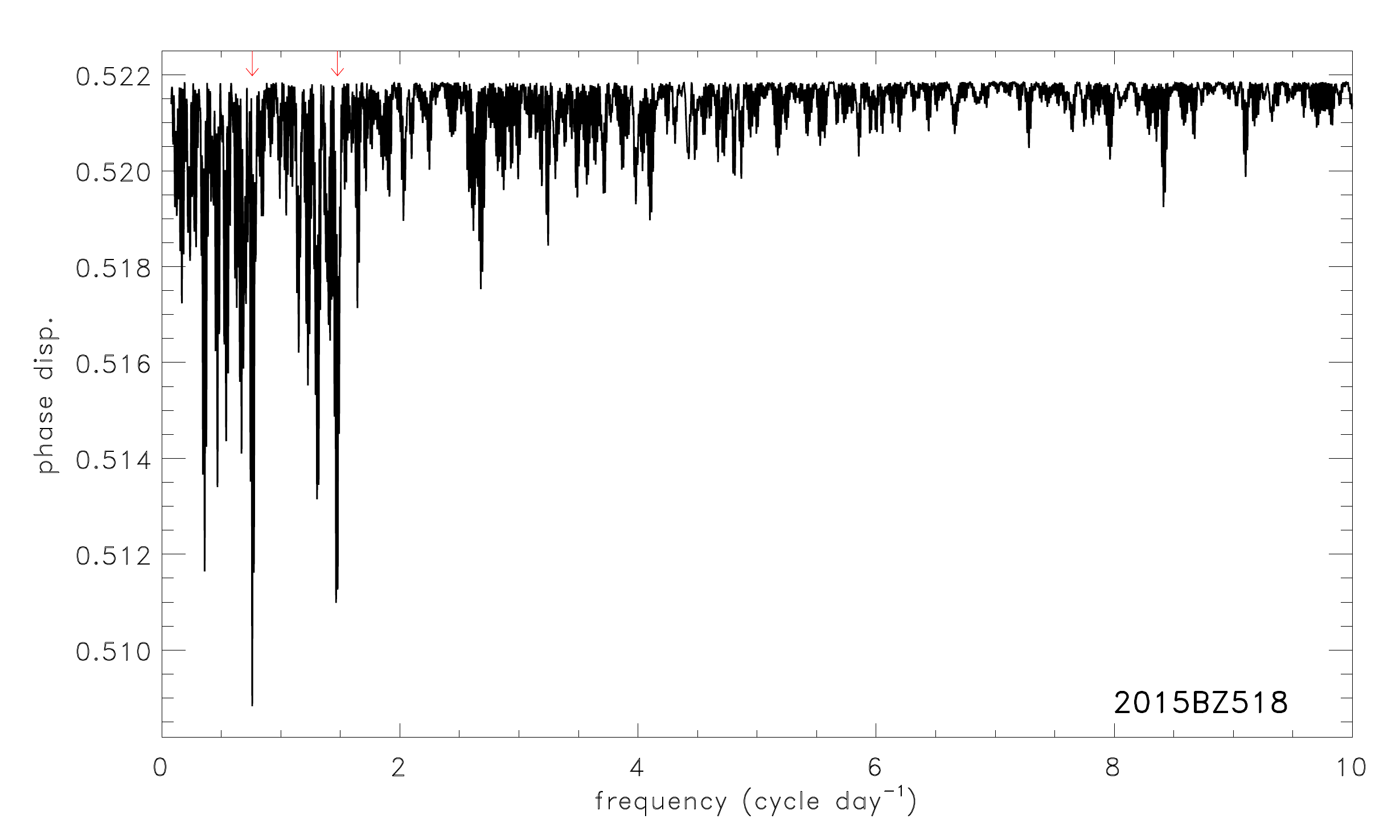}
\includegraphics[width=0.33\textwidth]{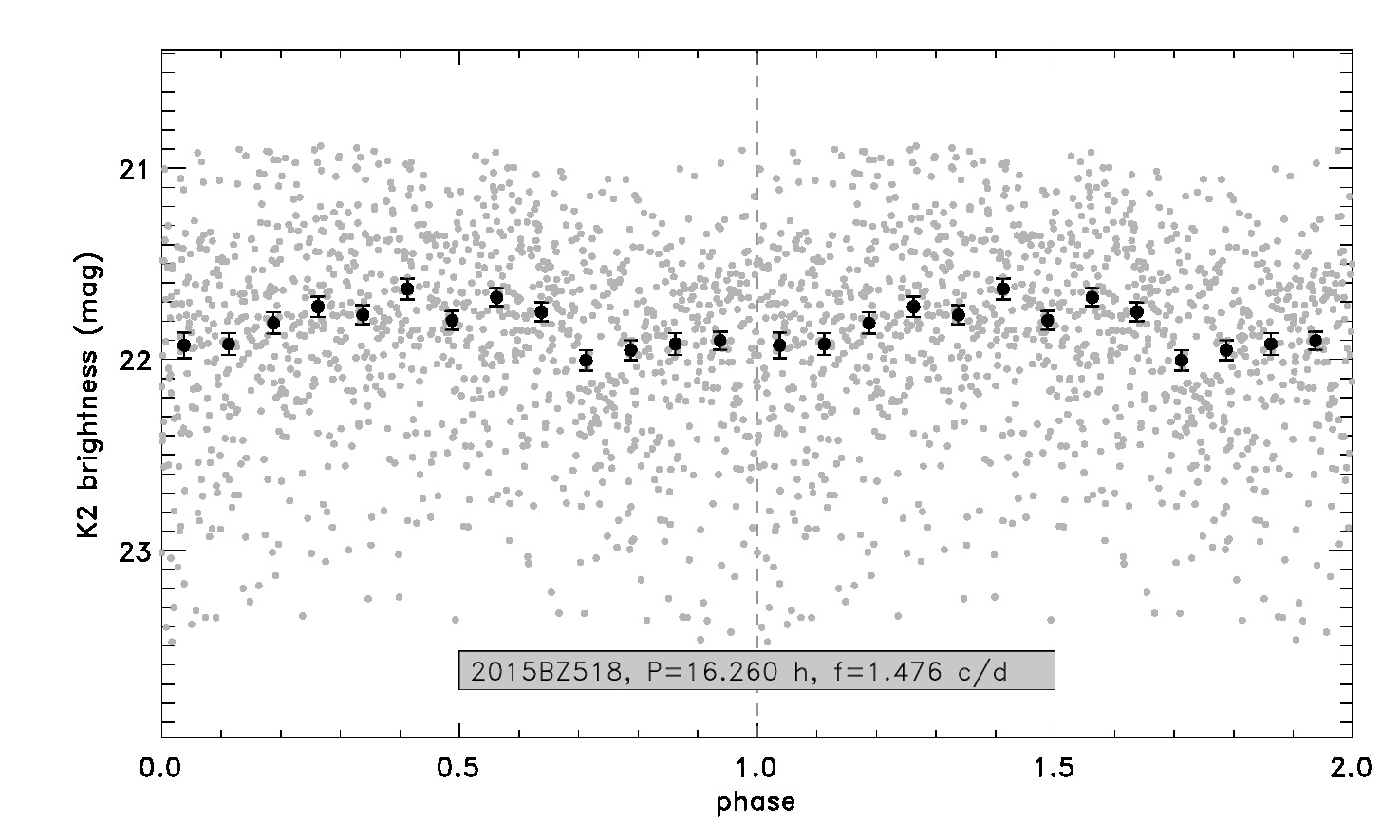}
\includegraphics[width=0.33\textwidth]{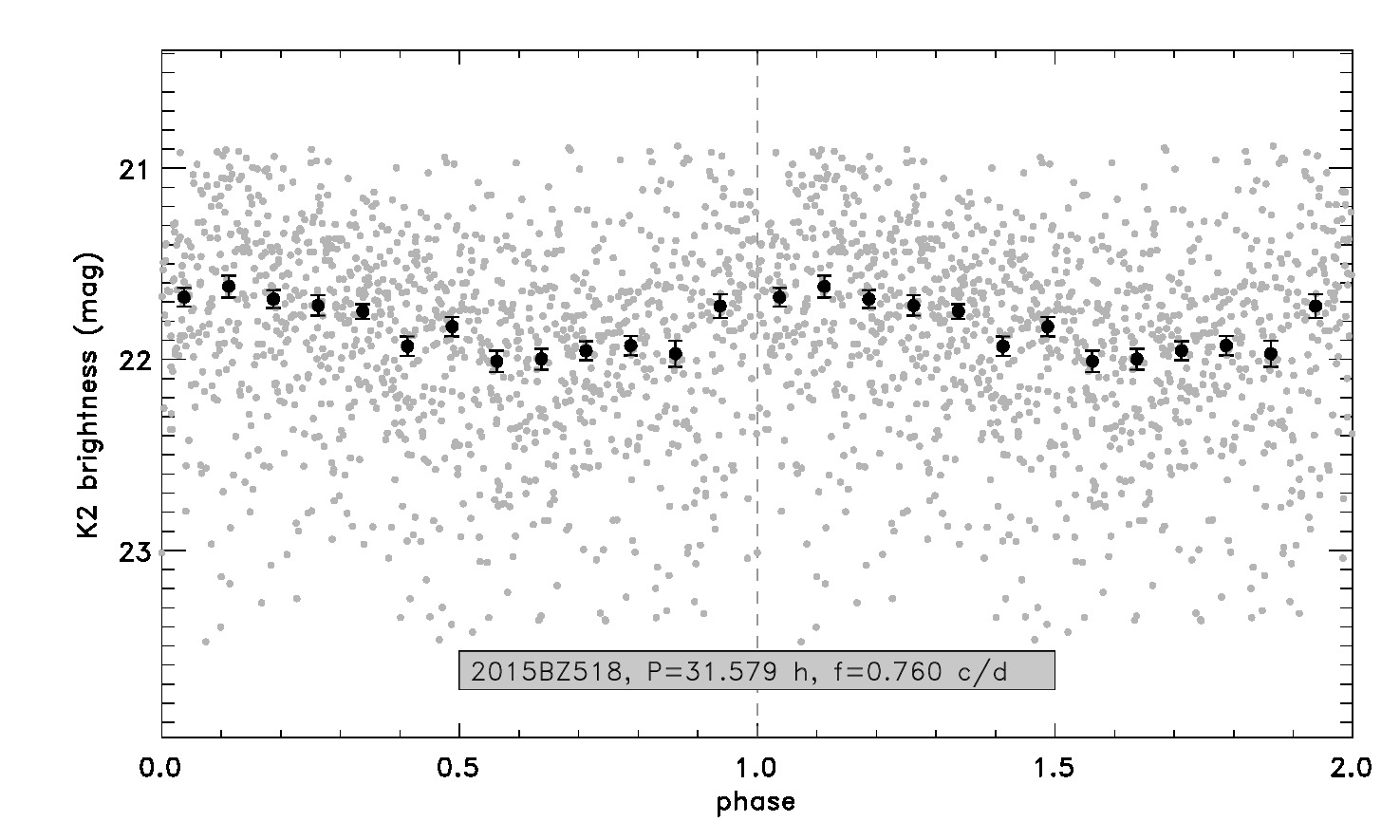}
}
\caption{
}
\end{figure}
\begin{figure}
\ContinuedFloat
\hbox{
\includegraphics[width=0.33\textwidth]{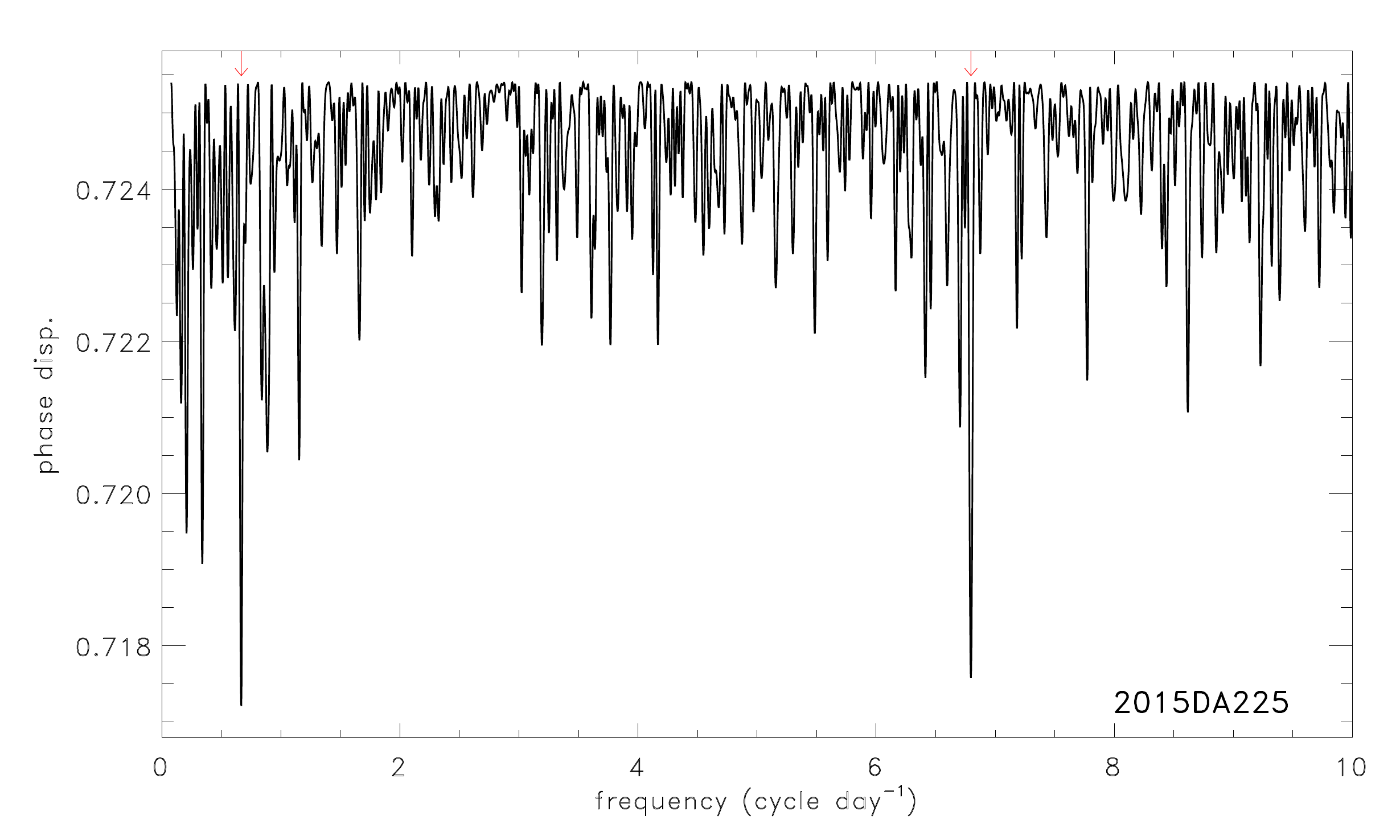}
\includegraphics[width=0.33\textwidth]{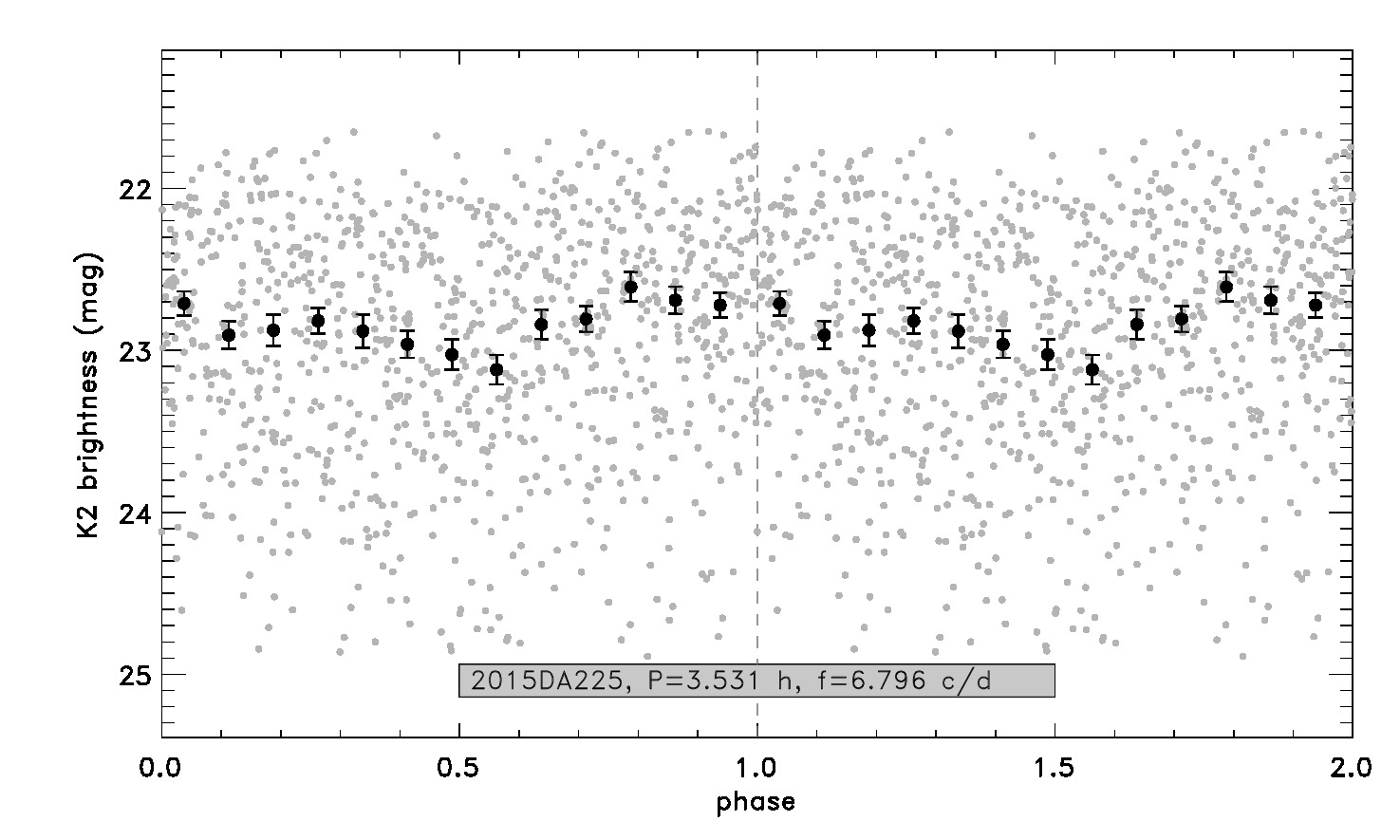}
\includegraphics[width=0.33\textwidth]{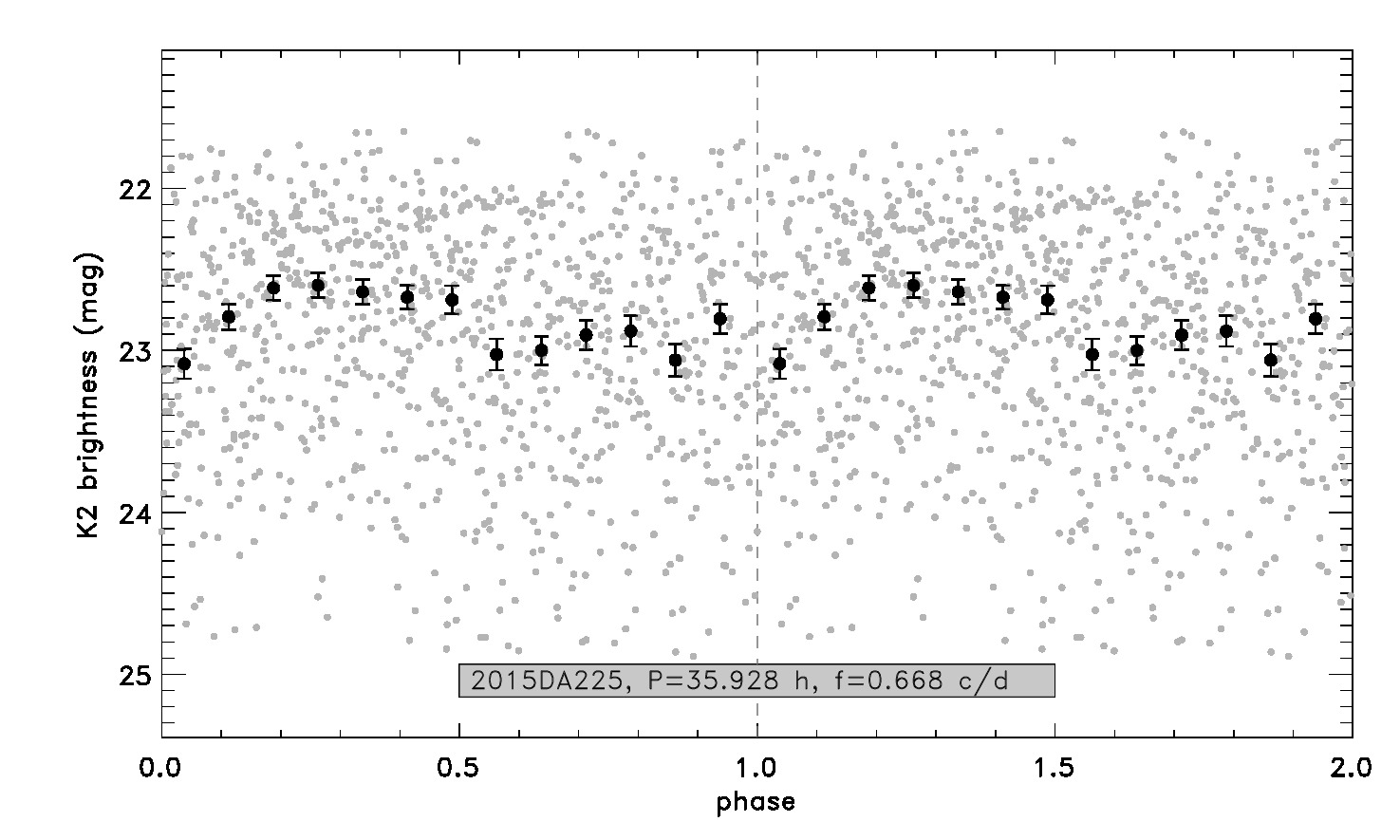}
}
\caption{
}
\end{figure}

\clearpage

\section{Individual objects without detected light curve periods \label{sect:nondetect}}

\begin{figure}[ht]
\centering
\hbox{
\includegraphics[width=0.33\textwidth]{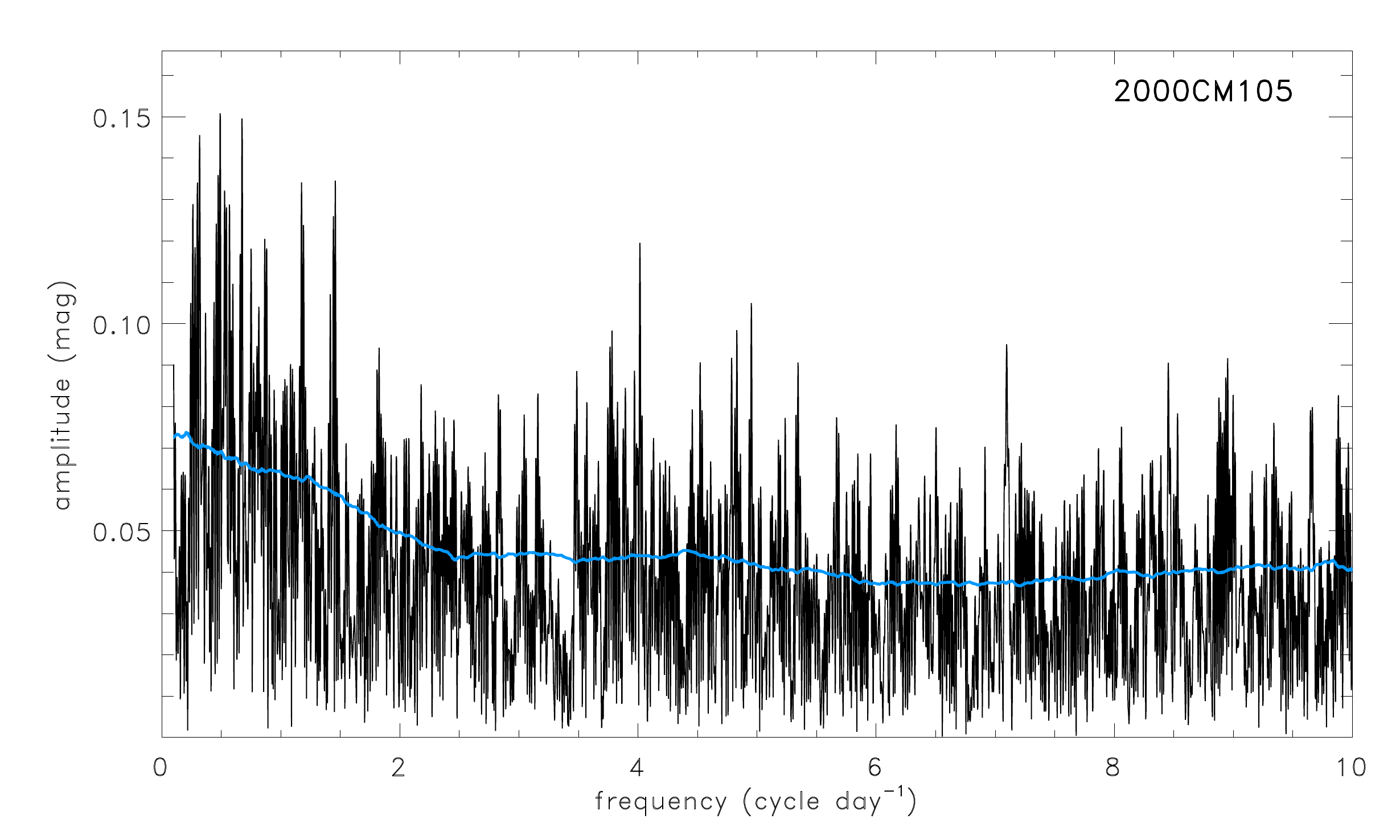}
\includegraphics[width=0.33\textwidth]{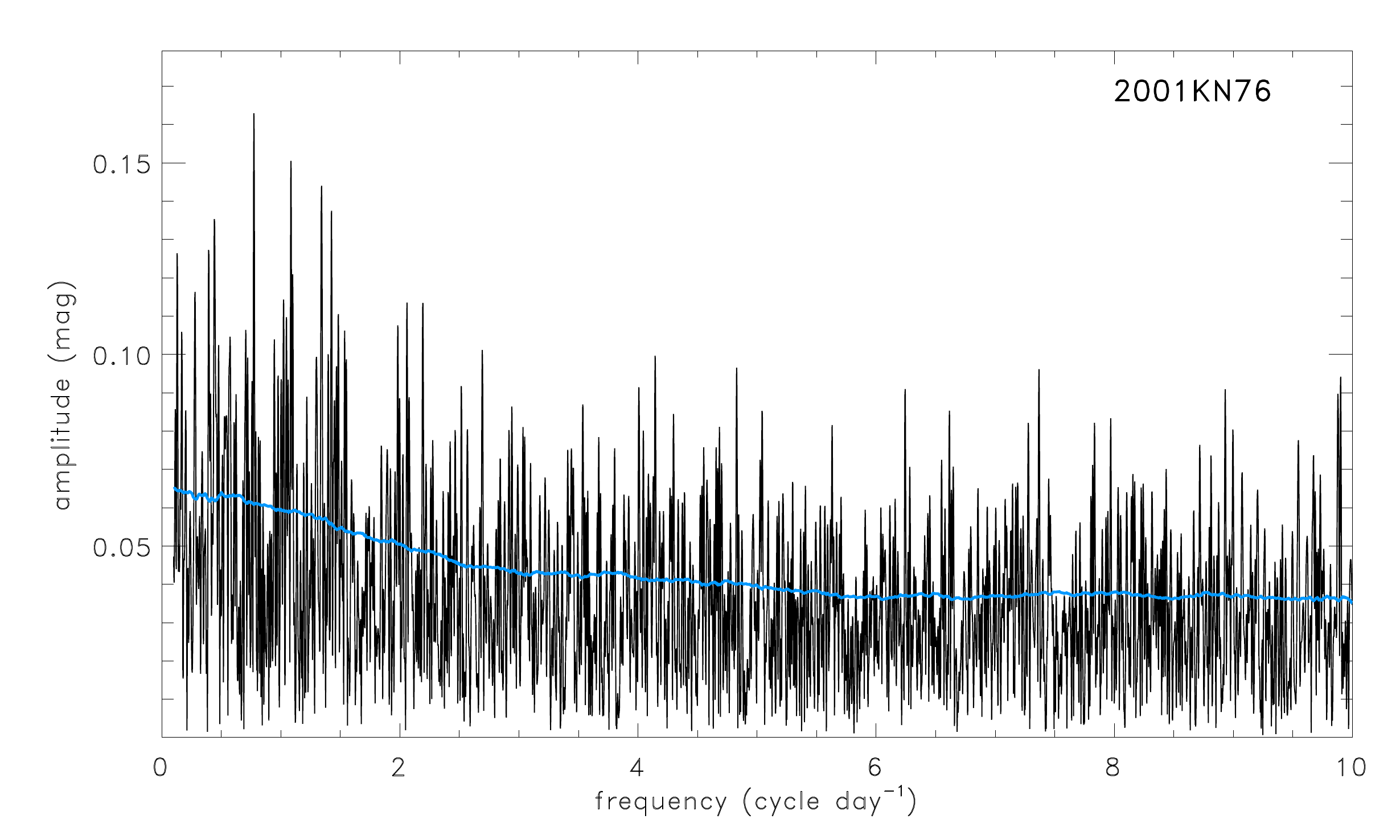}
\includegraphics[width=0.33\textwidth]{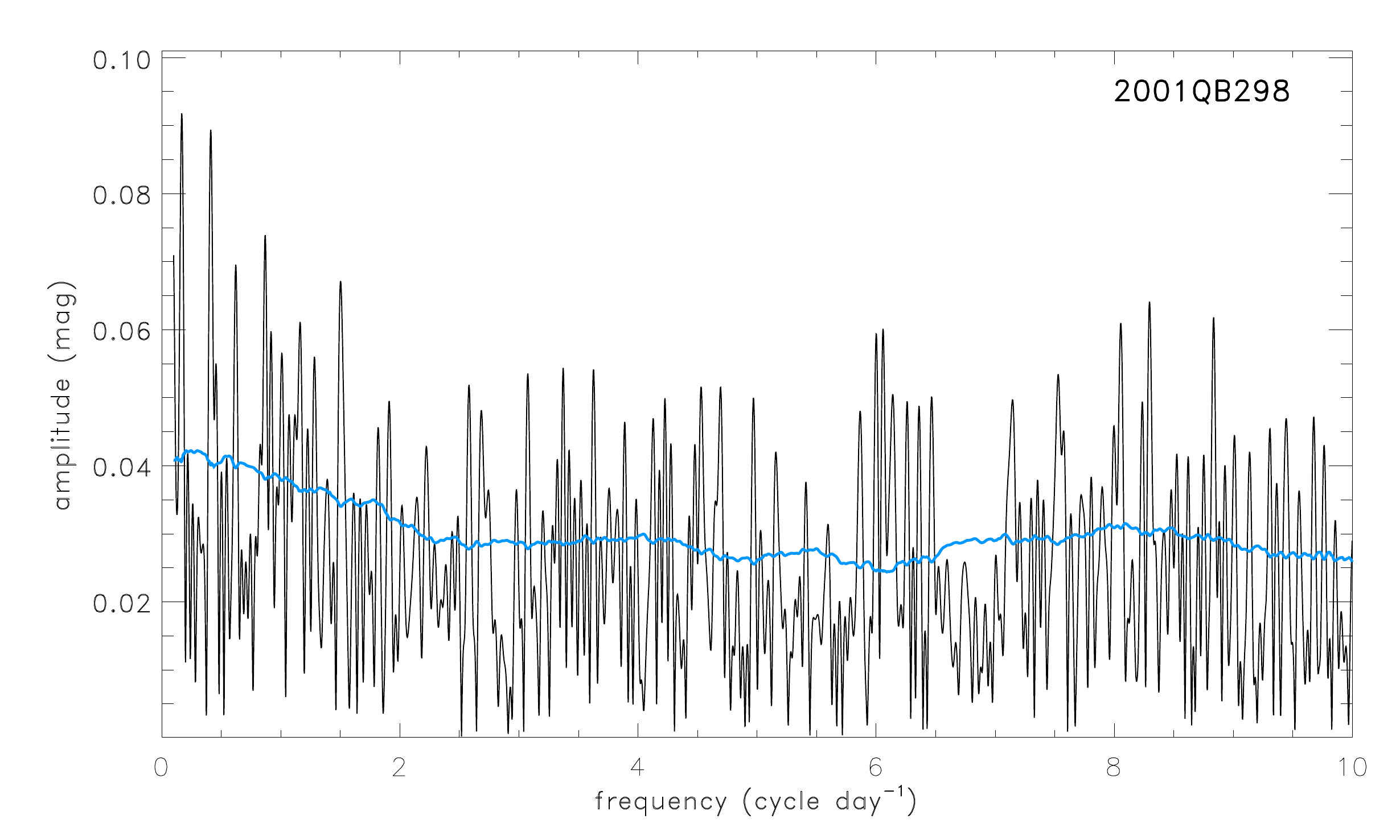}
}
\hbox{
\includegraphics[width=0.33\textwidth]{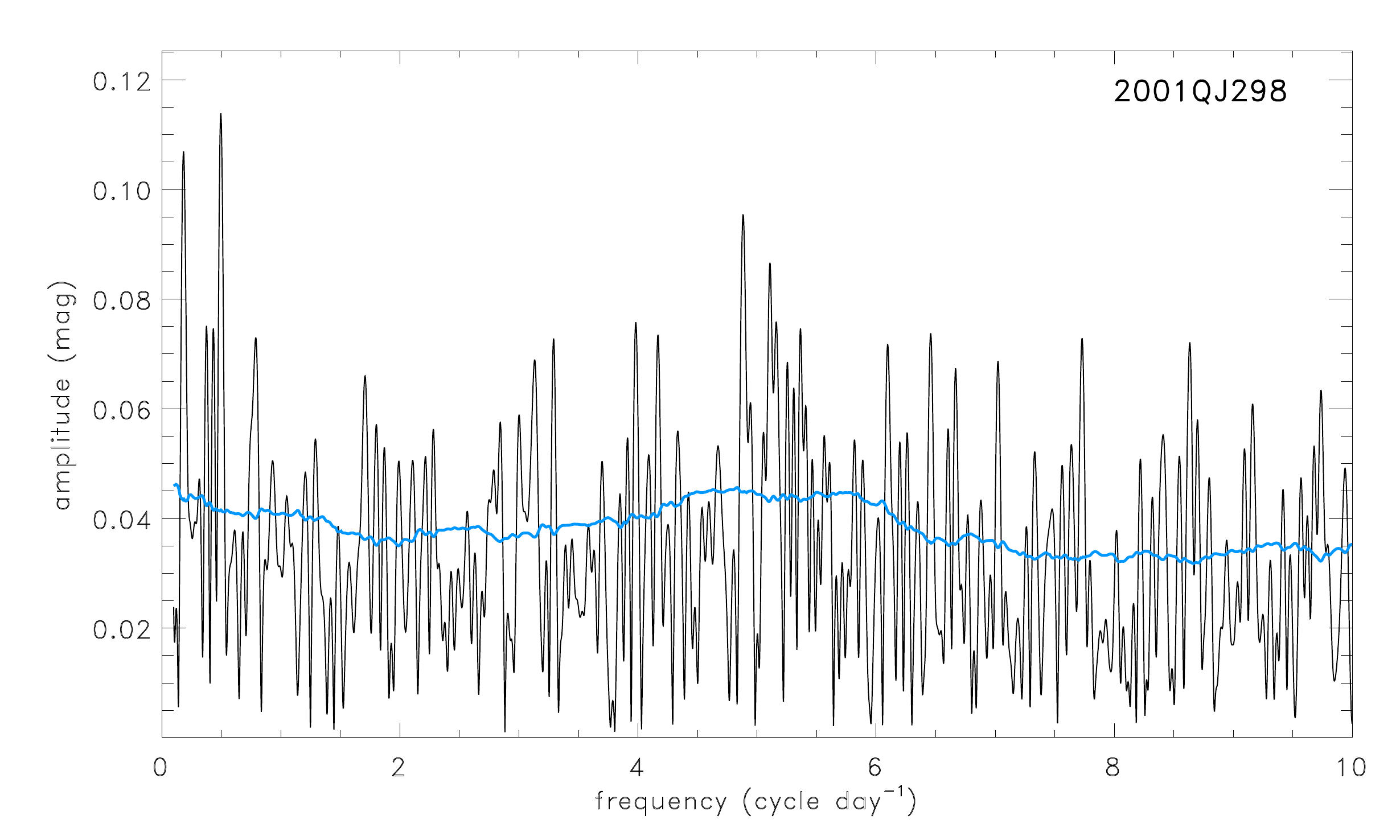}
\includegraphics[width=0.33\textwidth]{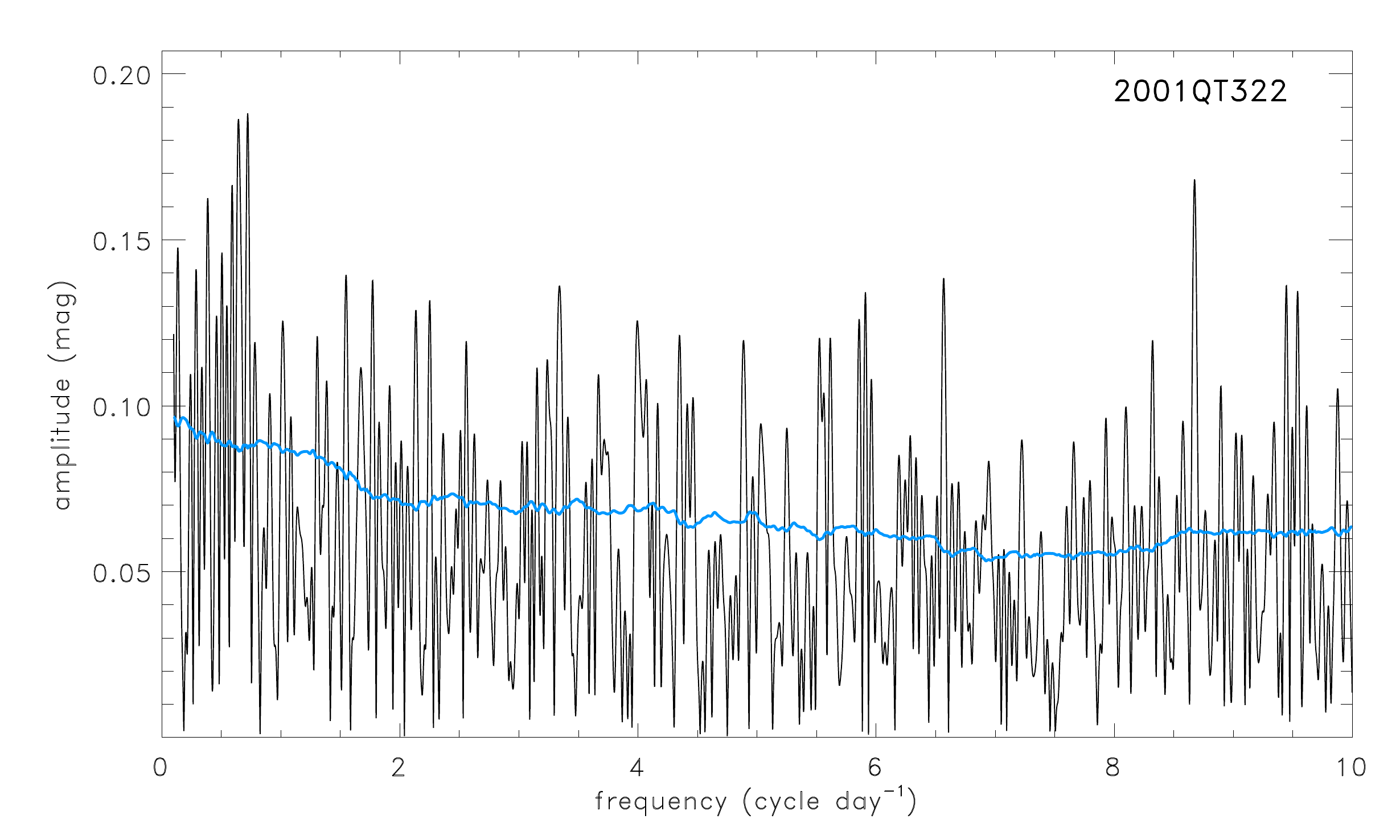}
\includegraphics[width=0.33\textwidth]{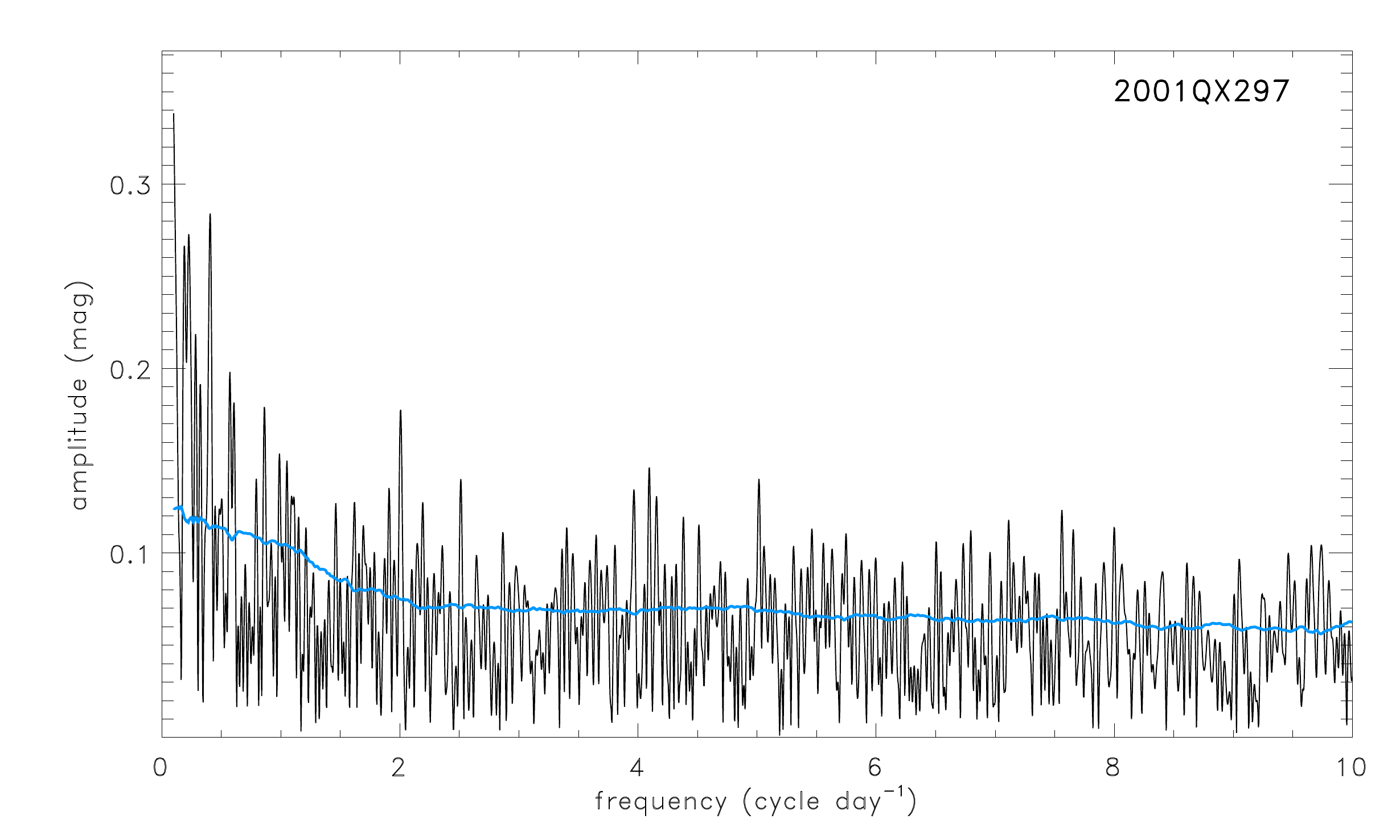}
}
\hbox{
\includegraphics[width=0.33\textwidth]{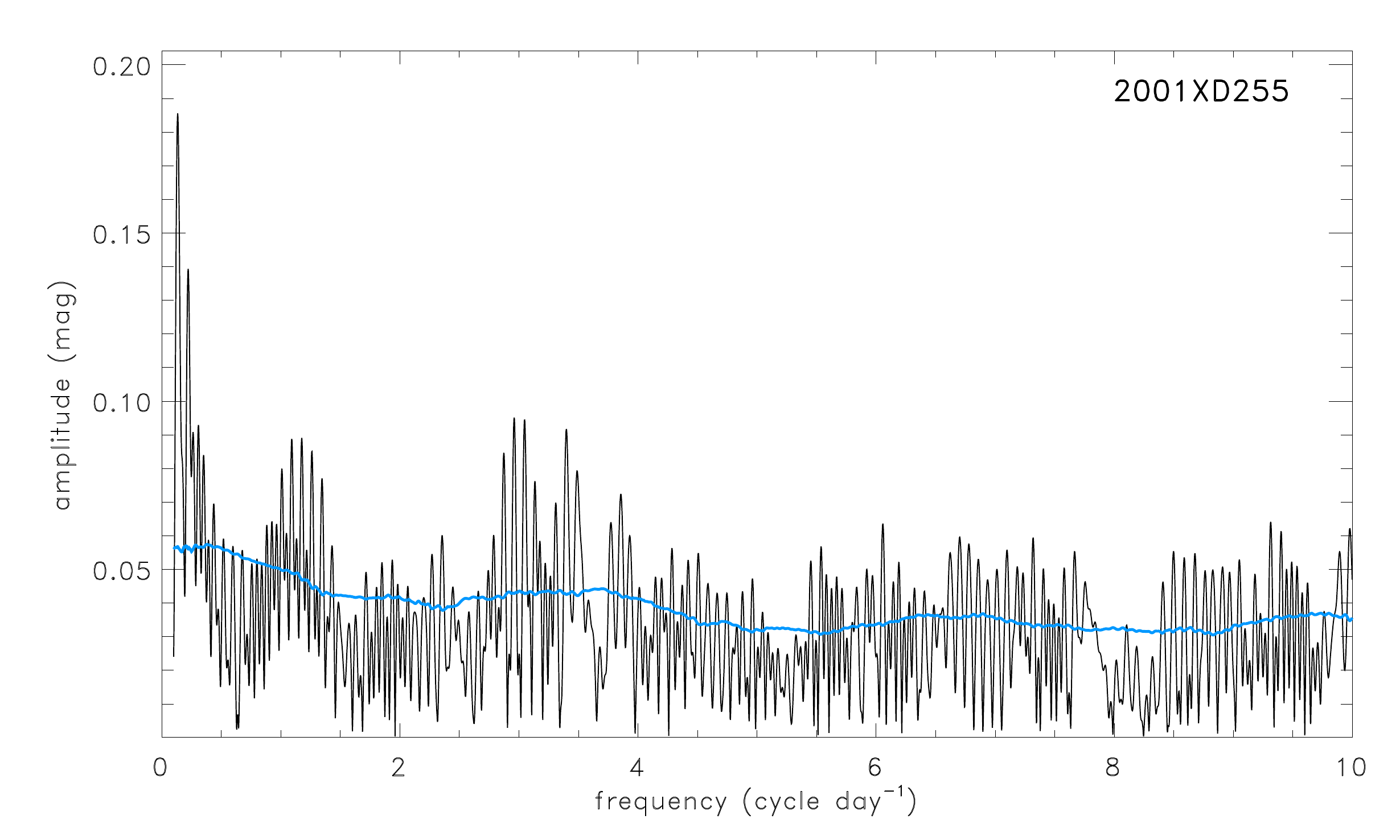}
\includegraphics[width=0.33\textwidth]{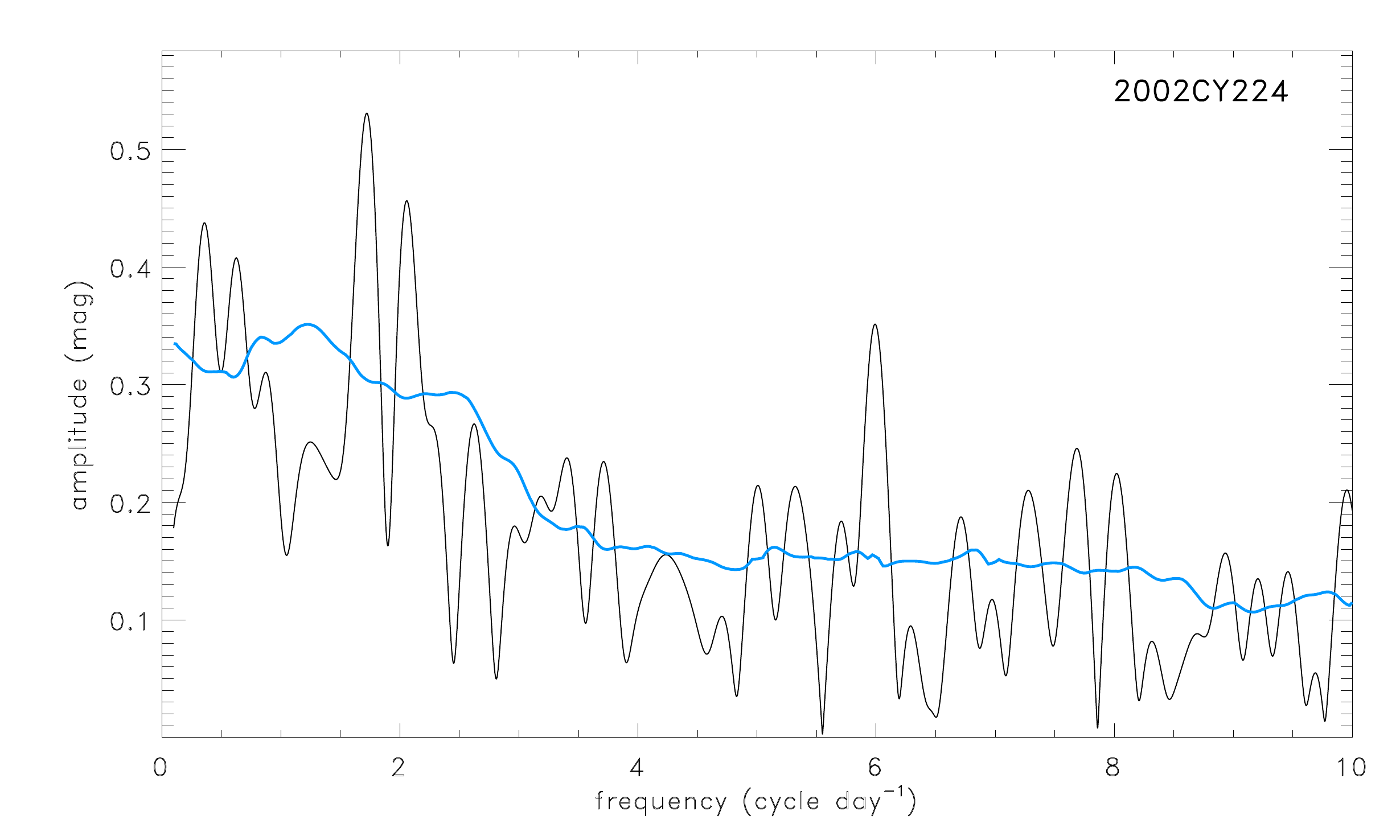}
\includegraphics[width=0.33\textwidth]{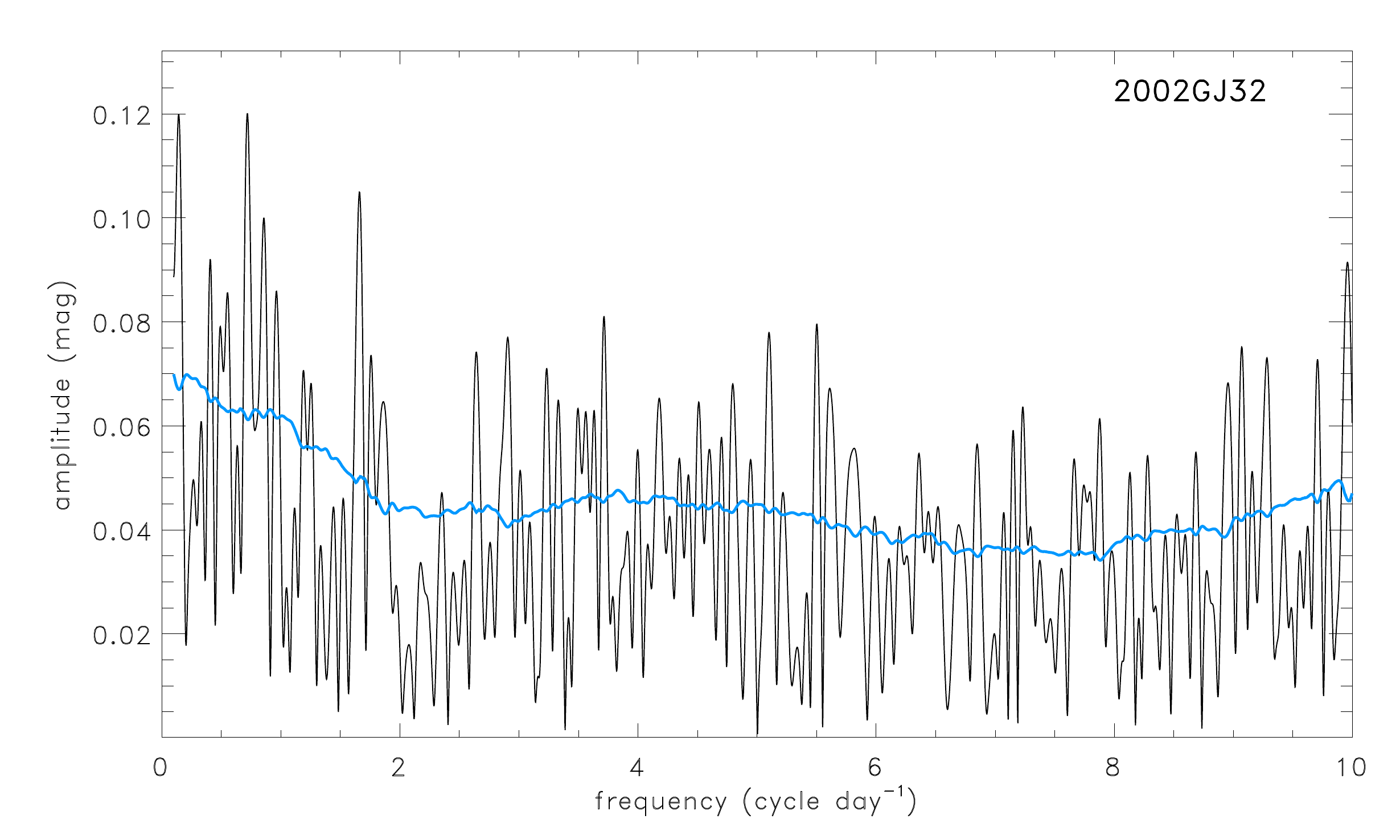}
}
\hbox{
\includegraphics[width=0.33\textwidth]{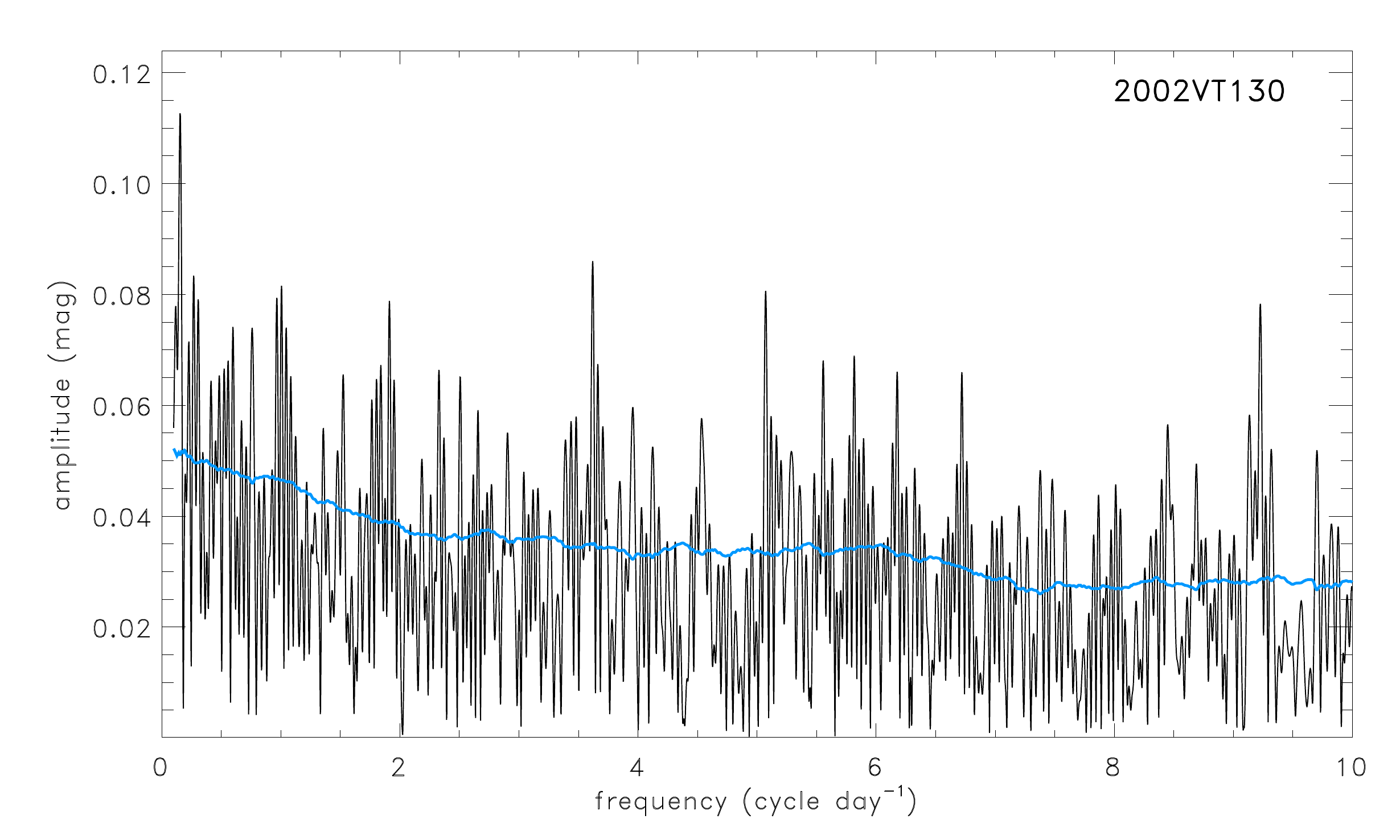}
\includegraphics[width=0.33\textwidth]{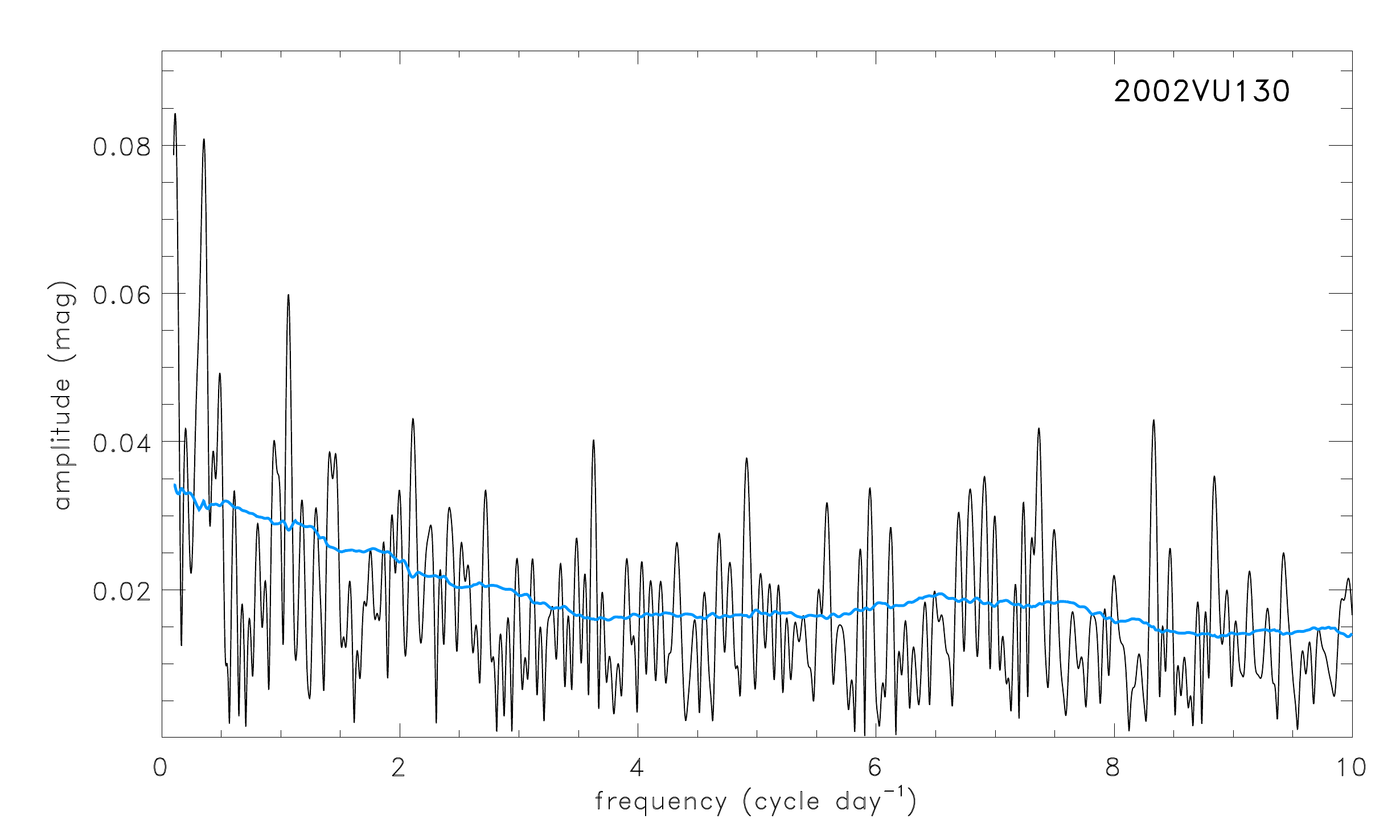}
\includegraphics[width=0.33\textwidth]{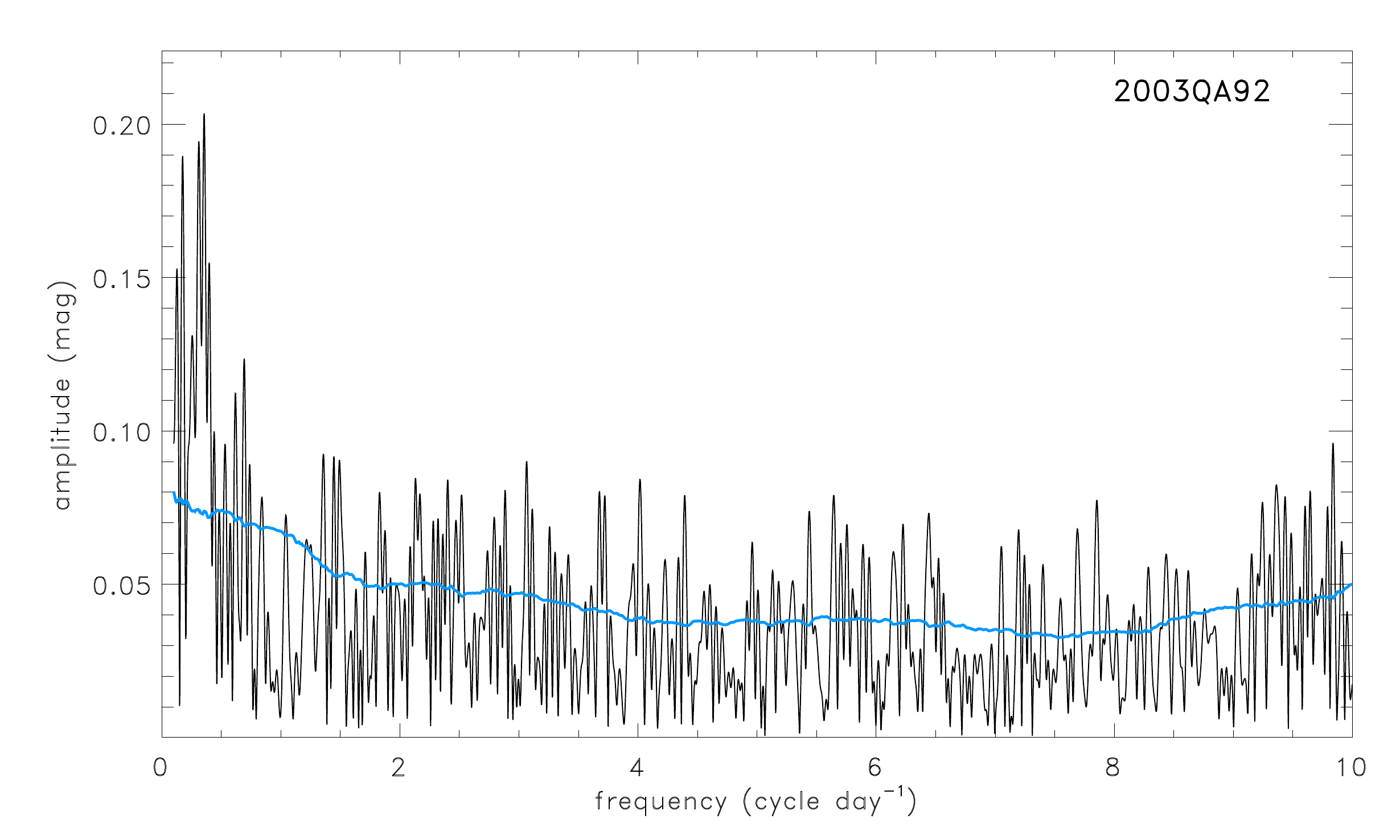}
}
\caption{In this figure we present the Fourier spectrum (amplitude vs. frequency) of the light curves of targets for which no peaks in the Fourier/residual spectra matched the detection criteria, as described in Sect.~\ref{sect:basicdatareduction}.
The blue curve represents the 1$\sigma$ r.m.s. of the frequency spectrum. The name of the target is indicated in each figure. 
\label{fig:nodetect0} }
\end{figure}


\begin{figure}[ht!]
\centering
\ContinuedFloat
\hbox{
\includegraphics[width=0.33\textwidth]{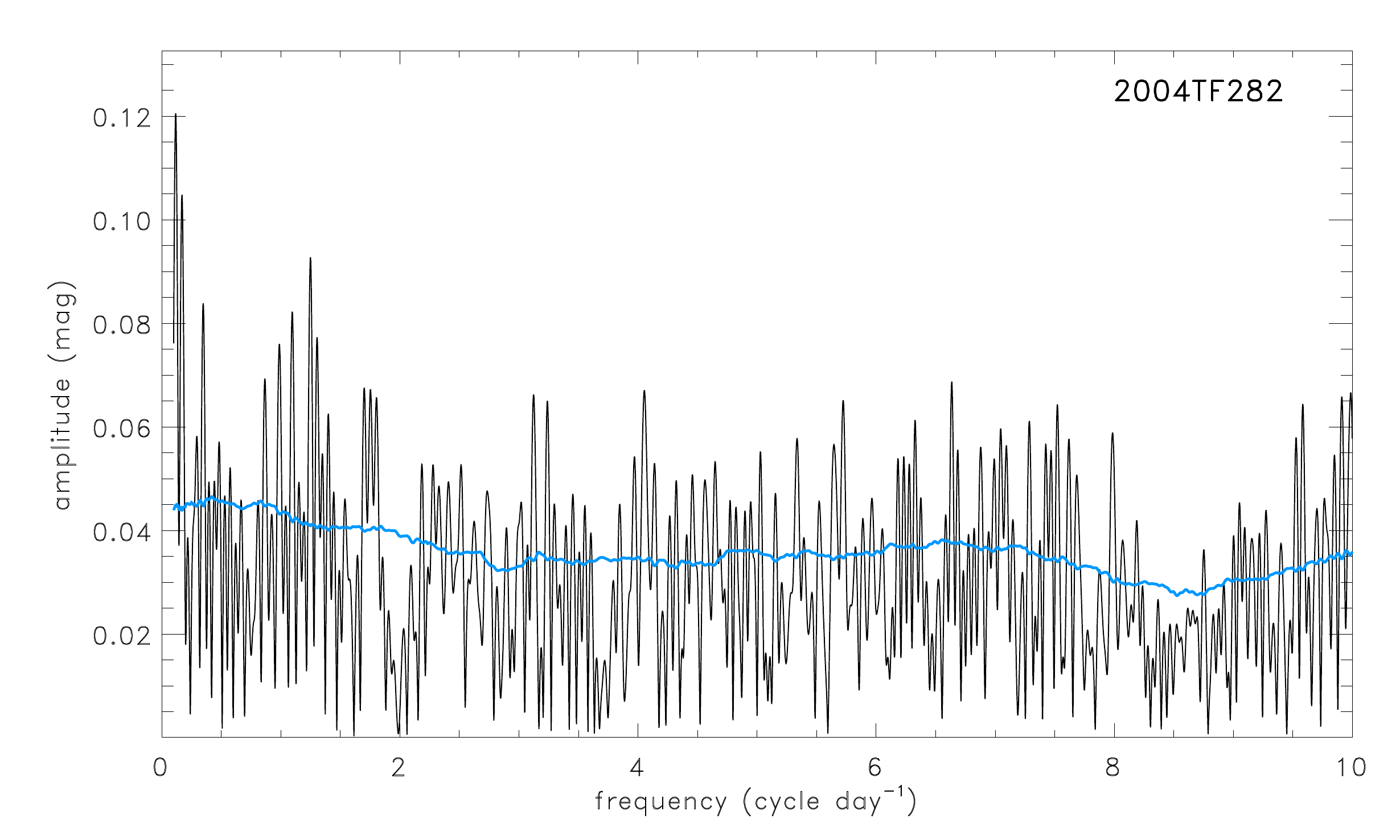}
\includegraphics[width=0.33\textwidth]{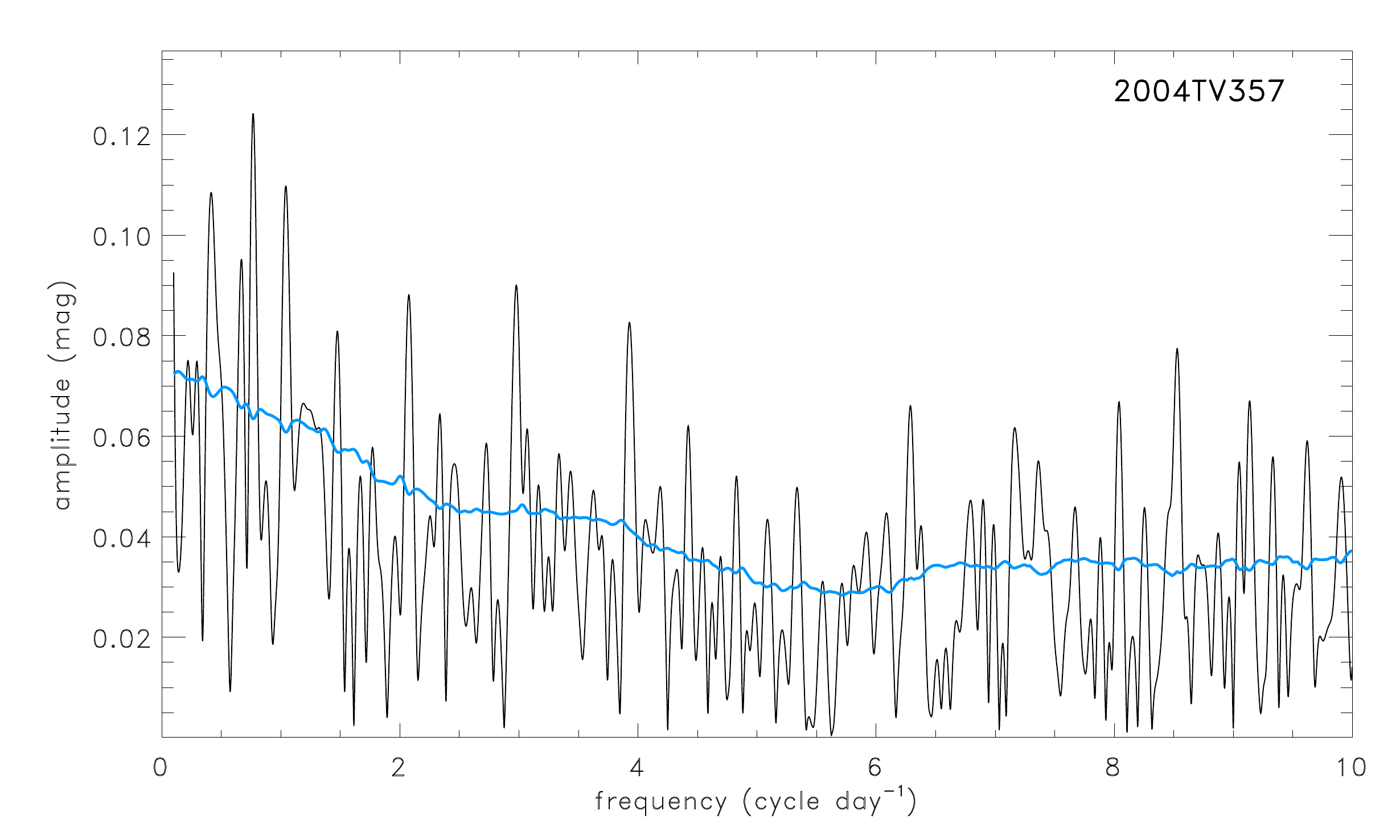}
\includegraphics[width=0.33\textwidth]{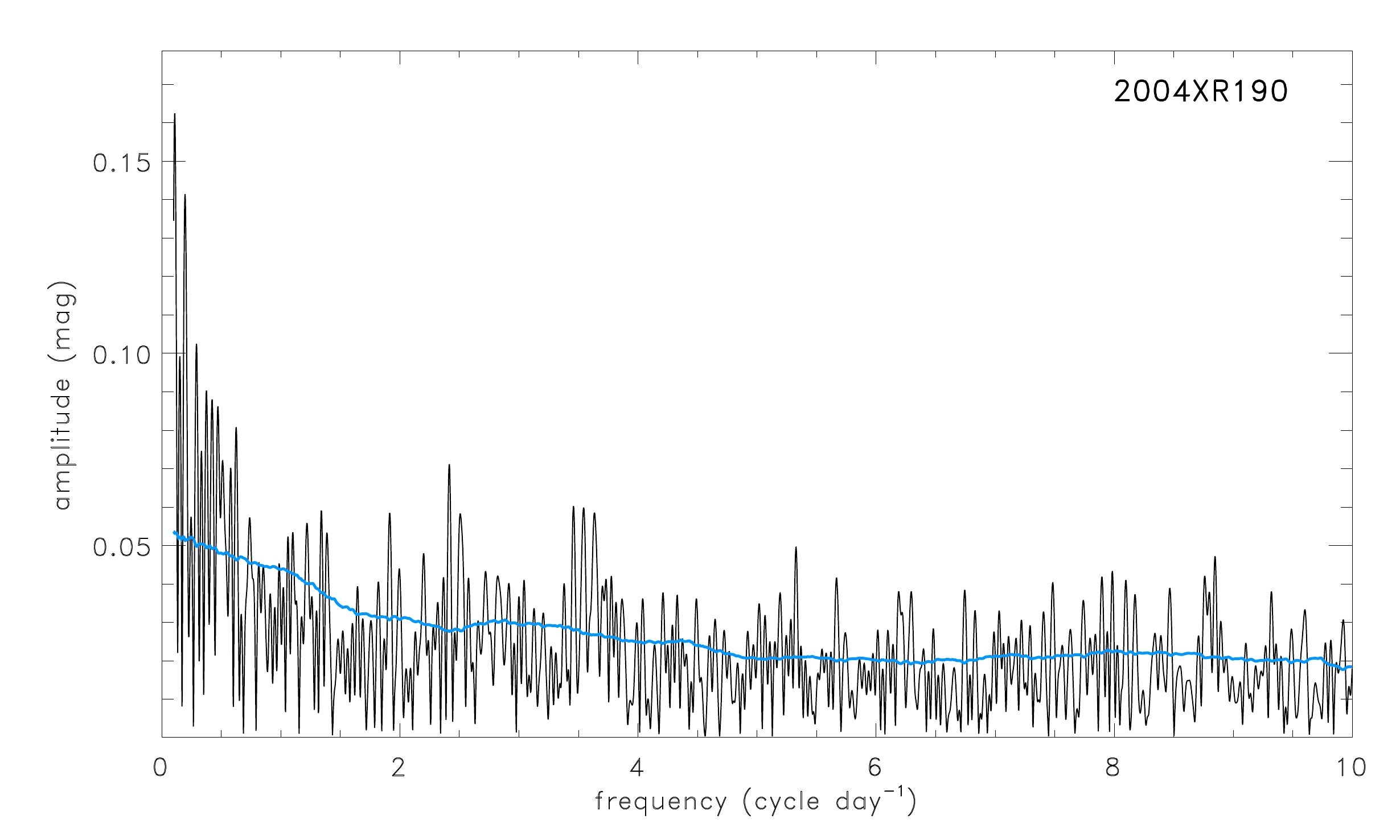}
}
\hbox{
\includegraphics[width=0.33\textwidth]{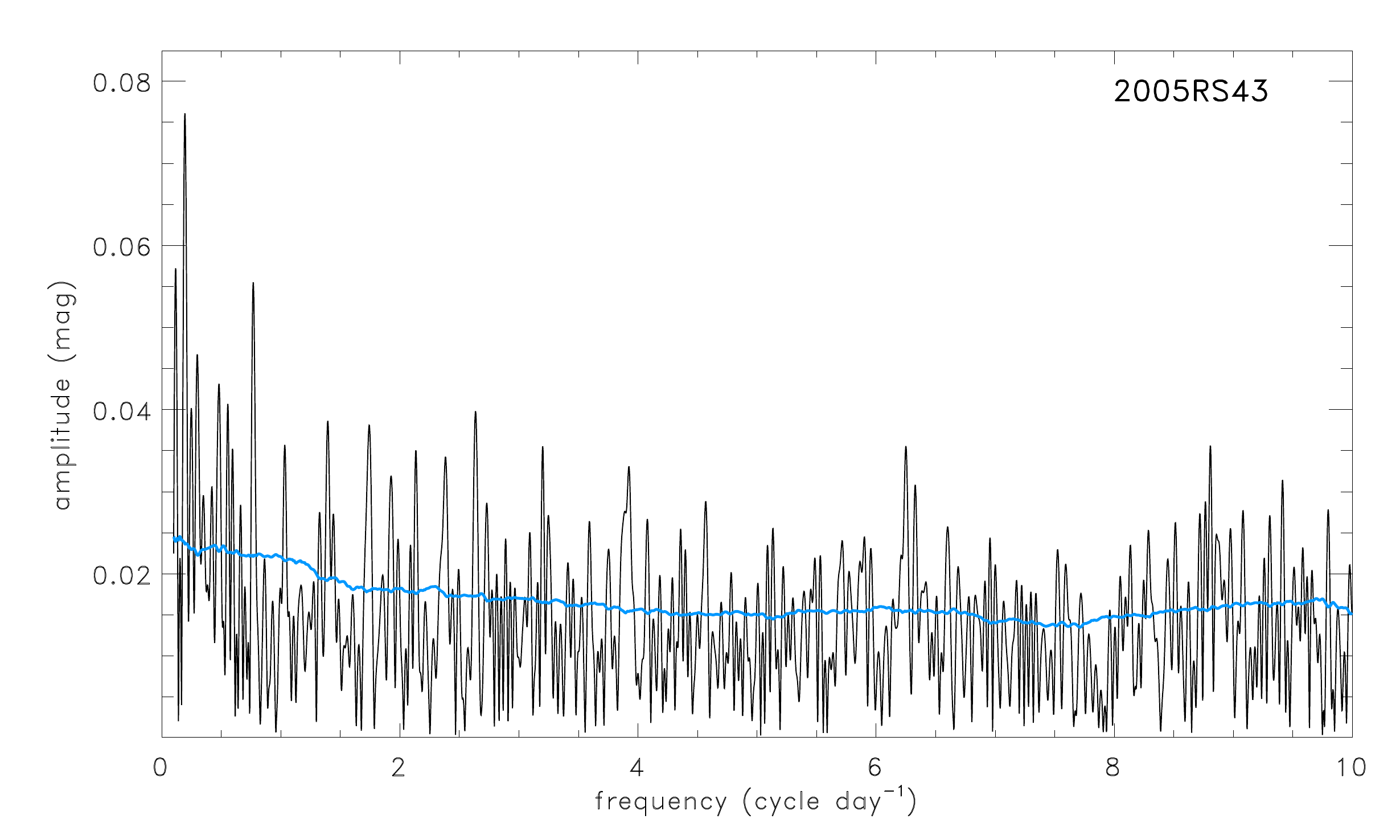}
\includegraphics[width=0.33\textwidth]{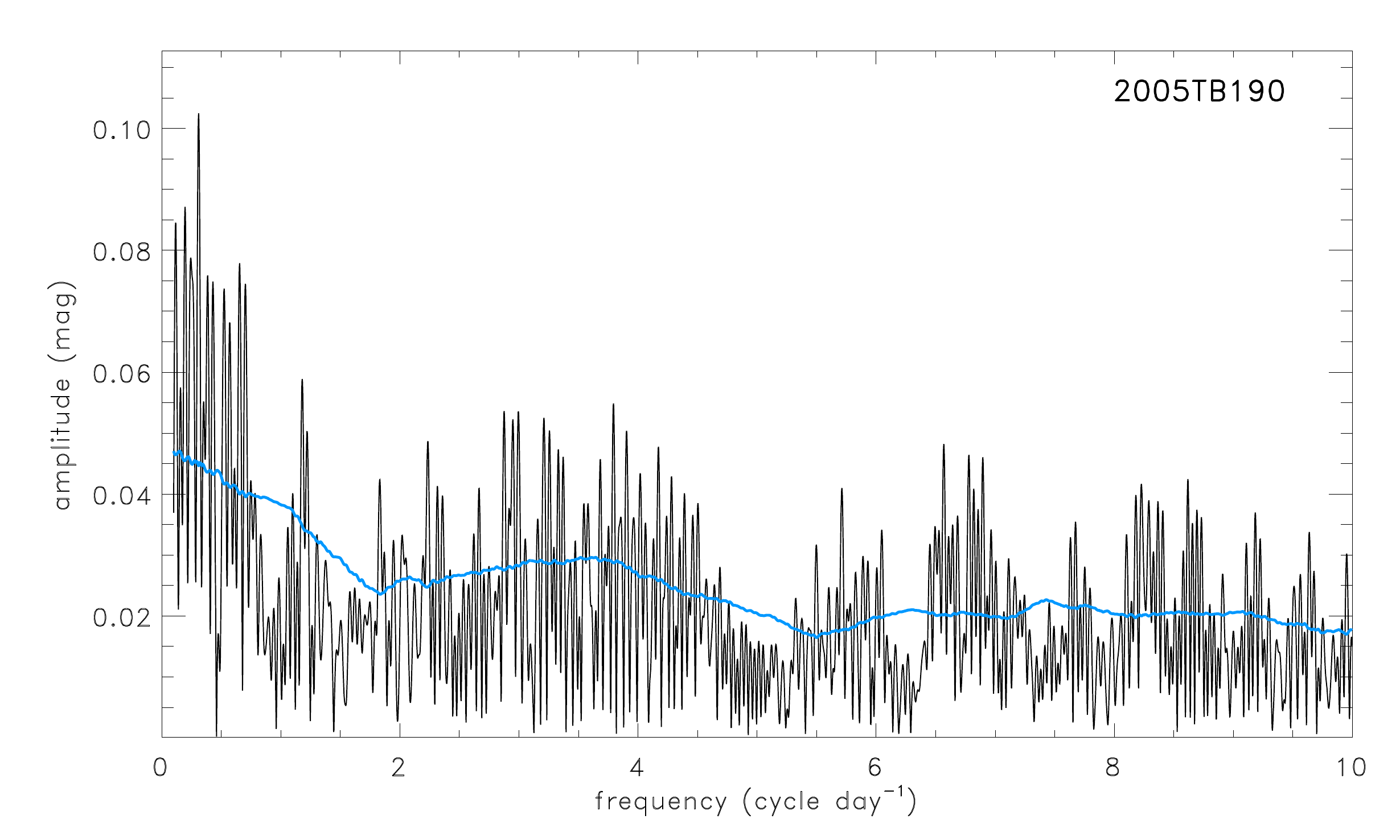}
\includegraphics[width=0.33\textwidth]{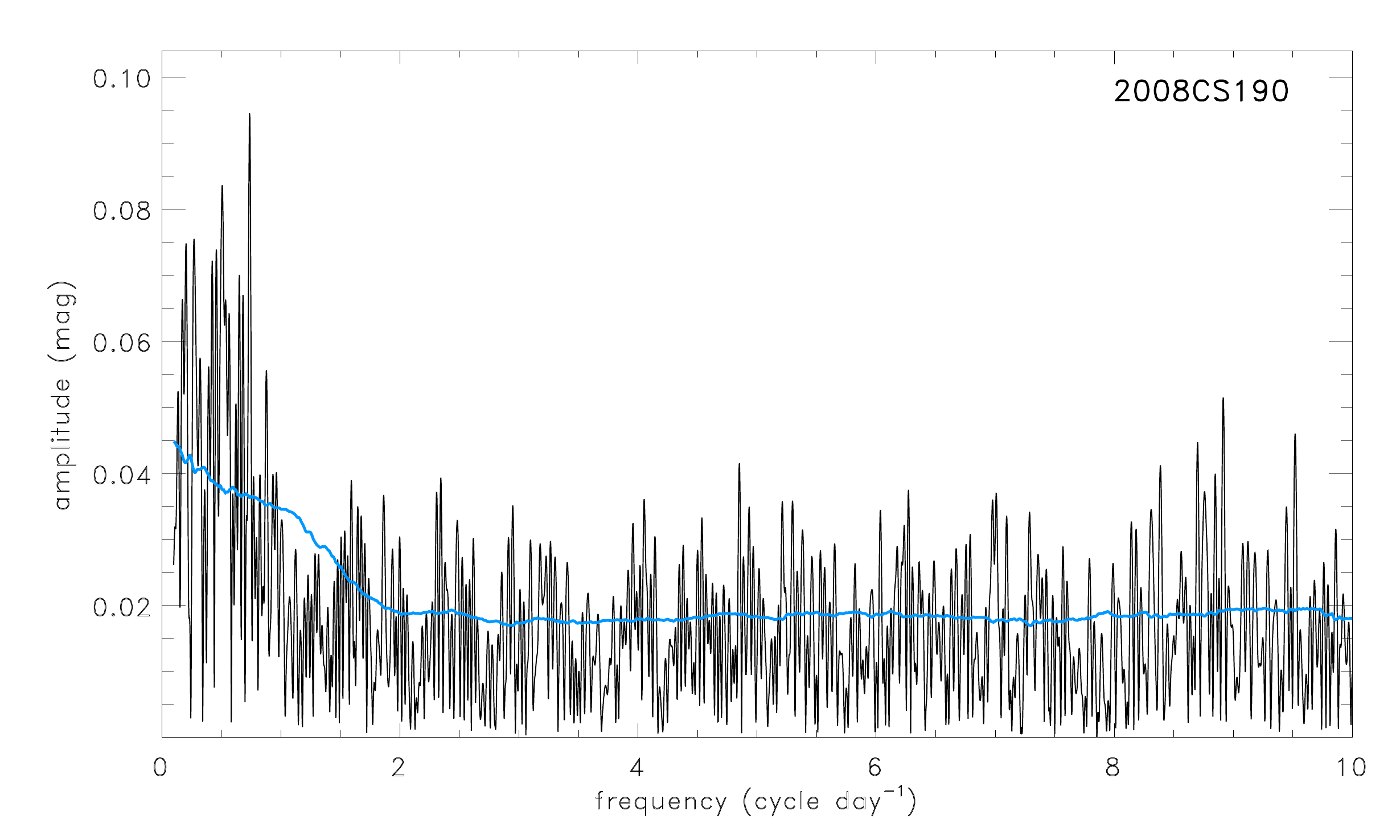}
}
\hbox{
\includegraphics[width=0.33\textwidth]{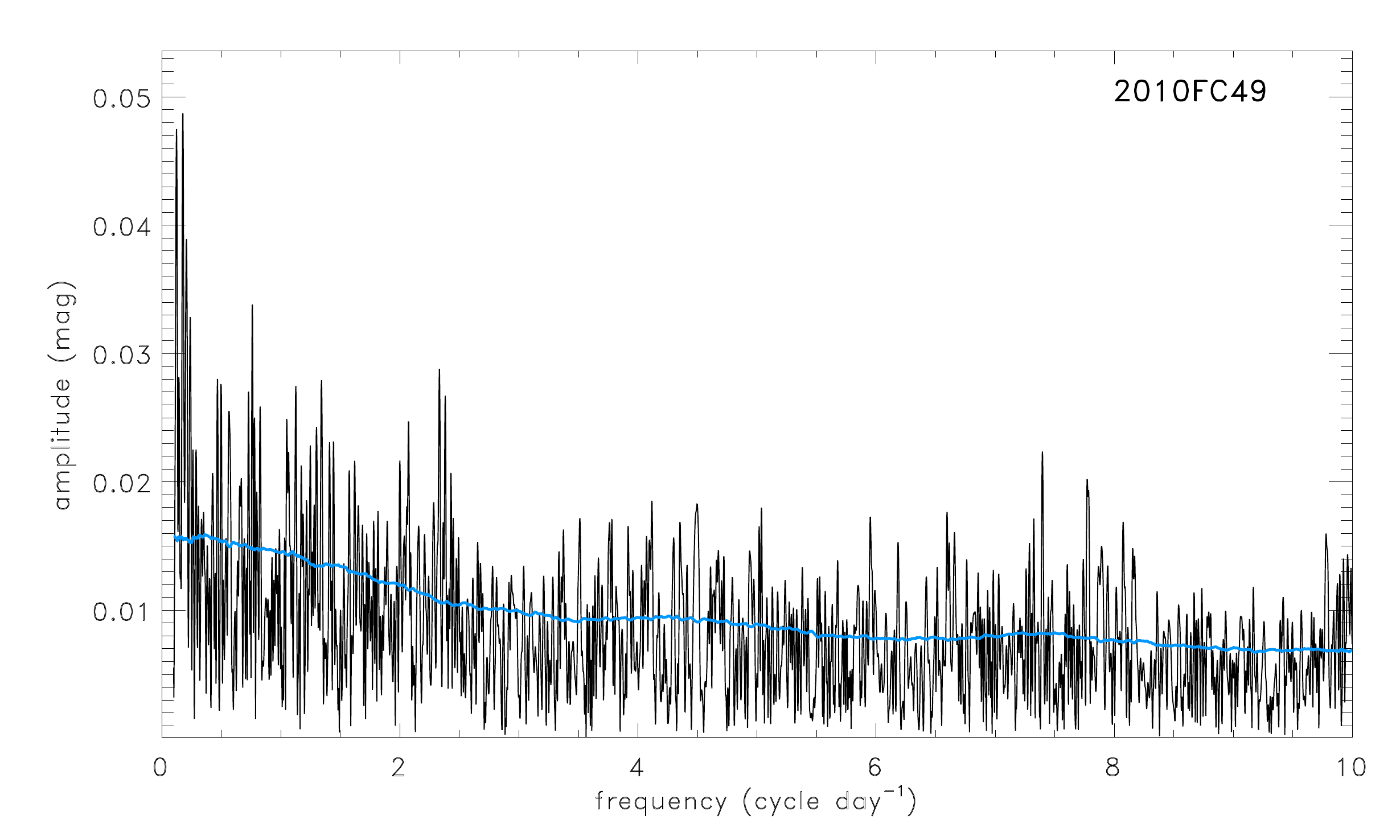}
\includegraphics[width=0.33\textwidth]{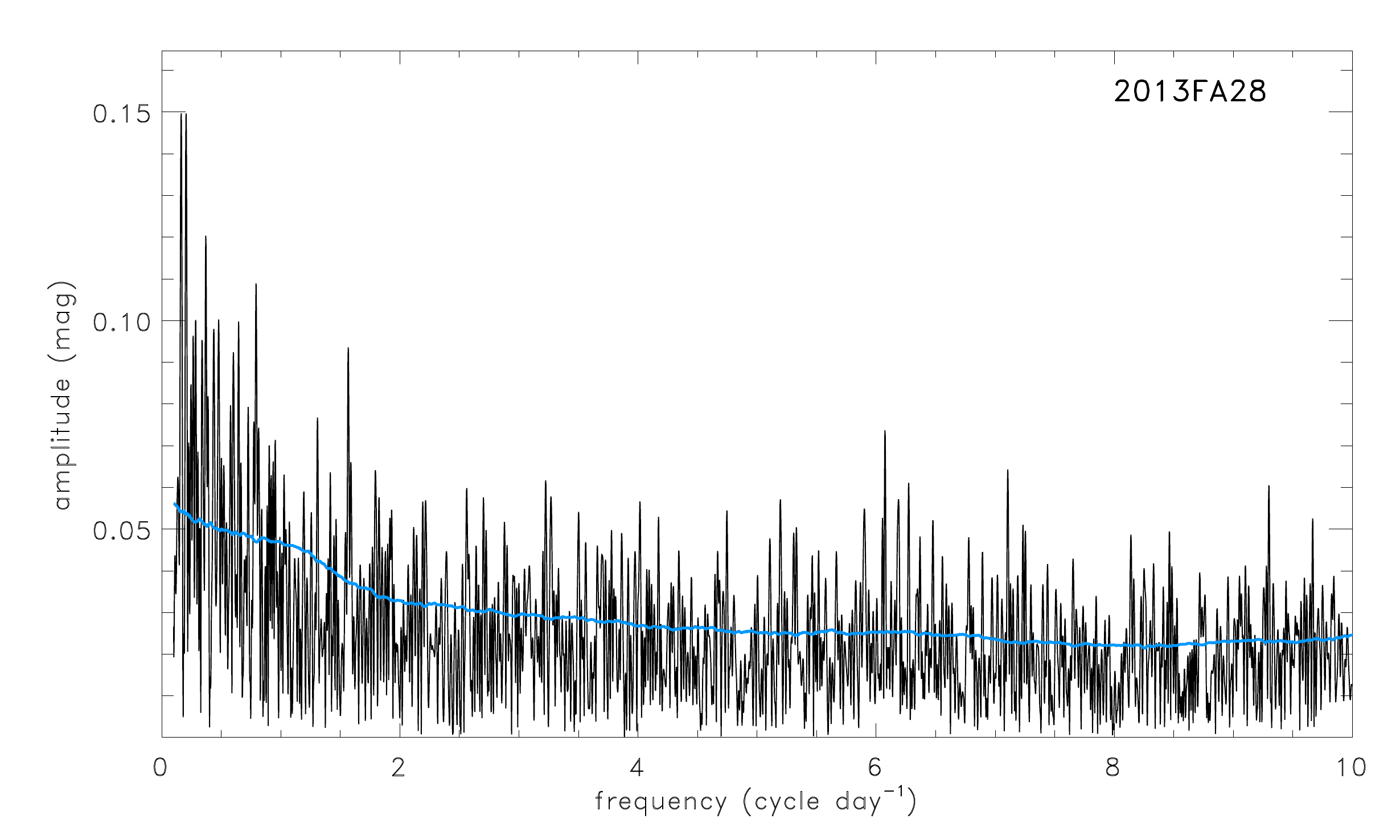}
\includegraphics[width=0.33\textwidth]{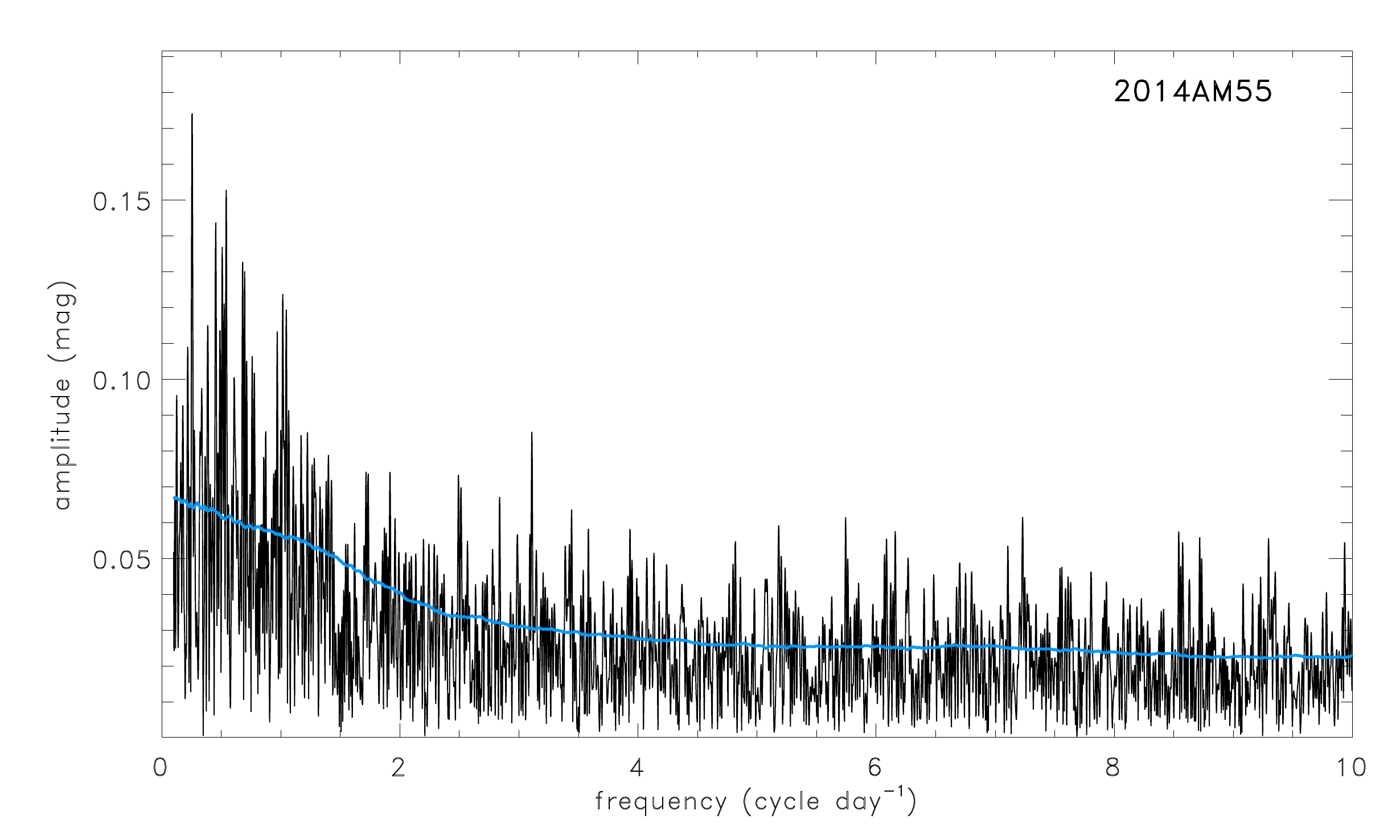}
}
\hbox{
\includegraphics[width=0.33\textwidth]{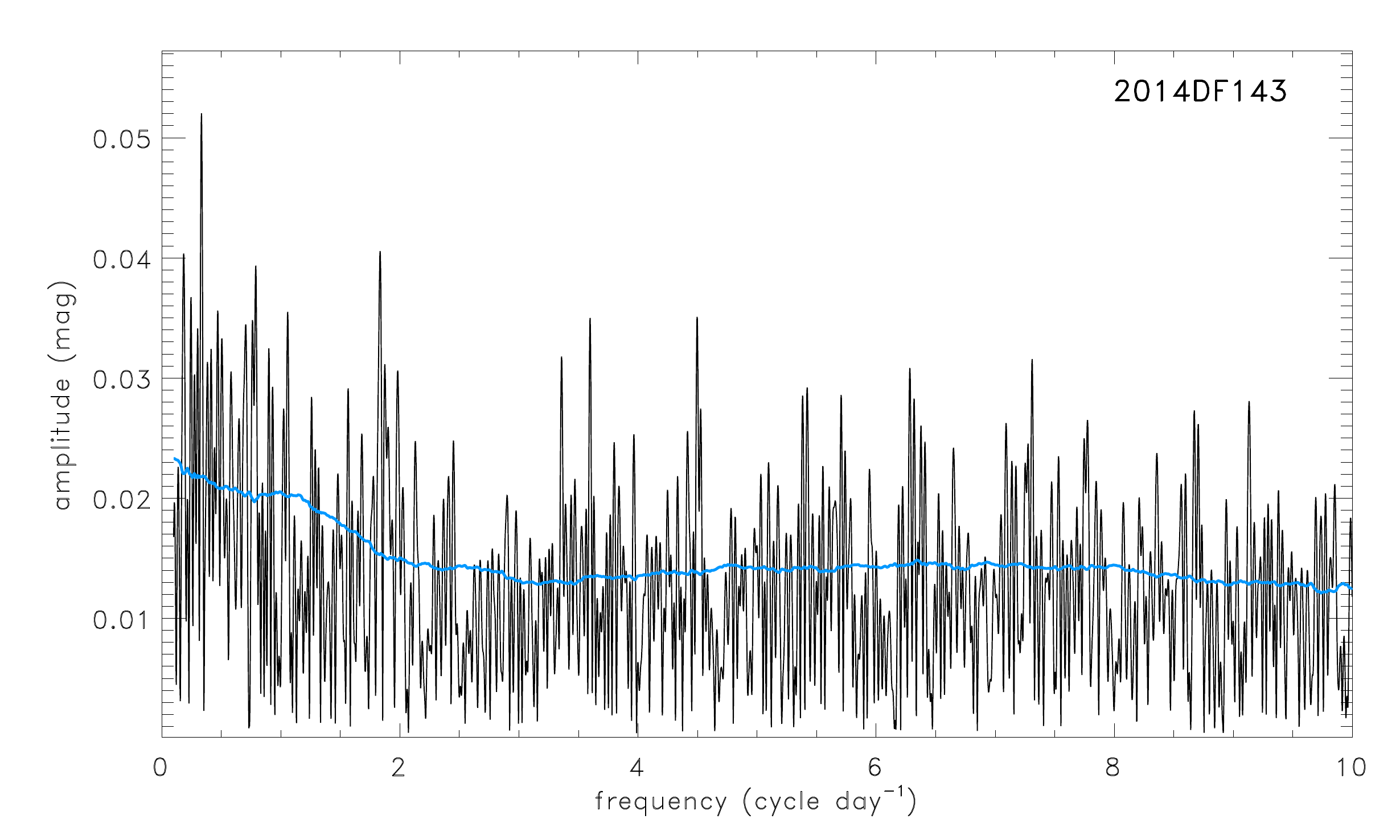}
\includegraphics[width=0.33\textwidth]{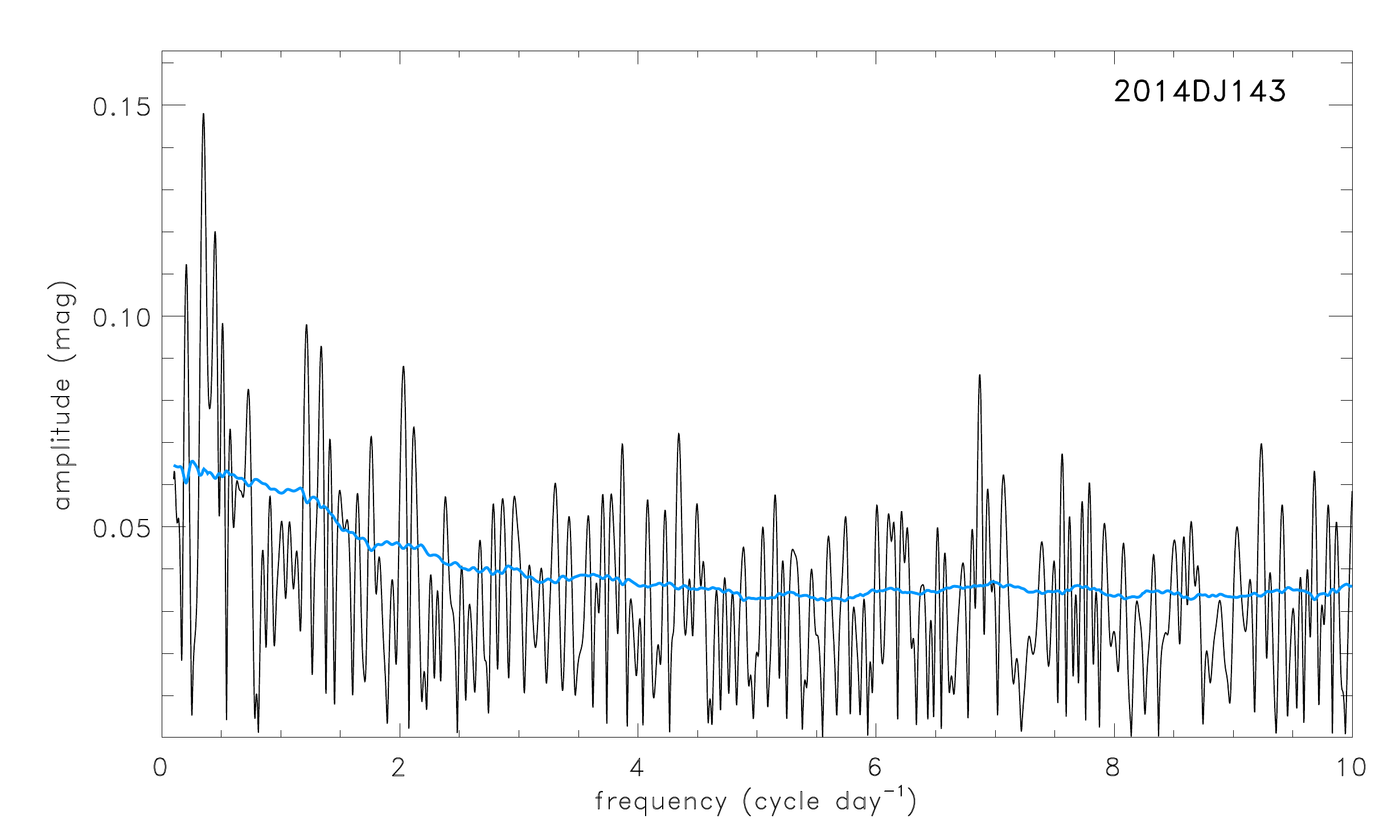}
\includegraphics[width=0.33\textwidth]{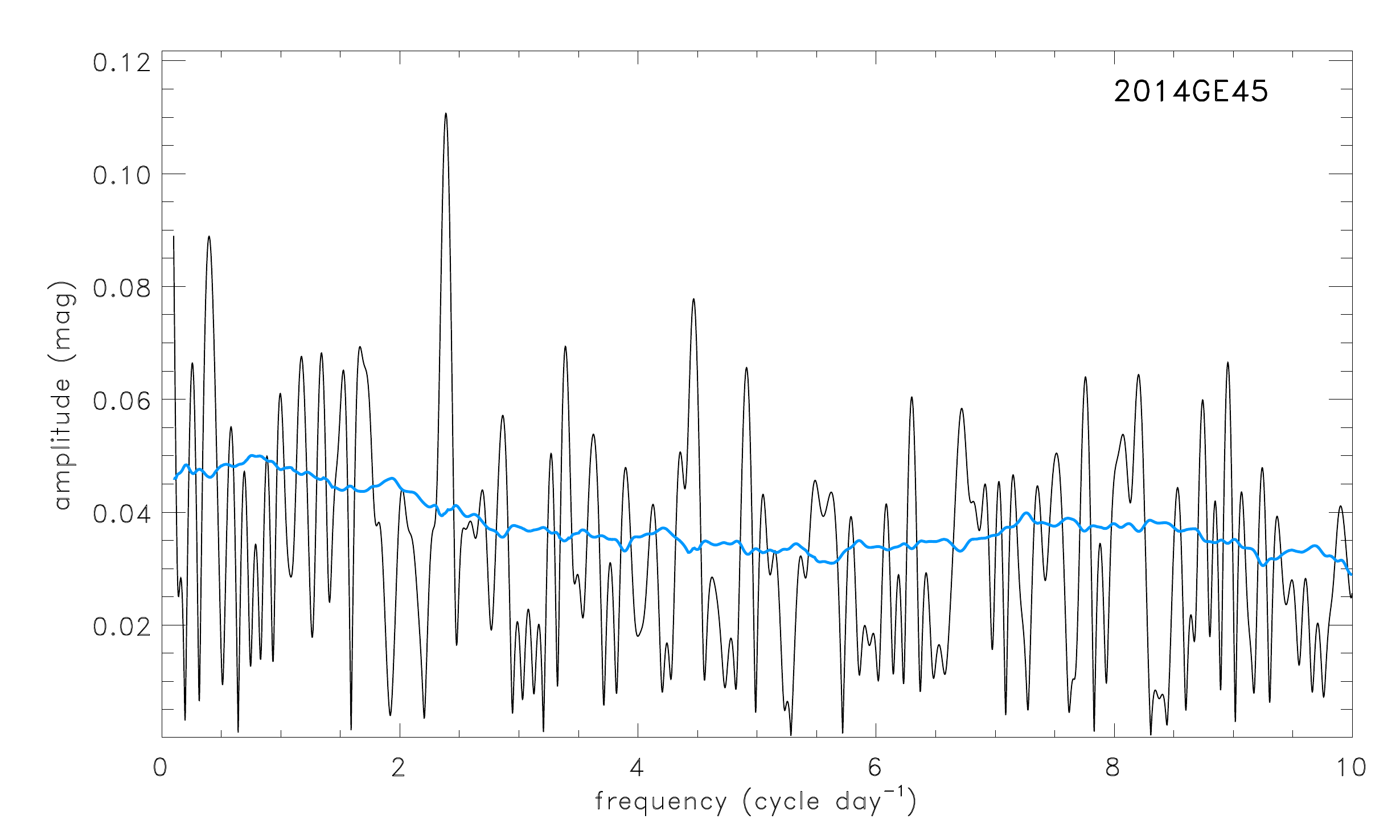}
}
\hbox{
\includegraphics[width=0.33\textwidth]{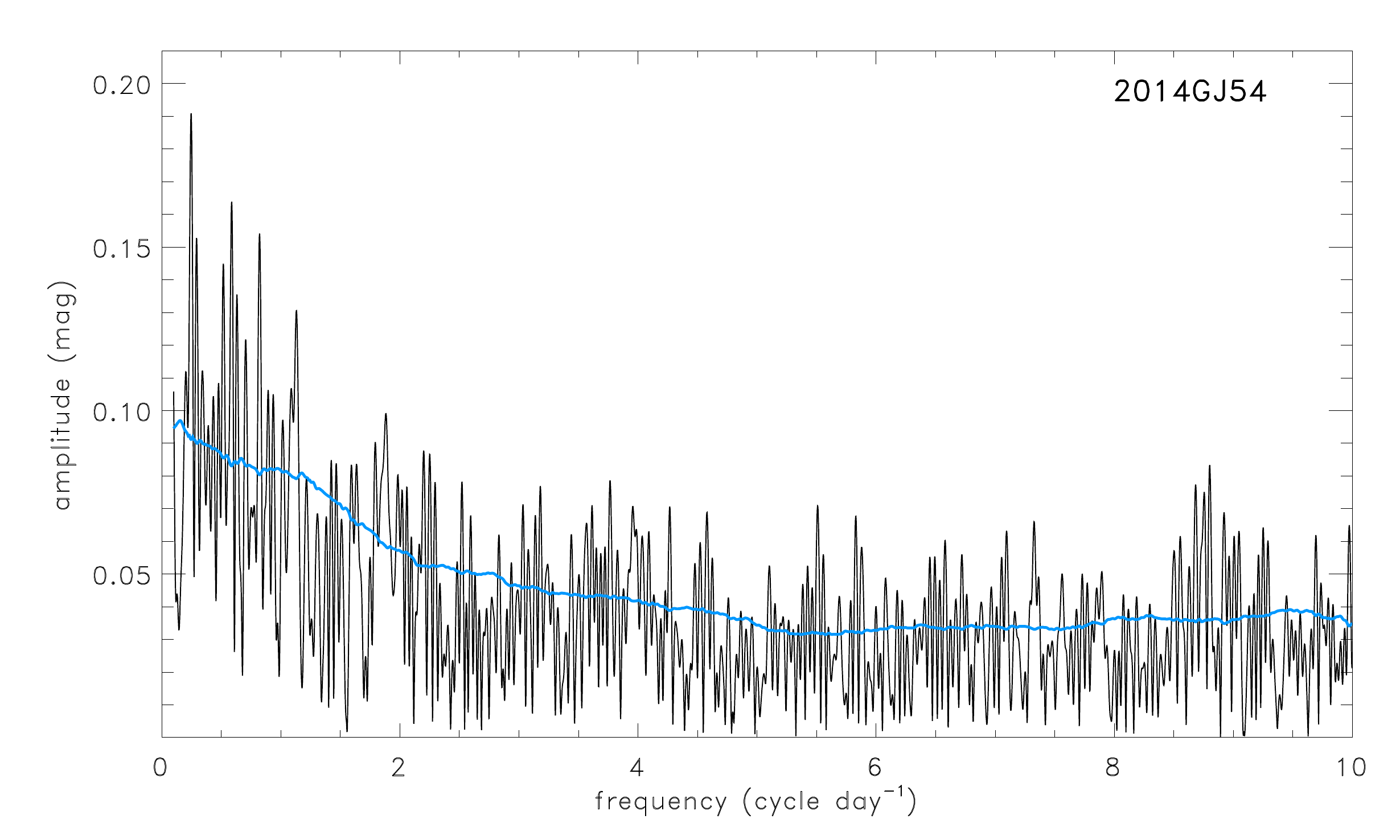}
\includegraphics[width=0.33\textwidth]{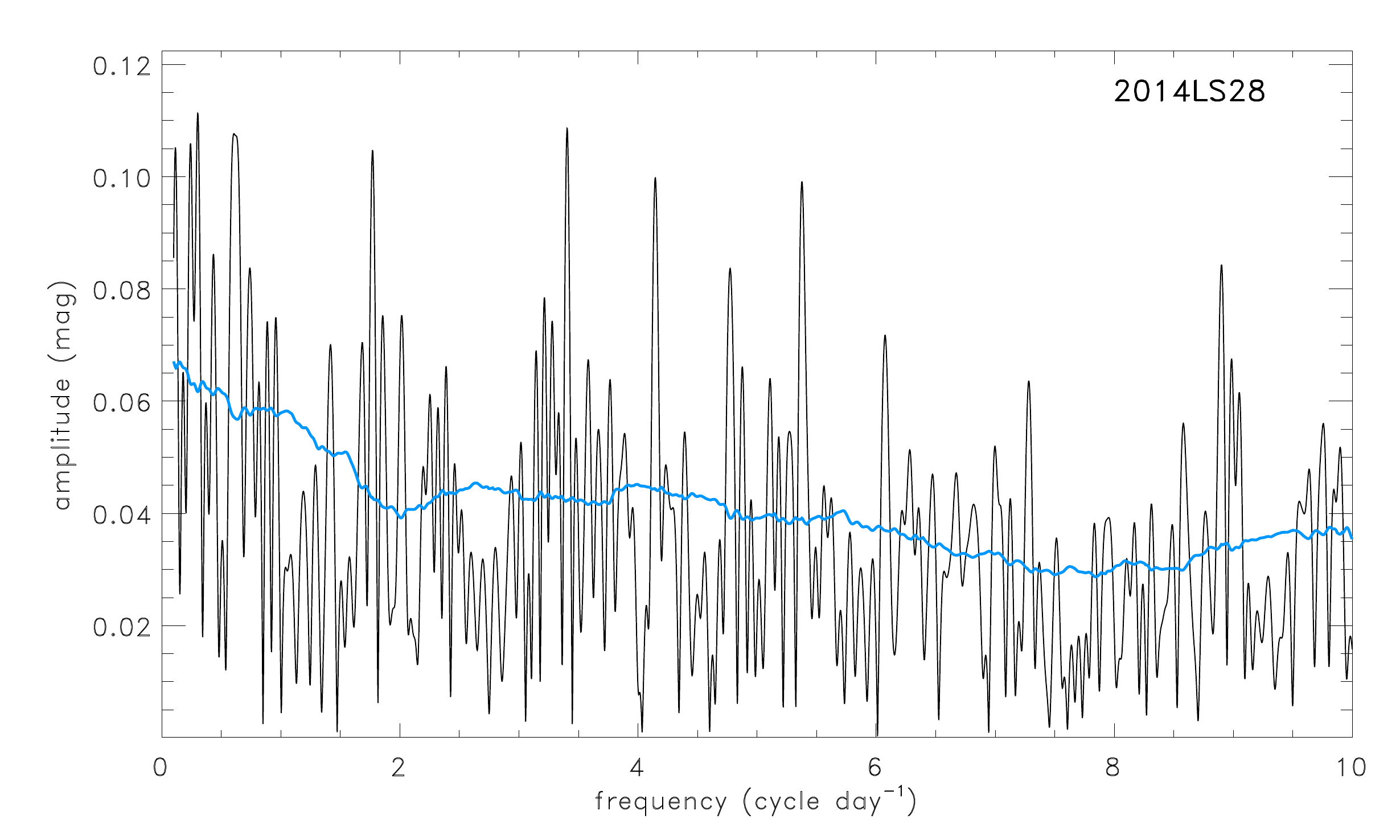}
\includegraphics[width=0.33\textwidth]{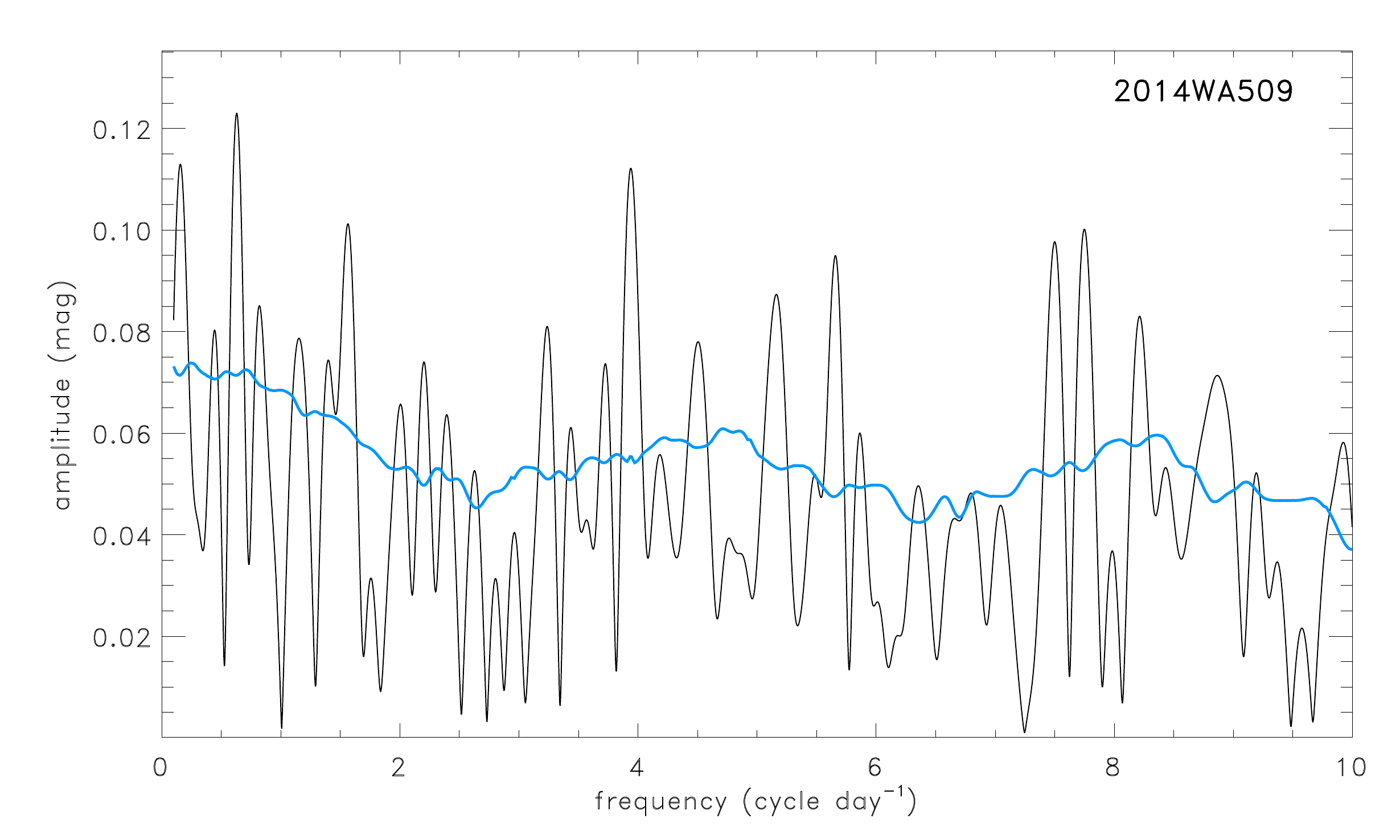}
}
\hbox{
\includegraphics[width=0.33\textwidth]{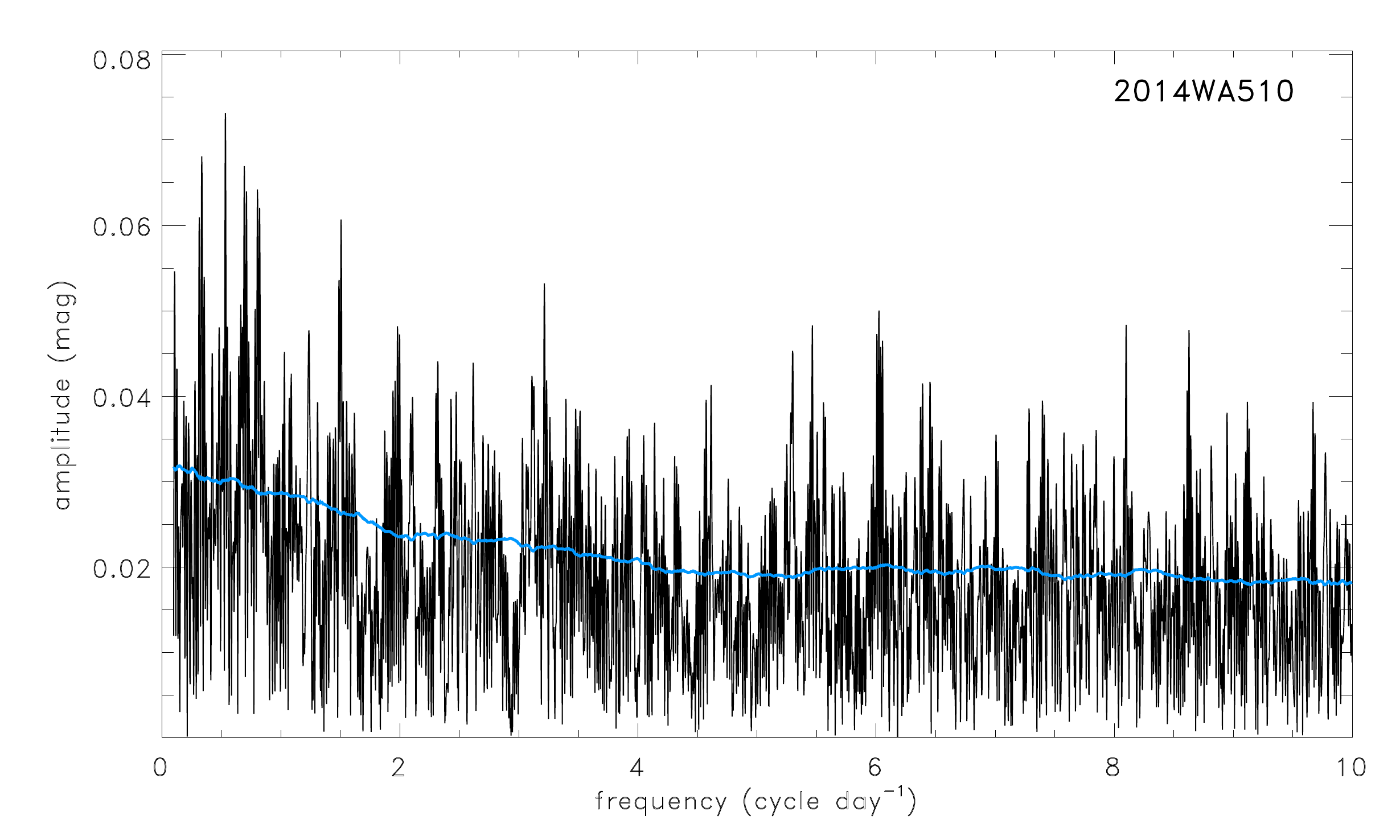}
\includegraphics[width=0.33\textwidth]{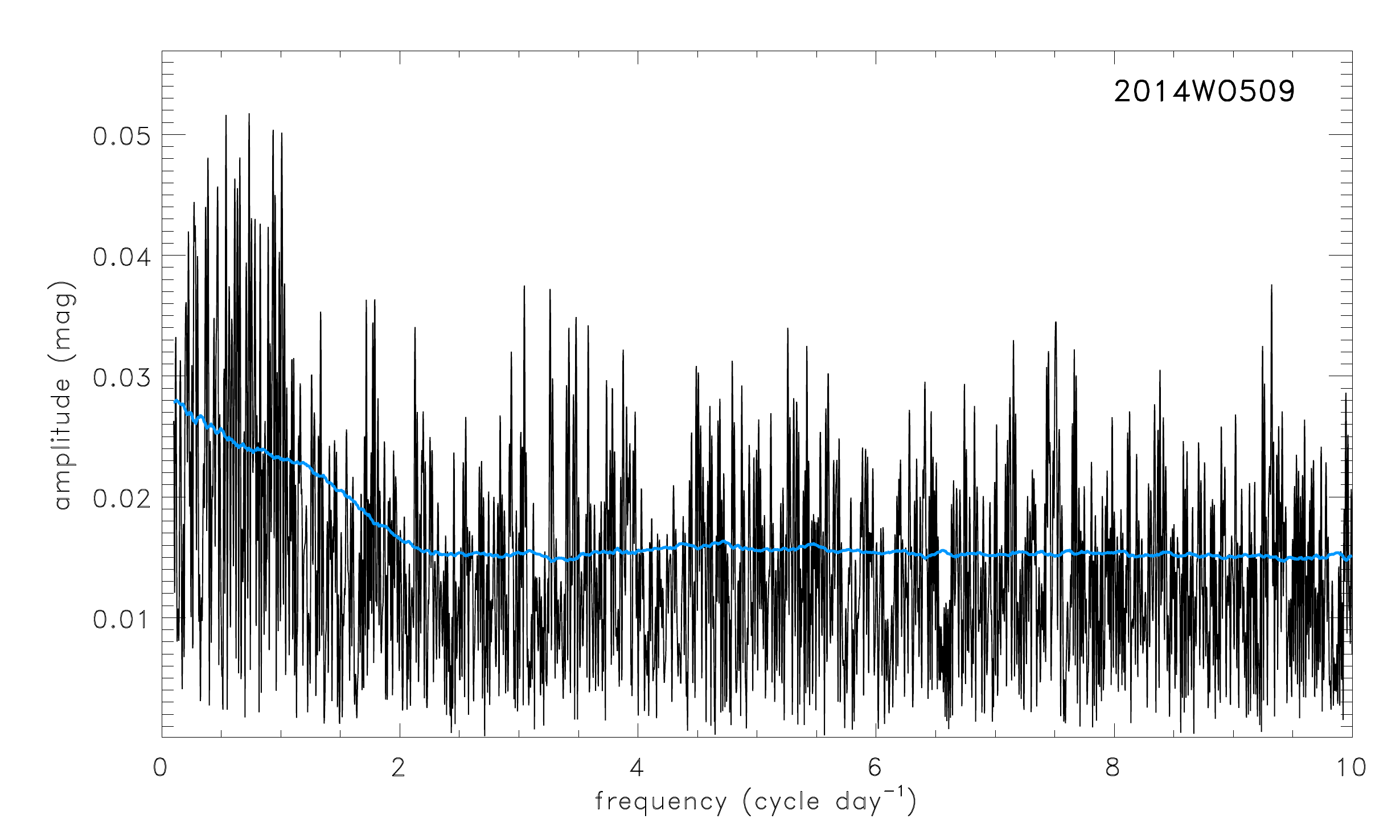}
\includegraphics[width=0.33\textwidth]{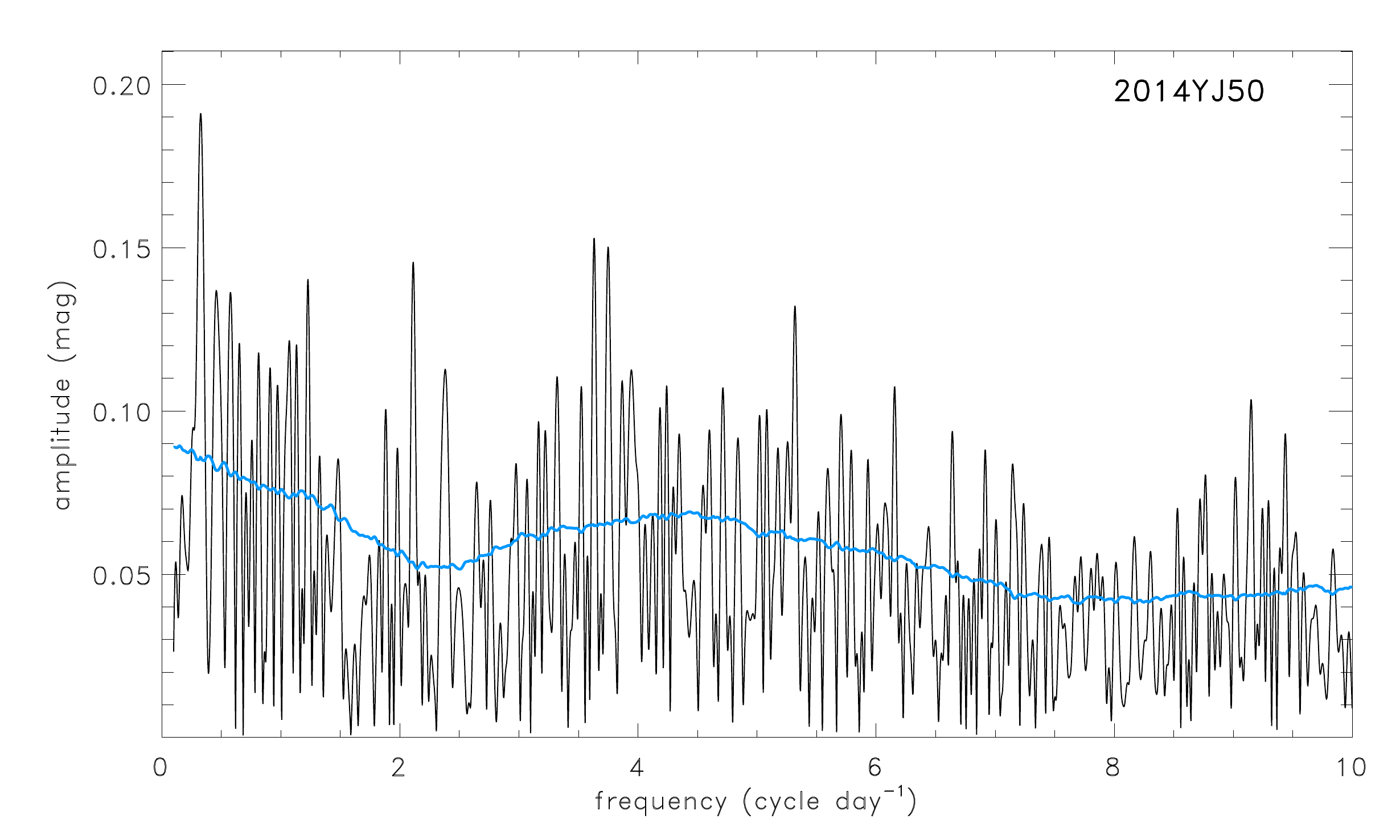}
}
\caption{
}
\end{figure}

\clearpage

\section{Light curve data}

The data of all light curves presented in this paper are available as a single ASCII data table. Note that all data presented are \emph{uncorrected} for heliocentric/observer distances and phase angle. { An example table is shown below}. 

\begin{table}[h!]
    \centering
    \begin{tabular}{cccccccc}
     \hline
        Desig.  & Name/ID             & Camp. &   JD          & m       &  $\delta$m &  R.A.      & DEC \\ \hline   
         26375   &   (26375) 1999 DE9 &  C10  & 2457584.09825  & 20.6242  &  0.1930  & 181.30961  &  -6.24154  \\
         26375   &   (26375) 1999 DE9 &  C10  & 2457584.13911  & 20.5693  &  0.1496  & 181.30934  &  -6.24136  \\
         26375   &   (26375) 1999 DE9 &  C10  & 2457584.15955  & 20.5848  &  0.1296  & 181.30921  &  -6.24127 \\
         26375   &   (26375) 1999 DE9 &  C10  & 2457584.20041  & 20.5265  &  0.1244  & 181.30893  &  -6.24108 \\
         26375   &   (26375) 1999 DE9 &  C10  & 2457584.22085  & 20.3055  &  0.0783  & 181.30880  &  -6.24099 \\
         26375   &   (26375) 1999 DE9 &  C10  & 2457584.24128  & 20.3238  &  0.0911  & 181.30866  &  -6.24090 \\
         26375   &   (26375) 1999 DE9 &  C10  & 2457584.26172  & 20.2942  &  0.1116  & 181.30852  &  -6.24080 \\
         26375   &   (26375) 1999 DE9 &  C10  & 2457584.28215  & 20.3690  &  0.0928  & 181.30839  &  -6.24071 \\
         26375   &   (26375) 1999 DE9 &  C10  & 2457584.30258  & 20.4051  &  0.0982  & 181.30825  &  -6.24062 \\
         26375   &   (26375) 1999 DE9 &  C10  & 2457584.32302  & 20.3264  &  0.1024  & 181.30812  &  -6.24053 \\
         \hline
    \end{tabular}
    \caption{Example table presenting the first lines of the ASCII data table of the K2 TNO photometry results. The columns of the table are: 
 (1) identification in Minor Planet Center's packed designation format\footnote{{https://www.minorplanetcenter.net/iau/info/PackedDes.html}} (characters 1-8);
 (2) target name or asteroid number, if available, or provisional designation (chars 9-30), in 'human-readable' format;
 (3) Campaign ID;
 (4) Julian date; 
 (5) K2 brightness (mag); 
 (6) uncertainty of K2 brightness (mag);
 (7)-(8) R.A. and DEC of the center of the measuring aperture, centered on the target on the K2 image (decimal degrees). These coordinates are taken from the object's ephemeris provided by NASA's Horizon system, generated as seen from the Kepler spacecraft at the time of the data reduction. Note that a later refinement in the object's orbit might result in (slightly) different R.A. and DEC values. }
    \label{tab:my_label}
\end{table}

\end{document}